\documentclass[ALICE,manyauthors]{cernphprep}

\usepackage[comma,square,numbers,sort&compress]{natbib}
\usepackage{siunitx}
\usepackage{hyperref}
\usepackage{graphicx}  
\usepackage{dcolumn}   
\usepackage{bm}        
\usepackage{amssymb}   
\usepackage{amsfonts}
\usepackage{graphics}
\usepackage{grffile}   
\usepackage{epsfig}
\usepackage{units}
\usepackage[usenames]{color}
\usepackage[normalem]{ulem} 
\usepackage[utf8]{inputenc}
\usepackage[T1]{fontenc}
\usepackage{subfigure}
\usepackage{placeins}

\usepackage{floatrow}
\DeclareFloatFont{footnotesize}{\footnotesize}
\begin{document}

\begin{titlepage}
\PHyear{2017}
\PHnumber{222}      
\PHdate{29 August}  

\hyphenation{FPGAs ALICE TRD MCM azi-muth  min-im-ise in-fra-struc-ture}

\title{The ALICE Transition Radiation Detector:\\
    construction, operation, and performance}
\ShortTitle{The ALICE Transition Radiation Detector}   

\Collaboration{ALICE Collaboration%
         \thanks{See Appendix~\ref{app:collab} for the list of collaboration members}}
\ShortAuthor{ALICE Collaboration}       
\begin{abstract}

The Transition Radiation Detector (TRD) was designed and built to enhance the capabilities of the ALICE detector at the Large Hadron Collider (LHC). While aimed at providing electron identification and triggering, the TRD also contributes significantly to the track reconstruction and calibration in the central barrel of ALICE. In this paper the design, construction, operation, and performance of this detector are discussed. A pion rejection factor of up to~410 is achieved at a momentum of 1~GeV/$c$ in p--Pb collisions and the resolution at high transverse momentum improves by about 40\% when including the TRD information in track reconstruction. The triggering capability is demonstrated both for jet, light nuclei, and electron selection.

\end{abstract}
\end{titlepage}
\setcounter{page}{2}

\newcommand{\checkit}{\TBD{(Is this correct?)}\xspace}
\newcommand{\dd}{\mathrm{d}}

\newcommand{\pp}{\mbox{pp}\xspace}
\newcommand{\ppb}{\mbox{p--Pb}\xspace}
\newcommand{\pbpb}{\mbox{Pb--Pb}\xspace}

\newcommand{\tb}{time bin\xspace}

\newcommand{\TBD}[1]      {}
\newcommand{\tbd}        {\TBD}

\newcommand{\strike}     {\sout}
\newcommand{\sqrts}      {\ensuremath{\sqrt{s}}\xspace}
\newcommand{\sqrtsnn}    {\ensuremath{\sqrt{s_\mathrm{NN}}}\xspace}

\newcommand{\gevc}       {~\ensuremath{\mathrm{GeV}\!/c}\xspace}
\newcommand{\gevcc}      {~\ensuremath{\mathrm{GeV}\!/c^2}\xspace}
\newcommand{\mevc}       {~\ensuremath{\mathrm{MeV}\!/c}\xspace}
\newcommand{\mevcc}      {~\ensuremath{\mathrm{MeV}\!/c^2}\xspace}
\newcommand{\lumi}       {\ensuremath{\mathcal{L}}\xspace}
\newcommand{\mum }       {~\ensuremath{\mu\mathrm{m}}\xspace}

%
%


%
%
\newcommand{\figs}{Figs.\xspace}
\newcommand{\Figs}{Figures\xspace}
\newcommand{\eqn}{Eq.\xspace}
\newcommand{\Eqn}{Equation\xspace}
\newcommand{\figref}[1]{Fig.~\ref{#1}}
\newcommand{\Figref}[1]{Figure~\ref{#1}}
\newcommand{\tabref}[1]{Tab.~\ref{#1}}
\newcommand{\Tabref}[1]{Table~\ref{#1}}
\newcommand{\appref}[1]{App.~\ref{#1}}
\newcommand{\Appref}[1]{Appendix~\ref{#1}}
\newcommand{\secs}{Secs.\xspace}
\newcommand{\Secs}{Sections\xspace}
\newcommand{\secref}[1]{Sec.~\ref{#1}}
\newcommand{\Secref}[1]{Section~\ref{#1}}
\newcommand{\chaps}{Chaps.\xspace}
\newcommand{\Chaps}{Chapters\xspace}
\newcommand{\chapref}[1]{Chap.~\ref{#1}}
\newcommand{\Chapref}[1]{Chapter~\ref{#1}}
\newcommand{\lstref}[1]{Listing~\ref{#1}}
\newcommand{\Lstref}[1]{Listing~\ref{#1}}

%
%
\newcommand{\hic}{heavy-ion collision\xspace}
\newcommand{\hics}{heavy-ion collisions\xspace}
\newcommand{\Hics}{Heavy-ion collisions\xspace}
\newcommand{\fee}{front-end electronics\xspace}
\newcommand{\readout}{read-out\xspace}

\newcommand{\run}[1]{Run\,#1}
\newcommand{\LS}[1]{LS~#1}

\newcommand{\Minbias}{Minimum bias\xspace}
\newcommand{\minbias}{minimum bias\xspace}

\newcommand{\ptjet}{\ensuremath{\pt^{\rm jet}}\xspace}
\newcommand{\pthadron}{\ensuremath{\pt^{\rm hadron}}\xspace}
\newcommand{\rr}{\ensuremath{R}}

%
%
\newcommand{\cLight}{\textit{c}\xspace}

\newcommand{\Ecorr}{\ensuremath{E_{\rm{corr}}}\xspace}
\newcommand{\Ereco}{\ensuremath{E_{\rm{reco}}}\xspace}
\newcommand{\sumpt}{\ensuremath{\sum\left(p_{\rm matched}^{\rm track}\right)}\xspace}
\newcommand{\DEcorr}{\ensuremath{\Delta\Ecorr}\xspace}
\newcommand{\DEmc}{\ensuremath{\Delta{E_{\rm{MC}}}}\xspace}
\newcommand{\Rcorr}{\ensuremath{R_{\rm{corr}}}\xspace}

\newcommand{\ptcorr}{\ensuremath{\pt^{\rm corr}}\xspace}
\newcommand{\ptraw}{\ensuremath{\pt^{\rm raw}}\xspace}
\newcommand{\Ajet}{\ensuremath{A_{\rm jet}}\xspace}

\newcommand{\kB}{kByte\xspace}
\newcommand{\MB}{MByte\xspace}
\newcommand{\GB}{GByte\xspace}
\newcommand{\TB}{TByte\xspace}
\newcommand{\PB}{PByte\xspace}
\newcommand{\EB}{EByte\xspace}

\newcommand{\MBs}{\ensuremath{\mathrm{\MB}\kern-0.05em/\kern-0.02em \mathrm{s}}\xspace}
\newcommand{\GBs}{\ensuremath{\mathrm{\GB}\kern-0.05em/\kern-0.02em \mathrm{s}}\xspace}
\newcommand{\TBs}{\ensuremath{\mathrm{\TB}\kern-0.05em/\kern-0.02em \mathrm{s}}\xspace}
\newcommand{\Mbits}{\ensuremath{\mathrm{Mbit}\kern-0.05em/\kern-0.02em \mathrm{s}}\xspace}
\newcommand{\Gbits}{\ensuremath{\mathrm{Gbit}\kern-0.05em/\kern-0.02em \mathrm{s}}\xspace}

\newcommand{\lumiUnit}{\ensuremath{{\rm cm}^{-2} {\rm s}^{-1}}}

%
%
\newcommand{\sqrtS}{\ensuremath{\sqrt{s}}\xspace}
\newcommand{\sqrtSnn}{\ensuremath{\sqrt{s_{\mathrm{NN}}}}\xspace}
\newcommand{\sqrtSE}[2][TeV]{$\sqrtS = #2\,\mathrm{#1}$\xspace}
\newcommand{\sqrtSnnE}[2][TeV]{$\sqrtSnn = #2\,\mathrm{#1}$\xspace}

\newcommand{\snn}{\sqrtSnn}
\newcommand{\sE}{\sqrtSE}
\newcommand{\snnE}{\sqrtSnnE}

\newcommand{\pt}{\ensuremath{p_{\mathrm{T}}}\xspace}

\newcommand{\dedx}{\ensuremath{\mathrm{d}E/\mathrm{d}x}\xspace}
\newcommand{\dndy}{\ensuremath{\mathrm{d}N/\mathrm{d}y}\xspace}
\newcommand{\dndeta}{\ensuremath{\mathrm{d}N_{\rm ch}/\mathrm{d}\eta}\xspace}

\newcommand{\Raa}{\ensuremath{R_\mathrm{AA}}\xspace}
\newcommand{\Taa}{\ensuremath{T_\mathrm{AA}}\xspace}
\newcommand{\ncoll}{\ensuremath{N_{\mathrm{coll}}}\xspace}
\newcommand{\npart}{\ensuremath{N_{\mathrm{part}}}\xspace}

\newcommand{\mee}{\ensuremath{\mathrm{m}_{\ee}}\xspace}

\newcommand{\rhosc}{\ensuremath{\rho_{\mathrm{sc}} }\xspace}
\newcommand{\ir}{\ensuremath{R_{\mathrm{int}}}\xspace}
\newcommand{\Tzero}{\ensuremath{t_{\mathrm{0}}}\xspace}
\newcommand{\TzeroI}{\ensuremath{t_{\mathrm{0},i}}\xspace}
\newcommand{\TzeroIplus}{\ensuremath{t_{\mathrm{0},i+1}}\xspace}
\newcommand{\Tzeroseed}{\ensuremath{t_{\mathrm{0}}^{\mathrm{seed}}}\xspace}
\newcommand{\Tdigit}{\ensuremath{t_{\mathrm{digit}}}\xspace}
\newcommand{\Textrapol}{\ensuremath{t_{\mathrm{extrapol}}}\xspace}
\newcommand{\zdigit}{\ensuremath{z_{\mathrm{digit}}}\xspace}
\newcommand{\Tdriftion}{\ensuremath{t_{\mathrm{d}}^{\mathrm{ion}}}\xspace}
\newcommand{\Tdrift}{\ensuremath{t_{\mathrm{d}}}\xspace}
\newcommand{\vdrift}{\ensuremath{v_{\mathrm{d}}}\xspace}
\newcommand{\vecVdrift}{\ensuremath{\vec{v}_{\mathrm{d}}}\xspace}
\newcommand{\zdrift}{\ensuremath{z_{\mathrm{d}}}\xspace}
\newcommand{\lorang}{\ensuremath{\Theta_{\mathrm{L}}}\xspace}

\newcommand{\Lpad}{\ensuremath{L_{\mathrm{pad}}}\xspace}
\newcommand{\Tint}{\ensuremath{T_{\mathrm{int}}}\xspace}
\newcommand{\zcls}{\ensuremath{z_{\mathrm{cls}}}\xspace}
\newcommand{\zvtx}{\ensuremath{z_{\mathrm{vtx}}}\xspace}
\newcommand{\zlength}{\ensuremath{z_{\mathrm{maxdrift}}}\xspace}
\newcommand{\zroc}{\ensuremath{z_{\mathrm{roc}}}\xspace}
\newcommand{\ExB}{\ensuremath{E\times B}\xspace}

\newcommand{\vecRvtx}{\ensuremath{\vec{r}_{\mathrm{vtx}}}\xspace}
\newcommand{\vecRcls}{\ensuremath{\vec{r}_{\mathrm{cls}}}\xspace}
\newcommand{\vecRro}{\ensuremath{\vec{r}_{\mathrm{ro}}}\xspace}

\newcommand{\rphi}{\ensuremath{r\varphi}\xspace}
\newcommand{\localy}{local-$y$\xspace}
\newcommand{\localx}{local-$x$\xspace}
\newcommand{\zpos}{$z$-position\xspace}
\newcommand{\zposs}{$z$-positions\xspace}
\newcommand{\xlab}{$x_{\rm lab}$}
\newcommand{\ylab}{$y_{\rm lab}$}
\newcommand{\zlab}{$z_{\rm lab}$}


\newcommand{\photon}{\ensuremath{\gamma}\xspace}
\newcommand{\wplus}{W$^{+}$\xspace}
\newcommand{\wminus}{W$^{-}$\xspace}
\newcommand{\zboson}{Z$^{0}$\xspace}

\newcommand{\eplus}{\ensuremath{{\rm e}^{+}}\xspace}
\newcommand{\eminus}{\ensuremath{{\rm e}^{-}}\xspace}
\newcommand{\epm}{\ensuremath{{\rm e}^{\pm}}\xspace}
\newcommand{\ee}{\ensuremath{{\rm e}^{+}{\rm e}^{-}}\xspace}

\newcommand{\mup}{\ensuremath{\mu^{+}}\xspace}
\newcommand{\mumi}{\ensuremath{\mu^{-}}\xspace}
\newcommand{\mupm}{\ensuremath{\mu^{\pm}}\xspace}
\newcommand{\mumu}{\ensuremath{\mu^{+}\mu^{-}}\xspace}

\newcommand{\rhop}{\ensuremath{\rho^{+}}\xspace}
\newcommand{\pip}{\ensuremath{\pi^{+}}\xspace}
\newcommand{\pim}{\ensuremath{\pi^{-}}\xspace}
\newcommand{\piz}{\ensuremath{\pi^{0}}\xspace}
\newcommand{\kap}{\ensuremath{{\rm K}^{+}}\xspace}
\newcommand{\kam}{\ensuremath{{\rm K}^{-}}\xspace}
\newcommand{\kashort}{\ensuremath{{\rm K}^{0}_{s}}\xspace}
\newcommand{\pbar}{\ensuremath{\rm\overline{p}}\xspace}
\newcommand{\ppbar}{\ensuremath{\rm p\overline{p}}\xspace}
\newcommand{\jpsi}{\ensuremath{{\rm J}\kern-0.02em/\kern-0.05em\psi}\xspace}
\newcommand{\psiP}{\ensuremath{\Psi^{\prime}}\xspace}
\newcommand{\upsi}{\ensuremath{\Upsilon}\xspace}
\newcommand{\upsiP}{\ensuremath{\Upsilon^{\prime}}\xspace}
\newcommand{\upsiPP}{\ensuremath{\Upsilon^{\prime\prime}}\xspace}
\newcommand{\kzerol}     {K\ensuremath{^0_L}\xspace}
\newcommand{\kzeros}     {K\ensuremath{^0_\mathrm{S}}\xspace}

\newcommand{\qbar}{\ensuremath{\rm\overline{q}}\xspace}
\newcommand{\ubar}{\ensuremath{\rm\overline{u}}\xspace}
\newcommand{\dbar}{\ensuremath{\rm\overline{d}}\xspace}
\newcommand{\cc}{\ensuremath{{\rm c}\bar{{\rm c}}}\xspace}


\newcommand{\ArCOtwo}{Ar-CO$_2$ (82-18)\xspace}
\newcommand{\XeCOtwo}{Xe-CO$_2$ (85-15)\xspace}
\newcommand{\Kr}{\ensuremath{^{83{\rm m}}{\rm Kr}}\xspace}

\section{Introduction}
\label{Chapterintro}

A Large Ion Collider Experiment (ALICE)~\cite{Aamodt:2008zz,Abelev:2014ffa} is the dedicated heavy-ion experiment at the Large Hadron Collider (LHC) at CERN. In central high energy nucleus--nucleus collisions a high-density deconfined state of strongly interacting matter, known as quark--gluon plasma (QGP), is supposed to be created~\cite{Borsanyi:2010bp,Bazavov:2014pvz,BraunMunzinger:2007zz}. ALICE is designed to measure a large set of observables in order to study the properties of the QGP. Among the essential probes there are several involving electrons, which originate, e.g. from open heavy-flavour hadron decays, virtual photons, and Drell-Yan production as well as from decays of the $\psi$ and $\rm \Upsilon$ families. The identification of these rare probes requires excellent electron identification, also in the high multiplicity environment of heavy-ion collisions. In addition, the rare probes need to be enhanced with triggers, in order to accumulate the statistics necessary for differential studies. The latter requirement concerns not only probes involving the production of electrons, but also rare high transverse momentum probes such as jets (collimated sprays of particles) with and without heavy flavour. The ALICE Transition Radiation Detector (TRD) fulfils these two tasks and thus extends the physics reach of ALICE.

Transition radiation (TR), predicted in 1946 by Ginzburg and Frank~\cite{Ginzburg:1945zz}, occurs when a particle crosses the boundary between two media with different dielectric constants. For highly relativistic particles ($\gamma \gtrsim 1000$), the emitted radiation extends into the \mbox{X-ray} domain for a typical choice of radiator~\cite{Garibian1958,Garibian1974,Garibian1975133}. The radiation is extremely forward peaked relative to the particle direction~\cite{Garibian1958}. As the TR photon yield per boundary crossing is of the order of the fine structure constant ($\alpha=1/137$), many boundaries are needed in detectors to increase the radiation yield~\cite{Dolgoshein:1993nb}. The absorption of the emitted X-ray photons in high-$Z$ gas detectors leads to a large energy deposition compared to the specific energy loss by ionisation of the traversing particle. 

Since their development in the 1970s, transition radiation detectors have proven to be powerful devices in cosmic-ray, astroparticle and accelerator experiments~\cite{Cherry:1974de,Cobb:1976bv,Appuhn:1987dj,DETOEUF1989310,Dolgoshein:1993nb,Beck:1995my,
Piekarz:1995np,Ackerstaff:1998av,vonDoetinchem:2006gy,Abat:2008zzb,Ave:2011zz}. The main purpose of the transition radiation detectors in these experiments was the discrimination of electrons from hadrons via, e.g.\ cluster counting or total charge/energy analysis methods. In a few cases they provided charged-particle tracking. The transition radiation photons are in most cases detected either by straw tubes or by multiwire proportional chambers (MWPC). In some experiments~\cite{Dolgoshein:1993nb,Appuhn:1987dj,Beitzel:1992iy,Piekarz:1995np} and in test setups~\cite{Ludlam:1980tm,WATASE1986379,Holder:1988je,O'Brien:1994tv}, short drift chambers (usually about \SI{1}{\centi\metre}) were employed for the detection. Detailed reviews on the transition radiation phenomenon, detectors, and their application to particle identification can be found in~\cite{Dolgoshein:1993nb,Favuzzi2001,Andronic:2011an,PDG2014}.

The ALICE TRD, which covers the full azimuth and the pseudorapidity range $-0.84<\eta<0.84$ (see next section), is part of the ALICE central barrel. The TRD consists of 522~chambers arranged in 6~layers at a radial distance from \SI{2.90}{\metre} to \SI{3.68}{\metre} from the beam axis. Each chamber comprises a foam/fibre radiator followed by a Xe-CO$_2$-filled MWPC preceded by a drift region of \SI{3}{\centi\metre}. The extracted temporal information represents the depth in the drift volume at which the ionisation signal was produced and thus allows the contributions of the TR photon and the specific ionisation energy loss of the charged particle \dedx to be separated. The former is preferentially absorbed at the entrance of the chamber and the latter distributed uniformly along the track. 
Electrons can be distinguished from other charged particles by producing TR and having a higher \dedx due to the relativistic rise of the ionisation energy loss. The usage of the temporal information further enhances the electron-hadron separation power. Due to the fast read-out and online reconstruction of its signals, the TRD has also been successfully used to trigger on electrons with high transverse momenta and jets (3~or more high-\pt tracks). Last but not least, the TRD improves the overall momentum resolution of the ALICE central barrel by providing additional space points at large radii for tracking, and tracks anchored by the TRD will be a key element to correct space charge distortions expected in the ALICE TPC in LHC \run{3}~\cite{ALICE:2014qrd}. A first version of the correction algorithm is already in use for~\run{2}.

In this article the design, construction, operation, and performance of the ALICE TRD is described. Section~\ref{Chaptertrdsystem} gives an overview of the detector and its construction. The gas system is detailed in Section~\ref{Chaptergas}. The services required for the detector are outlined in Section~\ref{infra}. In Section~\ref{Chapterreadoutelectronics} the read-out of the detector is discussed and the Detector Control System (DCS) used for reliable operation and monitoring of the detector is presented in Section~\ref{Chapterdcs}. The detector commissioning and its operation are discussed in Section~\ref{Chaptercomminbeam}. Tracking, alignment, and calibration are described in detail in Sections~\ref{Chaptertracking},~\ref{Chapteralign}, and~\ref{Chaptercalib}, while various methods for charged hadron and electron identification are presented in Section~\ref{Chapterpid}. The use of the TRD trigger system for jets, electrons, heavy-nuclei, and cosmic-ray muons is described in Section~\ref{Chaptertrigger}.

\section{Detector overview}
\label{Chaptertrdsystem}

\begin{figure}[bt]
   \centering
\includegraphics[width=.48\textwidth]{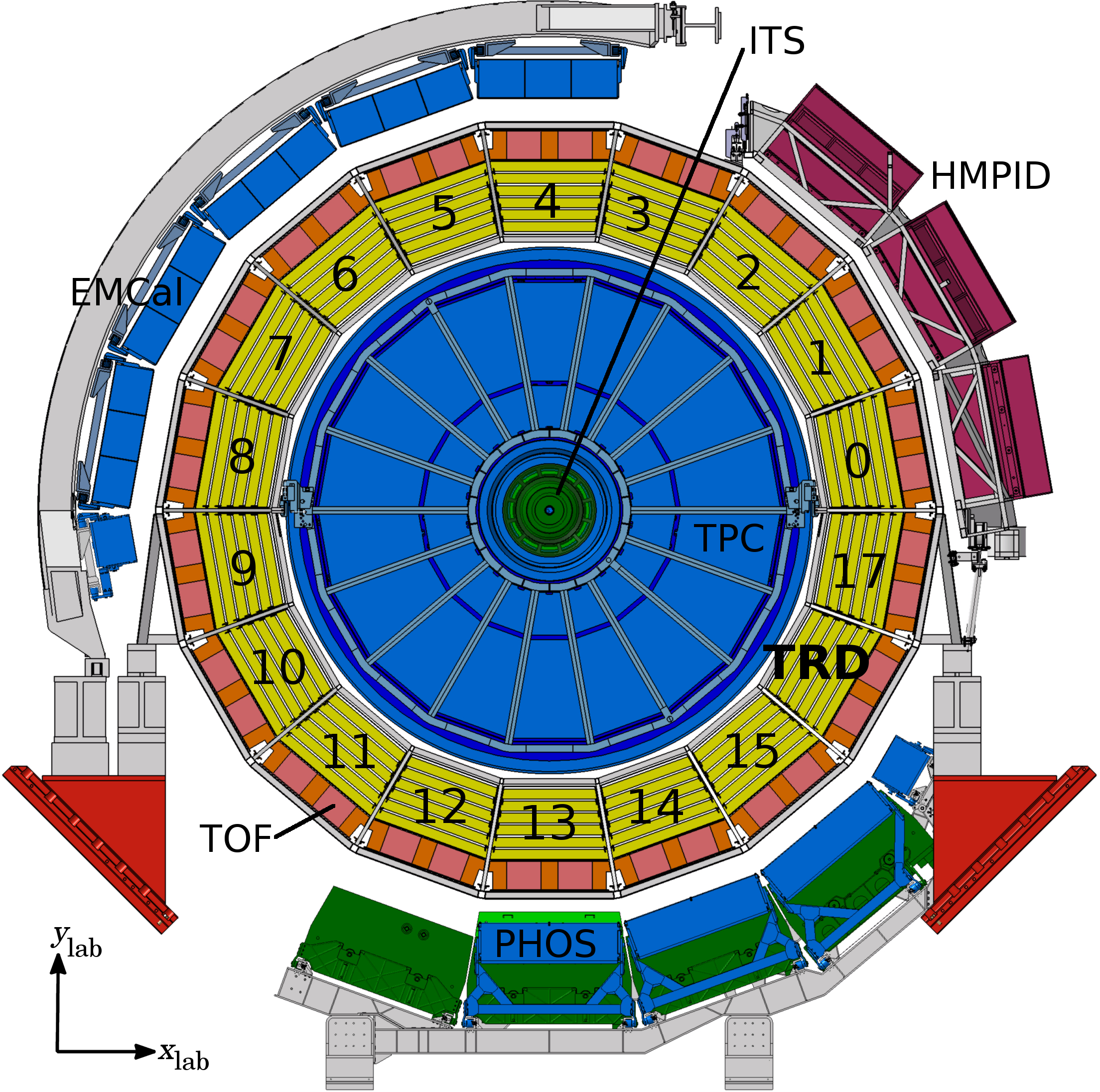}
  \caption[]{Schematic cross-section of the ALICE detector perpendicular to the LHC beam direction (status of the detector since the start of LHC \run{2}). The central barrel detectors cover the pseudorapidity range $| \eta | \lesssim 0.9 $ and are located inside the solenoid magnet, which provides a magnetic field with strength $B$~=~\SI{0.5}{\tesla} along the beam direction.}
   \label{Figure_ALICE}
\end{figure} 

A cross-section of the central part of the ALICE detector~\cite{Aamodt:2008zz,Abelev:2014ffa}, installed at Interaction Point~2 (IP2) of the LHC, is shown in Fig.~\ref{Figure_ALICE}. The central barrel detectors cover the pseudorapidity range $| \eta | \lesssim 0.9 $ and are located inside a solenoid magnet, which produces a magnetic field of $B$~=~\SI{0.5}{\tesla} along the beam direction. The Inner Tracking System (ITS)~\cite{Dellacasa:1999kf}, placed closest to the nominal interaction point, is employed for low momentum tracking, particle identification (PID), and primary and secondary vertexing. The Time Projection Chamber (TPC)~\cite{Alme:2010ke}, which is surrounded by the TRD, is used for tracking and PID. The Time-Of-Flight detector (TOF)~\cite{Cortese:2002kf} is placed outside the TRD and provides charged hadron identification. The ElectroMagnetic Calorimeter (EMCal)~\cite{Cortese:2008zza}, the PHOton Spectrometer (PHOS)~\cite{Dellacasa:1999kd}, and the High Momentum Particle Identification Detector (HMPID)~\cite{Beole:1998yq} are used for electron, jet, photon and hadron identification. Their azimuthal coverage is shown in Fig.~\ref{Figure_ALICE}. Not visible in the figure are the V0 and T0 detectors~\cite{Cortese:2004aa,Abbas:2013taa}, as well as the Zero Degree Calorimeters (ZDC)~\cite{Dellacasa:1999ke}, which are placed at small angles on both sides of the interaction region. 
These detectors can be employed, e.g.\ to define a minimum-bias trigger, to determine the event time, the centrality and event plane of a collision~\cite{Abelev:2014ffa,Adam:2016ilk,Abelev:2013qoq}. 
Likewise, the muon spectrometer~\cite{ALICE:1999aa,ALICE:muon2000} is outside the view on one side of the experiment, only, covering $-$4~$< \eta <$~$-$2.5.

Figure~\ref{Figure_ALICE} also shows the definition of the global ALICE coordinate system, which is a Cartesian system with its point of origin at the nominal interaction point (\xlab, \ylab, \zlab~=~0); the \xlab-axis pointing inwards radially to the centre of the LHC ring and the \zlab-axis coinciding with the direction of one beam and pointing in direction opposite to the muon spectrometer. According to the (anti-)clock-wise beam directions, the muon spectrometer side is also called C-side, the opposite side A-side.

\begin{figure}[tb]
   \centering
   \includegraphics[width=.6\textwidth]{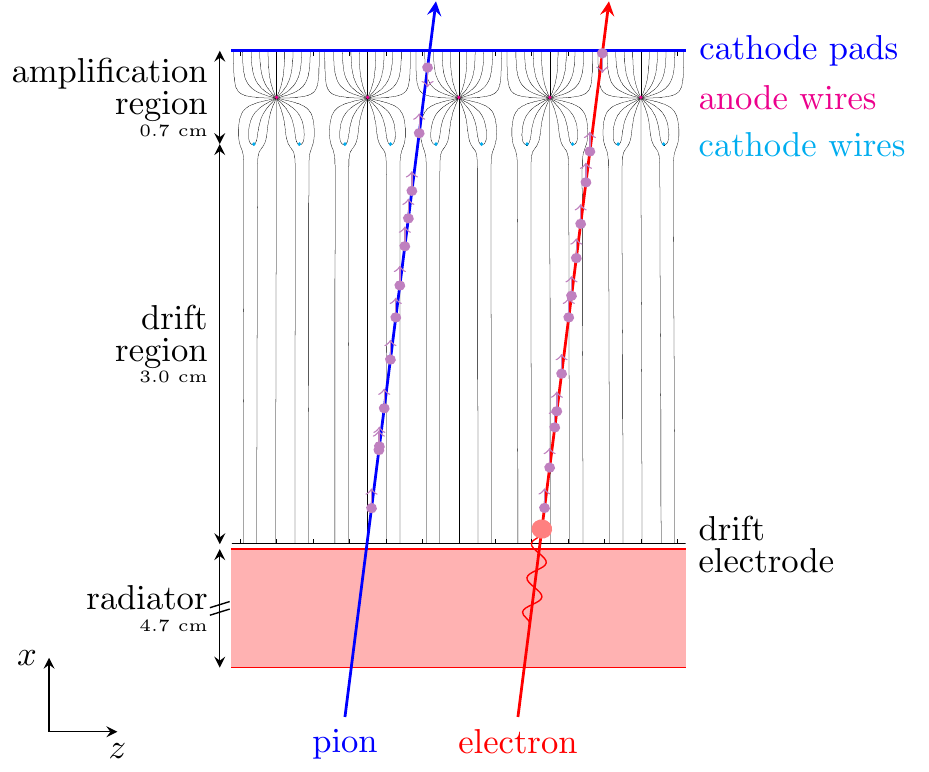}
   \caption[]{Schematic cross-section of a TRD chamber in the $x$-$z$ plane 
(perpendicular to the wires) with tracks of a pion and an electron to illustrate the ionisation energy deposition and the TR contribution. The large energy deposition due to the TR photon absorption is indicated by the large red circle in the drift region. The drift lines (solid lines) are calculated with Garfield~\cite{garfield} and correspond to the nominal voltage settings for chamber operation. The radiator is not drawn to scale.}
   \label{fig:roccross}
\end{figure}

\begin{figure}[tb]
   \centering
\includegraphics[trim=1.7cm 12.5cm 0cm 2.7cm,clip=true,width=1.0\textwidth]{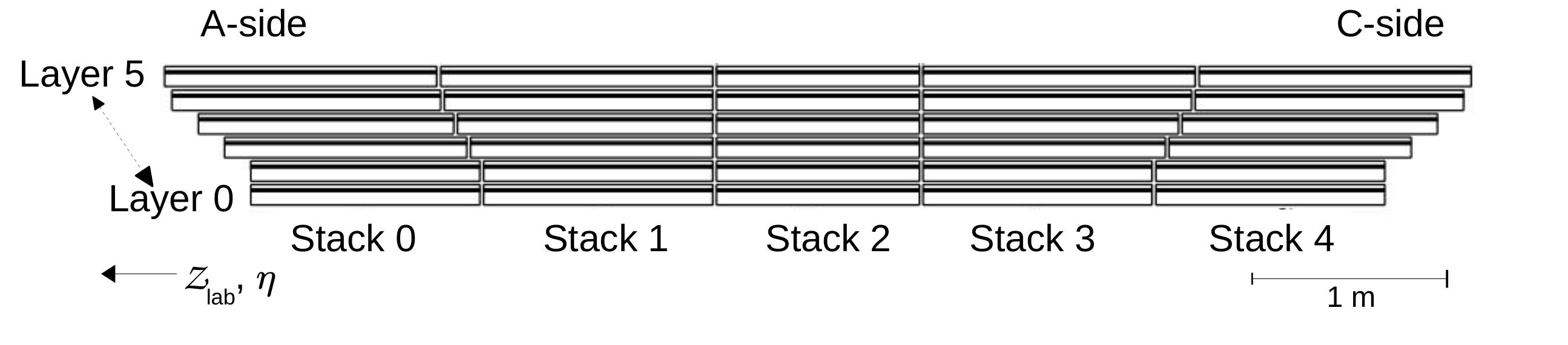}
  \caption[]{Cross-section (longitudinal view) of a supermodule.}
   \label{fig:crosssmlong}
\end{figure}

The design of the TRD is a result of the requirements and constraints discussed in the Technical Design Report~\cite{Cortese:519145}. It has a modular structure and its basic component is a multiwire proportional chamber (MWPC). Each chamber is preceded by a drift region to allow for the reconstruction of a local track segment, which is required for matching of TRD information with tracks reconstructed with ITS and TPC at high multiplicities. TR photons are produced in a radiator mounted in front of the drift section and then absorbed in a xenon-based gas mixture.
A schematic cross-section of a chamber and its radiator is shown in Fig.~\ref{fig:roccross}. The shown local coordinate system is a right-handed orthogonal Cartesian system, similar to the global coordinate system, rotated such that the $x$-axis is perpendicular to the chamber.
Six layers of chambers are installed to enhance the pion rejection power.
An eighteen-fold segmentation in azimuth ($\varphi$), with each segment called `sector', was chosen to match that of the TPC read-out chambers. In the longitudinal direction (\zlab), i.e.\ along the beam direction, the coverage is split into five stacks, resulting in a manageable chamber size. 
The five stacks are numbered from 0~to~4, where stack~4 is at the C-side and stack~0 at the A-side. Layer 0 is closest, layer 5 farthest away from the collision point in the radial direction. In each sector, 30~read-out chambers (arranged in 6~layers and 5~stacks) are combined in a mechanical casing, called a `supermodule' (see Fig.~\ref{fig:crosssmlong} and Section~\ref{secsupermodule}).

In total the TRD can host 540~read-out chambers (18~sectors $\times$~6~layers $\times$~5~stacks), however in order to minimise the material in front of the PHOS detector in three sectors (sectors 13--15, for numbering see Fig.~\ref{Figure_ALICE}) the chambers in the middle stack were not installed. This results in a system of 522~individual read-out chambers.
The main parameters of the detector are summarised in Table~\ref{tab:TRDParam}.

\begin{table}[tb]
\centering
\begin{tabular}{ l  l }
\hline
\textbf{Parameter} & \textbf{Value}  \\
\hline
Pseudorapidity coverage &  $-$0.84~$< \eta <$~+0.84 \\
Azimuthal coverage $\varphi$ & \ang{360} \\
Radial position  & \SIrange{2.90}{3.68}{\metre} \\
Length of a supermodule  & \SI{7.02}{\metre} \\
Weight of a supermodule  & \SI{1.65}{\tonne} \\
Segmentation in $\varphi$ & 18~sectors \\ 
Segmentation in \zlab & 5~stacks \\
Segmentation in $r$ & 6~layers \\
Total number of read-out chambers & 522 \\
Size of a read-out chamber (active area) & \SI{0.90}{\metre}~$\times$~\SI{1.06}{\metre} to \SI{1.13}{\metre}~$\times$~\SI{1.43}{\metre} \\
Radiator material & fibre/foam sandwich \\
Depth of radiator & \SI{4.7}{\centi\metre} \\
Depth of drift region & \SI{3.0}{\centi\metre} \\
Depth of amplification region & \SI{0.7}{\centi\metre}  \\ 
Number of time bins (\SI{100}{\nano\second})   & 30 (22--24) \\
Total number of read-out pads & \num{1\,150\,848}  \\
Total active area & \SI{673.4}{\square\metre} \\
Detector gas &  \XeCOtwo \\
Gas volume  & \SI{27}{\cubic\meter} \\
Drift voltage (nominal) & ${\sim}$\SI{2150}{\volt} \\
Anode voltage (nominal) & ${\sim}$\SI{1520}{\volt}\\
Gas gain (nominal)  & ${\sim} 3200$ \\
Drift field  & ${\sim}$\SI{700}{\volt/\centi\metre} \\
Drift velocity & ${\sim}$\SI{1.56}{\centi\metre/\micro\second} \\
Avg. radiation length along $r$ $\langle X/X_0 \rangle$  & 24.7\% \\
\hline
\end{tabular}\caption{General parameters of the TRD. The indicated weight corresponds to a supermodule with 30~read-out chambers; the length of the supermodule does not include the connected services. At maximum 30 time bins can be read out, typical values used in~\run{1} and~\run{2} are 22--24 (see Section~\ref{sec:fee}). }\label{tab:TRDParam}
\end{table}

At the start of the first LHC period (\run{1}) in 2009 the TRD participated with seven supermodules. 
Six further supermodules were built and integrated into the experiment during short winter shutdown periods of the accelerator, three in each winter shutdown period of 2010 and 2011. The TRD was completed during the Long Shutdown~1 (LS) of the LHC in 2013--2014. With all 18~supermodules installed, full coverage in azimuth was accomplished for the second LHC period (\run{2}) starting in 2015.

\subsection{Read-out chambers}\label{sec:readoutchamber}

The size of the read-out chambers changes radially and along the beam direction (see Fig.~\ref{fig:crosssmlong}). The active area per chamber thus varies from \SI{0.90}{\metre}~$\times$~\SI{1.06}{\metre} to \SI{1.13}{\metre}~$\times$~\SI{1.43}{\metre} ($x \times z$). The optimal design of a read-out chamber (see Fig.~\ref{fig:roccross}) was found considering the requirements on precision and mechanical stability, and minimisation of the amount of material.

\begin{figure}[tb]
  \centering
  \includegraphics[width=.5\textwidth]{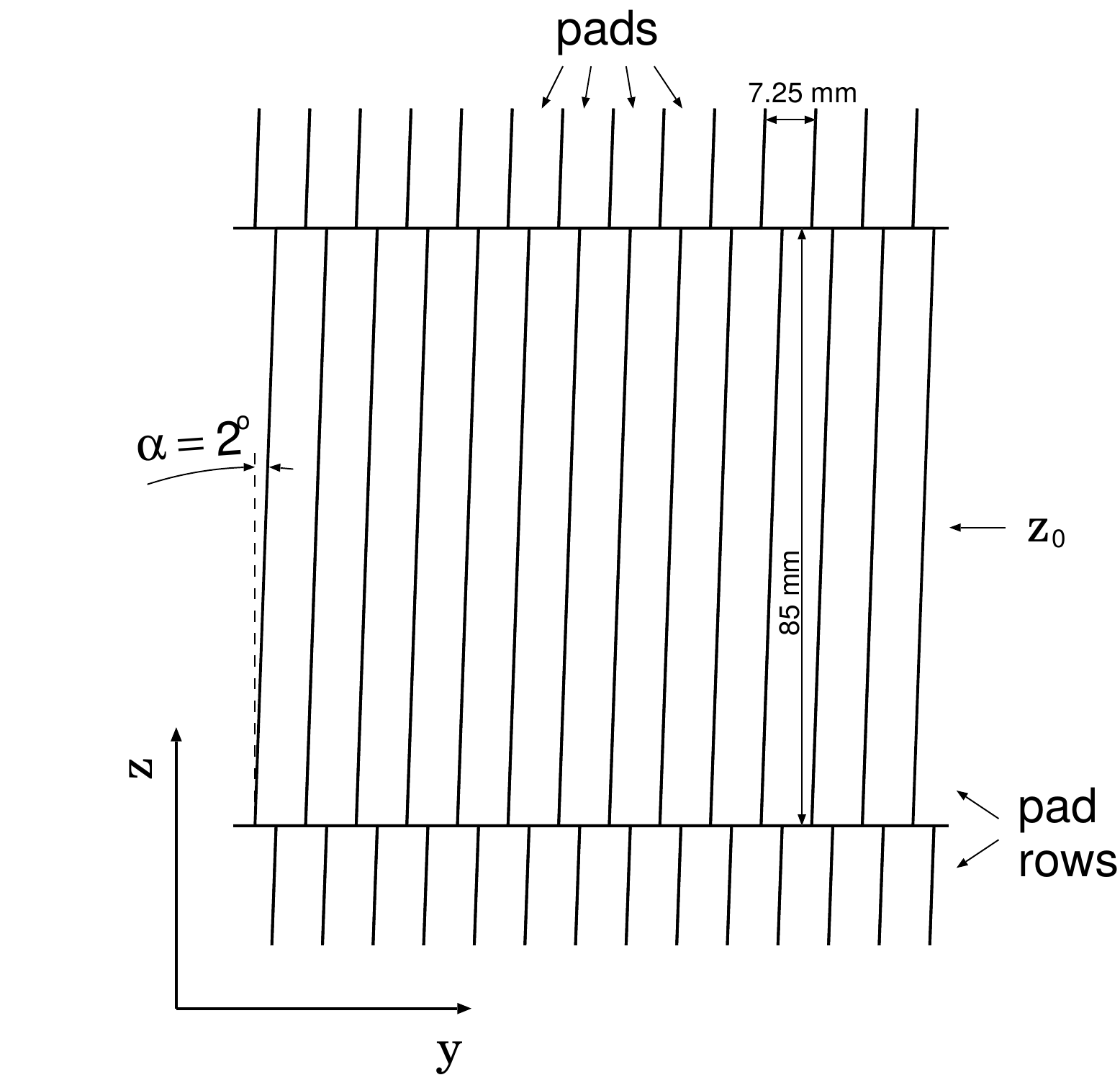}
  \caption[]{Pad geometry of a TRD read-out chamber in layer~3 (not stack~2).  
The pad tilt is $\pm$~\ang{2} with respect to the $z$-axis (along the beam direction), with the sign alternating between layers.  }\label{fig:tiltpad}
\end{figure}

The construction of the radiator, discussed in the following sub-section, is essential for the mechanical stability of the chamber. The drift electrode, 
an aluminised mylar foil (\SI{25}{\micro\meter} thick), is an integral part of the radiator.
To ensure a uniform drift field throughout the entire drift volume, a field cage with a voltage divider chain is employed~\cite{Cortese:519145}. The current at nominal drift voltage is about \SI{170}{\micro\ampere}.
The grounded cathode wires are made of Cu-Be and have a diameter of \SI{75}{\micro\meter}, while the anode wires are made of Au-plated tungsten with a diameter of \SI{20}{\micro\meter}.
The pitch for the cathode and anode wires are \SI{2.5}{\milli\metre} and \SI{5}{\milli\metre}, respectively; the tensions 
at winding were \SI{1}{\newton} and \SI{0.45}{\newton}~\cite{DEmschermann}. The wire lengths vary from \SI{1.08}{\metre} to \SI{1.45}{\metre}. The maximum deformation of the chamber frame was \SI{150}{\micro\meter} under the wire tension indicated, leading to a maximum 10\% loss in wire tension. Even with an additional \SI{1}{\milli\bar} overpressure in the gas volume (see Section~\ref{Chaptergas}), the deformation of the drift electrode can be kept within the specification of less than \SI{1}{\milli\metre}. The segmented cathode pad plane is manufactured from thin Printed Circuit Boards (PCB) and glued on a light honeycomb and carbon fibre sandwich to ensure planarity and mechanical stiffness.
The design goal of having a maximum deviation from planarity of \SI{150}{\micro\meter} was achieved with only a few chambers exceeding slightly this value. The PCBs of the pad plane were produced in two or three pieces.
The PCBs are segmented into 12 (stack~2) or 16~pads along the $z$-direction, and 144~pads in the direction of the anode wires (\rphi). The pad area varies from~\SI{0.635}{\centi\meter}~$\times$~\SI{7.5}{\centi\meter} to~\SI{0.785}{\centi\meter}~$\times$~\SI{9}{\centi\meter}~\cite{DEmschermann} to achieve a constant granularity with respect to the distance from the interaction point.
The pad width of~\SI{0.635}{\centi\meter} to~\SI{0.785}{\centi\meter} in the \rphi direction was chosen so that charge sharing between adjacent pads (typically three), which is quantified by the pad response function (PRF)~\cite{Adler:2005bc}, is achieved. As a consequence, the position of the charge deposition can be reconstructed in the \rphi-direction with a spatial resolution of $\lesssim$~\SI{400}{\micro\meter}~\cite{Adler:2005bc}. In the longitudinal direction, the coarser segmentation is sufficient for the track matching with the inner detectors. In addition, the pads are tilted by $\pm$~\ang{2} (sign alternating layer-by-layer) as shown in Fig.~\ref{fig:tiltpad}, which improves the $z$-resolution during track reconstruction without compromising the \rphi resolution. For clusters confined within one pad row, a $z$ position at the row centre is assumed, $z_{\rm cluster}=z_0$. The honeycomb structure also acts as a support for the read-out boards.
The pads are connected to the read-out boards by short polyester ribbon cables via milled holes in the honeycomb structure.

The original design of the TRD was conceived such that events with a multiplicity of d$N_{\rm ch}$/d$\eta$~=~\num{8000} would have lead to an occupancy of 34\% in the detector~\cite{Cortese:519145}.
The fast read-out and processing of such data on $1.15 \cdot 10^6$ read-out channels required the design and production of fully customised front-end electronics (see Section~\ref{Chapterreadoutelectronics}).

\begin{figure}[tb]
   \centering
\includegraphics[width=.5\textwidth]{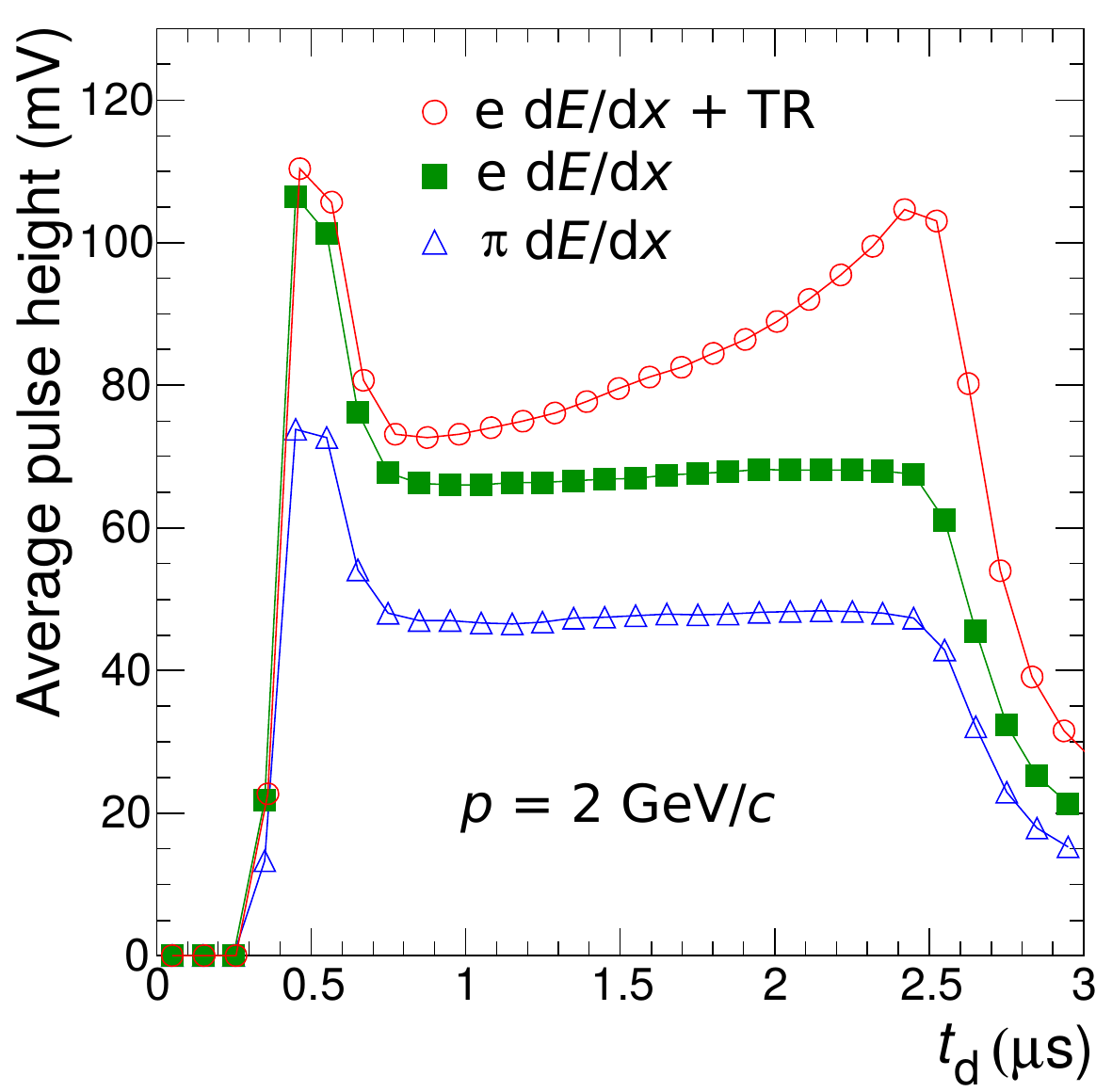}
  \caption[]{Average pulse height as a function of drift time for pions
and electrons (with and without radiator). The time axis is shown with an arbitrary offset of
\SI{0.3}{\micro\second}. The measurements were performed at the CERN PS with prototype read-out chambers that were smaller in overall size (active area~\SI{25}{\centi\meter}~$\times$~\SI{32}{\centi\meter}) but otherwise similar in construction to that of the final detector. Figure taken from~\cite{Andronic:2004uy}.}
   \label{fig:ph}
\end{figure}

The positive signal induced on the cathode pad plane is amplified using a 
charge-sensitive PreAmplifier-ShAper (PASA) (see Section~\ref{Chapterreadoutelectronics}) and the signals on the cathode pads are sampled in time bins of 100~ns inside the TRAcklet Processor (TRAP, see Section~\ref{Chapterreadoutelectronics}). For LHC~\run{1} and~\run{2} running conditions (see Section~\ref{Chapterhvinbeam}), the probability for pile-up events is small.
The averaged time evolution of the signal is shown in 
Fig.~\ref{fig:ph} for pions and electrons, with and without radiator.
In the amplification region (early times), the signal is larger, because the ionisation from both sides of the anode wires contributes to
the same time interval. The contribution of TR is seen as an increase in the measured average signal at times corresponding to the entrance of the chamber (around \SI{2.5}{\micro\second} 
in Fig.~\ref{fig:ph}), where the TR photons are preferentially absorbed.
At large times (beyond \SI{2.5}{\micro\second}), the effect
of the slow ion movement becomes visible as a tail. Various approximations of the time response function, the convolution of the long tails with the shaping of the PASA, were studied in order to optimally cancel the tails in data, see Section~\ref{Chaptertracking}.

\begin{figure}[tb]
   \centering
\includegraphics[width=.6\textwidth]{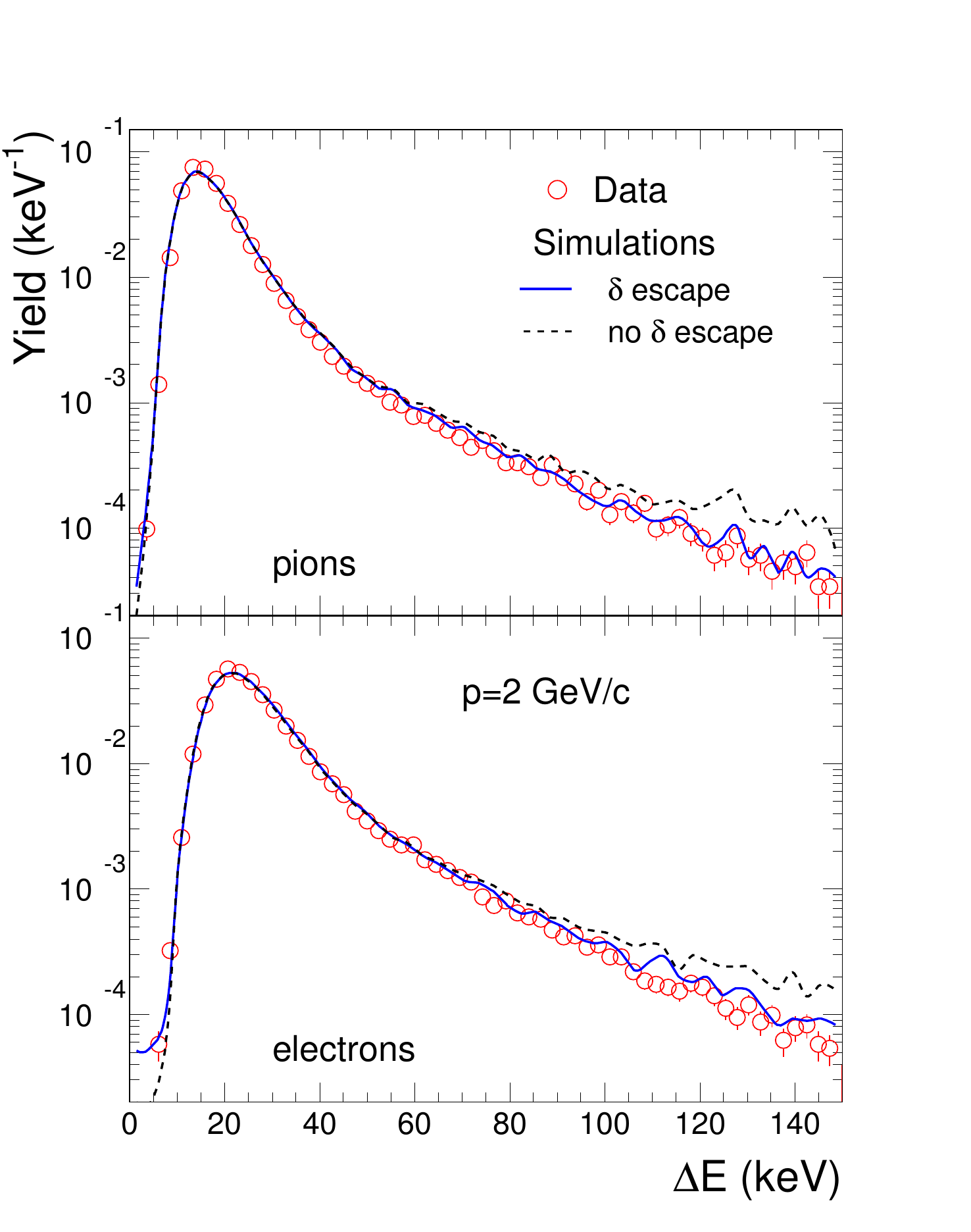} 
  \caption[]{Distributions of the ionisation energy loss of pions and electrons with momenta of 2\gevc. The symbols represent the measurements  obtained at the CERN PS with prototype read-out chambers that were smaller in overall size (active area ~\SI{25}{\centi\meter}~$\times$~\SI{32}{\centi\meter}) but otherwise similar in construction to that of the final detector. The lines are simulations accounting (continuous line) or not (dashed line) for the long range of $\delta$-electrons as compared to the chamber dimensions. Figure taken from~\cite{Andronic:2003qm}.  \TBD{(Replace with GEANT4 once available M.Voelkl)}}
   \label{fig:dedxwithsimtestbeam}
\end{figure}

The knowledge of the ionisation energy loss is important for the control
of the detector performance and for tuning the Monte Carlo simulations.
A set of measurements was performed with prototype read-out chambers with detachable radiators 
for pions and electrons at various momenta~\cite{Andronic:2003qm}.
An illustration of the measured data is shown in Fig.~\ref{fig:dedxwithsimtestbeam} for pions and electrons with a momentum of 2\gevc. The simulations describe the Landau distribution of the total ionisation energy deposition, determined from the calibrated time-integrated chamber signal.
A compilation of such measurements over a broad momentum range 
including data obtained with cosmic-ray muons and from collisions recorded with ALICE is shown 
in Section~\ref{Chapterpid}, Fig.~\ref{TRD_MPV}.

Measurements of the position resolution in the \rphi-direction ($\sigma_y$) and angular resolution $\sigma_{\varphi}$, conducted with prototype chambers, established that the required performance of the detector and electronics ($\sigma_y$~$\lesssim$\SI{400}{\micro\metre} and $\sigma_{\varphi}$~$\le$~\ang{1}) is reached for signal-to-noise values of about~40, which corresponds to a moderate gas gain of about 3500~\cite{Adler:2005bc}.

The production of a chamber was performed in several steps~\cite{BerndEDR} and completed in one week on average.
First, the aluminium walls of the chamber were aligned on a precision table and glued to the radiator panel. The glueing table was custom-built to ensure the required mechanical precision and time-efficient handling of the components.
For almost all junctions the two-component epoxy glue Araldite$\textsuperscript{\textregistered}$ AW~116 with hardener HV~953BD was used. In a few places, where a higher viscosity glue was needed, Araldite$\textsuperscript{\textregistered}$ AW~106 was applied.
In a second step, the cathode and anode wires were wound on a custom-made winding machine and glued onto a robust aluminium frame in order to keep the wire tension.
This aluminium frame was subsequently placed on top of the chamber body,
and the cathode and anode wires were transferred to the G10 ledges glued to the chamber body.
After gluing of the anode and cathode wire planes, the tension 
of each wire was checked by moving a needle valve with pressurised air
across the wires. The induced resonance frequency in each
wire was determined by measuring the reflected light of an LED~\cite{hg-dipl}. Afterwards the pad plane and honeycomb structure were placed on top of the chamber body. Following this production process, each chamber was subjected to a series of quality control tests with an Ar-CO$_2$ (70-30) gas mixture. 
The tests were performed once before the chamber was sealed with epoxy (closed with clamps) and repeated after chamber validation and glueing. In the following the requirements are described~\cite{fkramer}.
The anode leakage current was required not to exceed a value of 10~nA. 
The gas leak rate was determined by flushing the chamber with the Ar-CO$_2$ gas mixture and measuring the O$_2$ content of the outflowing gas. It was required to be less than \SI{1}{\milli\bar}~$\cdot$~\si{\litre/\hour}. In addition,
the leak conductance was measured at an underpressure of 0.4--\SI{0.5}{\milli\bar}
in the chamber. The underpressure test was only introduced at a later stage of the mass production after viscous leaks were found, see Section~\ref{Gas:viscous} for more details.
Comparisons of the anode current induced by a $^{109}$Cd source placed at 100 different positions
across the active area allowed determinations of the gain uniformity. The step size for this two-dimensional scan
was about 10~cm in both directions and the measured values were required to be within $\pm$~15\% of the median.
Electrically disconnected wires were detected by carrying out a one-dimensional scan perpendicular to the wires with a step size of \SI{1}{\centi\metre}. This scan clearly identified any individual
wire that was not connected due to the visible gas gain 
anomaly in the vicinity of this wire, and allowed for repair. 
For one position the absolute gas gain was determined by measuring the anode current and by 
counting the pulses of the $^{109}$Cd source. 
The long term stability was characterised by monitoring the gas gain in intervals 
of 15~minutes over a period of 12~hours.

\subsection{Radiator}

\begin{figure}[tb]
   \centering
\includegraphics[width=.6\textwidth]{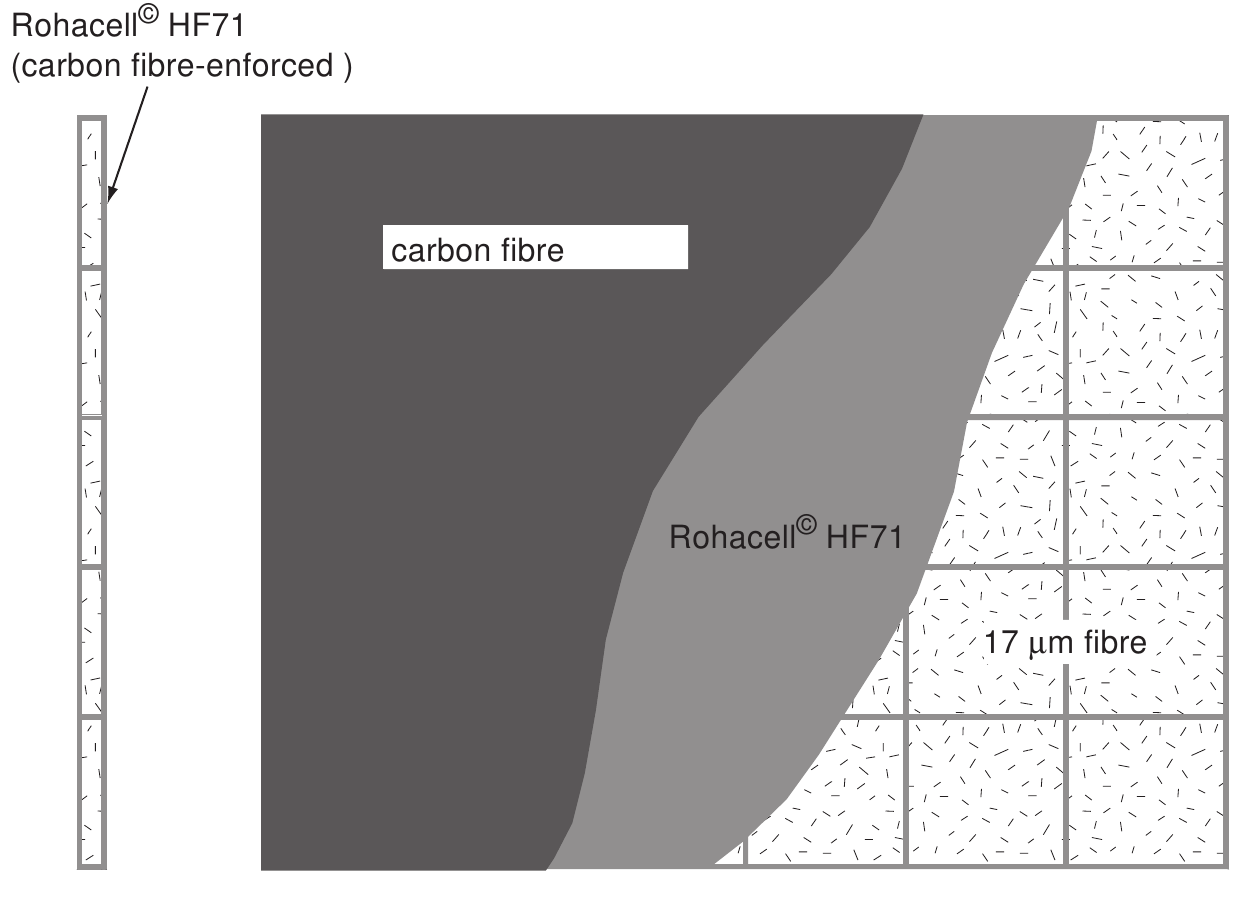}
  \caption[]{Side (left) and top (right) view of the design of the TRD sandwich radiator~\cite{Cortese:519145}.}
   \label{Figure_radcross}
\end{figure}

\begin{figure}[tb]
   \centering
\includegraphics[width=.6\textwidth]{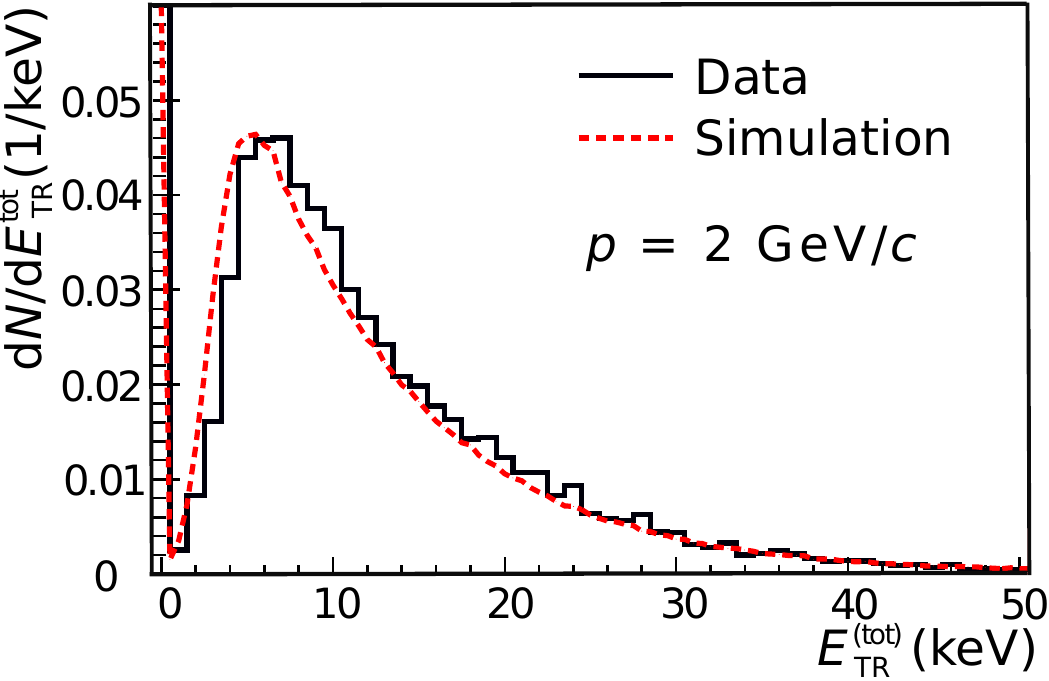} 
  \caption[]{Measured and simulated spectra of TR produced by electrons with a momentum of 2\gevc for the ALICE TRD sandwich radiator. Figure adapted from~\cite{Andronic:2005hw}. }
   \label{fig:trspectra}
\end{figure}

The design of the radiator is shown in Fig.~\ref{Figure_radcross}. 
Polypropylene fibre mats of \SI{3.2}{\centi\metre} total thickness are sandwiched between 
two plates of Rohacell$\textsuperscript{\textregistered}$ foam HF71, which are mechanically reinforced by lamination of carbon fibre sheets 
of \SI{100}{\micro\meter} thickness. 
Aluminised kapton foils are glued on top, to ensure gas tightness and to also serve as the drift electrode. For mechanical reinforcement, cross-bars of Rohacell$\textsuperscript{\textregistered}$ foam of 
\SI{0.8}{\centi\metre} thickness are glued between the two foam sheets of the sandwich, 
with a pitch of 20--\SI{25}{\centi\metre} depending on the chamber size.
After construction the transmission of the full radiator was measured using the K$_{\alpha}$ line of Cu at \SI{8.04}{\kilo \electronvolt} to ensure the homogeneity of the radiators~\cite{cb-dipl}. This line was chosen as its energy is close to the most probable value of the TR spectrum (see Fig.~\ref{fig:trspectra}).

Measurements with prototypes~\cite{Andronic:2005hw} indicated that such a 
sandwich radiator produces 30--40\% less TR compared to a regularly spaced foil radiator. However, constructing a large-area detector with radiators made out of 100~regularly spaced foils each is infeasible. The impact of various radiators constructed from fibres and/or foam on, e.g.\ particle identification is discussed in~\cite{Andronic:2005hw,Andronic:2004uy}. 
Based on these measurements the fibre/foam sandwich radiator design was chosen for the 
final detector.

The spectra of TR produced by electrons with a momentum of 2\gevc as measured with the ALICE TRD 
sandwich radiator is shown in Fig.~\ref{fig:trspectra}. Such a measurement is 
important for the tuning of simulations in the ALICE setup. As the production
of TR is not included in GEANT3~\cite{Brun:1994aa}, which is used to propagate generated particles through the ALICE apparatus for simulations, we have explicitly added it to our simulations in AliRoot~\cite{ALIROOT}, the ALICE offline framework for simulation, reconstruction and analysis. An effective parameterisation of the irregular radiator in terms of a regular foil radiator is employed as an approximation. The simulations describe the data satisfactorily including the momentum dependence~\cite{Andronic:2005hw}.

\subsection{Supermodule}\label{secsupermodule}

The detector is installed in the spaceframe (the common support
structure for most of the central barrel detectors) in
18~supermodules, each of which can host 30~read-out chambers arranged
in 5~stacks and 6~layers (see Fig.~\ref{fig:crosssmlong}). The overall
shape of the supermodule is a trapezoidal prism with a length of
\SI{7.02}{\metre} (\SI{8}{\metre} including services). Its height is
\SI{0.78}{\metre} and the shorter (longer) base of the trapezoid is
\SI{0.95}{\metre} (\SI{1.22}{\metre}). The weight of a supermodule
with 30~read-out chambers is about \SI{1.65}{\tonne}. Mechanical
stability is provided by a hull of aluminium profiles and sheets,
connected with stainless steel screws. The materials were chosen to
minimise the interference with the magnetic field in the solenoid
magnet. In front of PHOS, where minimal radiation length is required, 
the aluminium sheets of the short and long base of the trapezoid were replaced by
carbon-fibre windows.

All service connections must be routed internally to the end-caps of
the supermodule. Those that require materials with large radiation
length are placed at the sidewalls, outside the active area of the TRD
and most other detectors in ALICE. This includes the low-voltage power
distribution bus bars as well as other copper wires for the Detector
Control System (DCS) board power, network and high-voltage (HV)
connections between the fanout boxes and read-out chambers, and the
rectangular cooling pipes (see Section~\ref{infra} for more details).

Low-voltage (LV) power for the read-out boards is provided via copper
power bus bars (2~for each layer and voltage as described in Table~\ref{tab:lv_channels}) with a cross-section
of~\SI{6}{\milli \metre}~$\times$~\SI{6}{\milli \metre} (per channel)
running along the sidewalls of the supermodule. Each read-out board is
connected directly to the power bus bars. Heat generated by ohmic
losses in the power bus bars is partially transferred to the adjacent
cooling pipes (see Section~\ref{labelcooling}). The power bus bars
protrude about \SI{30}{\centi\metre} from each side of the supermodule
hull, where they are equipped with capacitors for voltage
stabilisation. On one end-cap of the supermodule the power-bus bars
are connected via a low-voltage patch panel to
the long supply lines to the power supplies outside of
the magnet.

Each read-out chamber is equipped with~6 or 8~read-out boards (see
Section~\ref{sec:readoutchamber}) and one DCS board (see
Section~\ref{InfraSCN}). Power is provided and controlled separately
for each DCS board by a power distribution box. The DCS boards
are connected via twisted-pair cables to Ethernet patch panels at the
end-caps and the boards of two adjacent layers are connected via
flat-ribbon cables in a daisy chain loop to provide low-level Joint
Test Action Group (JTAG) access to neighbouring boards.

For each chamber, three optical fibres are routed to the end-cap on
the C-side. Two fibres connect the optical read-out interfaces to a
patch panel, where they are linked via the Global Tracking Unit (GTU)
(see Section~\ref{Chapterreadoutelectronics}) to the Data AcQuisition
(DAQ) systems. One trigger fibre connects the DCS board to the trigger
distribution box (see Section~\ref{sec:pt_ov}), which receives the
trigger signals from the pretrigger system or its back-up system and
splits them into 30~fibres (+ 2 spares).

The supermodules were constructed from 2006 to 2014. In the following,
we discuss the sequence of required steps. After the construction of
the supermodule hulls, the power bus bars and patch panels for the
distribution of low voltage for the read-out boards and the cooling
bars for the water cooling were mounted on the sidewalls. Next the
power distribution box (DCS board power), the box for trigger signal
distribution, a patch panel for the optical read-out fibres, and the
high-voltage distribution boxes were installed at the end-caps.

Before integrating the read-out chambers into a supermodule, they were
equipped with electronics (read-out boards, DCS boards) and cooling
pipes. After a series of tests were performed to ensure stable
operation~\cite{ElectronicsInt1,ElectronicsInt2}, the chambers were
then inserted layer by layer. The first connection established during
the installation was the gas link between the chambers (using
polyether ether ketone connectors). The chambers were fixed to the
hull with three screws on each of the long sides after performing a manual 
physical alignment. As demonstrated by later measurements
(Section~\ref{Chapteralign}), the alignment in \rphi between the
chambers is of the order of 0.6--\SI{0.7}{\milli\metre} (r.m.s.).

The cables to and from the read-out boards used for JTAG, low-voltage
sensing, Ethernet, and DCS power were routed along one side of the
chambers. The cable lengths in the active area on top of the chambers
were minimised, avoiding cables from the read-out pads to cross. On the other side of the chambers,
only the high voltage cables were routed. They were soldered at two
separate HV distribution boxes for anode and drift voltage at one
end-cap of the supermodule. Each read-out board (38~per layer) was
connected to the power bus bars (low voltage) using pre-mounted
cables. The cooling pipes (4~per read-out board) were connected by
small Viton tubes. In the $z$-direction across the read-out chambers, only
optical fibres for the trigger distribution (1~per chamber) and data
read-out (2~per chamber) were routed.

In addition to layer-wise tests during installation, a final test
was done after completion. The test setup consisted of
low-voltage and high-voltage supplies, a cooling plant, a gas
system~\cite{hgr-dipl}, as well as a full trigger setup and read-out
equipment. Also a trigger for cosmic rays was built and
installed~\cite{bb-dipl,ja-dipl}. It was used for first measurements
of the gas gain and the chamber alignment, and to also study the zero
suppression during
assembly~\cite{sw-dipl,es-dipl,hg-dipl,mw-dipl,ba-dipl,cw-bach}.

After transport to CERN pre-installation tests were performed
(see Section~\ref{Chaptercommissioning} and~\cite{Klein:Diplom08}) and
the supermodules were installed in the space frame with a precision of
\SI{1}{\centi\metre} (r.m.s.) in \zlab-direction. The maximum tolerance in
$\varphi$ is \SI{2}{\centi\metre} due to constraints given by the space frame.

In addition to the sequential assembly and installation, four
supermodules were completely disassembled again in 2008 and 2009. The initial tests were not sensitive to viscous leaks of the read-out chambers and thus the supermodules were rebuilt after improving the gas tightness (see Section~\ref{Gas:viscous}). Furthermore, in 2013 during \LS{1}, one
supermodule was disassembled in order to improve the high-voltage
stability of the read-out chambers (see
Section~\ref{Chapterbeamoperation}).

\subsection{Material budget}

A precise knowledge of the material budget of the detector is important to obtain a precise description of the detector in the Monte Carlo simulations, which are used, e.g. to compute the track reconstruction efficiencies. 

\begin{figure}[tb]
\centering
\includegraphics[width=0.7\textwidth]{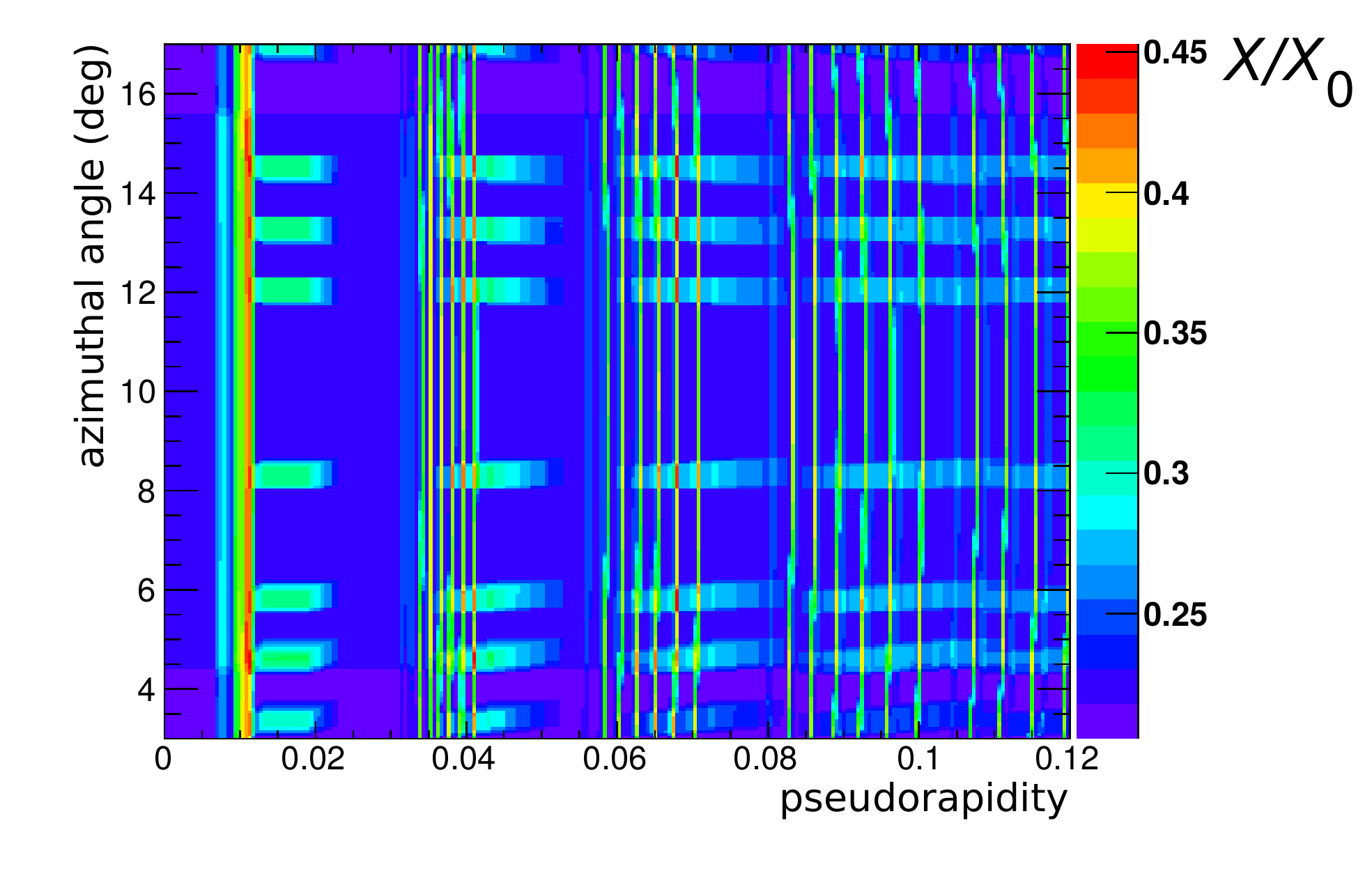}
\caption{
The radiation length map in units of $X/X_{0}$ in
a zoomed-in part of the active detector area as a function of the pseudorapidity
and the azimuthal angle, calculated from the geometry in
AliRoot (the colour scale has a suppressed zero). The positions of the MCMs and the cooling pipes are visible as hot spots. The radiation length was calculated for particles originating from the collision vertex. Therefore the cooling pipes of the six layers overlap for small, but not large $\eta$.
}
\label{FIG_GEO:roc_material}
\end{figure}

The TRD geometry, as implemented in the simulation part of AliRoot,
consists of the read-out chambers, the
services, and the supermodule frame. All these parts are placed
inside the space frame volume. The material of a read-out chamber is obtained including several material components. A general overview of the various components is given in Table~\ref{TAB_GEO:roc_geom}.

\begin{table}[tb]
\begin{center}
\begin{tabular}{ll}
\hline
\textbf{Description}    &  \textbf{$\textit{X}/\textit{X}_{0}$}   \textbf{(\%)}   \\
\hline
Radiator & 0.69 \\
Chamber gas and amplification region & 0.21 \\
Pad plane  & 0.77 \\
Electronics (incl. honeycomb structure) & 1.18 \\
\hline
Total & 2.85 \\
\hline
\end{tabular}
\caption{Parts of one read-out chamber, radiator, electronics, and their average contribution to the radiation length in the active area for particles with normal incidence.}
\label{TAB_GEO:roc_geom}
\end{center}
\end{table}

The material budget in the simulation was adjusted to match the
estimate based on measurements during the construction phase of the final detector.
The supermodule frames consist of the aluminium sheets on the sides,
top, and bottom of a supermodule together with the traversing support
structures, such as the LV power bus bars and cooling arteries.  
Additional electronics equipment is represented by
aluminium boxes that contain the corresponding copper layers to mimic the
present material. The services are also introduced, including, e.g.\ the gas distribution
boxes, cooling pipes, power and read-out cables, and power connection
panels. 

Figure~\ref{FIG_GEO:roc_material} shows the resulting radiation length
map, quantified in units of radiation length ($X/X_{0}$), in a zoomed-in part of the active detector area. It is clearly visible
that the Multi-Chip Modules (MCM)s on the read-out boards (see Section~\ref{Chapterreadoutelectronics}) and the cooling pipes introduce hot spots in $X/X_{0}$. After averaging over the shown area, the mean value is found
to be $\langle X/X_{0}\rangle =$~24.7\% for a supermodule with aluminium profiles and sheets and 30~read-out chambers (6 chambers per stack with the material budget as indicated in Table~\ref{TAB_GEO:roc_geom}). The reduced material budget of the supermodules in front of the PHOS detector (carbon fibre inserts instead of aluminium sheets and no read-out chambers in stack~2) is likewise modelled in the simulation. In regions directly in front of PHOS $\langle X/X_{0}\rangle$ is only 1.9\%.

The total weight of a single fully equipped TRD supermodule as described in the AliRoot geometry, 
including all services, is \SI{1595}{\kilogram}, which is about 3.3\% less than its real weight. This discrepancy can be attributed to material of
service components, such as the gas manifold (see Section~\ref{Gas:system}) and the patch panel, outside the active area, which were not introduced in the AliRoot geometry.

\section{Gas}\label{Chaptergas}

At atmospheric pressure, a total of \SI{27}{\cubic\meter} of a xenon-based gas mixture must be circulated through the TRD detector. This expensive gas cannot be flushed through, but rather has to be re-circulated in a closed loop by using a compressor and independent pressure and flow regulation systems.
The gas system of the TRD follows a pattern in construction,
modularisation, control, and supervision which is common to all LHC gaseous detectors, with emphasis on the regulation of a very small overpressure on the read-out chambers and on the minimisation of leaks. 
The basic modules such as mixer, purification, pump, exhaust, analysis, etc.,
are based on a set of equal templates applied to the hardware and the software. A Programmable Logic Controller (PLC)
controls each system and the user interacts with it through a supervision panel.
Upon a global command, the PLC executes a sequence
that configures all elements of the gas system for a given operation mode and continuously regulates the active elements of the system.
In this manner the modules and operational conditions can be customised to the specific requirements of each detector, from the control of the stability of the overpressure in the detectors, the circulation flow, and the gas purification, recuperation and distillation, to the monitoring of the gas composition and quality (\XeCOtwo, and as little O$_2$, H$_2$O and N$_2$ as possible).

\subsection{Gas choice}\label{Gas:choice}
\begin{figure}[tb]
\centering
\includegraphics[width=.45\textwidth]{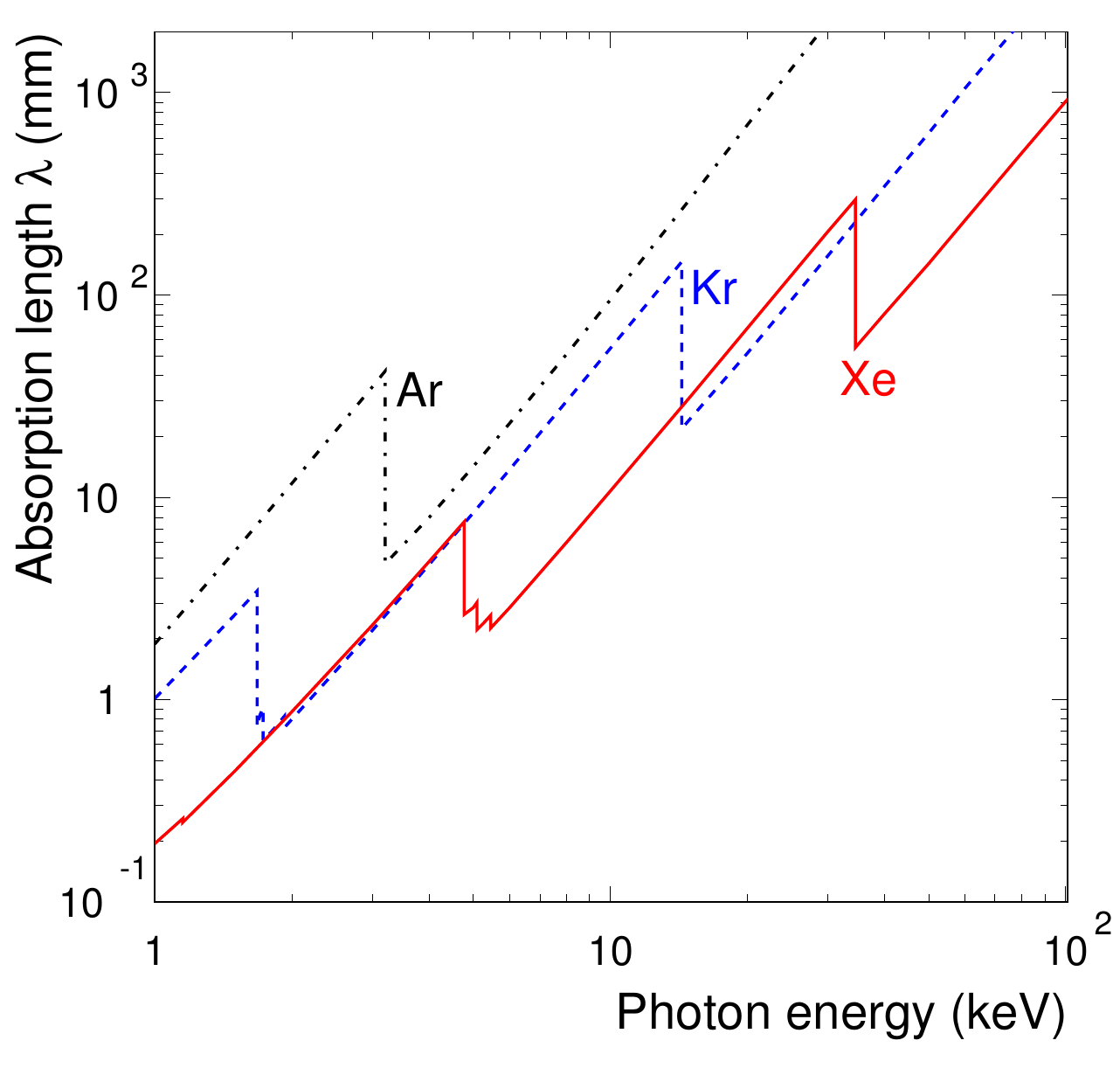} 
\caption{Absorption length of X-rays in noble gases in the relevant energy range of TR production.}
\label{fig:Xray_cross_sec}
\end{figure}

As well as being an array of tracking drift chambers, the TRD is an electron identification device, achieved through the detection of TR photons. In order to efficiently absorb these several keV photons, a high $Z$ gas is necessary. Figure~\ref{fig:Xray_cross_sec} shows, for three noble gases, the absorption length of photons of energies in the range of typical TR production. At around \SI{10}{\kilo\electronvolt} the absorption length in Xe is less than a \si{\centi\metre}, whereas for Kr it is several cm. This argues for the choice of Xe as noble gas for the operating mixture.
CO$_2$ is selected as the quenching gas, since hydrocarbons are excluded for flammability and ageing reasons. The choice of the exact composition
is in this case rather flexible, since the design of the wire chambers leaves enough freedom in the choice of the drift
field and anode potential. The best compromise for the CO$_2$ concentration corresponds to the mixture \XeCOtwo, which ensures a very good efficiency of TR photon absorption by Xe and provides stability against discharges to the detector. 

Furthermore, this mixture exhibits a nice stability of the drift velocity, 
at the nominal drift field, also with
the inevitable contamination of small amounts of N$_2$ that accumulates in the gas through leaks (see Section~\ref{Gas:specs}).
\begin{figure}[tb]
   \centering
\includegraphics[width=.7\textwidth]{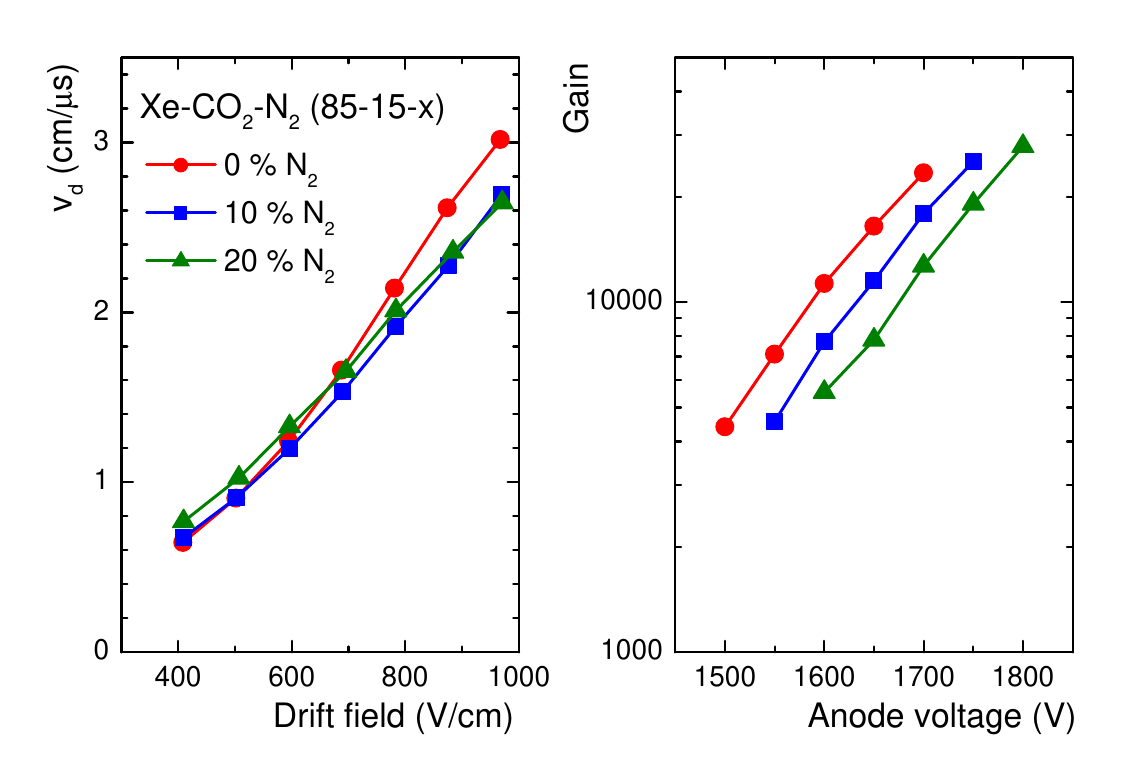} 
  \caption{Left: Drift velocity as a function of the drift field for the nominal gas mixture Xe-CO$_2$ and different admixtures of N$_2$. Right: Gain as a function of the anode voltage for the same gas mixtures.}
   \label{fig:Gas_properties}
\end{figure}
The drift velocity of the \XeCOtwo mixture, pure and with substantial admixtures of N$_2$, as a function of the drift field, is shown in Fig.~\ref{fig:Gas_properties} (left). The drift velocity does not depend on the N$_2$ contamination at the nominal drift field of \SI{700}{\volt/\centi\metre}. On the other hand, as illustrated in Fig.~\ref{fig:Gas_properties} (right), the anode voltage would need a \SI{50}{\volt} readjustment to keep the gain constant when increasing the concentration of N$_2$ by 10\% in the mixture. It should be noted that intakes of less than 5\% N$_2$ are typically observed in one year of operation. After 2--3 years of operation, the N$_2$ is cryogenically separated from the Xe (see Section~\ref{Gas:recuperation}).

The operation of the chambers in a magnetic field of \SI{0.5}{\tesla}, perpendicular to the electric drift field (\SI{700}{\volt/\centi\metre}), forces the drifting electrons on a trajectory, which
is inclined with respect to the electric field. The so-called Lorentz angle is about \SI{9}{\degree} for this gas mixture (see Section~\ref{Chaptercalib}).

For commissioning purposes, where TR detection is not necessary, the read-out chambers are flushed with \ArCOtwo, which is available in a premixed form at low cost.

\subsection{Requirements and specifications}\label{Gas:specs}
The TRD consists of read-out chambers with an area of about \SI{1}{\square\metre} which are built with low material budget. This poses a severe restriction on the maximum overpressure that the detector can hold. Therefore, while in operation, the pressure of each supermodule is regulated by the gas system to a fraction of a mbar above atmospheric pressure and the safety bubblers, installed close to the supermodules, are adjusted to release gas at about \SI{1.3}{\milli\bar} overpressure. The detector can hold an overpressure in excess of \SI{5}{\milli\bar}.

Another tight constraint arises from the highly disadvantageous surface-to-volume ratio of the detector, which enhances the challenge of keeping the gas losses through leaks to a minimum.
Cost considerations drive the criterion for the maximum allowable leak rate of the system: a reasonable target is to lose less than 10\% of the total gas volume through leaks in one year. This translates into a total leak conductance of \SI{1}{\milli \litre/\hour} per supermodule at \SI{0.1}{\milli\bar} overpressure. As a result, unlike in other gas systems, gas is not continuously vented out to the atmosphere. Furthermore, the filling and emptying of the system must be performed with marginal losses of xenon. Adequate gas separation and cryogenic distillation techniques are therefore implemented.
Furthermore, any pulse-height measuring detector must be operated with a gas free of electronegative substances, such as O$_2$, which is continuously removed from the gas stream. Precautions are taken by chromatographic analyses of both the supply xenon and of the air inside the volume of the solenoid magnet to avoid any SF$_6$ contamination of the gas through gas supply cylinders or from neighbouring detectors.

\subsection{Description of the gas system}\label{Gas:system}

\begin{figure}[tb]
\centering
\includegraphics[trim=0cm 4cm 0cm 4cm,clip=true,width=.5\textwidth]{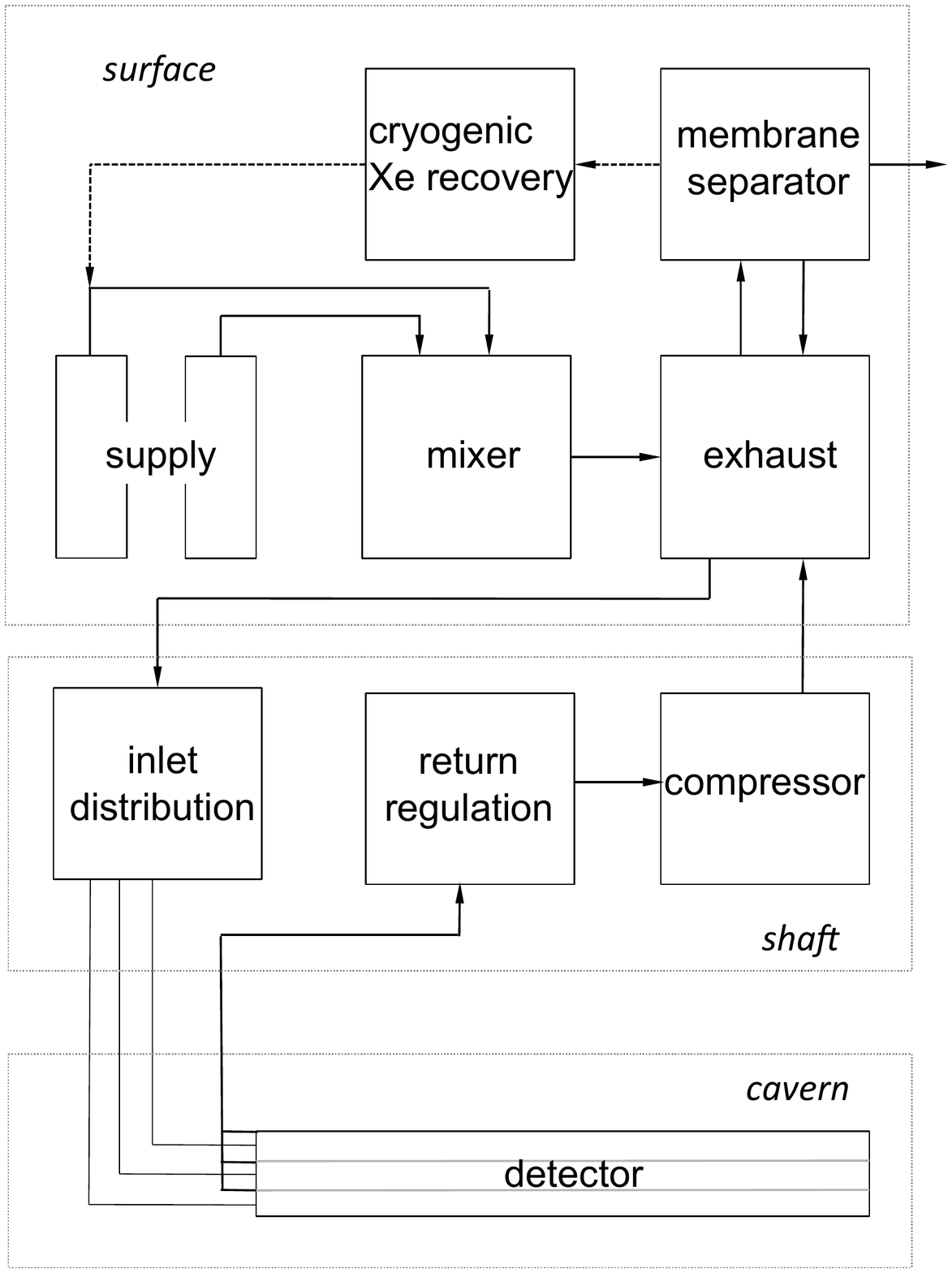} 
\caption[Gas system sketch]{Schematic view of the TRD gas system. The gas circulates in a closed loop pushed by a compressor. The flow for each supermodule is determined by the pressure set at individual pressure reducers in the inlet distribution modules. The overpressure is regulated with individual pneumatic valves at the return modules. The gas is purified at the surface and, when needed, supply gas is mixed and added to the loop. For the filling and the removing of the expensive xenon, semipermeable membranes are used to separate it from the CO$_2$. The recovered xenon can be treated in a cryogenic plant in order to remove accumulated N$_2$, prior to storage.}
\label{fig:Sketch_gas_system}
\end{figure}

The TRD gas system follows the general architecture of all closed loop systems of the LHC detectors, but is customised to meet the requirements specified above. The various modules of the gas system are distributed, as shown schematically in Fig.~\ref{fig:Sketch_gas_system}, on the surface, in a location halfway down the cavern shaft, and in the cavern.
The gas is circulated by compressors that suck the gas from the detector and compresses it to a high pressure value. This pumping action is regulated to keep the desired overpressure at the detector. In the high-pressure part of the system, at the surface, gas purification, mixing, and other operations are carried out. On its way to the cavern, the gas is distributed to individual supermodules using pressure regulators. The gas circulates through the detector and at the outlet of each sector a gas manifold is used to return the gas through a single line and to hold the pressure regulation hardware. Halfway to the surface, a set of pneumatic valves is used to regulate the flow from each supermodule in order to keep the desired overpressure. The gas is then compressed into a high pressure buffer prior to circulation back to the surface.

\subsubsection{Distribution}\label{Gas:distribution}
Xenon is a heavy gas; its standard condition density at ambient conditions is \SI{5.76}{\kilogram/\cubic\metre}, 4.7~times that of air. This means that over the \SI{7}{\metre} height-span of the TRD in the experiment, the total hydrostatic pressure difference between the top and the bottom supermodules would be about \SI{2.8}{\milli\bar}. In order to overcome this, gas is circulated separately through each supermodule (except the top three and the bottom three, which are installed at similar heights) and the pressure is thus individually regulated to equal values everywhere. In addition, due to the different heights of the supermodules, 
the gas, supplied from the surface, would flow unevenly through the different supermodules, the lower ones being favoured over the higher ones.
This second inconvenience is overcome by supplying the gas to each supermodule from the distribution area (half way down the cavern shaft) through \SI{4}{\milli\metre} thin lines over a length of about \SI{100}{\metre}. The pressure drop of the circulating gas in these lines, of several tens of mbar, is much larger than the difference in hydrostatic pressure between supermodules, and therefore nearly equal flow, at equal overpressure, is assured in all supermodules. 

The six layers of the supermodules are supplied from one side (A-side) with three inlet lines, each of them serving two consecutive layers. Small bypass bellows connect two consecutive layers on the opposite side. In the A-side, a manifold arrangement is used to connect the gas outlets and a common safety bubbler, pressure sensors and back-up gas. The return outlets in each supermodule are connected together into one line which returns to the pump module. The three top and three bottom supermodules are connected to one single return line each. This arrangement results into 14~independently regulated circulation loops. Each supermodule has its own two-way bubbler, which provides the ultimate safety against over- or underpressure.

\subsubsection{Pump}\label{Gas:pump}
In the distribution area, the flow through each return line is regulated by a pneumatic valve per loop driven by the pressure sensors located at the detector. In this area, the gas is kept at a pressure slightly below atmospheric pressure, and it is stored in a \SI{0.8}{\cubic\metre} buffer container before it is compressed by two pumps which operate at a constant frequency. The compressor module drives a bypass valve in order to maintain a calculated pressure set point at its inlet. In this manner, a dual regulation concept is used to handle the 14 loops. The role of the inlet buffer is to act as a damper of possible regulation oscillations. This pressure regulation system keeps the overpressure in the supermodule stable at \SI{0.1}{\milli\bar} above atmospheric pressure (set point) within \SI{0.03}{\milli\bar}. 

A \SI{0.93}{\cubic\metre} high pressure buffer at the compressor outlet is used as a storage volume. Its content varies according to the atmospheric pressure, either by providing gas to the detectors, or by receiving it from them. The overpressure in this buffer typically ranges between 0.8 and \SI{2}{\bar}. Knowledge of all the system volumes allows the pressure in the buffer to be predicted for any atmospheric pressure value. Gas leaks ultimately result in a reduction of this pressure, in that case the dynamic regulation of the high pressure triggers the injection of fresh gas from the mixer until the high pressure is restored.
From this buffer, the pressurised gas is circulated up to the gas building at the surface.

\subsubsection{Purifier}\label{Gas:purifier}
The purifier module consists of two 3~litre cartridges each filled with a copper catalyser which is efficient in chemically removing oxygen by oxidising the copper, and mechanically removing water by absorption. Upon saturation, the PLC switches between cartridges at the pre-defined frequency, and launches an automatic regeneration cycle where CuO$_2$ is reduced at high temperatures with a flow of H$_2$ diluted in argon. As the detector is rather gas tight, the O$_2$ intake through leaks is moderate, and the purifier keeps it between 0 and 3~ppm. However, H$_2$O diffusion, probably through the aluminised Mylar foil which constitutes the drift electrode of every read-out chamber, makes it necessary to switch between purifiers about every 3.5~days, in order to keep the H$_2$O content below a few hundred ppm.

\subsubsection{Recirculation}\label{Gas:exhaust}
The surface module is used to recirculate the gas at high enough pressure to the distribution modules in the cavern shaft area. It also contains provisions for extracting gas samples for analysis, and a bypass loop to allow for the installation of containers such as a krypton source for gain calibration (see Section~\ref{Chaptercalib}).

\subsubsection{Mixer}\label{Gas:mixer}
Under normal operation and since the gas is only exhausted through leaks, gas injection into the system happens only if the pressure in the high pressure buffer falls below a dynamic threshold, as explained above. On such occasions, the mixer is activated and injects the nominal gas mixture at a rate of a few tens of \si{\litre/\hour} until the high pressure buffer is replenished. The amount of gas injected by the mixer during a given period provides a direct measurement of the leak rate.

In addition, a second set of mass flow controllers provides flows in the \si{\cubic\metre/\hour} range and is used for filling and emptying the detector. 

\subsubsection{Backup system}\label{Gas:backup}
When the gas system is in stop mode, e.g.\ when there is a power failure, the safety bubbler installed on each supermodule ensures that the detector pressure always remains within about $\pm$\SI{1.3}{\milli\bar} relative to atmospheric pressure. In order to avoid that air, i.e.\ oxygen, enters the detector, the external side of the bubbler is connected to a continuous flow of neutral gas, in this case N$_2$, that flows through the bubbler in case of a large detector underpressure. The choice of N$_2$ is driven by the small influence on the gas properties that this admixture has (see Fig.~\ref{fig:Gas_properties}). The full TRD is served by three independent backup lines, each with connections to six supermodule bubblers, and arranged such that the flow points downwards. In this way, if the xenon mixture is exhausted through the bubblers, it falls down the back-up line, relieving its high hydrostatic pressure. A differential pressure transmitter measures the pressure difference between the detector and the backup gas.

\subsubsection{Analysis}\label{Gas:analysis}
The control of the gas quality is perhaps the most demanding aspect of running detectors where both signal amplitude and drift time information are important. This control is even more crucial for the ALICE TRD, where accurate and uniform drift velocity and gain values are needed for triggers based on online tracking and particle identification. Thus, in addition to effective tightness of the system and continuous removal of O$_2$ and H$_2$O, constant monitoring of the gas composition and in particular of the N$_2$ is necessary. Although for a large volume system such as that of the TRD the changes in composition are obviously slow, the precision and stability requirement of the measuring instruments are quite challenging.
Furthermore, constantly measuring analysers, such as O$_2$, H$_2$O and CO$_2$ sensors, must be installed in the gas loop, since xenon must not be exhausted. Therefore they must be free of outgassing of contaminants into the gas. 

The analysis module samples the return gas from individual supermodules in a bypass mode, before it is compressed. For this, a fraction of the gas is pushed through the analysis chain by a small pump, and returned to the loop at the compressor inlet. Usually, the PLC is programmed to continuously sample one supermodule after the other, for about 10~minutes each.

An external gas chromatograph is used to periodically measure the gas composition. This device is not in the gas loop; rather, the gas is exhausted while purging and sampling a small stream for a few seconds every few hours.

\subsubsection{Membranes}\label{Gas:membranes}
One system volume of xenon is injected for operation and, typically every two or three years, removed for cleaning and storage. This means that it must be possible to separate CO$_2$ from Xe.
This separation is achieved with a set of two semipermeable membrane cartridges. Each cartridge consists of a bundle of capillary polyimide tubes through which the mixture flows. The bundle is in turn enclosed in the cartridge case. While the CO$_2$ permeates through the polyimide walls, most of the xenon is contained and continues to flow into the loop. The permeating gas can be circulated through the second membrane cartridge to further separate and recover most of the Xe. 

During the filling, the detector is first flushed with CO$_2$ and then, in closed-loop circulation, the xenon is injected as the CO$_2$ is removed through the membranes. The reverse process is used for the recuperation of the xenon into a cryogenic plant.

\subsubsection{Recuperation}\label{Gas:recuperation}
N$_2$ inevitably builds up in the gas through small leaks and cannot be removed by the purifier cartridges. Therefore, after each long period (2--3 years) of operation, the N$_2$ is cryogenically separated from the Xe. A cryogenic buffer is filled with xenon after separating it from CO$_2$. At the same time, CO$_2$ is injected into the gas system in order to replace the removed gas.

The cryogenically isolated buffer is surrounded by a serpentine pipe with a regulated flow of liquid nitrogen (LN$_2$) in order to keep its temperature at \SI{-170}{\degreeCelsius}, just above the N$_2$ boiling point (\SI{-195.8}{\degreeCelsius}). At this temperature Xe (and CO$_2$) freezes whereas N$_2$ stays in the gaseous phase. Once the buffer is full, the stored gas is pumped away. After this, the buffer is heated up in a regulated way, and the evaporating Xe is compressed into normal gas cylinders. The resulting Xe has typically a N$_2$ contamination of \textless 1\%, and the total Xe loss (due to the efficiency of the membranes and the cryogenic recovery process) is about \SI{1}{\cubic\metre} for a full recovery operation.

\subsection{Operational challenges}\label{Gas:incidents}
The gas system has been operating reliably over several years in several modes, but mainly in so-called run mode. Aside from minor incidents, a number of important leaks have been dealt with, which deserve a brief description.
 
\subsubsection{Viscous leaks}\label{Gas:viscous}
As part of the standard quality assurance procedure, a leak test was performed on each chamber prior to installation in the supermodule. The leak test consisted of flushing the chamber with gas and measuring the O$_2$ contamination at the exhaust, where the overpressure was typically about \SI{1}{\milli\bar}. It was found, however, that a supermodule would lose gas even if the O$_2$ content was very low. The reason turned out to be the particular construction of the pad planes, which are glued to a reinforcement honeycomb panel with a carbon fibre sheet. Viscous leaks would develop between the glued surfaces and gas would find its way out through the cut-outs for the signal connections machined in the honeycomb sandwich. The impedance of this kind of leak is large enough that gas can escape the detector with no intake of air through back-diffusion. The concerned read-out chambers were then extracted and repaired, and the leak tests on subsequent chambers were modified such that the O$_2$ was measured both at over- and underpressure in the read-out chamber, resulting in a tight system.

\subsubsection{Argon contamination}\label{Gas:argon}
At one point, the routine gas analysis with the gas chromatograph showed increasing levels of Ar in the Xe-CO$_2$ mixture. This elusive leak came from a faulty pressure regulator which was pressurised with argon on the atmospheric side. Occasionally, depending on the pressure, the membrane of the regulator would leak and let Ar enter the gas volume. A total of 1\% Ar accumulated in the mixture and was removed by cryogenic distillation, together with N$_2$.

\subsubsection{Leak in pipe}\label{Gas:pipe}
The last major leak in the system was detected when suddenly the pressure at the high pressure buffer started to steadily decrease. Any leak of the system would appear, while running, as a decrease in the high pressure buffer, because the system always ensures the right overpressure at the read-out chambers. By stopping the system and isolating all of its modules, it was found that the source of the leak was a long, stainless steel pipe which connected the compressor module, half way down the cavern shaft, to the surface, where the gas, still at high pressure, is cleaned and recirculated. It was not possible to find the exact location of the leak. This was solved by replacing the pipe by a spare.

\section{Services}
\label{infra}

The supermodules installed in the space frame require service
infrastructure for their operation. To reduce the weight, the
connections (low and high voltage, cooling, gas, read-out, and control
lines) are routed via dedicated frames on the A- and
C-side, respectively. Both frames are \SI{2}{\metre} extensions of the space
frame with similar geometry, but mechanically independent except for the
flexible services.  Most of the equipment, such as the low-voltage power
supplies, is placed in the cavern underground and thus inaccessible
during beam operation. Some devices are situated in counting rooms in
the cavern shaft, which are supervised radiation areas but accessible.

\subsection{Low voltage}
\label{InfraLowV}

The low voltage system supplies power to various components of
the TRD. The largest consumer is the Front-End Electronics (FEE), i.e.\ the electronics of the Read-Out Boards (ROB) mounted on the
chamber (see Section~\ref{Chapterreadoutelectronics}). To minimise noise, separate
(floating) voltage rails are
used for analogue and digital components. The power supply channels
for analogue \SI{1.8}{\volt}, analogue \SI{3.3}{\volt}, and digital
\SI{1.8}{\volt} are grouped such that one power supply channel
supplies two layers of a supermodule. For the digital \SI{3.3}{\volt}
there is one channel per supermodule. For each supermodule, this
results in the supply channels listed in
Table~\ref{tab:lv_channels}. The DCS boards (see
Section~\ref{InfraSCN}) are powered by a power distribution box (PDB), two of which (in two adjacent supermodules) are supplied by a
dedicated channel. The PDBs are controlled by Power Control Units
(PCU) over a redundant serial interface.

\begin{table}[tb]
\centering
  \begin{tabular}{lrr}
  \hline
    \textbf{Channel} & \textbf{~~~$\mathbf{U_\mathrm{nom} (\mathrm{V})}$}
    & \textbf{~~~$\mathbf{I_\mathrm{typ} (\mathrm{A})}$}\\
    \hline
    3 x Analogue \SI{1.8}{\volt} & 2.5 & 125\\
    3 x Analogue \SI{3.3}{\volt} & 4.0 & 107\\
    3 x Digital \SI{1.8}{\volt} & 2.5 & 95--150\\
    Digital \SI{3.3}{\volt} & 4.0 & 110\\
    \hline
    DCS boards & 4.0 & 2~$\times$~30\\
    \hline
  \end{tabular}
  \caption{Number of low voltage channels, nominal voltages and
    typical currents for the electronics on the chamber-mounted
    read-out boards of one supermodule. The current for the TRAP cores
    (digital 1.8~V) increases with the trigger rates. The current for
    the DCS boards is 2~$\times$~\SI{30}{\ampere} for two adjacent
    sectors.}
  \label{tab:lv_channels}
\end{table}

Because of the high currents, the intrinsic resistances of the cables
and connections are critical and are constantly monitored by measuring
the voltage drop between the power supply unit (terminal voltage) and the patch
panel at each supermodule (sense voltage). Typical values are
6--\SI{8}{\milli\ohm}, depending on the cable length. In addition, the
voltages at the end of each power bus bar are monitored.

The Global Tracking Unit (GTU) (see Section~\ref{sec:gtu}) uses
additional power supplies which are shared with the PCUs. The pretrigger system (see Section~\ref{sec:pt_ov}) is
powered by separate power supplies, laid out in a fail-safe redundant
architecture.

Different customizations of the Wiener PL512 power supply units are used. The
power supplies feeding the FEE are connected to a PLC-based interlock
based on the status of the cooling. Power is automatically cut in
case of a cooling failure.

During the \run{1} operation, several low-voltage connections on the
supermodules showed increased resistivity resulting in excessive
heat dissipation, which in some cases required to switch off part of
the detector until the problem could be fixed during an
access. Later, during \LS{1}, the affected supermodules were pulled out of the
experiment and the connections were reworked
in the cavern. The supermodules were re-inserted and re-commissioned
immediately after the rework. The complete procedure took about one
day per supermodule.

\subsection{Cooling}\label{labelcooling}

The complexity of the cooling system, whose cooling medium is deionised water, is driven by the large amount of
heat sources (more than \SI{100000}{}) distributed over the complete
active area of the detector. Heat is produced by the MCMs and the Voltage Regulators (VR) on the read-out boards, the
DCS boards, and the power bars. The total heat dissipation in a
supermodule amounts to about \SI{3.3}{\kilo\watt}, of which about
\SI{2.6}{\kilo\watt} are produced in the FEE, the remaining
\SI{700}{\watt} originate from the voltage regulators and the bus
bars. The DCS boards contribute with about \SI{130}{\watt} per
supermodule. Overall, the rate of heat to be carried away during detector
operation amounts to \SI{55}{\kilo\watt} and \SI{70}{\kilo\watt} in
\pbpb and pp collisions, respectively, due to different read-out rates.  
Apart from the power bus bars, the heat
sources are positioned on top of the read-out boards.

In the cooling system the pressure is kept below atmospheric pressure. Thus a leak leads to air entering in the system but no
water is spilled onto the detector. The cooling plant~\cite{Santos2003} consists of a
\SI{1500}{\litre} storage tank positioned at the lowest point outside
the solenoid magnet, which is able to contain all the water of the
installation, the circulation pump, the 18~individual circuits that
supply cooling water to the 6~layers of each supermodule, and the heat
exchanger connected to the CERN chilled water network. The reservoir
is kept at 300--\SI{350}{\milli\bar} below atmospheric pressure by
means of a vacuum pump that also removes any air collected through
small leaks. In addition, the pressure of the circulation pump
(\SI{1.8}{\bar}) and the diameter of all pipes are chosen such that a
sub-atmospheric pressure is maintained in all places of the detector,
despite a difference in height of about \SI{7}{\metre} between the
lowest and the highest supermodule. Each circuit is equipped with
individual heaters and balancing valves in order to control the
temperature and the flow in each loop separately. The heaters are
regulated by a proportional-integral-derivative controller.  A
temperature stability in the cooling water of
$\pm$\SI{0.2}{\degreeCelsius} is achieved. The typical water flow is
about \SI{1300}{\litre/\hour} per supermodule. To avoid corrosion a
fraction of the total water flow is passed by a deioniser to keep the
water conductivity low. As the water is in contact with similar materials (stainless steel and
aluminium), the TRD cooling system also supplies the water to the
cooling panels of the thermal screening between TPC and
TRD~\cite{Alme:2010ke}.

The loop regulations and cooling plant control is done by a
PLC. Warnings and alarms are issued by the PLC if the parameters are
outside the allowed intervals and read out by the Detector Control
System (see Section~\ref{Chapterdcs}).  Two independent security
levels were implemented in each loop. The first continuously
monitors the pressure of each loop and stops the water
circulation of the cooling plant if any value reaches atmospheric
pressure.  Secondly, large safety valves were installed at the
entrance to each supermodule. They will open in case an overpressure of
\SI{50}{\milli\bar} is reached, providing a low resistance path for
the water evacuation in case of emergency.

The cold water is supplied in the lowest point of each supermodule and
the warm water is collected on the highest point in order to have more
homogeneous water flow in all pipes. A water manifold at one end-cap
of the supermodule distributes the water in parallel to the 6~layers
inside each supermodule, and on the opposite side a similar manifold
collects the warm water.  In each layer, two rectangular pipes along
the $z$-direction
(65~$\times$~8~$\times$~\SI{7500}{\cubed\milli\metre}) supply
(collect) water to (from) the meanders, 76~individual cylindrical
aluminium pipes (\SI{3}{\milli\metre} in diameter) running across the
$y$-direction where the heat sources are. A total of 17~meander types
were designed for the system. To bring the water from the rectangular
pipes to the individual meanders, the rectangular pipe has small
stainless steel pipes (\SI{3}{\milli\metre} diameter and
\SI{5}{\centi\metre} length) soldered at the proper position for each
MCM row. A Viton tube of about~\SI{2}{\centi\metre} length is used to
connect the small stainless steel pipes and the meanders as well as
for the connections between the two meanders (one per ROB) in
$y$-direction. A total of about 25000~Viton tube connectors were used
in the system. This kind of connector was previously used in the
CERES/NA45 leakless cooling system \cite{Adamova:2008mk} because of
its low price and reliability.

The cooling pad mounted on top of the heat source consists of an
\SI{0.4}{\milli\metre} thick aluminium plate. The meander is glued on
top of the pad by aluminium-filled epoxy
(aluminium powder: Araldite$\textsuperscript{\textregistered}$ 130:100
by weight) to increase the thermal conductivity.  In order to maximise the heat transfer, the longest
possible path was chosen. The choice of aluminium was driven by the
necessity of keeping the material budget as low as possible in the
active area of the detector.

\subsection{High voltage}\label{InfraHVsection}
The high voltage distribution for the drift field and the anode-wire plane
is made separately for each chamber, reducing the affected area to one chamber in case of failure. 
The power supplies for the drift channels and anode-wires were purchased from ISEG~\cite{pap:iseg} (variants of the model EDS~20025). Each module has 32~channels, which are grouped in independent 16-channel boards. Each channel is independently controllable in terms of the voltage setting and current limit as well as monitoring of current and voltage. Eight modules are placed into each crate and remotely controlled via CANbus (Controller Area Network) from DCS (see Section~\ref{Chapterdcs}). The HV crates are placed in one of the counting rooms in the cavern shaft, which allows access even during beam operation. 

For each of the 30~read-out chambers in a supermodule one power supply is needed for the
drift field and one for the anode-wire plane. A multiwire HV cable connects the 32~channel HV module
with a 30~channel HV fanout box (patch box) located at one end of the
supermodule, where the output is redistributed to single wire HV cables (see
Section~\ref{secsupermodule}).  The individual HV cables are then connected to a HV
filter box, mounted along the side of the read-out chamber. The HV filter
box supplies the HV to the 6~anode segments and the drift cathode of
the read-out chamber, and in addition it allows connection of the HV ground
to the chamber ground. It consists of a network of a resistor and capacitors (\SI{2.2}{\nano\farad}
and \SI{4.7}{\nano\farad}) to suppress
load-induced fluctuations of the voltages in the chamber.

The HV crates are equipped with an Uninterruptible Power Supply (UPS) and a battery to bridge short term power failures.  In case of a longer power failure (>~\SI{10}{\second}) a controlled ramp-down is initiated, i.e.\ the HV of the individual drift and anode-wire channels is slowly ramped down. Details on maximum applied voltages, channel equalisation, ramp speed as well as high-voltage instability observed during data taking are discussed in Sections~\ref{Chapterdcs} and~\ref{Chapterbeamoperation}.

\subsection{Slow control network}\label{InfraSCN}

The slow control of the TRD is based on Detector Control Systemboards~\cite{KrawDCS}. They communicate with the DCS (see Section~\ref{Chapterdcs}) by a \SI{10}{\mega \bit/\second} Ethernet
interface, mostly using Distributed Information Management (DIM) as
protocol for information exchange. The use of Ethernet allows the use of standard network equipment, but a dedicated network restricted to
the ALICE site is used. The DCS boards are used as end points for the
DCS to interact with subsystems of the detector. Later sections will
discuss how the DCS boards are used as interface to the various
components, e.g.\ the front-end electronics or the GTU.

The DCS boards were specifically designed for the control of the detector
components and are used by several detectors in ALICE. At the core,
the board hosts an Altera Excalibur EPXA1 (ARMv4 core + FPGA), which hosts a Linux operating system on the processor and user logic
in the FPGA fabric depending on the specific usage of the
board. The DCS board also contains the Trigger and Timing Control receiver (TTCrx) for clock
recovery and trigger reception. The Ethernet interface is implemented
with a hardware PHY (physical layer) and a soft-Media Access Controller (MAC) in the FPGA fabric. In case of the
boards mounted on the detector chambers, the FPGA also contains the
Slow Control Serial Network (SCSN) master used to configure the front-end electronics. Further
general purpose I/O lines are, e.g.\ used for JTAG and I$^2$C
communication.

Since the Ethernet connections are used for configuration and
monitoring of the detector components, reliable operation is crucial.
All DCS boards are connected to standard Ethernet switches installed
in the experimental cavern outside of the solenoid magnet. Because of
the stray magnetic field and the special Ethernet interface of the DCS
board (no inductive coupling), there are limitations on the usable switches. Since the failure of
an individual switch would result in the loss of connectivity to a large
number of DCS boards, a custom-designed Ethernet multiplexer was installed in front of the
switches in the second half of \run{1}. This allows the connection of each DCS board to be remotely switched between two different switches with separate uplinks to the DCS
network. The multiplexers themselves are implemented with fully
redundant power supplies and control interfaces.

\section{Read-out}
\label{Chapterreadoutelectronics}

\begin{figure*}[bt]
   \centering
   \includegraphics[width=.7\textwidth]{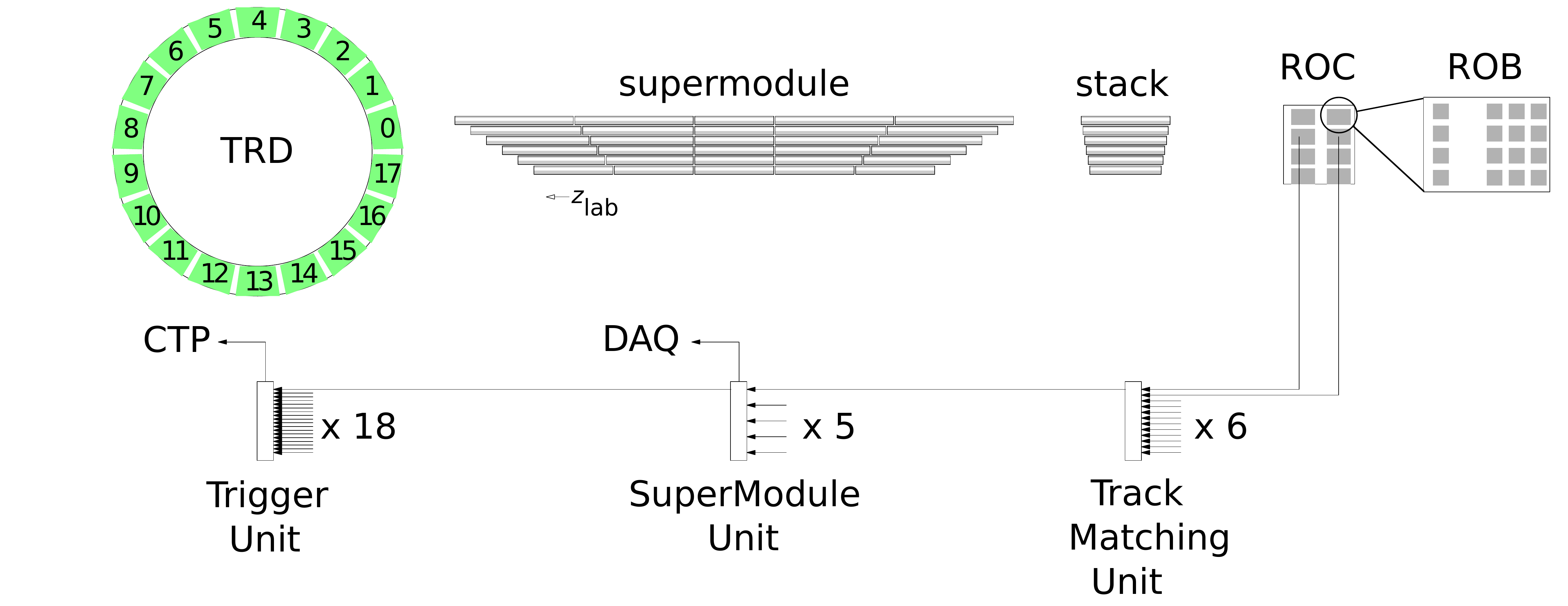}
   \caption[]{Detector structures and corresponding read-out
     stages~\cite{Klein:2014rxa}. The top row of the figure represents the detector and the bottom row the GTU components. The dimensions are not to scale. }
   \label{fig:blockreadout}
\end{figure*}

The read-out chain transfers both raw data and condensed information
for the level-1 trigger. While the former requires sufficient
bandwidth to minimise dead time, the latter depends on a low latency,
i.e.\ a short delay of the transmission. The data from the detector
are processed in a highly parallelised read-out
tree. Figure~\ref{fig:blockreadout} provides an overview and relates
entities of the read-out system to detector components. In the
detector-mounted front-end electronics, the data are processed in
Multi-Chip Modules grouped on Read-Out Boards (ROB) and eventually
merged per half-chamber. Then, they are transmitted optically to the
Track Matching Units (TMU) as the first stage of the Global Tracking
Unit (GTU). The data from all stacks of a supermodule are combined on
the SuperModule Unit (SMU) and eventually sent to the Data AcQuisition
system (DAQ) through one Detector Data Link (DDL) per supermodule.

The read-out of the detector is controlled by trigger signals distributed to both the FEE and the GTU. The ALICE trigger system is
based on three hardware-level triggers (level-0, 1, 2) and a High
Level Trigger (HLT)~\cite{Fabjan:684651} implemented as a computing farm. In addition to
these levels, the FEE requires a dedicated wake-up signal as described in the next subsection.

\subsection{Pretrigger and LM system}
\label{sec:pt_ov}

\begin{figure}[tb]
  \centering
  \includegraphics[width=.3\textwidth]{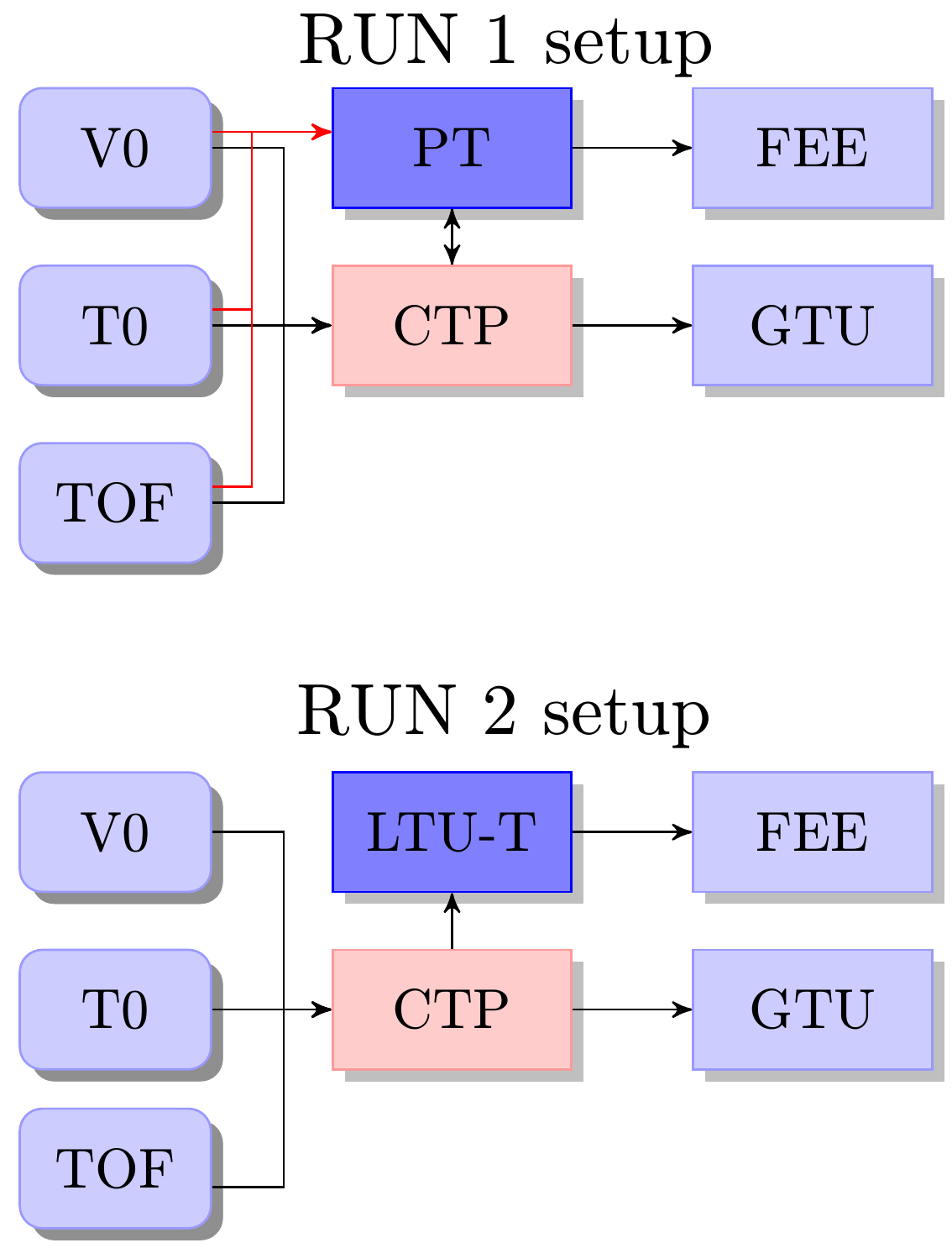}
  \caption{In \run{1}, the wake-up signal required for the front-end
    electronics was generated by a dedicated pretrigger system. In
    \run{2}, the functionality was implemented in the central trigger
    processor and the LTU-T serves as an interface to the TRD FEE.}
  \label{fig:pt_upgrade}
\end{figure}

Both FEE and GTU must receive clock and trigger signals, which are
provided by the Central Trigger Processor (CTP)~\cite{Fabjan684651}
using the Trigger and Timing Control (TTC) protocol over optical
fibres. While the GTU only needs the level-0/1/2 and is directly
connected to the CTP, the FEE requires a more complicated setup. To
reduce power consumption, it remains in a sleep mode when idle and
requires a fast wake-up signal before the reception of a level-0
trigger to start the processing. During \run{1}, an intermediate
pretrigger system was installed within the solenoid
magnet~\cite{sz-dipl,sts-dipl}. Besides passing on the clock and
triggers received from the CTP, it generated the wake-up signal from
copies of the analogue V0 and T0 signals (reproducing the level-0
condition) and distributed it to the front-end electronics. In
addition, the signals from TOF were used to generate a pretrigger and
level-0 trigger on cosmic rays. Because of limitations of this setup,
the latencies of the contributing trigger detectors at the
CTP were reduced for \run{2} (also by relocating the respective detector
electronics) such that the functionality of the pretrigger system
could be integrated into the CTP. The latter now issues an LM (level
minus~1) trigger for the TRD before the level-0 trigger. An interface
unit (LTU-T) was developed for protocol
conversion~\cite{Klewin:Master15} in order to meet the requirements of
the TRD front-end electronics. A comparison of the two designs is
shown in Fig.~\ref{fig:pt_upgrade}. The new system has been used since
the beginning of collision data taking in \run{2}.

\subsection{Front-end electronics}
\label{sec:fee}

\begin{figure*}[tb]
  \centering
  \vspace{0.5cm}
  \includegraphics[width=1.0\textwidth]{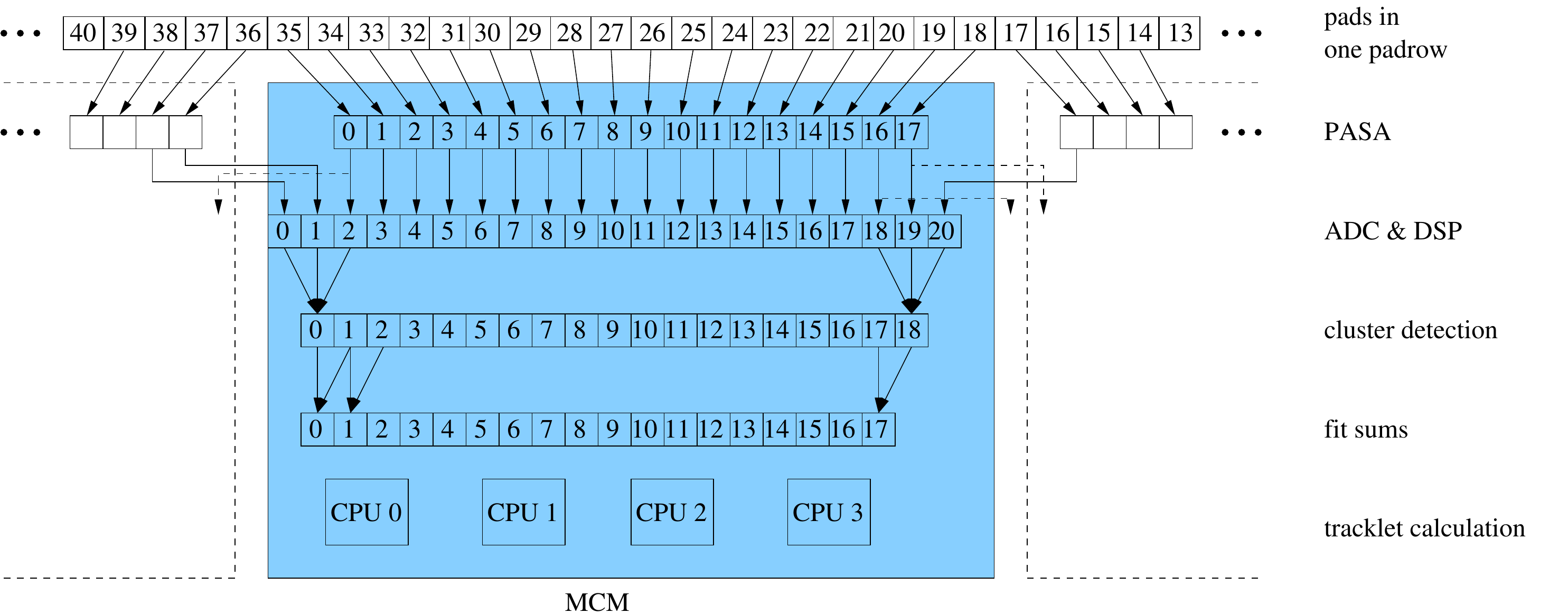}
  \caption{Connections in one MCM~\cite{Klein:Diplom08}.}
  \label{fig:mcm_trkl}
\end{figure*}

The FEE is mounted on the back-side
of the read-out chamber. It consists of MCMs which
are connected to the pads of the cathode plane with flexible flat
cables. An MCM comprises two ASICs, a PASA
and a TRAP, which feature a large number of
configuration settings to adapt to changing operating conditions. The
signals from 18~pads are connected to the charge-sensitive inputs of
the PASA on one MCM. An overview of the connections is shown in
Fig.~\ref{fig:mcm_trkl}.

The very small charges induced on the read-out pads (typically
\SI{7}{\micro\ampere} during \SI{1}{\nano\second}) are not amenable to
direct signal processing. Therefore, the signal is first integrated
and amplified by a Charge Sensitive Amplifier (CSA). Its output is a
voltage signal with an amplitude proportional to the total charge. The
CSA has a relatively long decay time, which makes it vulnerable to
pile-up. A differentiator stage removes the low frequency part of the
pulse. The exponential decay of the CSA feedback network, in
combination with the differentiator network, leads to an undershoot at
the shaper output with the same time constant as the CSA feedback
network. A Pole-Zero network is used to suppress the undershoot. A
shaper network is required to limit the bandwidth of the output signal
and avoid aliasing in the subsequent digitisation process. At the same
time the overall signal-to-noise ratio must be optimised. These
objectives are achieved by a semi-Gaussian shaper, implemented with
two low-pass filter stages. Each stage consists of two second-order
bridged-T filters connected in cascade. The second shaper consists of
a fully differential amplifier with a folded cascode configuration and
a common-mode feedback circuit. This circuit network was implemented
to prevent the output of the fully differential amplifier from
drifting to either of the two supply voltages. It establishes a stable
common-mode voltage. The last stage in the chain comprises a pseudo-differential amplifier
with a gain of~2. This stage
adapts the DC voltage level of the PASA output to the input DC-level
of the TRAP ADC~\cite{Soltveit:2012jp}.

\begin{table}[tb]
  \centering
  \begin{tabular}{ll}
  \hline
    \textbf{Parameter} & \textbf{Value} \\
    \hline 
    PASA gain & \SI{12}{\milli\volt/\femto\coulomb}\\
    PASA power & \SI{15}{\milli\watt/channel }\\
    PASA pulse width (FWHM) & \SI{116}{\nano\second}\\
    PASA noise (equivalent charge) & $1000~\mathrm{e}$\\
    \hline
    TRAP power & \SI{12.5}{\milli\watt/channel }\\
    TRAP ADC depth & 10~bit\\
    TRAP sampling frequency & \SI{10}{\mega\hertz}\\
  \end{tabular}
  \caption{Achieved PASA and TRAP characteristics.}
  \label{tab:spec_pasa}
  \label{tab:spec_trap}
\end{table}

The differential PASA outputs are fed into the ADCs of the TRAP, the
second ASIC on the same MCM. The PASA and TRAP parameters are listed in Table~\ref{tab:spec_trap}. The TRAP is a custom-designed digital chip produced in the
UMC \SI{0.18}{\micro\meter} process. The TRAP comprises cycling 10-bit
ADCs for 21~channels, a digital filter chain, a hardware preprocessor,
four two-stage pipelined CPUs with individual single-port,
Hamming-protected instruction memories (IMEM, 4k x 24~bit), about 400
configuration registers usable by the hardware components, a quad-port
Hamming-protected data memory (DMEM, 1k x 32~bit), and an arbitrated
Hamming-protected data bank (DBANK, 256 x
32~bit)~\cite{Angelov:2006hu}. Three excess ADC channels are fed with
the amplified analogue signal from the two adjacent MCMs to avoid
tracking inefficiencies at the MCM boundaries. The signals of all
21~channels are sampled and processed in time bins of
\SI{100}{\nano\second}. The number of time bins to be read out, can be configured in the FEE. At the beginning of \run{1} 24~time bins were conservatively read out. 
At a later stage the number of time bins was reduced to~22 in order to
reduce the readout time and the data volume.

The first step in the TRAP is the digitisation of the incoming
analogue signals. In order to avoid rounding effects, the ADC outputs
are extended by two binary digits and fed into the digital filter
chain. First, the pedestal of the signal is equilibrated to a
configurable value. Then, a gain filter is used to correct for local
variations of the gain, arising either from detector imperfections or
the electronics themselves. A tail cancellation filter can be used to
suppress the ion tails. The filtered data are fed into a pre-processor
which contains hardware units for the cluster finding. The four CPUs
(MIMD architecture) are used for the further processing. The local
tracking procedure is discussed in detail in
Section~\ref{sec:local_online_tracking}.

The MCMs are mounted on the ROB. On each board, 16~chips
are used to sample and process the detector signals. A full detector
chamber is covered by 8~ROBs (6 for chambers in stack~2). The read-out
is organised in a multi-level tree. First, the data from four chips
are collected by so-called column merger chips. The latter, in
addition to processing the data from their own inputs, receive the
data from three more MCMs. The data are merged and forwarded to the board
merger, which combines the data from all chips of one ROB. One ROB per
half-chamber carries an additional MCM which acts as half-chamber
merger (without processing data of its own). It forwards the data to
the Optical Read-out Interface (ORI) from where it is transmitted
through an optical link (DDL) to the GTU. The link is
operated at 2.5~Gbit/s and is implemented for uni-directional
transmission without handshaking, i.e.\ the receiving side must be
able to handle the incoming data for a complete event as it
arrives. As the FEE does not provide multi-event buffering, the
detector is busy until the transmission from the FEE is finished. The
slowest half-chamber determines the contribution to the dead time of
the full detector.

\subsection{Global Tracking Unit}\label{sec:gtu}
\begin{figure}
  \centering
  \includegraphics[]{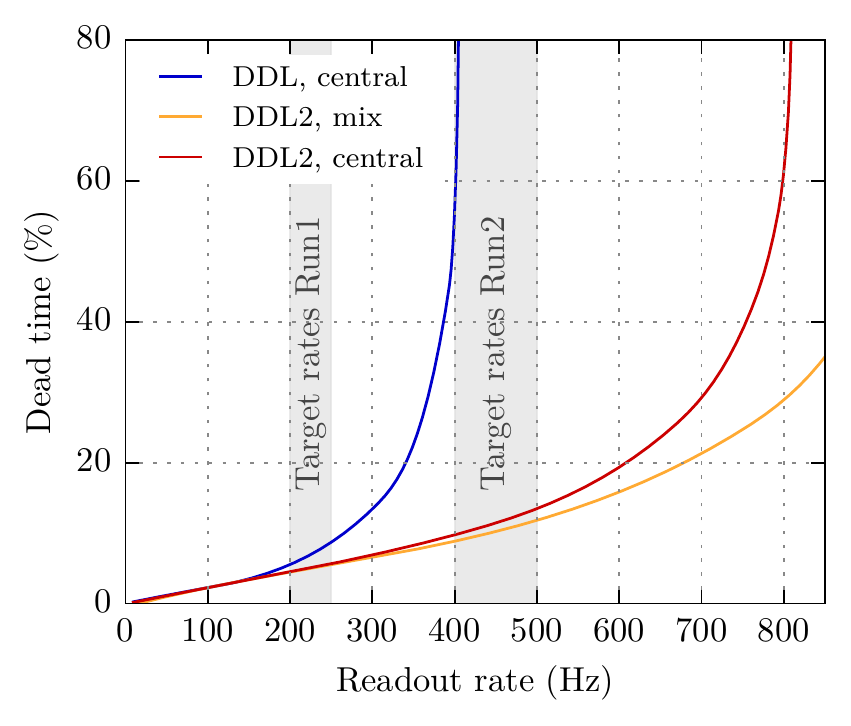}
  \caption{Simulation of the dead time as function of the read-out
    rate in Pb--Pb collisions for the effective DDL bandwidth in
    \run{1} (DDL) and \run{2} (DDL2). The simulation assumes a
    5\% L1/L0 accept ratio and no L2 rejects. 
    The scenarios central and mix correspond to an event size of 470~kB and 310~kB per supermodule, respectively, mimicking thus different event multiplicities.
}
  \label{fig:gtu-readout-plot}
\end{figure}

The GTU receives data via 1044 links from the
FEE. The aggregate net bandwidth amounts to
\SI{261}{\giga\byte/\second}. The two main tasks of the GTU are the
calculation of level-1 trigger contributions from a large number of
track properties in about \SI{2}{\micro\second} and the preparation of
the event data for read-out. Accordingly, the data processing on the
GTU features a trigger path, which is optimised for low latency, and a
data path, which equips the detector with the capability to buffer up to
4~events (multi-event buffering, MEB). The derandomisation of the incoming data rate fluctuations with
multiple event buffers minimises the read-out related dead time. The
data transfer from the GTU to the DAQ contributes to the dead time
only when the read-out rate approaches the rate which saturates the output
bandwidth as shown in Fig.~\ref{fig:gtu-readout-plot}.

\begin{figure*}[bt]
\centering
\includegraphics[width=.8\textwidth]{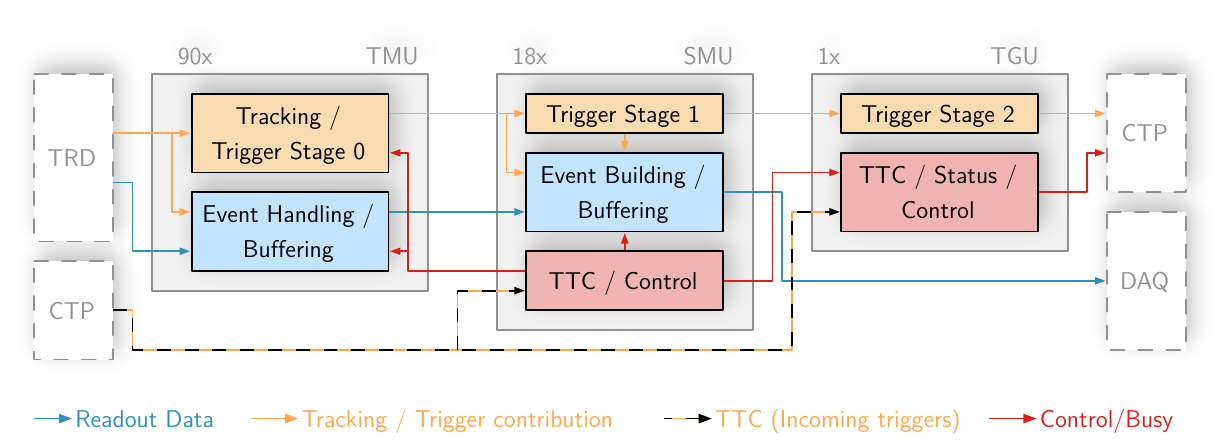}
\caption{
Major design blocks of the TMU and SMU stages of the GTU and
data flow. The busy and trigger logic information are combined on
the TGU before transmission to the CTP.
}
\label{fig:gtu-data-flow}
\end{figure*}

The GTU consists of three types of FPGA-based processing nodes
organised in a three-layer hierarchy (see
Fig.~\ref{fig:gtu-data-flow}). The central component of all nodes is
a Virtex-4$^\copyright$ FX100 FPGA, supplemented by a
\SI{4}{\mega\byte} source-synchronous DDR-SRAM, \SI{64}{\mega\byte}
DDR2-SDRAM and optical transceivers. Depending on the type, the nodes
are equipped with different optical parts and supplementary
modules. 90 TMUs and 18 SMUs are organised in 18 segments of 5+1 nodes (corresponding to the 18
sectors). The TMUs and SMU of a segment are interconnected using a custom
LVDS backplane, which is optimised for high-bandwidth transmissions at
low latency. A single top-level Trigger Unit (TGU) is connected to
the SMUs of the individual segments via LVDS transmission lines.

The data from one stack is received by the corresponding TMU.
Each TMU implements the global online tracking, which combines pre-processed
track segments to tracks traversing the corresponding detector stack,
as first stage of the trigger processing (see
Section~\ref{Chaptertrigger}). The TMUs furthermore implement the
initial handling and buffering of incoming events as a pipelined data
push architecture. Input shaper units monitor the structural
integrity of the incoming data and potentially restore it to a form
that allows for stable operation of all downstream entities.
Dual-port, dual-clock BRAMs in the FPGA are utilised to compactify
data of the 12 incoming link data streams to dense, wide lines
suitable for storage in the SRAM. The SRAM provides buffer space for
multiple events and its controller implements the required
write-over-read prioritisation to ensure that data can be handled at
full receiver bandwidth.  On the read side, a convenient interface is
provided to read out or discard stored events in accordance to the
control signals generated by the segment control on the SMU.

Via its DCS board the SMU receives relayed trigger data issued by the
CTP to synchronise the operation of the experiment. The trigger
sequences are decoded, and converted to suitable control signals and
time frames to steer the operation of the segment. The segment
control on the SMU supports operation with multiple, interlaced
trigger sequences in order to support the concurrent handling and
buffering of multiple events. Upon reception of a level-2 trigger,
the SMU requests the corresponding event data from the event buffers
and initiates the building of the event fragment for read-out. The
built fragment contains, in addition to the data originating from the detector,
intermediate and final results from tracking and triggering relevant for
offline verification, as well as checksums to quickly assess its
integrity.  The SMU implements the read-out interface to the DAQ/HLT with
one DDL. The endpoint of the DDL is a Source
Interface Unit (SIU), which in \run{1} was a dedicated add-on card
mounted on the SMU backside that operates at a line rate of
\SI[round-precision=3]{2.125}{\giga\bit/\second}.
The read-out upgrade for \run{2} integrates the functionality of the SIU into the SMU
FPGA and employs a previously unused transceiver on the SMU at a line
rate of \SI{4}{\giga\bit/\second}. The elimination of the interface
between SMU and SIU add-on card, the higher line rate as well as data
path optimisations resulted in an increase of the effective DDL output
bandwidth from \SI{189}{\mega\byte/\second} to
\SI{370}{\mega\byte/\second} in
\run{2}. Figure~\ref{fig:gtu-readout-plot} illustrates the performance
improvement for the assumed data taking scenarios.
With the upgrade the read-out-related dead time can be kept at an acceptable level. 
The almost linear increase at low rates is due to the dead time associated with the
L0--L1 interval and the FEE-GTU transmission. The typical aggregate output bandwidth for all 18~supermodules is \SI{126}{\mega\byte/\second}, \SI{202}{\mega\byte/\second}, and \SI{1260}{\mega\byte/\second} in \pp, \ppb, and \pbpb collisions (see also Section~\ref{Chapterbeamoperation}).

The top-level TGU consolidates the status of the segments, which
operate independently in terms of read-out, as well as the
segment-level contributions of the triggers.  It constitutes the
interface to the CTP, to which it communicates the detector busy status and
the TRD-global trigger contributes for various signatures (see Section~\ref{Chaptertrigger}).

\section{Detector Control System}\label{Chapterdcs}

The purpose of the DCS is to ensure safe detector conditions, to allow fail-safe, 
reliable and consistent monitoring and control of the detector, and to provide calibration data for offline reconstruction. 
In addition it provides detailed information on subsystem conditions and full functionality for expert 
monitoring and detector operation. Tools were implemented to reduce the operational complexity 
and the information on detector conditions to a level that allows operators to monitor and 
handle the detector in an intuitive and safe way. The TRD DCS is integrated with the rest of the ALICE detector control systems into one system which is operated by one operator.

\subsection{Architecture}
\label{DCS:Architecture}

\begin{figure*}[t]
\centering
\includegraphics[width=0.75\textwidth]{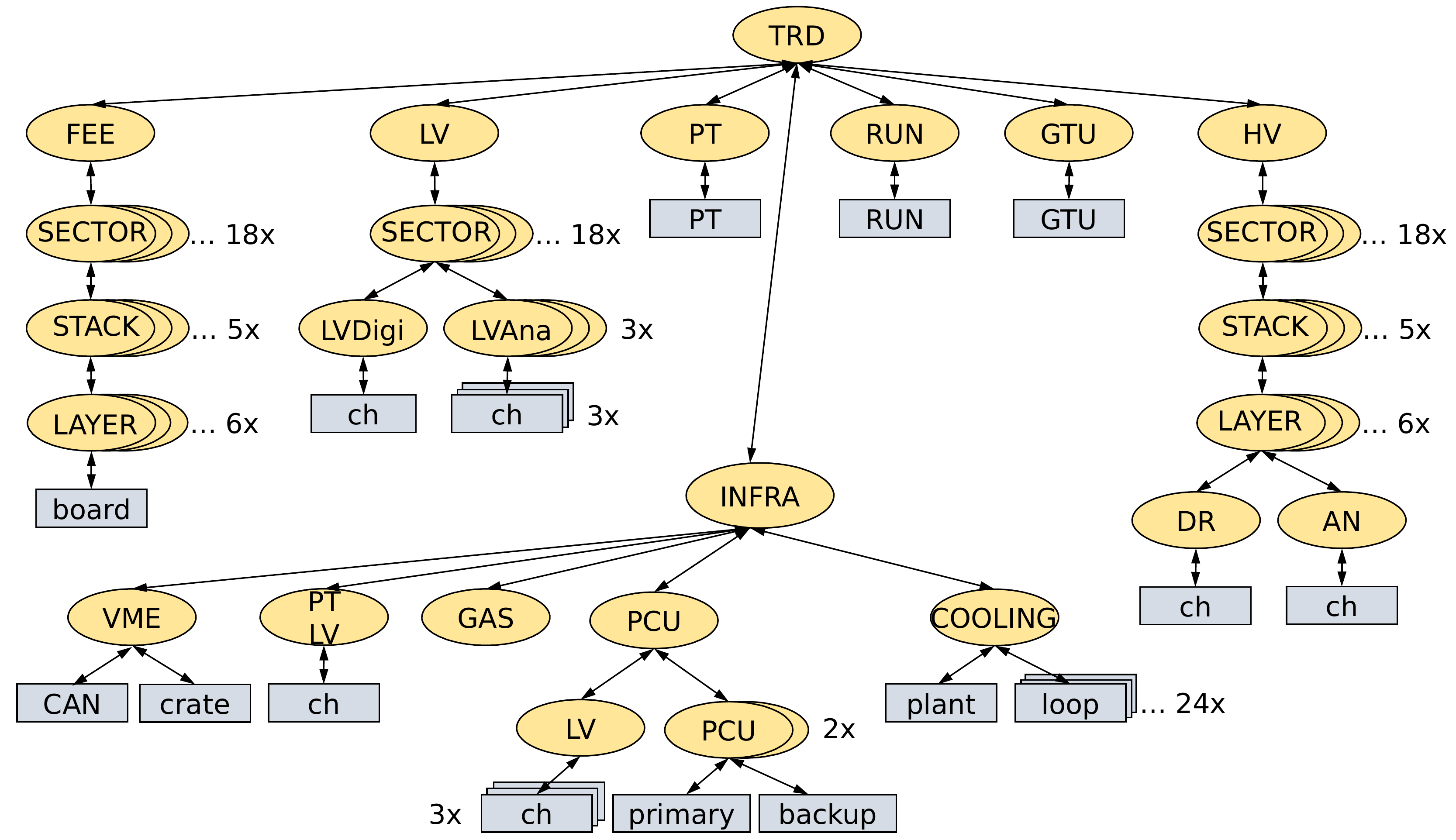}
\caption{Overview of the DCS software architecture. The tree structure consists of device units (boxes) and logical control 
units (ellipses). The abbreviations PT, DR and AN correspond to the pretrigger, the drift and anode channels, respectively.}
\label{fig:FSMtree}
\end{figure*}

The hardware architecture of the DCS can be divided into three functional layers. The field layer contains 
the actual hardware to be controlled (power supplies, FEE, etc). The control layer consists of 
devices which collect and process information from the field layer and make it available to the 
supervisory layer. Finally, the devices of the control layer receive and process commands from the supervisory 
layer and distribute them to the field layer. 

The software on the supervisory layer is distributed over 11~server computers. It is based on the  
commercial Supervisory Control and Data Acquisition (SCADA) system PVSS~II from the company ETM~\cite{ETM}, now called Symatic WinCC~\cite{WINCCref}. 
The implementation uses the CERN JCOP control framework~\cite{JCOP}, shared by all major LHC experiments. This framework 
provides high flexibility and allows for easy integration of separately developed components in combination with 
dedicated software developed for the TRD, including Linux-based processes. 

The software architecture is a tree structure that represents (sub-)systems 
of the detector and its devices, as shown in Fig.~\ref{fig:FSMtree}. The entities at the bottom of the hierarchy 
represent the devices (device units), logical entities are represented by control units. 
The DCS system monitors and controls 89 low voltage (LV) power supplies with more 
than 200~channels, and 1044~high voltage channels. The system also monitors the electronics configuration of more 
than one million read-out channels, the GTU, and the cooling and gas systems.

\subsection{Detector safety}
\label{DCS:DetectorSafety}

To ensure the safety of the equipment, nominal operating conditions are maintained by a 
hierarchical structure of alerts and interlocks. 
Whenever applicable, internal mechanisms of devices (e.g.\ power supply trip) are used to guarantee 
the highest level of reliability and security. Thresholds and status of the interlocks are controlled 
by the system, but the functioning of the device is independent of the communication between hardware and 
software. The possible range of applied settings (e.g.\ anode channel high voltage) is limited to a nominal 
range to prevent potential damage due to operator errors. 

In addition, the system employs a three-level alert system, which is used to warn operators and detector 
experts of any unusual detector condition.

On the control and supervisory layer, cross system interlocks protect the devices and ensure consistent 
detector operation. These are a few examples:  

\begin{itemize}

\item In case of a failure of the cooling plant for the FEE, a 
PLC-based interlock disables the LV power supplies.

\item The temperature of the FEE is monitored at the control and supervisory level and 
interlocked with the PCU to switch off the devices in case of overheating or 
loss of communication to the SCADA system. 

\item In case of a single LV channel trip, the corresponding FEE channels are consistently switched 
off.  

\item Unstable LHC beam conditions, e.g.\ during injection or adjustment of the beam optics, pose a potential 
danger to gas-filled detectors. Therefore the HV settings are adapted to the LHC status (see Section~\ref{Chapterhvinbeam}). At injection, the 
anode voltages are decreased automatically to an intermediate level to reduce the chamber gain. Restoring the nominal gain is inhibited until the LHC operators declare stable beams via a data interchange protocol. 

\end{itemize}

\subsection{High voltage}
\label{DCS:HV}

The HV system comprises 36~HV modules in 5~crates. The 1044~HV channels,  
1~of each polarity providing anode and drift voltage to each chamber, are controlled via a
250~kbit/s CAN bus through a dedicated 
Linux-based DIM server~\cite{Gaspar2001102}. The published DIM services, commands and remote procedure calls (RPC) resemble the 
logical structure of items used in commercial process control servers: the command to change a setting is confirmed by the server via 
a read back setting. In addition, the actual measured value from the device is published. Update rates for 
different services can be adjusted independently. 

The HV gain and drift velocity are equilibrated for each chamber individually to compensate for 
small differences in the chamber geometry. Changes of environmental 
conditions (atmospheric pressure and temperature) as well as small variations of the 
gas composition cause changes in gas gain and drift velocity. To ensure stable conditions for the level-1  
trigger (see Section~\ref{Chaptertrigger}), these dynamic variations are compensated by automatic adjustments of the anode and 
drift voltages which are performed in between runs. These and other automatic actions on the HV are described in Section~\ref{Chapterhvinbeam}.

\subsection{Detector operation}
\label{DCS:DetectorOperation}

The DCS employs a dedicated Graphical User Interface (GUI) and a Finite State Machine (FSM). The FSM allows experts and operators intuitive monitoring and operation of the detector. The FSM hierarchy 
reflects the structure of subsystems and devices shown in Fig.~\ref{fig:FSMtree}. Detector conditions 
are mapped to FSM states, and these are propagated from the device level upwards to the FSM top node. Standard 
operational procedures (configuration of read-out and trigger electronics, ramping voltages etc.) are 
carried out via FSM commands which propagate down to the devices and cause a transition to a 
different state. 

\begin{figure}[tb]
\centering
\includegraphics[width=0.45\textwidth]{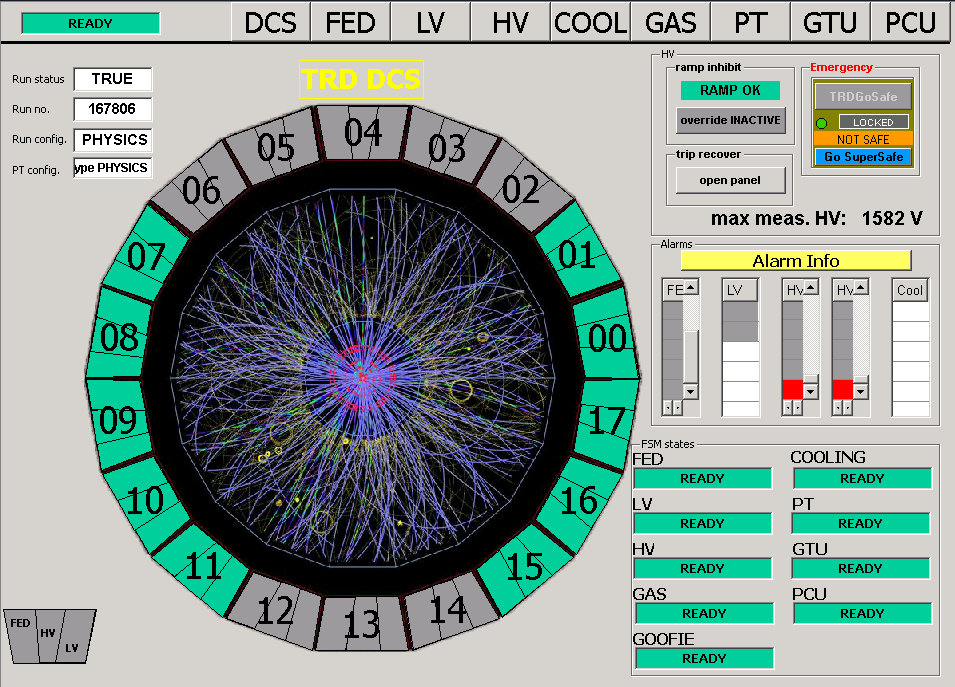}
\caption{Graphical User Interface: example panel (FSM top node) representing the status in the year 2010 (i.e.\ with 10 installed TRD sectors). The FSM state of the main systems of each 
sector is represented by the corresponding colour; run status and alarm summary are displayed. 
The panel gives quick 
access to emergency actions and detailed monitoring panels via single mouse clicks. }
\label{fig:panel}
\end{figure}

The GUI for detailed monitoring and expert operation comprises a dedicated panel for each node in the 
FSM tree. An example is shown in Fig.~\ref{fig:panel}. 
Detector subsystem `ownership', i.e.\ the right 
to execute FSM commands and change the detector state, is only granted to a single operator at a time, 
and is represented by symbolic `locks'. 
Operators can work on-site or access the DCS system remotely through appropriate gateways. 

The monitoring data acquired by the DCS system are stored in dedicated databases. Dedicated trending GUIs allow the experts to visualise 
the time dependence of the detector conditions. During data taking, the monitoring data needed for 
detector calibration is queried and made available for offline analysis (see Section~\ref{Chapterbeamoperation}). 

\section{Operation}\label{Chaptercomminbeam}

In this section, first the commissioning steps for the
detector and the required infrastructure and then the
operation and performance for different collision systems are described.

\subsection{Commissioning}\label{Chaptercommissioning}

The service connections in the cavern were prepared and tested in parallel to the construction of the supermodules. The low-voltage connections were tested with dummy loads and the leak tightness of the cooling loops was verified. The Ethernet connections were checked using both
cable testers and stand-alone DCS boards. The optical fibres for the
read-out were controlled for connectivity and mapping. These tests were
crucial in order to identify connection problems prior to the detector
installation when all connections were still well accessible.

The supermodules were installed in different installation blocks as described in Section~\ref{Chaptertrdsystem}.
Prior to the installation the supermodules were tested at the
surface site. They were rotated along the \zlab-axis to
the orientation corresponding to their foreseen installation position
(e.g.\ relevant for cooling). A test setup provided all relevant
services (low/high voltage, cooling, Ethernet, read-out, ...) to allow a full system test of each supermodule. The testing procedure included basic
functionality tests, such as water and gas tightness, front-end
electronic stress tests, read-out tests as well as checks of the noise
level~\cite{Klein:Diplom08}.

After successful surface testing, the supermodules were installed into
the space frame in the cavern (see
Section~\ref{secsupermodule}). Subsequently, the services were connected
and the basic tests described above repeated to verify operation in the final
setup. At this stage, also the full read-out of the detector with the
experiment-wide trigger and data acquisition systems was
commissioned. To check the data integrity of the read-out chain, test
pattern data, generated either in the FEE or in the GTU, were
used. Errors observed during those tests, e.g.\ bitflips on individual
connections on a read-out board, were cured by switching to spare
lines or by masking channels from the read-out if a correction was not
possible. After establishing the read-out, pedestal runs (without zero
suppression) were recorded to determine the baseline and noise of each
channel. If needed, further data were recorded to perform a Fourier
analysis in order to identify and fix noise sources,
e.g.\ caused by missing ground connections. In addition, these runs
were used to identify inactive channels which cannot be read out.

After each installation block of new supermodules, a dedicated
calibration run was performed before the actual data taking. The
detector is read out with radioactive $^{\rm 83m}$Kr distributed
through the gas system (see Section~\ref{sec:cal:kr}). Since this was
usually the first high-rate data taking after the end-of-year shutdown
(and installation), these runs and the preparations for them were an
important step to get ready for the real data taking.

Before each physics production run, periods of cosmic-ray data taking
were scheduled to study the performance of the detector system, to align
individual detector components (see Section~\ref{sec:ali:int}) and to
provide reference spectra for particle identification (see
Section~\ref{Chapterpid}). Data were obtained with and without magnetic
field. A two-level trigger condition was used to ensure sufficient
statistics in the detector acceptance, even when only the first
supermodules were installed in the horizontal plane (see
Section~\ref{sec:cosmic_trg}).

\subsection{High voltage operation}\label{Chapterhvinbeam}

To avoid HV trips during the critical phases of beam injection
(e.g.\ a possible kicker failure), the anode voltages are reduced to
values with very low gain. After the injection is completed, the anode
voltages are ramped up from ${\sim}$\SI{1030}{\volt} (gain of about a factor ${\sim}
100$ lower than nominal) to an intermediate voltage of ${\sim}$\SI{1230}{\volt} (gain ${\sim} 6.5\%$ of nominal). The ramp speed is \SI{6}{\volt/\second}. After the declaration of stable beams, the anode
voltages are ramped to the nominal voltages ($U_\mathrm{anode} \simeq$\SI{1520}{\volt}) for data taking. The drift voltages always remain at nominal settings.

To equalise the gain and drift velocity of all chambers, the results
from the calibration (see Section~\ref{Chaptercalib}) are used. The
nominal voltages and r.m.s.\ variations for drift and anode voltages are
2150$\pm$\SI{22}{\volt} and 1520$\pm$\SI{14}{\volt}, respectively.

Based on measurements in \pp, \ppb and \pbpb collisions in \run{1}, it
has been estimated that the chambers had a time averaged current of about 200~nA. This led to a total accumulated charge of less than \SI{0.2}{\milli\coulomb} per \si{\centi\metre} of wire for \run{1}. As the chambers were validated for charges above \SI{10}{\milli\coulomb/\centi\metre}, it is expected that no ageing effect occurs during the time the TRD is going to be operated. Up to now, in fact no deterioration in
the performance of tracking, track matching and energy resolution was
observed.

The average anode current as a function of the interaction rate as measured by the T0 detectors used for the ALICE luminosity measurement has a linear dependence with a slope of \SI{1/200}{\nano\ampere/Hz} for \ppb collisions at \sqrtsnn~=~5.02~TeV. The slope parameter was obtained from different LHC fills
ranging from minimum-bias data taking up to high rate interaction
running, where the LHC background conditions can be different. Under the vacuum conditions in \run{1}, about 1/3 of the current was due to the background rate, which is nearly negligible in \run{2}.

The expected dependence of the measured current on detector occupancy was found.
The probability for pile-up events in, e.g.\ \ppb collisions at
\sqrtsnn~=~5.02~TeV at \SI{200}{\kilo\hertz} interaction rate is about 14\% when averaged over time,
with a maximum of ${\sim}$24\% as calculated from the
bunch spacing and the number of bunch crossings in the LHC filling
scheme~\cite{Abelev:2014ffa,Bailey:691782} as well as the integration time of the read-out chamber (drift
length/drift velocity).

For the level-1 trigger it is crucial to reduce the time dependence of the drift velocity and the gain to a minimum. The former impacts the track matching, the latter
the electron identification. To ensure the required stability, the
anode and drift voltages are adjusted to compensate for pressure
changes (the temperature is sufficiently stable). The parameters for the correction were obtained by
correlating the calibration constants with pressure (see Section~\ref{Chaptercalib}). A relative pressure change d$p/p$ results in a change of gain of d$G/G = -6.76 \pm 0.04$ and drift velocity of d\vdrift/\vdrift~=~$-1.41 \pm 0.01$~\cite{lbergmann}. In addition, the dependences of the gain and drift velocity on the anode and drift voltage, respectively, as obtained from test beam measurements~\cite{pap:Andronic2004302} were used (from \run{2} onwards the dependence of gain on voltage was taken from the krypton calibration runs). This results in voltage changes of about \SI{0.83}{\volt} and \SI{1.4}{\volt} for a pressure change of \SI{1}{\milli\bar}. During \run{1} the gain and the drift velocity could be kept constant within about 2.5\% and 1\%, respectively. These values include the precision of the determination of the calibration constants (see Section~\ref{Chaptercalib}). The variations can be further reduced by measuring and correcting for the gas composition using a gas chromatograph installed during \LS{1}. 

During \run{1}, 10\% of the anode and 5.5\% of the drift channels
turned out to be problematic (see Fig.~\ref{fig:cali-qa}). The respective channels had to either be
reduced in anode voltage or switched off. As the detector is segmented
into 5~stacks along the beam direction and 6~layers in radial
direction, the loss of a single chamber in a stack is tolerable and
excellent performance is still achieved for tracking and particle
identification (see Sections~\ref{Chaptertracking} and \ref{Chapterpid}). Most
of the problematic chambers showed strange current behaviours
(trending vs time). The de-installation of a supermodule and disassembly of
the individual read-out chambers followed by detailed tests revealed
that the inspected problematic anode and drift channels had broken
filter capacitors (\SI{4.7}{\nano\farad}/\SI{3}{\kilo\volt}). Thus, the \SI{4.7}{\nano\farad} capacitors (see Section~\ref{InfraHVsection}) were removed from the resistor chain in the last supermodules built and
installed during the \LS{1} (5~supermodules).

\subsection{In-beam performance}\label{Chapterbeamoperation}

After commissioning with cosmic-ray tracks and krypton calibration runs in
2009, the detector went into operation and worked reliably during the first collisions at the LHC on December 6$^{\rm th}$ 2009. Since then, the detector has participated in data taking for all collision systems and energies
provided by the LHC~\cite{Abelev:2014ffa}:
\begin{itemize}
\item \pp collisions from \sqrts~=~0.9 to 13~TeV at low interaction rates (minimum-bias data taking) and high intensities (minimum-bias data taking and rare triggering) with a maximum interaction rate of 200--\SI{500}{\kilo\hertz}. During the rare trigger periods, the detector 
contributed level-1 triggers on high-\pt electrons and jets (see Section~\ref{Chaptertrigger}).
\item \ppb collisions at \sqrtsnn~=~5.02~TeV and 8~TeV with interaction rates at the level of \SI{10}{\kilo\hertz} (minimum-bias data taking) and at maximum \SI{200}{\kilo\hertz} (rare triggering). The detector contributed the same triggers as in the pp running scenario.
\item \pbpb collisions at \sqrtsnn~=~2.76~TeV and 5.02~TeV with maximum interaction rates of up to \SI{8}{\kilo\hertz} (minimum-bias and rare triggering).
\end{itemize}

At the beginning of a fill, once all detectors within ALICE are ready for
data taking, a global physics run is started. A run is defined in ALICE
as an uninterrupted period of data taking, during which the conditions
(trigger setup, participating detectors, etc.)\ do not change. A run can
last from a few minutes to several hours until either the experimental setup or conditions have to be changed or the beam is dumped. An additional end-of-run (EOR) reason is given by the occurrence of a problem related to a given detector or system.
The detector parameters measured
during a run, such as the voltages and currents of the anode
and drift channels as well as temperatures of the FEE, are dumped at the EOR to
the Offline Conditions 
Database (OCDB) via the Shuttle
framework~\cite{Grosse-Oetringhaus:1269925,calib:shuttlenote}. The relevant parameters
can then be used in the offline reconstruction and analysis. 

In order to ensure sufficiently stable conditions during a run, any
change, such as the failure of a part of the detector, e.g.\ due to
a LV/HV trip, triggers the ending of the run. In order to avoid too
frequent interruptions, the failure of a single chamber within a stack
is ignored. Technically, this is realised using the so-called
Majority Unit within DCS.

All subcomponents of the TRD detector (infrastructure and gas system)
are monitored via DCS (see Section~\ref{Chapterdcs}). In case any
entity deviates from nominal running conditions by pre-defined
thresholds a warning is issued. The single entity is either recovered
by the DCS operator in the ALICE Run Control Centre or by an expert
intervention. During \run{1} data taking, most interventions were
related to the recovery of single event upsets (SEU) and HV trips of
problematic channels by re-configuration of the FEE or ramping up of the
anode/drift channels. For \run{2} an automatic recovery of the FEE and HV was
put in place.

\subsubsection{Read-out performance}

\begin{figure}[tb]
  \centering
  \includegraphics[width=.45\textwidth]{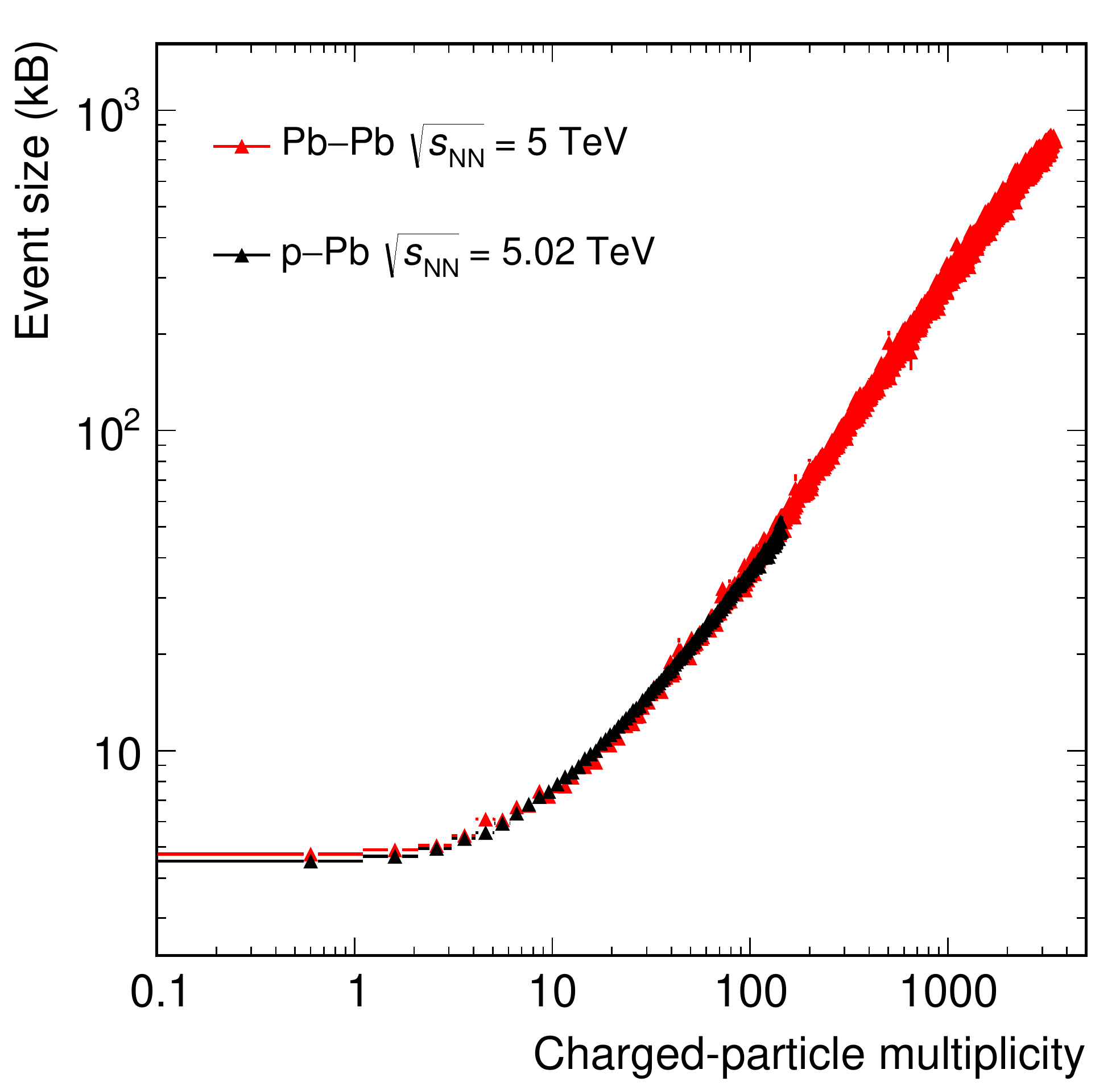}
  \caption{Event size vs charged-particle multiplicity for various collision systems for one supermodule. To obtain the charged-particle multiplicity, global tracks (see Section~\ref{Chaptertracking}) fulfilling minimum tracking quality criteria were counted on an event-by-event basis. }
  \label{fig:evsize}
\end{figure}

The event size depends on the charged-particle multiplicity. It is
therefore influenced by the collision system and the background
conditions of the LHC. The event size vs.\ charged-particle
multiplicity is shown for various collision systems for one
supermodule in Fig.~\ref{fig:evsize}. For the most central \pbpb
collisions an event size of 800~kB per supermodule is found. 

The dead time per event is composed of the front-end processing and
transmission time to the GTU and a potential contribution from the
shipping to DAQ. On average the former scales approximately linearly
with the event size and rate, the latter is suppressed by the MEB as
long as the read-out data rate stays sufficiently below the effective
link bandwidth. The typical event sizes of
\SI{7}{\kilo\byte}, \SI{14}{\kilo\byte},
\SI{200}{\kilo\byte} in minimum-bias data taking for \pp, \ppb, and \pbpb collisions result in
front-end contributions of \SI{20}{\micro\second},
\SI{25}{\micro\second}, \SI{50}{\micro\second}, respectively. This
does not include the read-out induced part. However, as
illustrated by the \pbpb case shown in
Fig.~\ref{fig:gtu-readout-plot}, the detector is typically operated in
the linear range of the curve, indicating that input rate fluctuations
are absorbed by the MEB and that the read-out does not contribute
significantly to the dead time.

The read-out rate during \run{1} and until now in \run{2} ranged from
about \SI{100}{\hertz} in rare trigger periods to about
\SI{850}{\hertz} in minimum-bias data taking in \pp and \ppb
collisions. In \pbpb collisions, the read-out rate was about
\SI{100}{\hertz} and \SI{350}{\hertz} for minimum-bias data taking in
\pbpb collisions in \run{1} and up to now in \run{2}, respectively.

\subsubsection{Radiation effects}

The radiation on the TRD was for \run{1} and \run{2} (until the end of 2016) rather low both in terms of flux and dose. The following radiation calculations for the inner radius of the TRD are based on simulations obtained using the FLUKA transport code~\cite{Ferrari:2005zk} and taking into account the measured multiplicities of \pbpb, \ppb and \pp collisions~\cite{Aamodt:2010ft,ALICE:2012xs,Aamodt:2010pb,Adam:2015pza,Adam:2015gka,Adam:2015gda} as well as the running scenarios (luminosities, running time, and interaction rate). For the indicated time range the Total Ionisation Dose (TID) and the Non-Ionising Energy Loss (NIEL), quoted in 1-MeV-neq fluence, were $7 \cdot 10^{-3}$~krad and $2 \cdot 10^{9}\rm{cm^{-2}}$, respectively. 
The flux of hadrons is highest in \pbpb collisions, because it is proportional to the product of the interaction rate and the particle multiplicity. For \pbpb collisions at \sqrtsnn~=~5.02~TeV, the flux of hadrons with $> 20$~keV energy and charged particles is about $3.8 \cdot 10^{-2}\rm{kHz/cm^{2}}$ and $2.5 \cdot 10^{-2}\rm{kHz/cm^{2}}$, respectively. The radiation load in terms of flux and dose are far below the values, for which the experiment was designed for~\cite{Aamodt:2008zz}.  

\begin{figure}[tb]
  \centering
  \includegraphics[width=.7\textwidth]{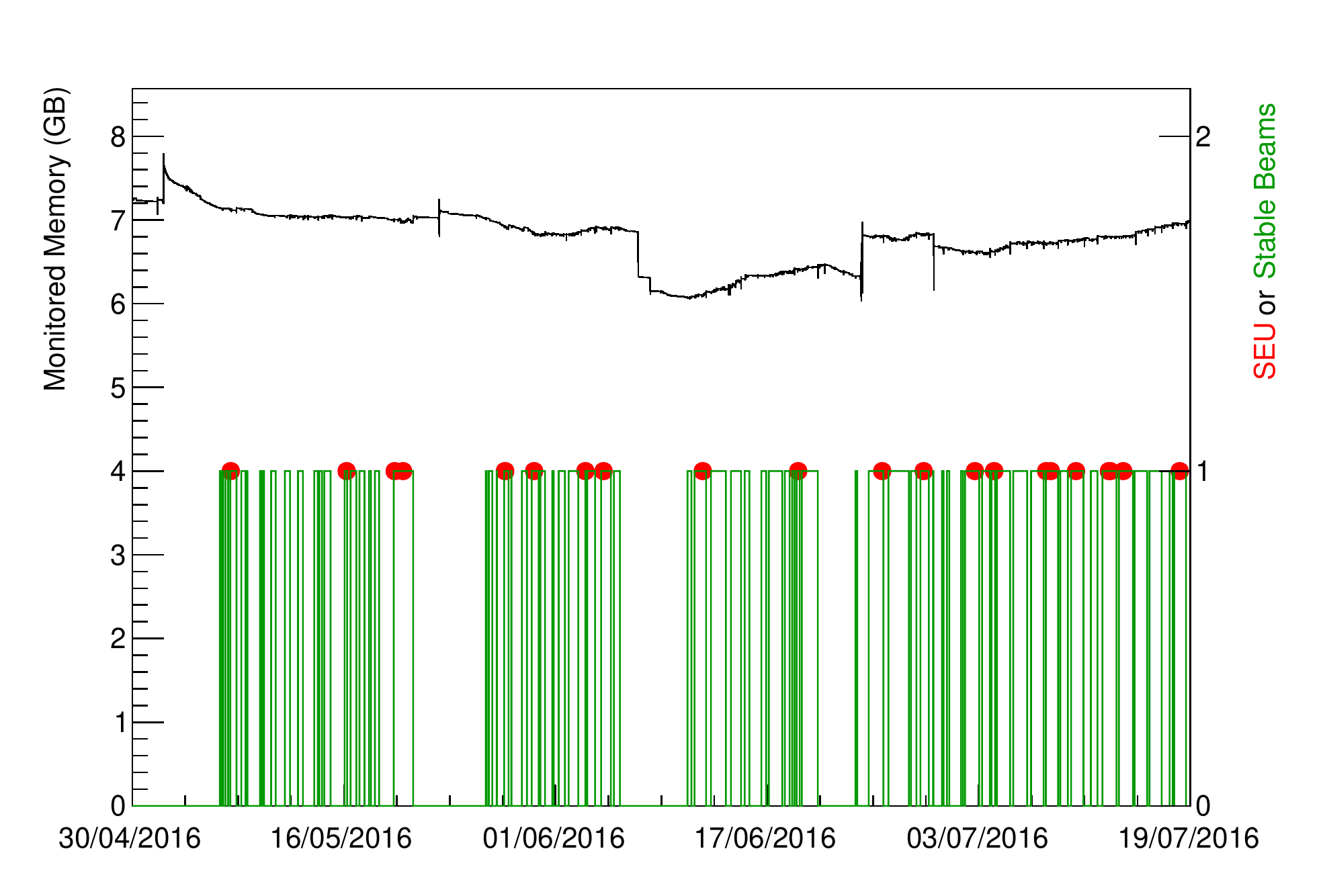}
  \caption{Monitored external DCS memory and occurrences of SEUs as a function of time. The periods of stable beam are indicated as well.}
  \label{trd:seu_ram}
\end{figure}

In the radiation environment described above, very few SEUs are observed in the electronics. The most affected device is the DCS
board, for which SEUs result in occasional reboots (a few DCS boards
per LHC fill). The DCS board is needed for control and monitoring but
is not part of the read-out chain meaning that the reboots do not affect
the data taking. The external RAM on the DCS board can be monitored
for SEUs by writing and verifying known patterns in unused areas of
the $\sim$13~MB memory per chamber. During 2.5~months of pp data taking at LHC luminosities of about $5 \cdot 10^{30}\rm{cm^{-2}s^{-1}}$, 20~SEUs as shown in Fig.~\ref{trd:seu_ram} were observed in the external RAM, i.e. a negligible amount compared to the occasional reboots of a few DCS boards. 

The memories of the TRAPs are Hamming-protected and, thus, resilient
to SEUs. However, the configuration registers are not protected and
can be affected by radiation. Therefore, the configuration is
compressed and written to a Hamming-protected memory area. In this
way, the registers can be checked (and corrected) against the
compressed configuration.

\subsubsection{Data quality assurance}\label{Chapterdqm}

The Data Quality Monitoring framework (DQM) provides online feedback
on the data and allows problems to be quickly spotted and identified during
data taking. The Automatic MOnitoRing Environment (AMORE) was
developed for ALICE~\cite{Haller:2014lya} and allows run-based, detector-specific analyses on the raw data. The results are visualised in a dedicated user interface. The monitored observables,
such as noise level, event size per supermodule, trigger timing, FEE
not sending data, are compared with reference values or diagrams
(depending on the data taking scenario). Deviations from the references
indicate a problem to the operator. Based on the information
obtained from the online DQM all runs are directly marked with a quality
flag, both globally and for the individual ALICE subdetectors. For the
offline physics analyses, lists of runs are selected based on these
flags according to the physics case under study.

\subsubsection{Pretrigger performance}

A dedicated wake-up signal is required for the FEE (see
Section~\ref{sec:pt_ov}). It should reflect
the level-0 trigger condition as closely as possible. However, as it needs to be generated
before the actual level-0 trigger, it cannot use the same information. This introduces some
inefficiency into the TRD read-out. In the early \run{1} LHC filling schemes
(e.g.\ during the LHC ramp-up in 2009) with only a few colliding bunches per
orbit, it was possible to send a wake-up signal for all of the bunch crossings with
potential interactions. This resulted in a fully efficient
operation~\cite{Klein:2014rxa}. During this time, the pretrigger
system was commissioned to use the V0 and T0 signals as inputs. They could then
also be used for filling schemes with many bunches. The trigger
condition was configured as closely as possible to the ALICE level-0
interaction trigger, i.e.\ a coincidence of either the V0 or the
T0 detectors (simultaneous signals in the A- and C-side, see Section~\ref{Chaptertrdsystem}). The
efficiency of the V0- and T0-derived wake-up signals depends on the
discrimination thresholds used for those detectors and on the inherent
dead time between pretrigger and the abort or end of the read-out
(see Section~\ref{Chapterreadoutelectronics}). The latter is
particularly important when subsequent collisions are close in time,
e.g.\ in LHC filling schemes that have bunch trains with 25 or 50 ns
bunch spacing~\cite{Jowett99}. For runs taken at low interaction rates
the pretrigger efficiency is above 97\%; for higher rates the efficiency
depends on the colliding bunch structure of the filling scheme and
reaches average values down to about
83\% in \run{1}~\cite{Klein:2014rxa}. These inefficiencies were avoided with the LM system used in \run{2} (see Section~\ref{sec:pt_ov}).

The analysis of electrons from heavy-flavour hadron decays in \ppb
collisions at \sqrtsnn~=~5.02~TeV in events satisfying the
pretrigger condition showed no bias compared to
results from events triggered with the ALICE level-0 minimum-bias
interaction trigger~\cite{mfleck}.

\section{Tracking}
\label{Chaptertracking}

The charged particle tracking in the ALICE central barrel is based on a Kalman filtering~\cite{Billoir:1983mz}. Track finding and fitting are performed simultaneously~\cite{Abelev:2014ffa}.
The algorithm operates on clusters of track hits from the individual detectors. The clusters carry position
information and, depending on the detector, the amount of charge from
the ionisation signal. The cluster parameters are calculated locally from the raw
data, implying that the cluster finding can be parallelised.

The global tracking starts from seed clusters at the outer radius of
the TPC (see Fig.~\ref{Figure_ALICE}). During the first inward propagation of the tracks previously
unassigned TPC clusters are attached while updating the track
parameterisation at the same time. If possible, the track is further propagated to the
ITS. Subsequently, an outward propagation adds information from TRD,
TOF, and HMPID. A second inward propagation is used to obtain the
final track parameters, which are stored at a few important detector
positions, most importantly at the primary vertex. 

The TRD contributes to the tracking in various ways. First, it adds
roughly \SI{70}{\centi\metre} to the lever arm, which improves significantly the
momentum resolution for high-$\pt$ tracks. Second, it increases the
precision and efficiency of assigning clusters from the detectors at
larger radii, in particular the TOF, to propagated
tracks. In addition, the TRD is used as reference to obtain correction
maps for distortions in the TPC, which arise from the build up of
space charge at high interaction rates. For this the TRD and ITS track segments are
reconstructed using as seeds the TPC tracks (with relaxed tolerances
accounting for potential distortions). Then, the estimate of the real
track position is built as a weighted average of the ITS and TRD
refitted tracks (without TPC information). The TPC
distortions are deconvoluted from the residuals between these
interpolations and the measured TPC cluster positions.

The tracking in the TRD can be subdivided into the formation of
tracklets (track segments within one read-out chamber) from clusters and the updating of the global tracks based on the tracklets. These steps are performed layer-by-layer. The chambers
within a layer can be treated in parallel. For each layer, a seed
track is prepared by propagation from the TPC and used to calculate
the intersection with a chamber. Based on this information a tracklet
is formed from the clusters in the vicinity of this intersection and then the track
parameterisation is updated accordingly. In the
following, details of the individual steps will be given.

\subsection{Clusterisation}

Primary ionisation in the detector gas leads to a signal that spreads
over several pads. Because of the slower ion drift, the charge carries
over into subsequent time bins, resulting in a correlation between time
bins (see Section~\ref{sec:readoutchamber}). The cluster algorithm combines
the data from adjacent pads in the same time bin, producing clusters
with information on position and total charge. The former is
calculated from the weighted mean of the charge shared between
adjacent pads (up to 3). Look-Up Tables
(LUT) are used to relate the measured charge distribution to the
actual position. These LUTs are the result of calculations for the
different pad width sizes, based on measurements in a test
beam~\cite{Adler:2005bc}. The
cluster position can deviate from the LUT values because of detector
parameters which are subject to calibration (see
Section~\ref{Chaptercalib}), most importantly the drift velocity
$v_\mathrm{d}$ and the time offset $t_{0}$ (time corresponding to the
position of the anode wires, see Fig.~\ref{fig:cali-pars}). In addition, a
correction for the $E \times B$ effect is applied.
The complete position characterisation
also includes the estimated uncertainty, which determines the weight
for updating the global track. The uncertainties are derived from
differential analyses of Monte Carlo simulations. Cluster properties
such as the deposited energy, time bin, and reconstructed position
relative to the pad with the maximum charge are taken into account as
well as particle level characteristics such as electrical charge and
incident angle. A linear model relates all uncertainties with
parameters being defined by all conditions determining a
cluster.

\begin{figure}[tb]
  \centering
  \includegraphics[width=0.6\textwidth]{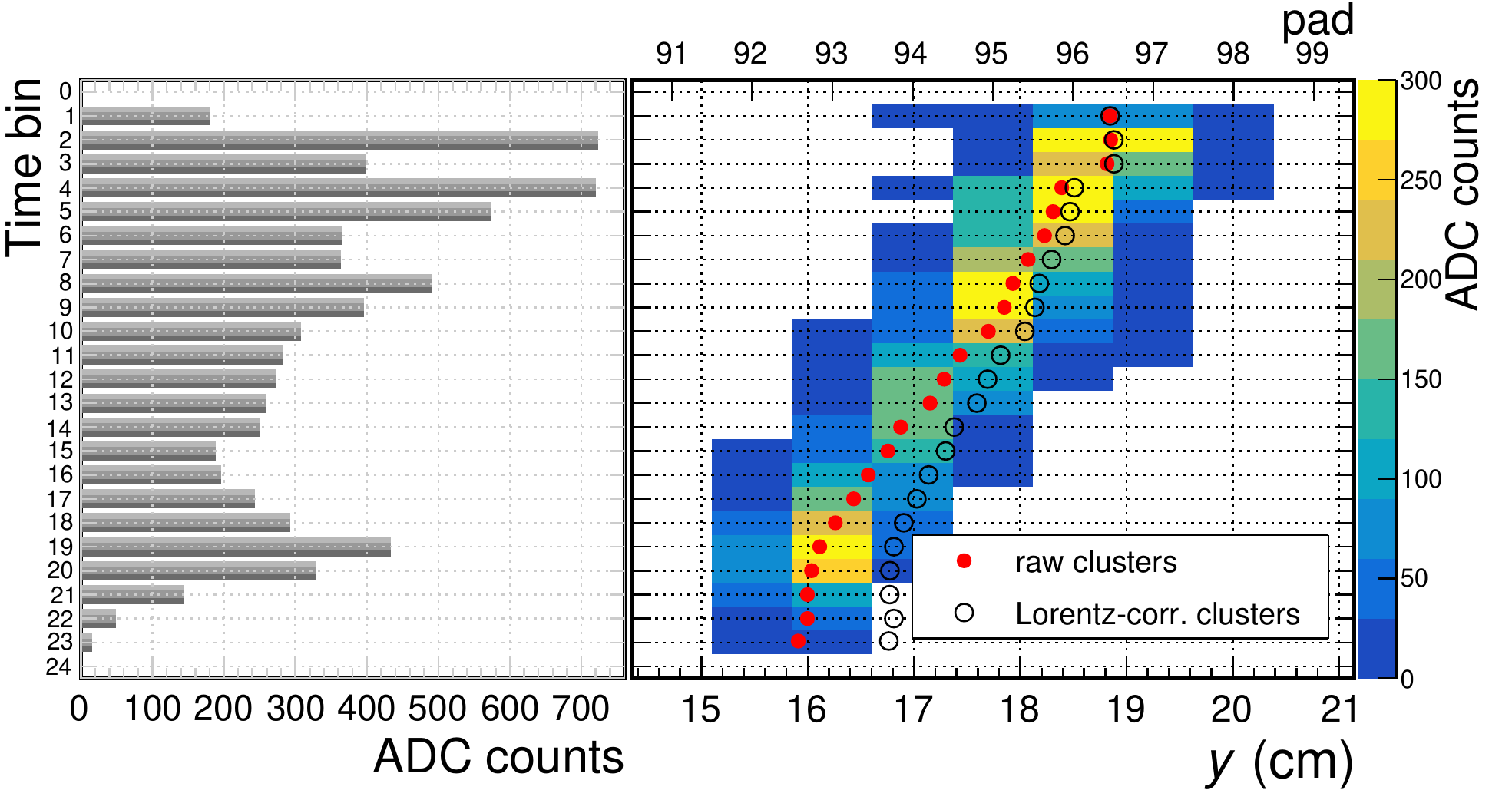}
  \caption{Signal produced by a positively charged particle ($\pt =
    0.5\gevc$). Left: Total
    charge per time bin used for particle identification.
    Right: Ionisation signal vs. pad number and time bin. The cluster positions are shown as reconstructed
    from the charge distribution (raw clusters) and after correction
    for the $E \times B$ effect (Lorentz-corr. clusters). }
    \label{fig.trklt.segm}
\end{figure}

\subsection{Track reconstruction}
\label{Chaptertracklet}

For the preparation of the TPC-based track seed used to match with the
TRD clusters, the Kalman parameterisation (at the outer radius of the
TPC) is propagated to the radial position of the anode wires of a
given chamber. At this radius the position is least affected by
variations in calibration parameters. If a chamber is rotated with
respect to the tracking frame, the radial position of the anode wires
depends on the intersection point of the track in the $y$-$z$
plane. As this is only known after the propagation, the preparation of
the track seed is an iterative process.

The clusters that are assigned to the seed track in a given layer are
combined into tracklets. A straight
line fit is sufficient for their description since the negligible
sagitta of the trajectory is only of the order of tens of microns.

Since in the read-out chamber the electrons drift in the radial direction, that is approximately parallel to the track, and due to the long ion tails, the signals pile up.
The measured charges, sampled in time intervals of \SI{100}{\nano\second},
are therefore correlated between different time samples. Since such
correlations degrade the angular resolution, a tail cancellation
correction is applied~\cite{Adler:2005bc}. It subtracts an exponential
tail proportional to the current signal from the subsequent samples
for each read-out pad.

The number of pads on the the read-out plane onto which a track is projected depends on the track incident angle. For decreasing transverse momentum, more pads will carry a signal. The Lorentz angle also affects this spread. For negatively (positively) charged particles
the Lorentz drift is along (opposite to) the track inclination, 
independent of the polarity of the magnetic field. On average, negatively charged particles are thus spread over fewer pads than positively charged ones. In the right panel of
Fig.~\ref{fig.trklt.segm}, an example of a positively charged particle
of $\pt = 0.5\gevc$ (worst case) is shown. Its projection spans over
6~pads.

\begin{figure}[tb]
  \centering
  \includegraphics[width=0.58\textwidth]{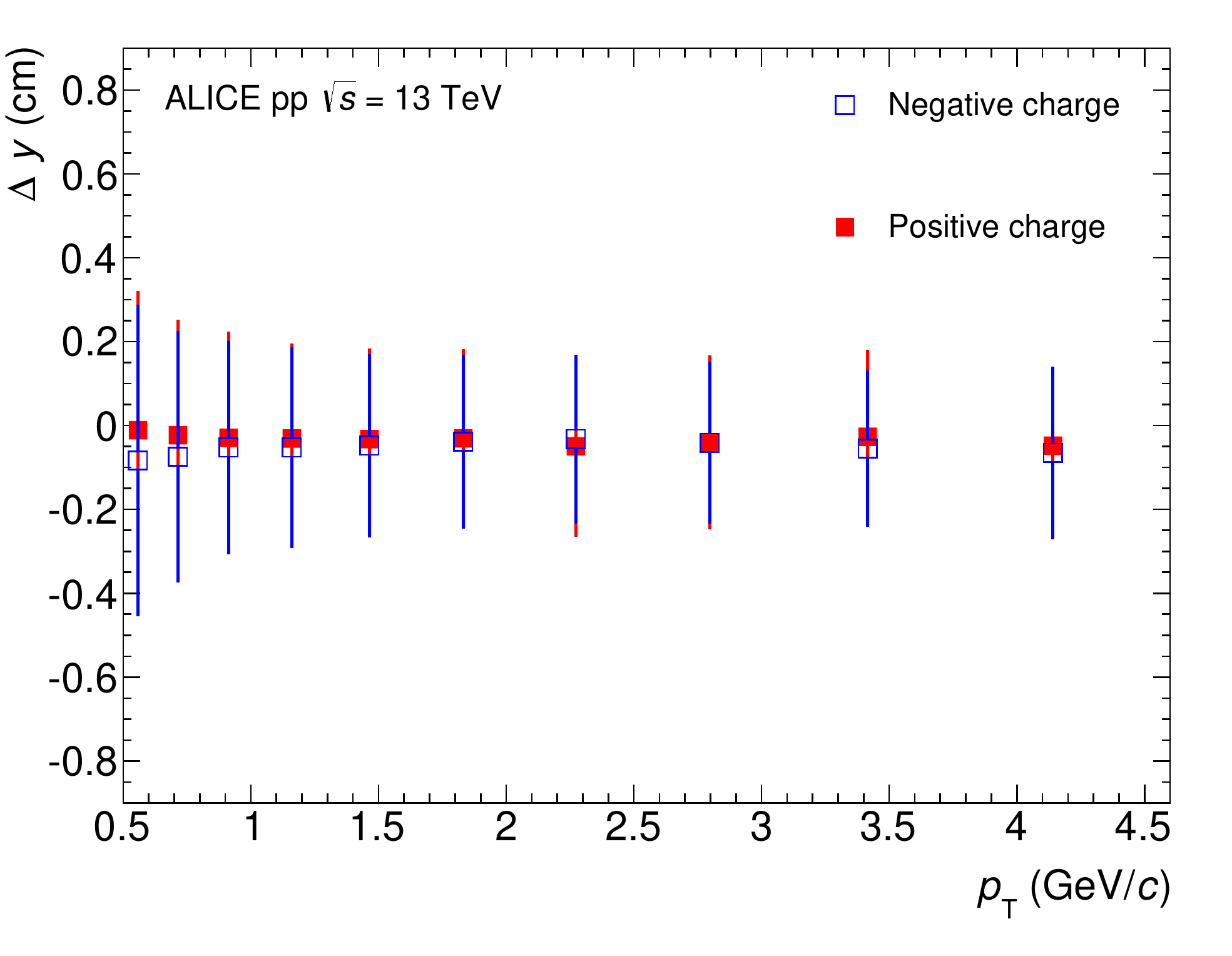}
  \caption{Residuals in $\Delta y$ of tracklets with respect to global tracks as a function of
    $\pt$ in pp collisions at $\sqrt{s} = 13~\mathrm{TeV}$. For every bin the mean (marker) and r.m.s.\ width (error bar)
    of the distribution are shown.}
  \label{fig.likelihood}
\end{figure}

The procedure to find candidates for seeds involves a preliminary
stage in which clusters are searched in the neighbourhood of the
propagated seed. In Fig.~\ref{fig.likelihood} the mean and width of
the residuals are shown for the arising tracklets in $\Delta y$ in
layer~0 as a function of the seed $\pt$. The imperfect tail
cancellation results in different position biases for tracklets from
positive and negative tracklets, the signal spreading over more pads
for the former.

\subsection{Performance}

\begin{figure}[tb]
  \centering
  \includegraphics[width=0.6\textwidth]{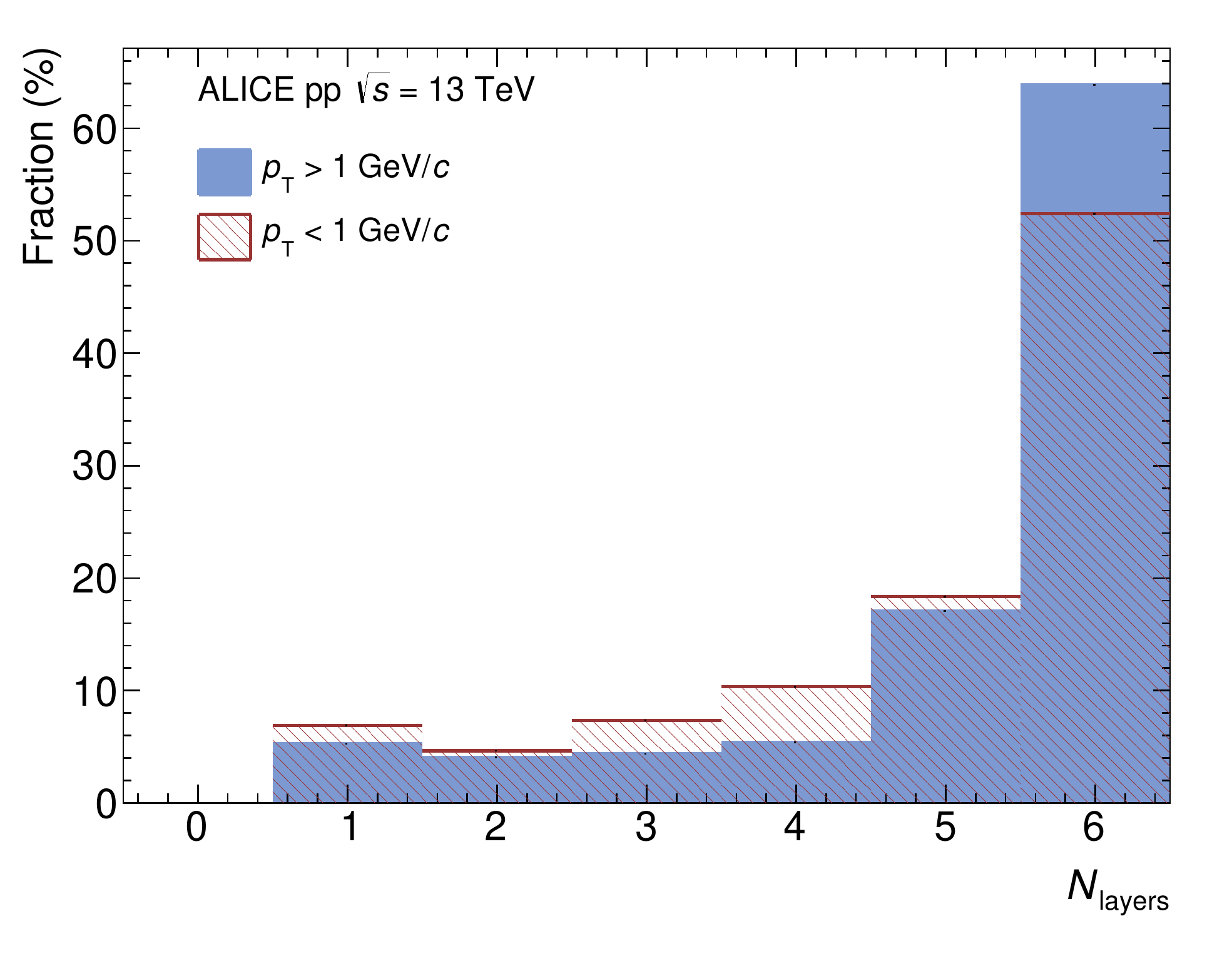}
  \caption{Fraction of tracks, originating from the primary vertex, consisting of a given number of
    layers in pp collisions at $\sqrt{s} = 13~\mathrm{TeV}$.}
  \label{fig:trk_comp}
\end{figure}

The relative frequencies of the number of tracklets assigned to a track are shown in Fig.~\ref{fig:trk_comp} for pp collisions at $\sqrt{s} = 13~\mathrm{TeV}$. Tracks consisting of 6~layers account
for more than 50\% (60\%) for $\pt < \SI{1}{\giga\electronvolt / c}$
($\pt > \SI{1}{\giga\electronvolt / c}$). Tracks with 4~and
5~layers are mainly produced by particles crossing dead areas of
the detector.

\begin{figure}[tb]
  \centering
  \includegraphics[width=0.6\textwidth]{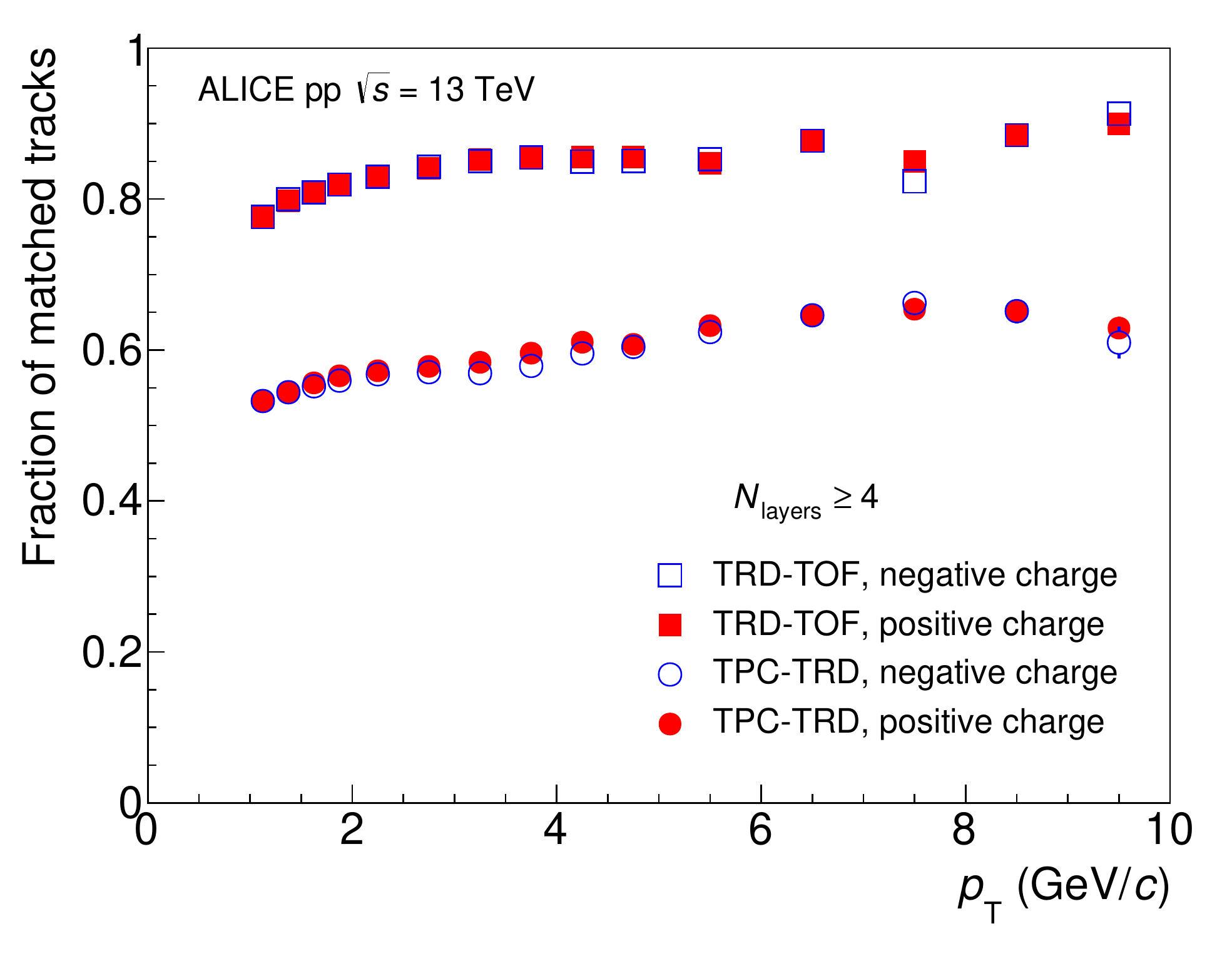}
  \caption{Fraction of tracks matched between
the TPC and the TRD (TPC-TRD) and further the TOF detector (TRD-TOF) as a function of transverse momentum in pp
    collisions at $\sqrt{s} = 13~\mathrm{TeV}$.}
  \label{fig.efficiency}
\end{figure}

A crucial figure of merit for the tracking is the fraction of global
tracks matched to the TRD. This includes acceptance effects, between
the TPC and the TRD as well as the TRD and the TOF detector. The momentum
dependence is shown in Fig.~\ref{fig.efficiency} for tracks with at
least 4~layers (about 75\% of all tracks). For positively charged particles,
the Lorentz drift of the electrons is opposite to the track
inclination, which (together with the tail cancellation) results in a
slightly higher efficiency.

\begin{figure}[tb]
  \centering
  \includegraphics[width=0.8\textwidth]{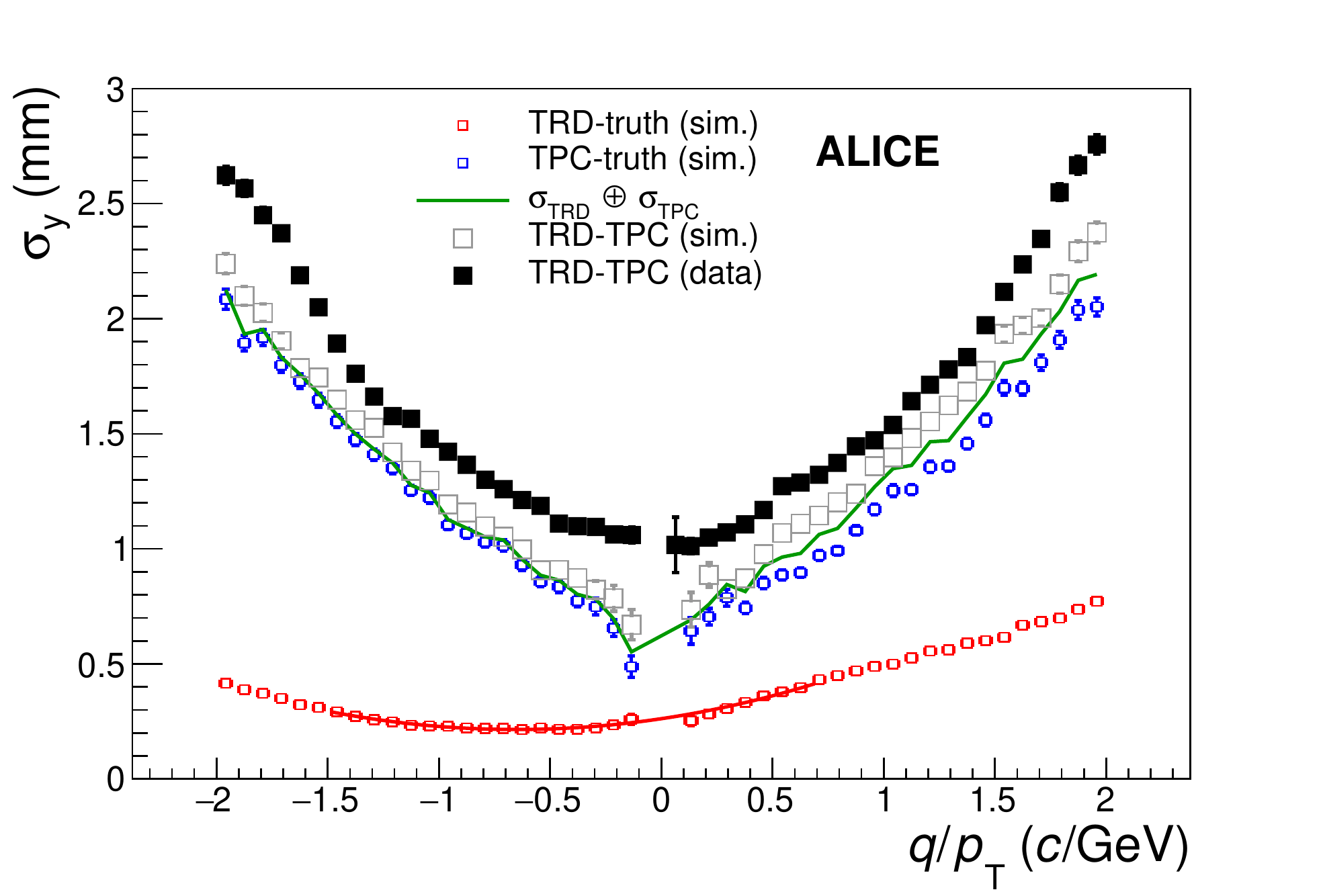}
  \caption{Dependence of the position resolution on charge over
    transverse momentum for simulated tracks in the TRD (red) and in
    the TPC (blue), reconstructed global tracks from simulation (gray)
    and from \pp collisions at $\sqrt{s} = 13~\mathrm{TeV}$ (black). The label TRD-TPC indicates global tracks reconstructed with the ITS and TPC that were extrapolated to the TRD. The green line represents the theoretical value for the combined resolution of TRD and global tracks. The red line shows a parabolic
    fit to the corresponding points.}
  \label{fig.pos.resolution}
\end{figure}

A systematic analysis of the position resolution in the bending plane
($r\varphi$) is presented in Fig.~\ref{fig.pos.resolution}. The
resolution ($\sigma_{\Delta y}$) is expressed as the width of a Gaussian fit to the
difference between the position reconstructed via tracklets and
different references ($\Delta y$). It is shown as a function of the
inverse transverse momentum scaled with the particle charge
($q/\pt$). First, the ideal position resolution is derived from
Monte Carlo simulations by comparing the reconstructed tracklet position with the
true particle position at the reference radial point (anode wire plane
of the read-out chamber). This is shown as the red curve in
Fig.~\ref{fig.pos.resolution}, calculated in local chamber coordinates
to decouple residual misalignment effects from the result. A parabolic
best fit is performed for which the parameters show the best
position resolution of close to \SI{200}{\micro\metre} at
\pt~=~1.8\gevc. The best performance is achieved for tracks where the
inclination angle cancels the $E\times B$ effect. In the case of real
data, the comparison can be performed only against a measured
estimator, i.e.\ against the reconstructed global (ITS + TPC)
track. The black curve shows the distribution for \pp collisions at
$\sqrt{s}=8~\mathrm{TeV}$. The combined position resolution of the TRD
and global tracks is around \SI{700}{\micro\metre} at very large
transverse momentum. In order to bridge the two results, observables
at the level of reconstruction and simulation are compared. The blue
curve shows the position resolution of the global tracks as
reconstructed against the true position from the Monte Carlo simulation.
The green line represents the theoretical value for the combined
resolution for TRD and global tracks, given by the quadratic sum of
the dependencies described by the red and the blue distributions. These
tracks from simulation yield a slightly worse resolution because the
theoretical limit does not consider the pad tilting. It is worth
noting that the simulated position resolution describes the measured
dependency reasonably well. Effects of remaining miscalibration and
misalignment of all central barrel detectors lead to a degradation of about
\SI{500}{\micro\metre} for the resolution in the TRD.

\begin{figure}[tb]
  \centering \includegraphics[width=0.5\textwidth]{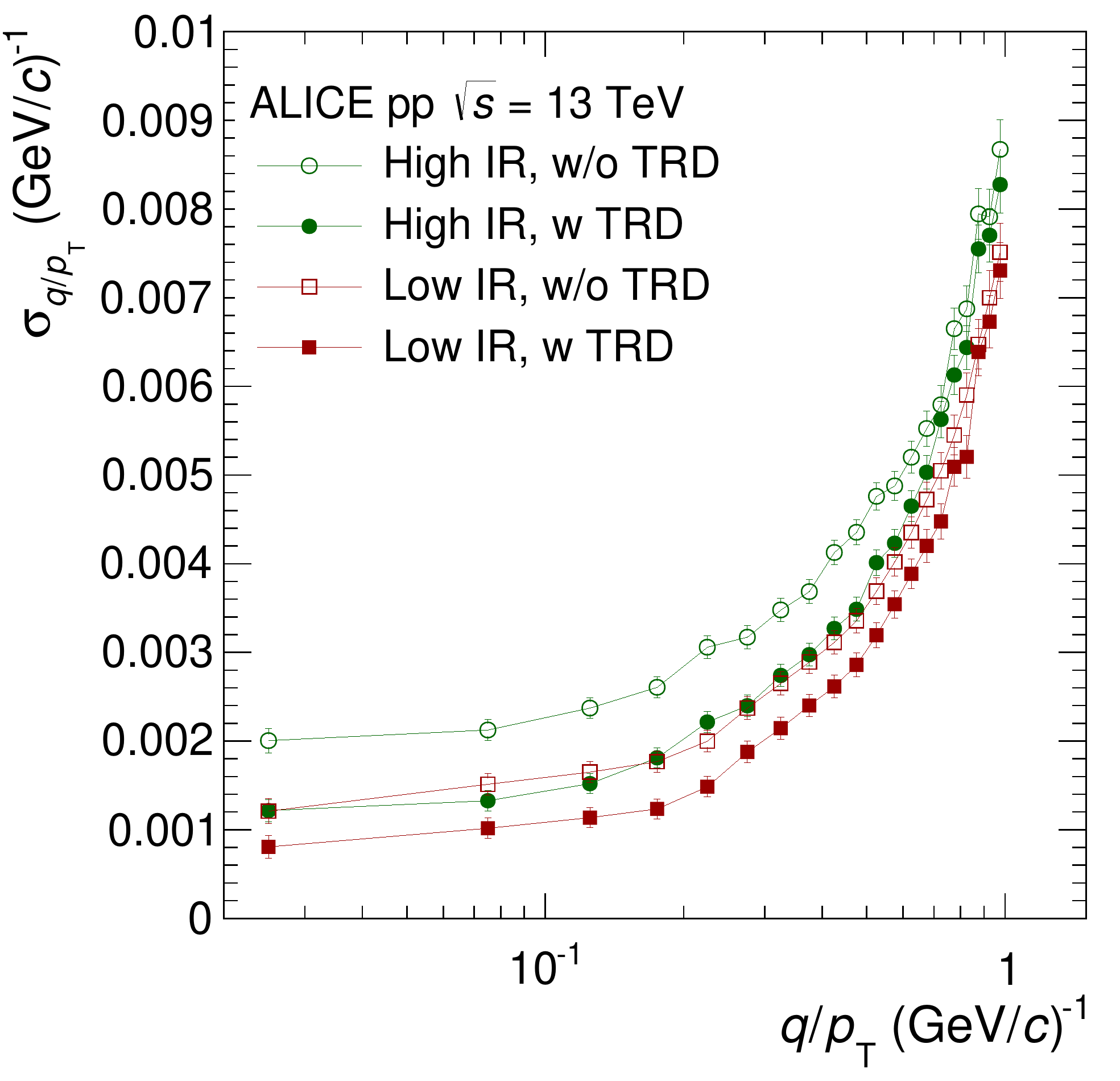}
  \caption{Improvement of the $q/$\pt resolution in data when TRD
    information is included as compared with the performance of tracking without TRD information for various running scenarios. The labels low and high IR indicate interaction rates of 12 and 230~kHz, respectively. }
  \label{fig.pt.resolution}
\end{figure}

The good position resolution capabilities demonstrated by the TRD
detector can be used in the central barrel tracking of ALICE to
improve the transverse momentum resolution of reconstructed
particles. Figure~\ref{fig.pt.resolution} shows the $q/$\pt resolution of
the combined ITS-TPC tracking with and without the TRD for various running scenarios.
In all considered cases the TRD was also used as reference to obtain the correction
maps for the distortions in the TPC. The inclusion of the TRD in tracking in addition improves the resolution
by about 40\% at high transverse momentum for \pp collisions recorded at both low (12~kHz) and high interaction (230~kHz) rates. For example in the low interaction scenario of pp collisions, the achieved \pt resolution is 3\% at 40~GeV. In addition the inclusion of the TRD in the track reconstruction improves the impact parameter resolution and the reconstruction of tracks that pass at the edges of the TPC sectors, i.e.\ increasing the acceptance of the experiment.

\section{Alignment}\label{Chapteralign}

The physical alignment of the detectors during installation (see 
Section~\ref{secsupermodule}) has a finite precision of the order of \SI{1}{\milli\metre} for 
chambers within a supermodule and of \SI{1}{\centi\metre} for supermodules in the spaceframe. 
The subsequent software alignment, i.e.\ 
accounting for the actual positions of supermodules and chambers 
in the reconstruction and simulation software, is the subject of this section. 
The alignment parameters (three shifts and three rotation parameters per alignable volume) are deduced 
from optical survey data and/or from reconstructed tracks. In the latter 
case, the obtained values have to be added to those already used during the 
reconstruction. The obtained alignment sets are stored in the OCDB and used in the subsequent reconstructions. 

The different alignment steps are described in the following subsections.
The alignment is checked and, if necessary, redone after shutdown periods and/or 
interventions that may affect the detector positions, e.g.\ installations 
of new supermodules.

\subsection{Internal alignment of chambers with cosmic-ray tracks}
\label{sec:ali:int}
\begin{figure}[bt]
\centering
\includegraphics[width=.45\textwidth]{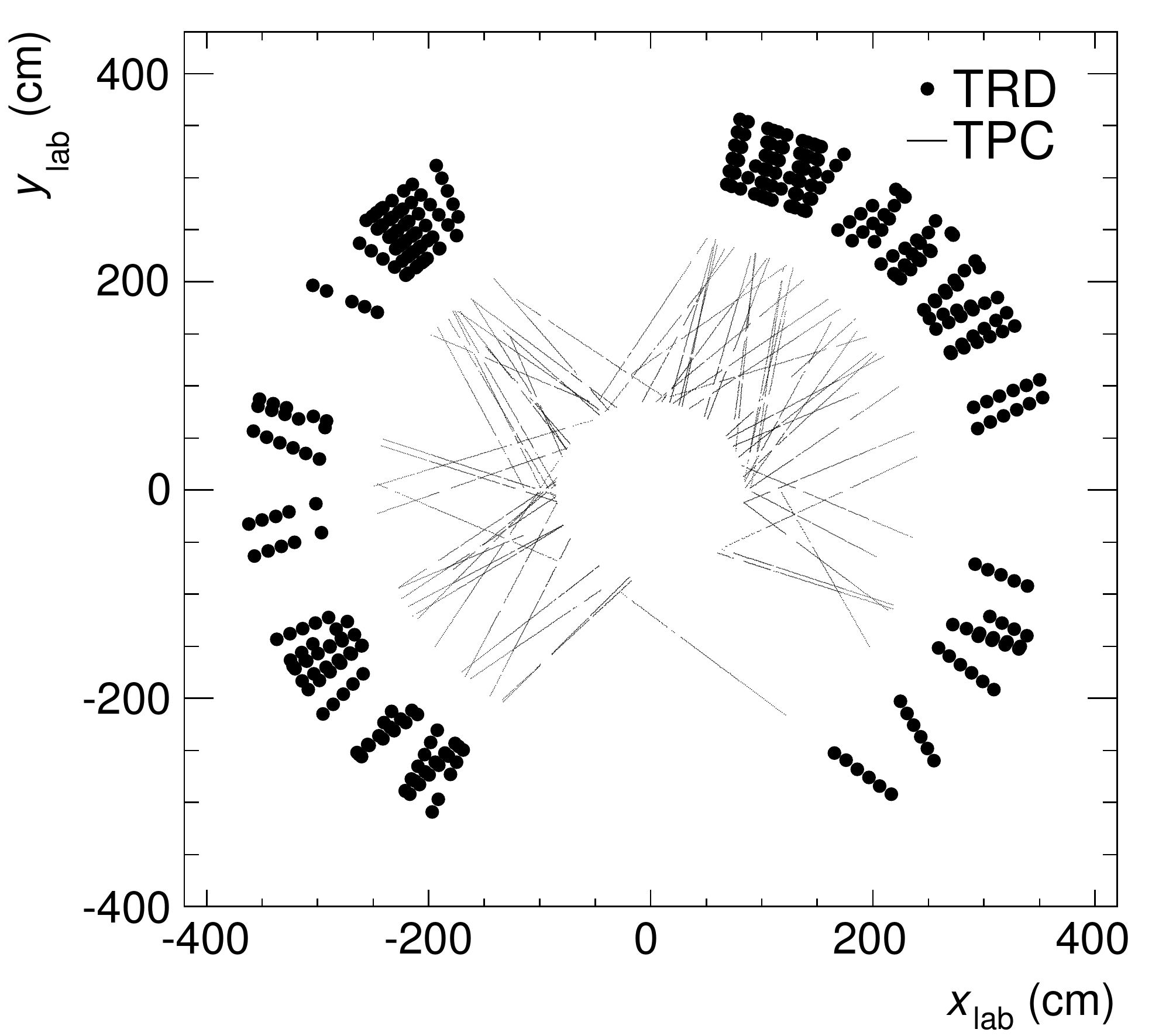} 
\includegraphics[width=.45\textwidth]{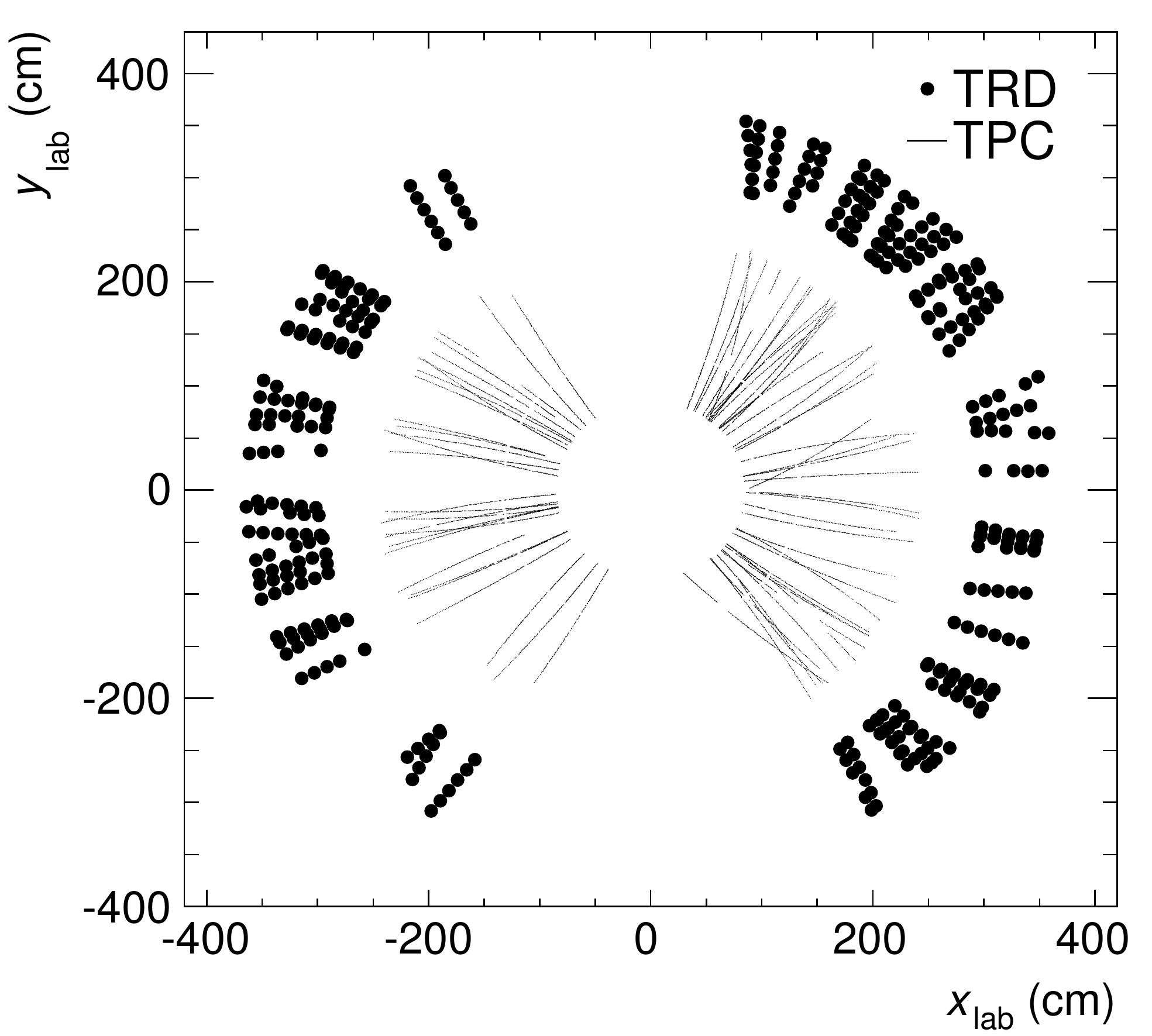} 
\caption{Left: Cosmic-ray tracks with at least 100~TPC clusters and 5~TRD layers, 
recorded without magnetic field, used for the relative $r\varphi$ alignment of 
the TRD chambers within stacks (internal alignment). 
Right: Charged-particle tracks with at least 100~TPC clusters and 4~TRD 
layers from \pp collisions at $\sqrts=8$~TeV, used for the alignment of the TRD 
with respect to the TPC (external alignment). Both figures show data from 2012 (setup 
with 13~supermodules).}
\label{fig:align-tracks} 
\end{figure}

The internal detector alignment, i.e.\ the relative alignment of the read-out chambers 
within one stack, is performed with cosmic-ray tracks recorded without magnetic 
field (Fig.~\ref{fig:align-tracks}, left). 
The local $y$ coordinates (see Section~\ref{Chaptertrdsystem}) of the chambers
of the intermediate layers L1--L4 (tracklet) are varied to minimise the $\chi^2$ of
straight tracks calculated from the hits in layers L0 and L5. The coordinates of the first and last chamber, L0 and L5, are
kept constant. Any misalignment of a stack, such as a tilt, possibly resulting from this constraint is removed
later during the stack alignment. Chamber tilts are neglected. The typical
spread (Gaussian $\sigma$) of the residual between tracklet and straight track is about
\SI{1}{\milli\metre} for a single chamber (see Table~\ref{tab:ali-resolution-int}). The initial chamber
misalignments of 0.6--\SI{0.7}{\milli\metre} are
reduced to 0.2--\SI{0.3}{\milli\metre} (r.m.s.). The minimum required statistics
is $O(10^3)$ tracks per read-out chamber (i.e.\ per stack). For a few stacks, located
around $\varphi=0$ and $\varphi$~=~\SI{180}{\degree}, with low statistics of
cosmic-ray tracks, charged tracks from \pp collisions taken without magnetic
field are used instead.
\begin{table}[bt]
\begin{center}
\begin{tabular}{lcc}
\hline
\begin{tabular}{@{}l@{}} \textbf{Alignment} \\ \textbf{volumes}  \end{tabular}    & \textbf{Input data set}           & \begin{tabular}{@{}c@{}} \textbf{Residual} \\ \textbf{width ($\sigma$)} \end{tabular} \\
\hline 
L0 & cosmics         & \SI{2}{\milli\metre} \\
L2 & cosmics         & \SI{1}{\milli\metre}\\
L5 & cosmics         & \SI{2}{\milli\metre} \\
L0 & \pp collisions  & 2--\SI{3}{\milli\metre} \\
L2 & \pp collisions  & 1--\SI{2}{\milli\metre} \\
L5 & \pp collisions  & 2--\SI{3}{\milli\metre} \\
\hline

\end{tabular}
\caption{Typical width of the tracklet-to-track residuals in $y$ observed during 
the internal alignment procedure. 
The residuals are between a tracklet (measured by a single chamber) and track (defined 
by the remaining chambers of the stack). L0--L5 refer to the six TRD chambers within a stack. 
The L0 and L5 resolutions are given only for comparison purposes as the positions 
of these two chambers are fixed during the minimisation.\label{tab:ali-resolution-int}} 
\end{center}
\end{table}

The internal $y$~alignment sets deduced from cosmic-ray tracks and from pp collisions 
agree within 0.18~mm (Gaussian $\sigma$). From this, the accuracy of the 
internal alignment is estimated to be about \mbox{$\Delta y = 0.18~{\rm mm}/\sqrt{2}=0.13~{\rm mm}$}. Similar agreement exists between cosmic-ray runs taken in different periods.

\subsection{Survey-based alignment of supermodules}
\label{sec:ali:survey}
The supermodules are subject to an optical survey after installation and, subsequently, 
after every hardware intervention that may affect the geometry of the detector. 
For this measurement, survey targets are inserted into precision holes existing at 
each end of every supermodule.

Because of poor accessibility of the muon-arm side, the supermodules are only surveyed on one side (A-side).
Four of the six alignment parameters, $x$, $y$, $z$ shifts and the rotation around 
the $z$-axis, are then 
determined for each supermodule by fitting the survey results. The typical 
survey precision is \SI{1}{\milli\metre}. The survey-based alignment procedure reduces the 
supermodule misalignment from its initial value of 1--\SI{2}{\centi\metre} to a few~\si{\milli\metre}.

\subsection{External alignment with tracks from beam--beam collisions}
\label{sec:ali:ext}
\begin{figure}[tb]
\centering
\includegraphics[width=0.6\textwidth]{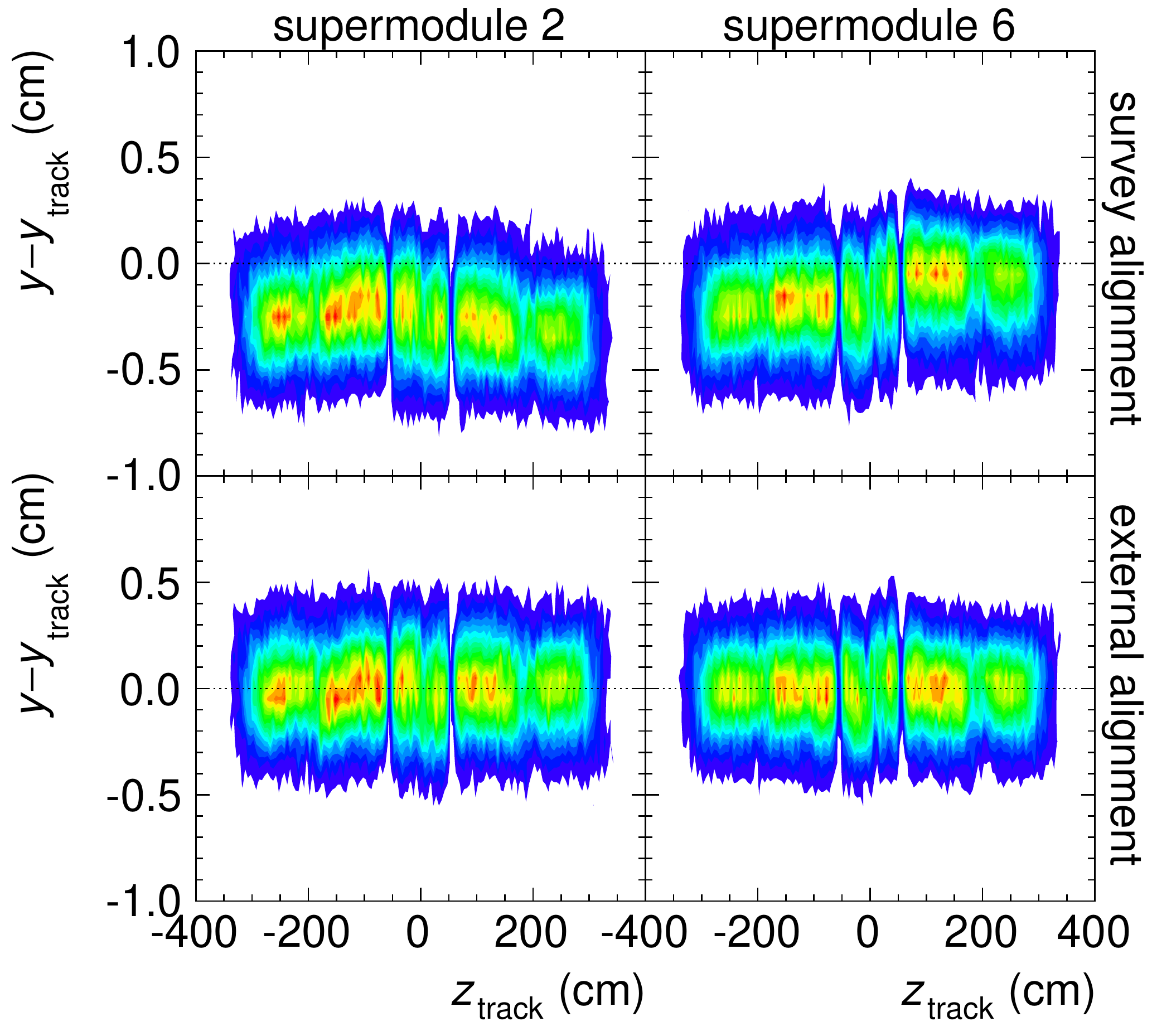}
\caption{TRD tracklet to TPC track residuals in $y$ as a function of the $z$~coordinate of the TPC track ($z_{\rm track}$)
for supermodules 2~(left) and 6~(right). The colour code is linear in the number 
of tracks. The upper and lower panels show the situation with the survey alignment 
and with in addition the external alignment, respectively. The data are from a 
2012 run of \pp collisions with B~=~$-$\SI{0.5}{\tesla}. The alignment set used for the lower 
plots was deduced from the same run. The internal alignment is applied in all 
four cases.}
\label{fig:align-befaft}
\end{figure}

The external alignment, i.e.\ the alignment of TRD volumes with respect 
to the TPC, is performed with charged-particle tracks recorded with magnetic 
field (Fig.~\ref{fig:align-tracks}, right). 
Only tracks with \mbox{$\pt>1.5\gevc$} are used. 
First, all six alignment parameters of each TRD supermodule are varied to minimise the residuals. 
Subsequently, the alignment of each stack is refined by adjusting its $x$ 
and $y$~positions and its rotation around the $z$-axis. 
The tracklet-to-track residuals in $y$ before and after alignment 
are shown in Fig.~\ref{fig:align-befaft} for two supermodules. 
As can be seen, the initial misalignment and the degree of improvement 
vary supermodule by supermodule. 
The typical width of the residuals (Gaussian $\sigma$) is about \SI{2}{\milli\metre}
(see Table~\ref{tab:ali-resolution-ext}). 
In the limit of low number of tracks per stack $N_{\rm track}$, the alignment precision 
is statistical: $\sigma/\sqrt{N_{\rm track}}$. With $N_{\rm track}=O(10^3)$, systematic 
effects start to dominate. 
\begin{table}[tb]
\begin{center}

\begin{tabular}{lc}

\hline
\textbf{Alignment volumes} & \textbf{Residual width ($\sigma$)} \\
\hline
L0 chamber          & \SI{1}{\milli\metre} \\
L5 chamber          & \SI{3}{\milli\metre} \\
stack               & \SI{2}{\milli\metre}\\
supermodule         & \SI{2}{\milli\metre}\\
\hline
\end{tabular}
\caption{Typical width of the tracklet-to-track residuals observed during the 
external alignment procedure with $\pt>1.5$\gevc tracks from \pp collisions. 
The residuals are between a TRD chamber, stack, or supermodule and the TPC track. 
L0 and L5 denote the first and the last (radially) TRD chambers within a stack. 
\label{tab:ali-resolution-ext}}
\end{center}
\end{table}

Figure~\ref{fig:align-befaft} shows the effect of an alignment procedure applied to the same data set from which it was deduced. However, one single alignment set is used for runs of a complete year.
This raises the question of the universality and temporal stability of the alignment,  
which can be addressed by comparing alignment sets deduced from various portions 
of data. Separate analyses of positive and negative tracks yield two alignment 
sets that agree within \SI{1}{\milli\metre} (r.m.s.\ of the $y$~shifts). 
A larger difference (\SI{2}{\milli\metre}) is seen between the two magnetic field polarities. 
Such differences can result from mechanical displacements and/or from the fact that 
the TPC calibration is performed separately for the two polarities. The presence 
of a step in the middle of the central TRD stack, at $z=0$, in 
Fig.~\ref{fig:align-befaft} indicates the latter. 
Several iterations of the TRD to TPC alignment and the TPC calibration with respect 
to the TRD are needed to achieve the best possible precision. In order to address the entanglement of the alignment and calibration of the central barrel detectors, an alternative approach was developed during \LS{1}. It is based on a combined alignment and calibration fit performed using the Millepede algorithm~\cite{Blobel:2002ax}.
The new method allows for a simultaneous alignment and calibration of the
ITS, TRD, and TOF, followed by the calibration of the TPC. The procedure is being 
used successfully in \run{2}.

\section{Calibration}\label{Chaptercalib}

\begin{figure}[tb]
\centering
\includegraphics[width=.5\textwidth]{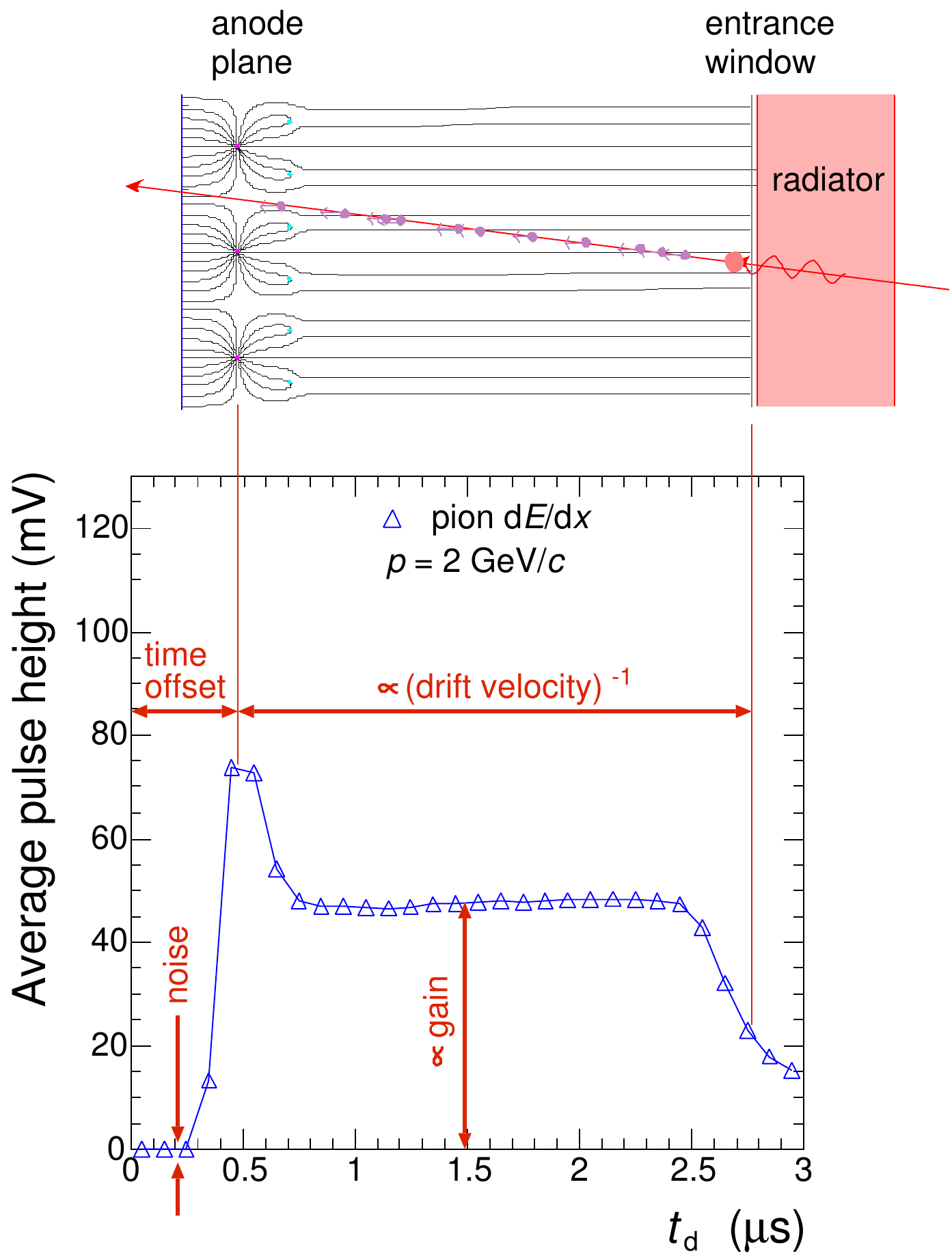}
\caption{Average pulse height vs.\ drift time plot (derived from Fig.~\ref{fig:ph}) 
illustrating the main calibration parameters. 
For better understanding, a sketch of the chamber cross-section with field lines 
from Fig.~\ref{fig:roccross} is shown at the top. 
The peak at the left and the edge on the right of the drift time spectrum correspond to 
the anode wires and the chamber entrance window. 
The temporal difference between them depends on the drift velocity. 
The anode-peak position defines the time offset. 
The mean pulse height and the pedestal width are related to the gain and the pad noise, 
respectively. 
\label{fig:cali-pars}}
\end{figure}

The ALICE calibration scheme is explained in~\cite{Abelev:2014ffa}. 
Here the calibration procedures for the TRD are described. 
The four basic calibration parameters for the TRD -- time offset, drift 
velocity, gain, and noise -- are illustrated in Fig.~\ref{fig:cali-pars}. 
The position of the anode wires and the entrance window are visible in the measured drift time spectrum 
as a peak (around \SI{0.5}{\micro\second}, caused by charges coming from both sides of the anode wires) 
and an edge (around \SI{2.8}{\micro\second}), respectively. 
Since the calibrated time represents the distance from the anode wires, 
the position of the anode peak provides the time offset. 
The time span between the anode peak and the entrance-window edge is inversely proportional 
to the drift velocity. 
The mean pulse height is proportional to the gain and the width of the pedestal 
is proportional to the pad noise. 

While ionisation electrons are attracted to the anode wires by an electric field 
$E$, the presence of a magnetic field perpendicular to it, $|E \times B|>0$, 
leads to a Lorentz angle of about 9$^{\circ}$ between the electron drift direction and 
the direction of the electric field. 
Knowledge of the Lorentz angle is necessary for the reconstruction of the tracklets, 
described in Section~\ref{Chaptertracklet} (see Fig.~\ref{fig.trklt.segm} and 
Fig.~\ref{fig:trkl_reco}). 

The complete list of the calibration parameters, organised according to the source from which they are determined, 
is given in Table~\ref{tab:cali}. 
\begin{table}[tb]
\begin{center}
\begin{tabular}{ll}
\hline
\textbf{Input data} & \textbf{Parameters} \\
\hline
pedestal runs & pad noise, pad status\\
runs with \Kr in the gas & relative pad gain  \\
physics runs (cpass0/1) & chamber status, time offset, drift velocity, Lorentz angle, gain \\
\hline
\end{tabular}
\caption{Sources of input data and the derived calibration parameters.
\label{tab:cali}}
\end{center}
\end{table}
Once determined for a given run, the calibration parameters are stored in the OCDB and 
used in the subsequent reconstructions. 
In the following, the methods used to determine the values of the calibration parameters
are discussed. 

\subsection{Pad noise and pad status calibration using pedestal runs}
\label{sec:cal:pede}

Short pedestal runs are taken roughly once per month during data taking. In these runs, events are triggered at random instants and the data are recorded without zero suppression. 
At the end of the run, an automatic analysis of the pedestal data is performed 
on the computers of the DAQ system~\cite{Chapeland:2010zz}. 
Hundred events are sufficient to calculate the position of the baseline of the analogue pre-amplifier and shaper output (pedestal) and its fluctuation (noise) for all electronics channels. The results are subsequently collected by the Shuttle system~\cite{calib:shuttlenote} and transported to the OCDB. The mean noise is 1.2~ADC counts, corresponding to an equivalent of 1200 electrons. The pad-by-pad r.m.s.\ value is 0.17~counts. 
The precision of the measurement is 0.015 counts (r.m.s.). 
Pads that have a faulty connection to the FEE, are connected to a non working FEE channel, 
have excessive noise, or are bridged with a neighbour 
are marked in the OCDB and treated correspondingly during the data taking and 
reconstruction chain (pad status).

\subsection{Pad gain calibration using \Kr decays}
\label{sec:cal:kr}

Pad-by-pad gain calibration of the TRD chambers is performed after every 
installation of new supermodules. It is done by injecting radioactive gas 
into the chambers and measuring the signals of the decay electrons. 
The method, developed by ALEPH~\cite{Decamp:1990jra,Blum:227125} and 
DELPHI~\cite{DeMin:1995tk}, is also used to calibrate the ALICE 
TPC~\cite{Alme:2010ke}. 

Solid $^{83}$Rb decays by electron capture into gaseous Kr and populates, among
others, the isomeric state \Kr with an excitation energy of
\SI{41.6}{\kilo\electronvolt} and a half-life of 1.8~hours.  The radioactive
krypton is injected into the gas circulation system and is distributed
over the sensitive volumes of all installed chambers. The krypton nuclei decay to their ground state by electron emission.
The decay energy, comparable to the energy lost by a minimum-ionising particle traversing the sensitive volume
of a read-out chamber (20--\SI{30}{\kilo\electronvolt}), gets deposited within
\SI{1}{\centi\metre} from the decay point. For each decay, the total signal
is calculated by integrating over $y$ (pad column), $z$ (pad row) and $x$ (drift
time), and filled into the histogram associated with the pad of maximum signal.

\begin{figure}[tb]
\centering
\includegraphics[width=.5\textwidth]{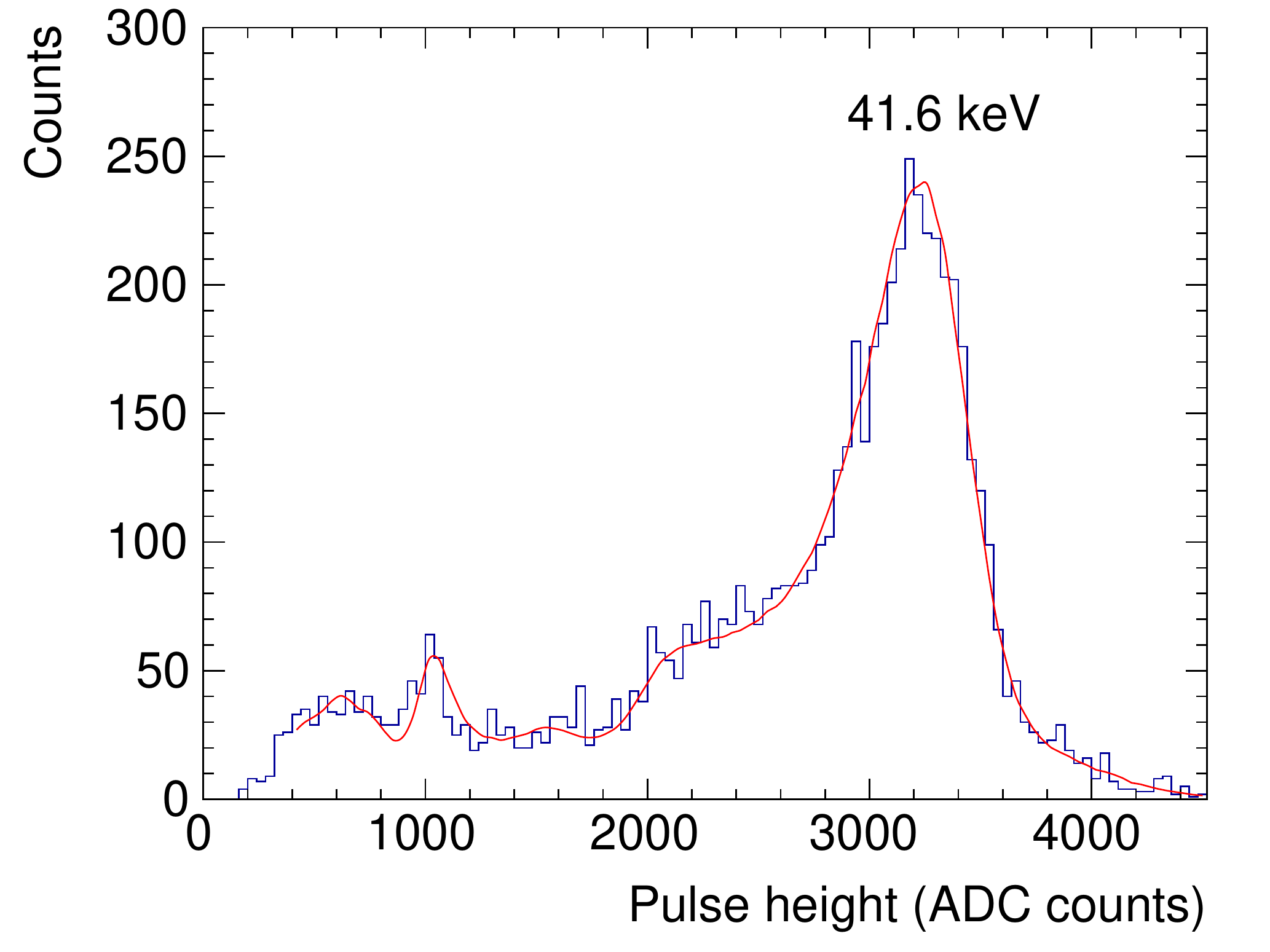}
\caption{Pulse height spectrum accumulated for one pad during the Kr-calibration run~\cite{mh-dipl,js-dipl}. 
The smooth solid line represents the fit from which the gain is extracted. 
\label{fig:cali-kr-spectrum}}
\end{figure}

With three gas inlets to each supermodule (see Section~\ref{Chaptergas}), 
groups of 10 chambers are connected in series. The difference between the decay 
rates seen in the first and last chamber of the chain was reduced to a factor of 
${\sim}$3 by increasing the gas flow during the krypton calibration run. 
With an $^{83}$Rb source intensity of \SI{5}{\mega\becquerel} and 
a measurement time of one week, the collected statistics is of the order of 
thousand counts per pad. This is sufficient to identify the expected decay lines 
in the distribution. An example is shown in Fig.~\ref{fig:cali-kr-spectrum}. 
The histogram of each pad is fitted by stretching horizontally the reference distribution. The
stretching factor is the measure of the pad gain. The energy resolution at \SI{41.6}{\kilo\electronvolt} is 10\%. 

\begin{figure}[tb]
\centering
\includegraphics[width=.6\textwidth]{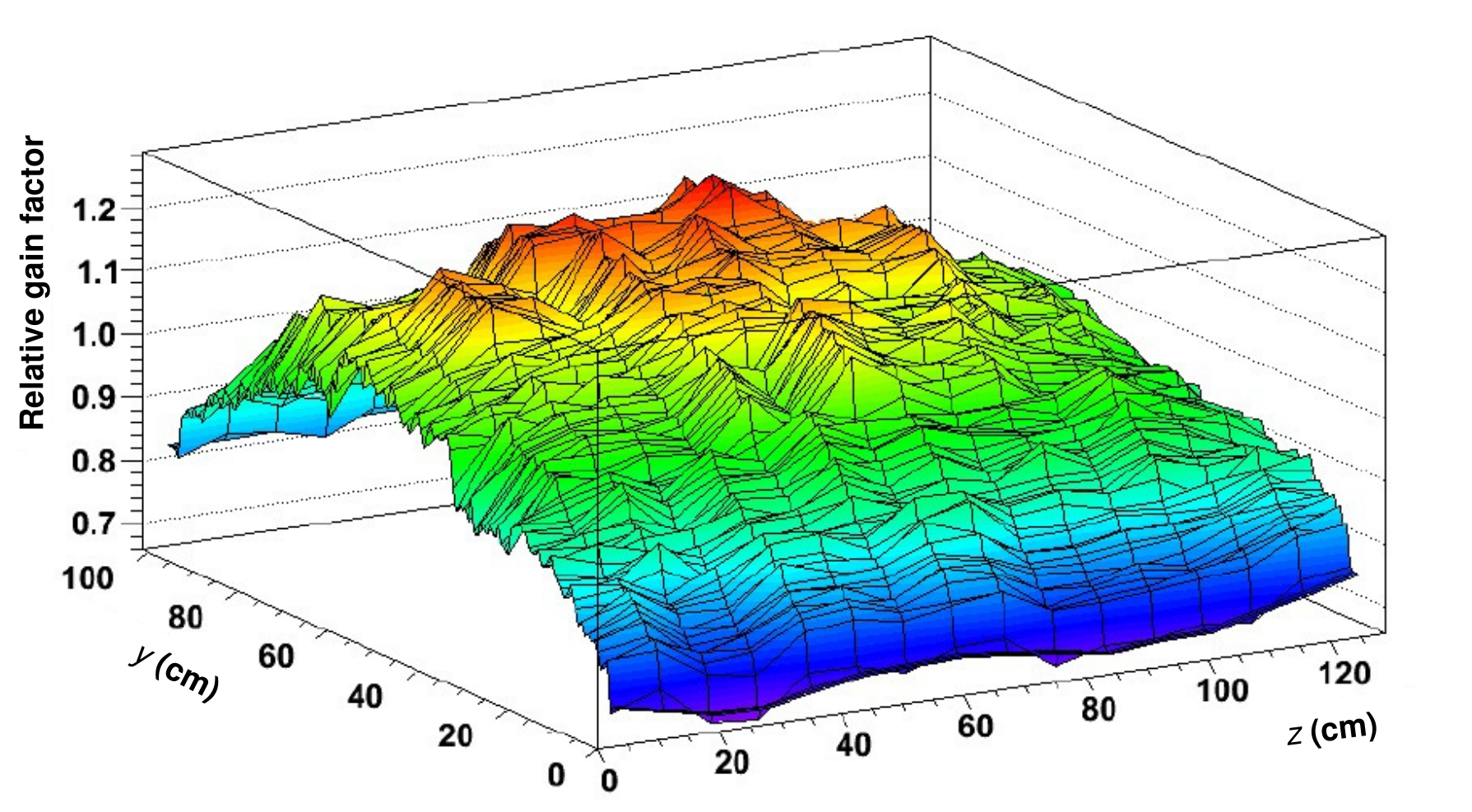}
\caption{Relative pad gains for one chamber calibrated with electrons from \Kr decays. 
\label{fig:cali-kr-pads}}
\end{figure}

The resulting pad gain factors for one particular chamber are shown in 
Fig.~\ref{fig:cali-kr-pads}. 
The short-range variations of up to 10\% reflect the differences between 
electronics channels. 
The long-range inhomogeneities originate from chamber geometry and are typically 
within $\pm$15\% (peak to peak). A detailed description of the krypton calibration 
can be found in~\cite{mh-dipl} and \cite{js-dipl}. 

\begin{figure}[tb]
\centering
\includegraphics[width=.5\textwidth]{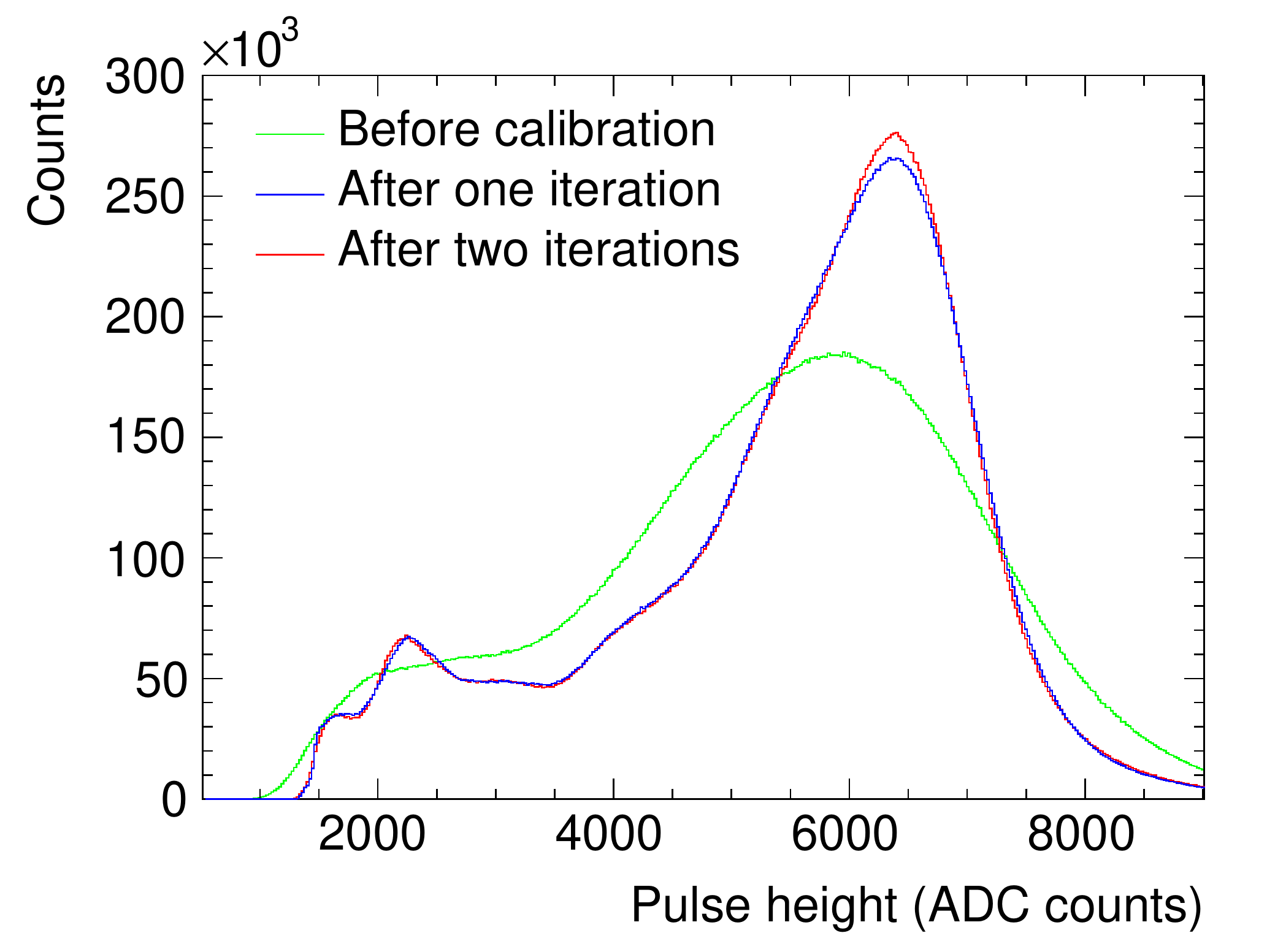}
\caption{Pulse height spectrum before the krypton-based calibration, after one 
and after two iterations (calibrations performed in consecutive years) for one read-out chamber.
\label{fig:cali-kr-spectrum2}}
\end{figure}

The improvement of the chamber resolution achieved by the krypton-based pad-by-pad 
calibration is presented in Fig.~\ref{fig:cali-kr-spectrum2}. The histograms show the pulse height spectrum before calibration, after one 
and after two iterations (calibrations performed in consecutive years), respectively. 

\subsection{Chamber calibration using physics data}
\label{sec:cal:cpass}

The anode and drift voltages of the individual chambers are adjusted
periodically (once a year) to equalise the chamber gains and drift
velocities.  Moreover, an automatic procedure is in place that continuously
adjusts the voltages depending on the atmospheric pressure, compensating the
impact of the environment on the gas properties (see
Section~\ref{Chapterhvinbeam}). This is important because the pulse height and
the tracklet angle are used for triggering (see Section~\ref{Chaptertrigger}).

\begin{figure}[tb]
\centering
\includegraphics[width=.5\textwidth]{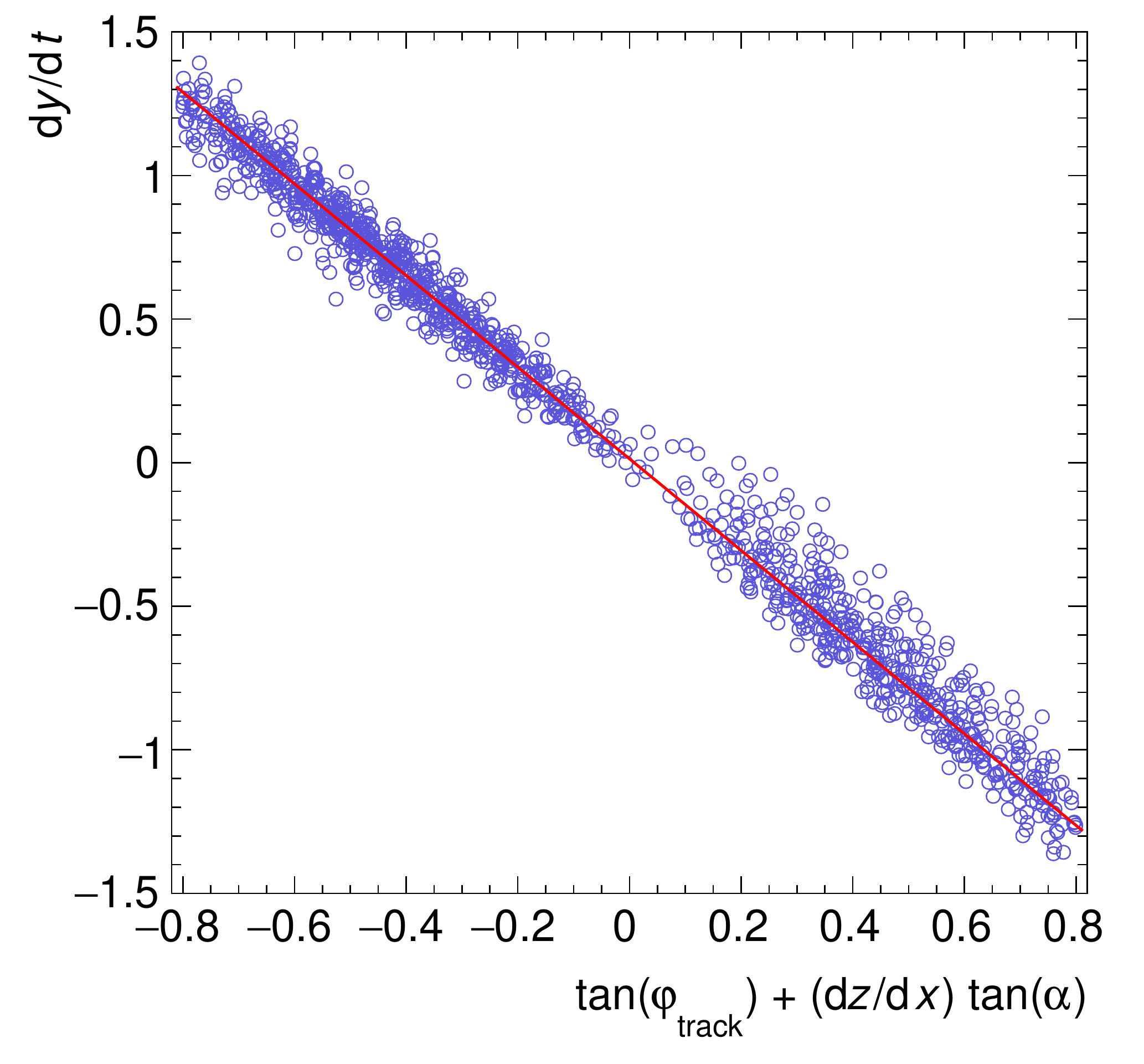}
\caption{
The derivative of the local tracking $y$ coordinate with respect to the drift time $t$
vs. the tangent of the azimuthal track inclination angle from global tracking. 
The slope and the offset of the fit (red line) give the drift velocity and the 
Lorentz angle, respectively. 
\label{fig:cali-drift}}
\end{figure}

In order to achieve the ultimate resolution for physics data analysis, 
the chamber status, time offset, drift velocity, Lorentz angle, and gain are 
calibrated run-by-run offline, using global tracks from physics runs. 
A sample of events of each run is reconstructed for this purpose. 
The required statistics is equivalent to $10^5$ \pp interaction events. 
The first reconstruction pass (cpass0) provides input for the calibration. 
The second pass (cpass1) applies the calibration and the reconstructed events 
are used as input for the data quality assurance analysis, and for the second 
iteration of the calibration. 
The read-out chamber status and the chamber-wide time offset, drift velocity, Lorentz 
angle, and gain values are extracted from cpass0 and updated after cpass1. 
The time offset is obtained as indicated in Fig.~\ref{fig:cali-pars}. 
The drift velocity and the Lorentz angle are derived from the correlation between 
the derivative of the local tracking $y$ coordinate with respect to the drift time,  
and the azimuthal inclination angle of the global track (see Fig.~\ref{fig:cali-drift}). 
The former represents the uncalibrated estimate of the tracklet angle. The latter 
is obtained from the extrapolation of the global track to the TRD. The correlation is fitted by a straight 
line. The effect of the pad tilt ($\dd y/\dd z = \tan(\alpha)$, $\alpha = \pm$ \SI{2}{\degree}, see Section~\ref{Chaptertrdsystem}) is taken into account by adding the respective term to the global track inclination. 
The slope and the offset parameters give the drift velocity and the Lorentz 
angle, respectively. 

The gain calibration factor is determined by histogramming, for each chamber, 
the deposited charge divided by the path length and taking the mean of this 
distribution. 
The last stage of the chamber calibration is to identify chambers for 
which a satisfactory calibration cannot be obtained or whose parameter values 
are very different from the mean. 
These chambers are masked in the data analysis and in the respective simulation. 

The typical mean values, chamber-by-chamber variations, stability, and precision 
of the calibration parameters are shown in Table~\ref{tab:calivar}. 
\begin{table}[tb]
\begin{center}
\begin{tabular}{lllll}
\hline
\textbf{Parameter}           &  \textbf{Mean}           & \textbf{Variations}             & \textbf{Stability}        & \textbf{Precision}    \\
\hline
\Tzero          &  \SI{145.2}{\nano\second}    & \SI{2.7}{\nano\second}               &  $\pm$\SI{3.4}{\nano\second}     &   \SI{1}{\nano\second}       \\
\vdrift         &  \SI{1.56}{\centi\metre/\micro\second} & 1--14\%              &  $\pm$3\%        &   0.4\%      \\
$\Psi_{\rm L}$    &  \SI{8.8}{\degree}    & \SI{0.3}{\degree}--\SI{0.5}{\degree}   &  $\pm$\SI{0.4}{\degree} &   \SI{0.05}{\degree}\\
gain            &  1.0 (a.u.)     & 3--16\%              &  $\pm$7\%        &   1.4\%      \\
\hline
\end{tabular}
\caption{
The typical mean values, chamber-by-chamber variations, stability (in the 
second half of 2012), and precision of the chamber calibration parameters 
\Tzero (drift time offset), \vdrift (drift velocity), $\Psi_{\rm L}$ (Lorentz 
angle for $B$~=~\SI{0.5}{\tesla}), and gain. 
For the chamber-by-chamber variations, which are subject to 
equalisation by adjusting the voltages, ranges are indicated. 
\label{tab:calivar}}
\end{center}
\end{table}
The chamber-by-chamber variation is quantified by the r.m.s.\ of the chamber 
distribution within one run. 
The stability is described via the maximum variations observed in one read-out chamber during 
half a year of running. 
The precision is defined as 1/$\sqrt{2}$ of the r.m.s.\ difference between the calibration 
parameters deduced from two high-statistics data sets taken under identical conditions. 

\subsection{Quality assurance}
\label{sec:cal:qa}

\begin{figure}[tb]
\centering
\includegraphics[width=.47\textwidth]{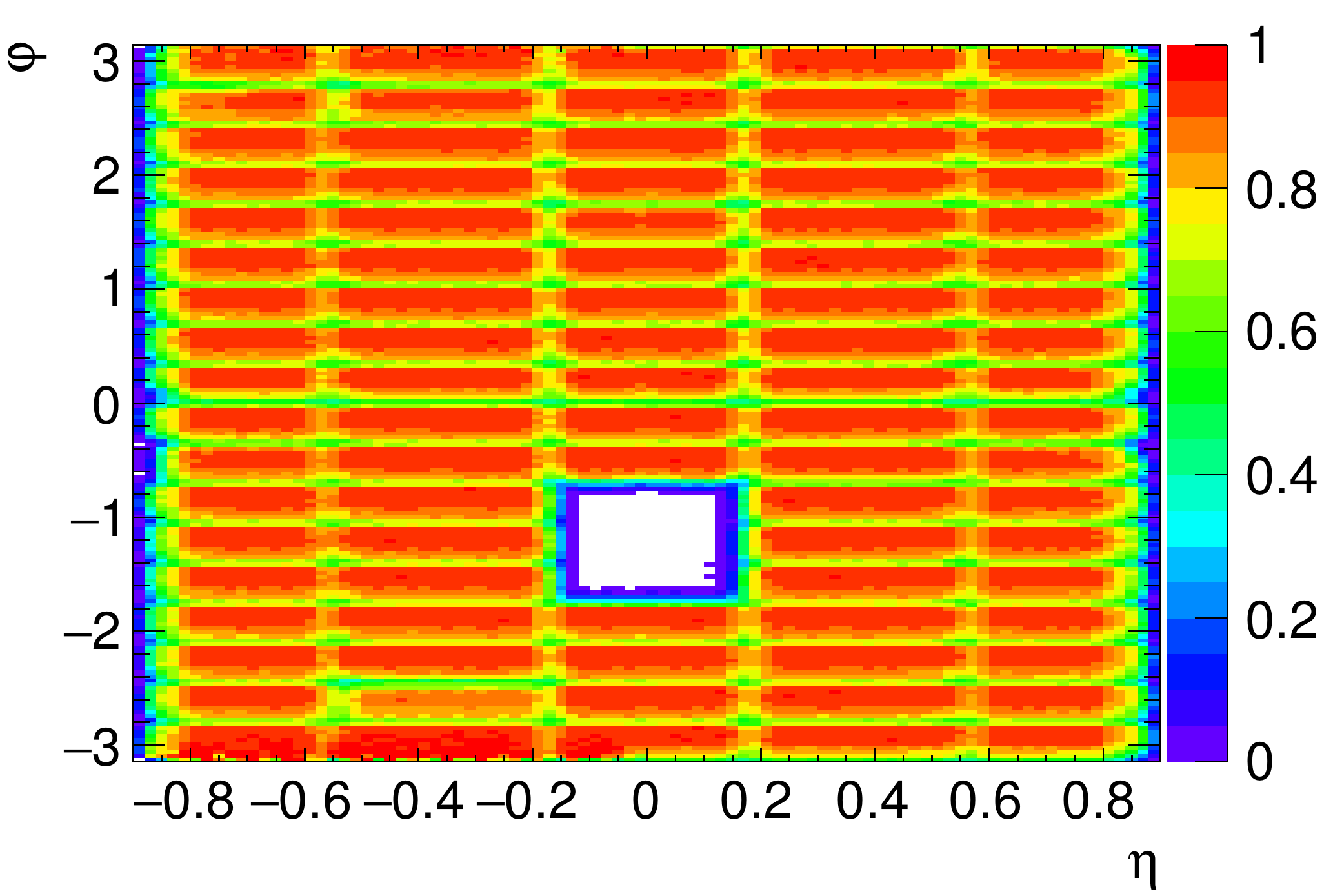}
\includegraphics[width=.47\textwidth]{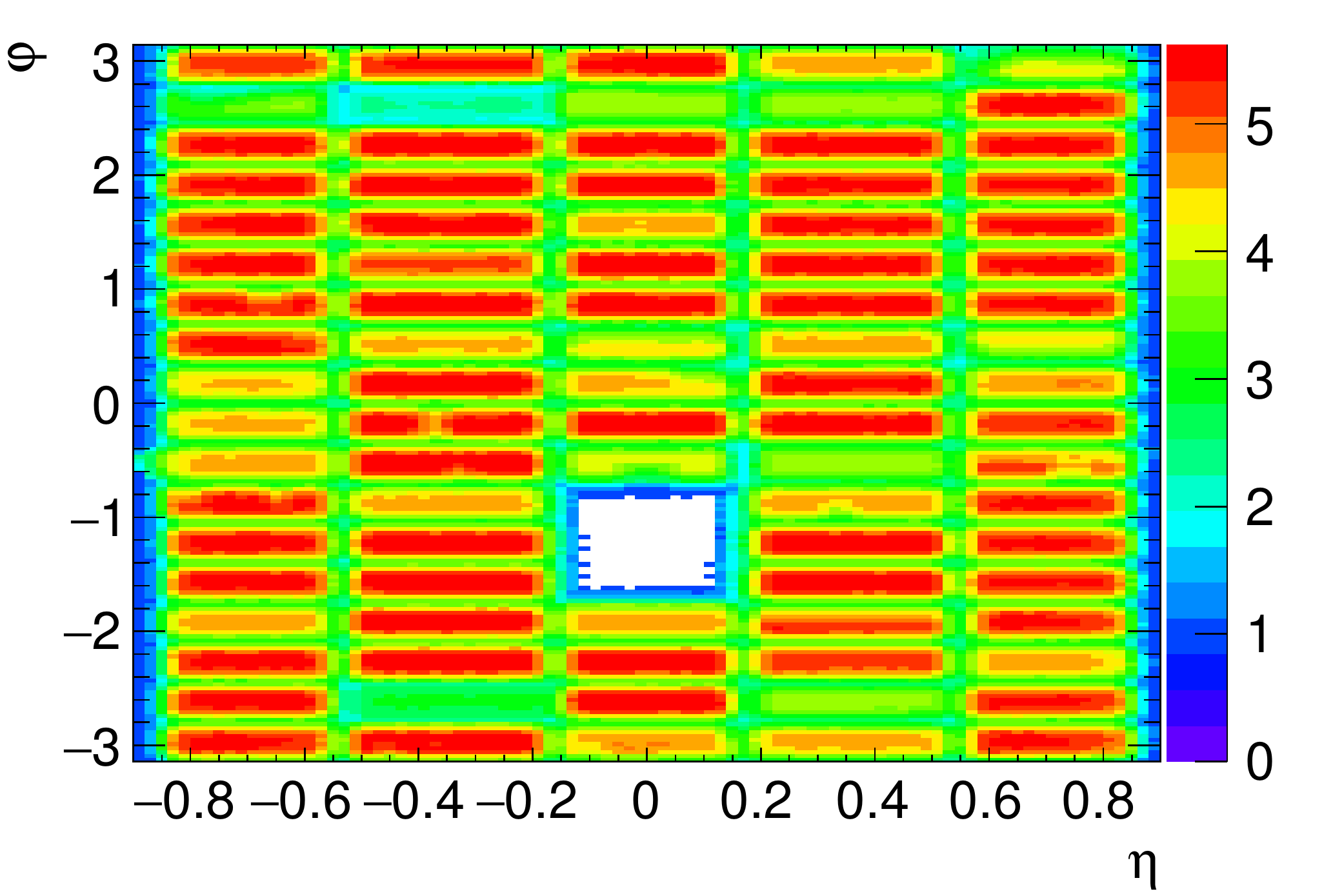}
\caption{Two quality-assurance plots (data from \pp collisions recorded in 2015 with 
all supermodules installed, tracks with at least 70~TPC clusters and $\pt>0.5$\gevc). Left: 
Efficiency of matching tracklets to TPC tracks. Right: Mean number of layers per track in each 
stack (cf.\ the discussion of inactive chambers in Section~\ref{Chapterhvinbeam}). 
\label{fig:cali-qa}}
\end{figure}

As described before, during cpass1 reconstructed events are subject to a quality assurance (QA) analysis in which 
control histograms monitoring the quality of the calibrated data are filled.  
The analogous monitoring of raw data, performed online, is described in 
Section~\ref{Chapterbeamoperation}. As an example, two such QA histograms, representing the efficiency and the mean number 
of layers in each stack (equivalent to the number of active layers) in one particular run of the 
\pp data taking in 2015, are shown in Fig.~\ref{fig:cali-qa}.
The efficiency drops at stack boundaries and the window in correspondence of the detector coverage of the PHOS 
detector are visible.

\begin{figure}[hbt]
   \centering
\includegraphics[width=.5\textwidth]{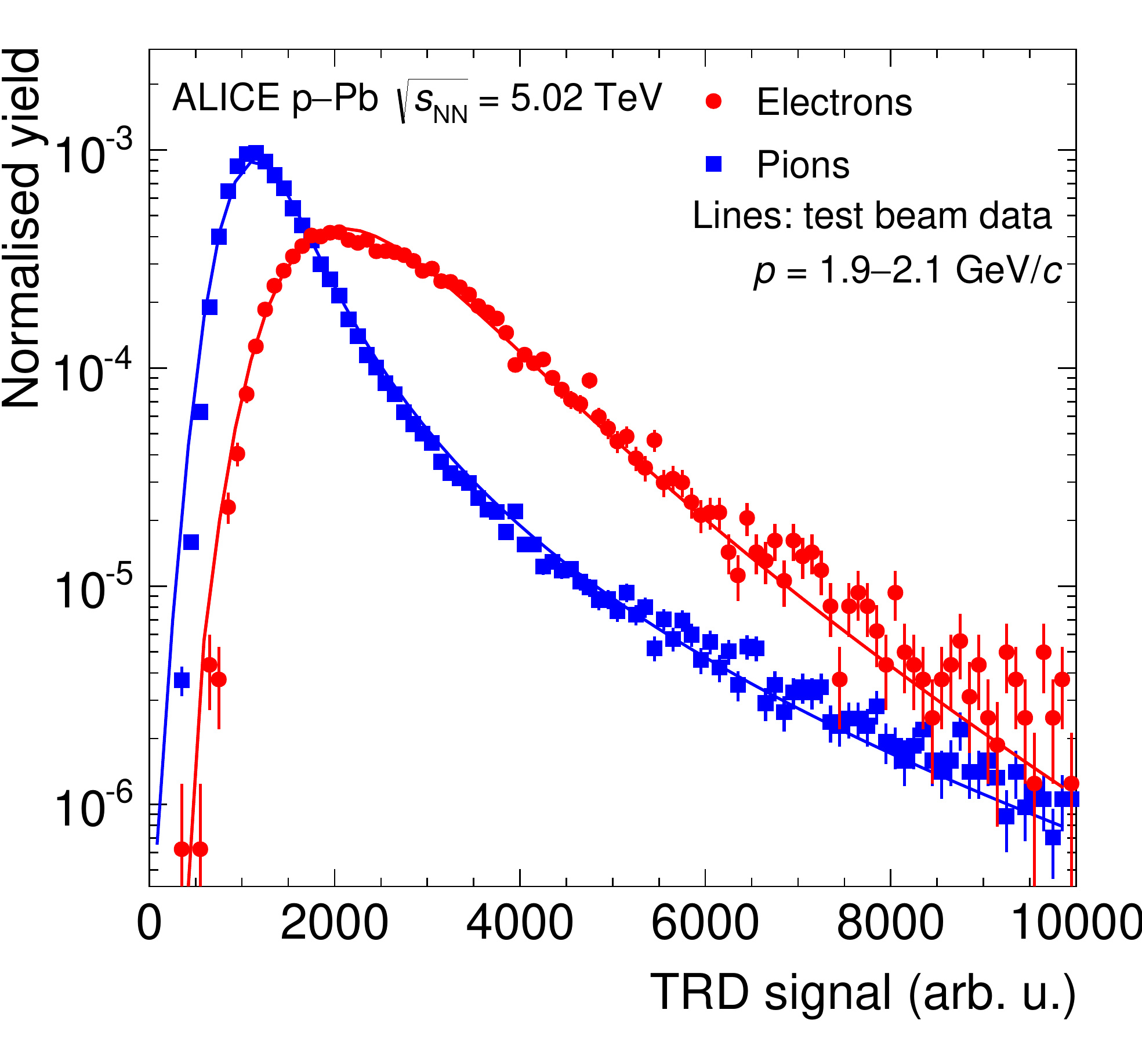} 
  \caption[]{Total integrated charge, normalised to the tracklet length, measured in a single read-out chamber for electrons and pions in \ppb collisions at \mbox{\sqrtsnn~=~5.02~TeV}, in comparison with results from test beam measurements (solid lines)~\cite{Andronic:2004uy,Bailhache:2006hs}. The electrons and pions from test beam measurements were scaled by one common factor to compensate the difference in gain of the two data sets. }
   \label{Figure_trackletcharge}
\end{figure}

\section{Particle identification}\label{Chapterpid}

The TRD provides electron and charged hadron identification based on the measurement of the specific energy loss and transition radiation. The total integrated charge measured in a tracklet~\cite{MFasel}, normalised to the tracklet length, is shown in  Fig.~\ref{Figure_trackletcharge} for electrons and pions in \ppb collisions at \mbox{\sqrtsnn~=~5.02~TeV}. The electron and pion samples were obtained by selecting tracks originating from $\gamma \rightarrow e^+ e^-$ conversions in material and from the decay K$^{0}_{\rm s} \rightarrow \pi^+ \pi^-$ via topological cuts and particle identification (PID) with the TPC and the TOF. The obtained electron sample has an impurity of less than 1\%.
Due to the larger specific energy loss and transition radiation, the average charge deposit of electrons is higher than that of pions. 
Charge deposit distributions recorded in test beam measurements at CERN PS in 2004 for electrons and pions in the momentum range 1 to 10 \gevc~\cite{Andronic:2004uy,Bailhache:2006hs} describe the results from collision data well (see Fig.~\ref{Figure_trackletcharge}), and can thus also be used as references for particle identification.

\begin{figure}[tb]
   \centering
\includegraphics[width=0.7\textwidth]{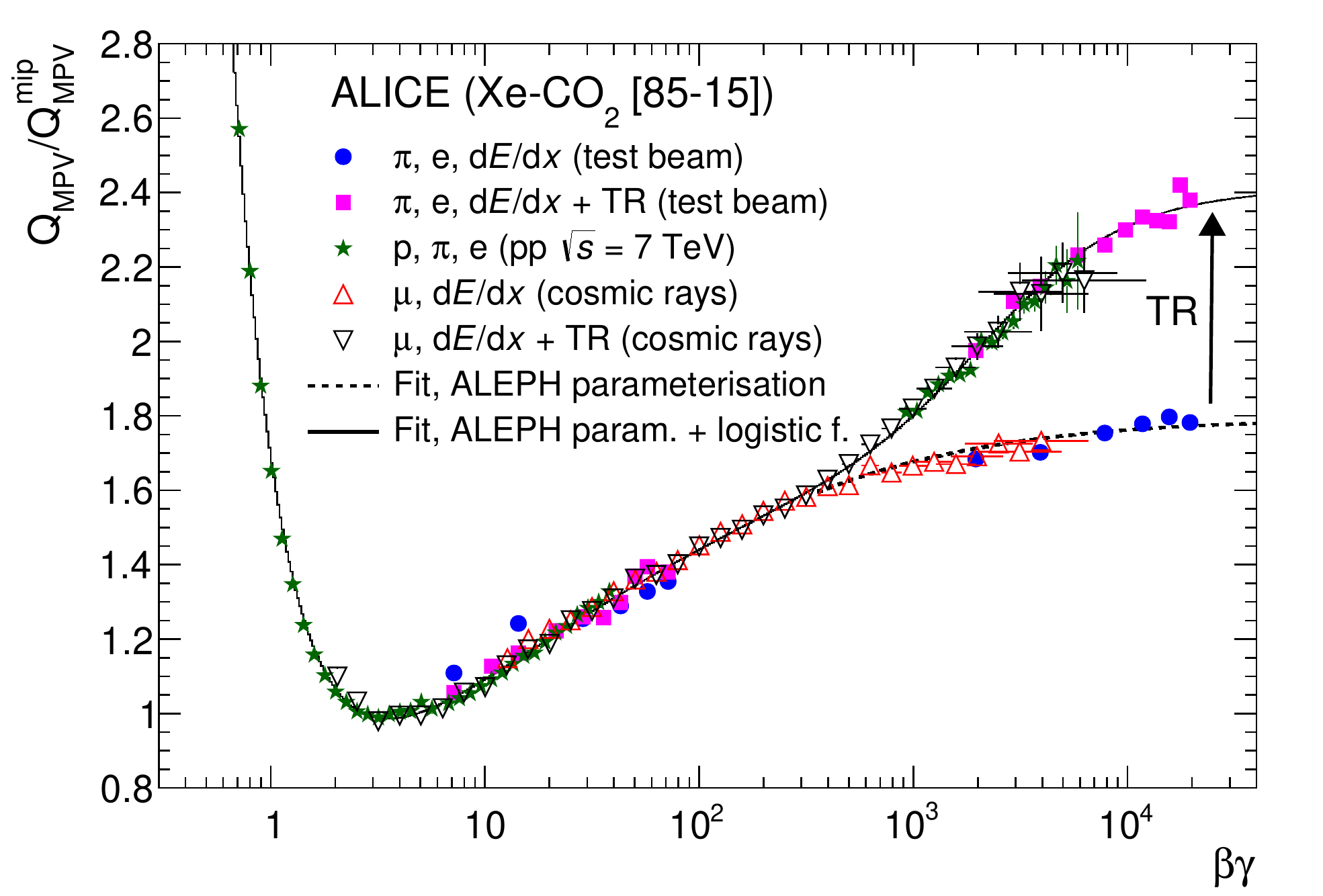} 
  \caption[]{Most probable charge deposit signal normalised to that of
minimum ionising particles as a function of $\beta\gamma$. 
  The data are from measurements performed in test beam runs, \pp collisions at 
  \mbox{\sqrts~=~7~TeV}, and cosmic-ray runs. Uncertainties in momentum and thus $\beta\gamma$ determination are drawn as horizontal and statistical uncertainties as vertical error bars. The shown fits correspond to the Equations~\ref{equ:aleph} and~\ref{equ:logistic} described in the text.}
   \label{TRD_MPV}
\end{figure}

The measured charge deposit distributions can be fitted by a modified Landau-Gaussian convolution: (Exponential~$\times$~Landau)~$\ast$~Gaussian~\cite{xlu,Collaboration:2012zsa}, where the Landau distribution is weighted by an exponential dampening (Landau$(x)~\rightarrow~e^{kx}$~Landau$(x)$). This function describes the specific charge deposit distributions for pions (\dedx) and electrons (\dedx~+~TR) well and can thus be used to extract the most probable energy loss.
The dependence of the most probable signal versus $\beta\gamma$ is shown in Fig.~\ref{TRD_MPV}. 
The data have been extracted from measurements (i) in a beam test at CERN PS in 2004 (pions and electrons)~\cite{Bailhache:2006hs}, (ii) with pp collisions at $\sqrt{s} = 7$~TeV (protons, pions and electrons)~\cite{MFasel} and (iii) with a cosmic-ray trigger in the ALICE setup (muons)~\cite{xlu}. The selection of the flight direction of the cosmic-ray muons allows only the specific energy loss (\dedx) or the summed signal (\dedx + TR) to be measured by selecting muons that first traverse the drift region and then the radiator, and vice versa~\cite{xlu,Collaboration:2012zsa}. To improve the momentum reconstruction of very high \pt cosmic-ray muons, a dedicated track fitting algorithm~\cite{xlu,Collaboration:2012zsa} was developed, combining the clusters of the two individual tracks in the two hemispheres of the TPC. This yields a better momentum resolution by about a factor of~10, e.g.\ at 1~TeV$/c$ the 1/\pt resolution is 8.1$\cdot 10^{-4}$ (GeV/$c$)$^{-1}$~\cite{xlu,Collaboration:2012zsa}.

The onset of the TR production is visible for $\beta\gamma \gtrsim 800$, both for electrons and high-energy (TeV scale) cosmic-ray muons. The signals for muons are consistent with those from electrons at the same $\beta\gamma$. The most probable signal (MPV) of the energy loss due to ionisation only, normalised to that of
minimum ionising particles (mip), is well described by the parameterisation proposed by the ALEPH Collaboration~\cite{Decamp:1990jra,blumrolandi} (shown in Fig.~\ref{TRD_MPV}):
\begin{equation}
\label{equ:aleph}
\left( \frac{Q_{\rm MPV}}{Q_{\rm MPV}^{\rm mip}} \right) = 0.2 \cdot \frac{4.4 - \beta^{2.26} - \ln \left[ 0.004 - \frac{1}{\left(\beta\gamma\right)^{0.95}} \right] }{\beta^{2.26}}.~~ 
\end{equation}
Minimum ionising particles are at a $\beta\gamma$ value of~3.5 and the \dedx in the relativistic limit is 1.8~times the minimum ionisation value. To describe the \dedx~+~TR signal, a parameterised logistic function is needed in addition. The formula, normalised to the signal for minimum ionising particles, is as follows:

\begin{equation}
\label{equ:logistic}
\frac{\rm TR}{\rm TR^{\rm mip}} = \frac{0.706}{1 + \exp(-1.85 \cdot (\ln \gamma - 7.80))}. 
\end{equation}
 The saturated TR yield in the relativistic limit is 0.7~times the minimum ionisation value. At $\beta\gamma$~=~$2.4 \cdot 10^3$ the logistic function reaches half its maximum value.

\subsection{Truncated mean method}
The TRD can provide electron (described in the next section) and hadron identification. For the hadron identification, the truncated mean is calculated from the energy loss (+TR) signal stored in the clusters (see Section~\ref{Chaptertracking})~\cite{xlu}. For the particle identification, the deviation from the expected most probable signal for a given species is then used after normalisation to the expected resolution of the truncated mean signal for the track under study.  

In order to obtain an approximately Gaussian shape, the long tail of the Landau distribution needs to be eliminated or at least strongly suppressed, which can be realised through a truncated-mean procedure.
The PID signal of a charged hadron passing through the detector is calculated using all $M$ clusters along the up to six layers (see Section~\ref{Chaptertracking}). The truncated mean is then calculated as the average over the $N$ lowest values: $N = f \cdot M$. The truncation fraction $f$~=~0.55 was chosen in order to maximise the separation power between minimum ionising pions with $p$~=~0.5\gevc and electrons with $p$~=~0.7\gevc. The different momenta were chosen to maximise the statistics of the electron sample~\cite{xlu}.
However, the cluster signal strength depends on the radial position of the cluster within the read-out chamber (see Fig.~\ref{fig:ph}). Therefore, the cluster amplitudes are first weighted with time-bin dependent calibration factors, found and applied during the cpass0/cpass1 calibration steps (see Section~\ref{Chaptercalib}). For example, for the cosmic-ray data sample, the weights are determined for tracks within the interval $1.65 \le {\rm log}_{10} (\beta\gamma) \le 2.5$ to eliminate kinematic dependences. These $\beta\gamma$ are far below the onset of TR. 
After applying this procedure, some non-uniformity over time bins remains ($\pm$15\%), which is due to the TR component~\cite{xlu}. 

\begin{figure}[htb]
\centering
\includegraphics[width=0.55\textwidth]{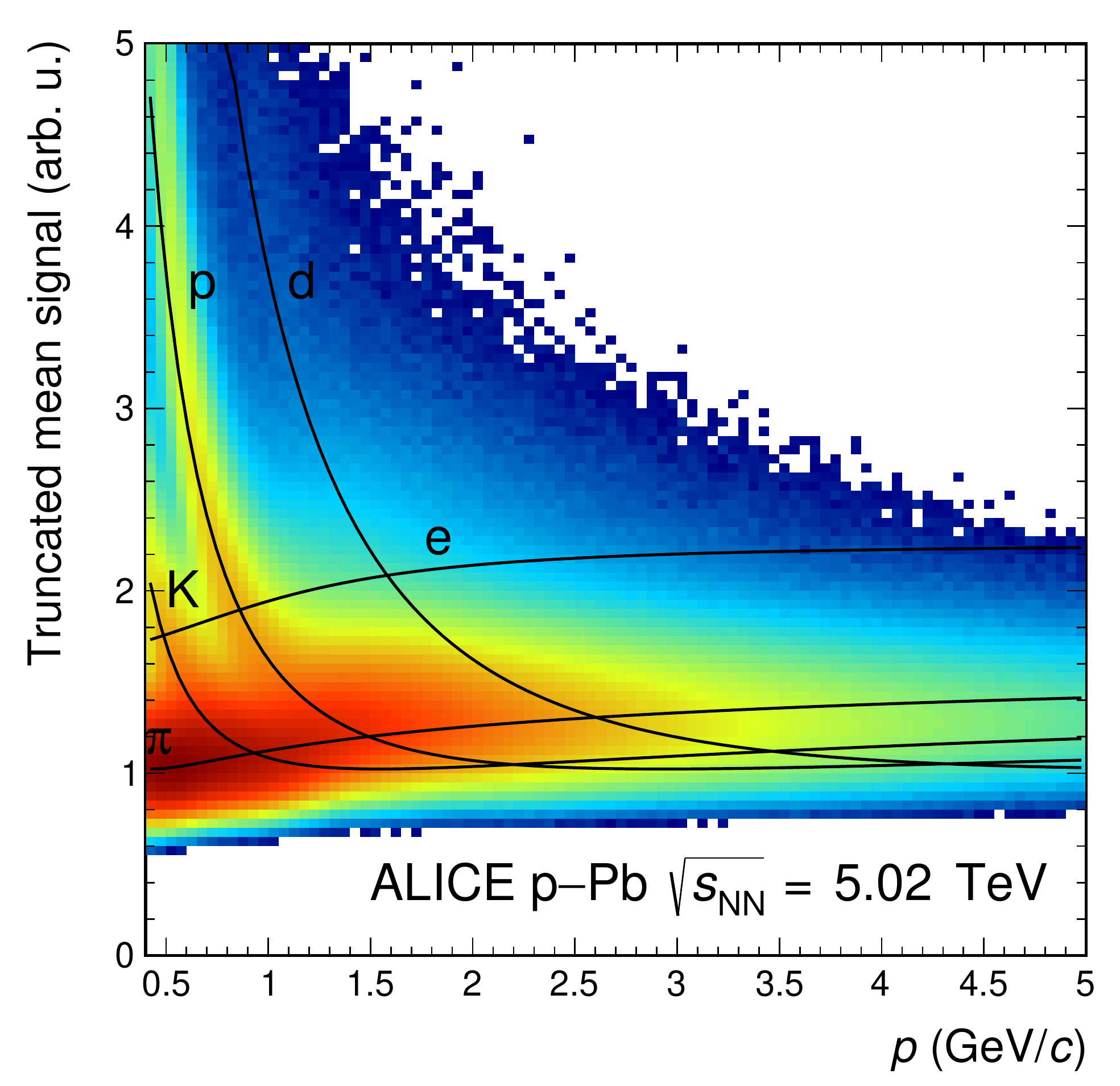}
\caption{Truncated mean signal as a function of momentum for \ppb collisions at \mbox{\sqrtsnn = 5.02 TeV}. The solid lines represent the expected signals for various particle species.}
\label{fig:TRD2dplotall} 
\end{figure}

Figure~\ref{fig:TRD2dplotall} shows the truncated mean signal as a
function of momentum for \ppb collisions at \sqrtsnn~= 5.02~TeV. The curves represent the expected signals for various particle species. These parameterisations were obtained by fitting the truncated mean signal (\dedx~+~TR) of electrons from conversion processes, pions from K$^{0}_{\rm s}$ and protons from $\Lambda$ decays
as a function of $\beta\gamma = \frac{p}{m}$ with a sum of the ALEPH parameterisation (Eq.~\ref{equ:aleph}) and logistic function (Eq.~\ref{equ:logistic}), see above.

\begin{figure}[tb]
   \centering
\includegraphics[width=.5\textwidth]{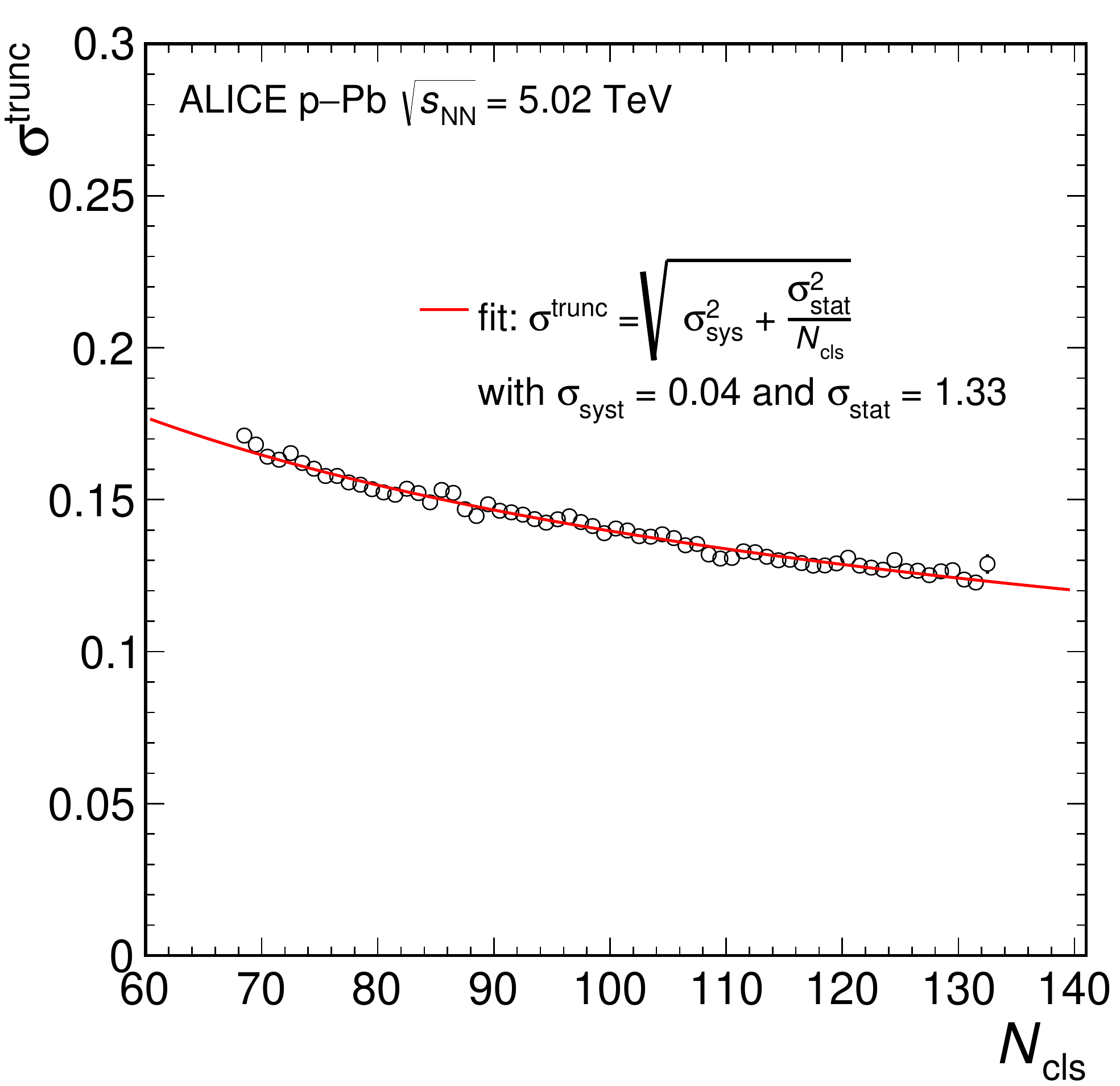} 
  \caption[]{Resolution of the truncated mean signal as a function of the number
of clusters in \ppb collisions at \mbox{\sqrtsnn~=~5.02~TeV}.}
   \label{Figure_truncmeanres}
\end{figure}

The resolution of the truncated mean signal is shown in Fig.~\ref{Figure_truncmeanres} as a function of the number of clusters ($N_{\rm cls}$), which is described by the function 
\begin{equation}
\sigma^{\rm trunc} = \sqrt{\sigma_{\rm sys}^2 + \frac{\sigma_{\rm stat}^2}{N_{\rm cls}}} ,
\end{equation}
where $\sigma_{\rm sys}$ describes systematic uncertainties due to, e.g.\ residual calibration effects.
The fit shows that the resolution is, as expected, mainly driven by a statistical scaling according to the law $\sigma^{\rm trunc} \propto 1/\sqrt{N_{\rm cls}}$. The results demonstrate a resolution of the truncated mean signal of 12\% for tracks with signals in all six layers. It should be noted that the resolution is, in parts, limited 
by the ion tails in the late time bins leading to a correlation between individual time bins (see Section~\ref{Chaptertracking}).

\begin{figure}[tb]
   \centering
\includegraphics[width=.5\textwidth]{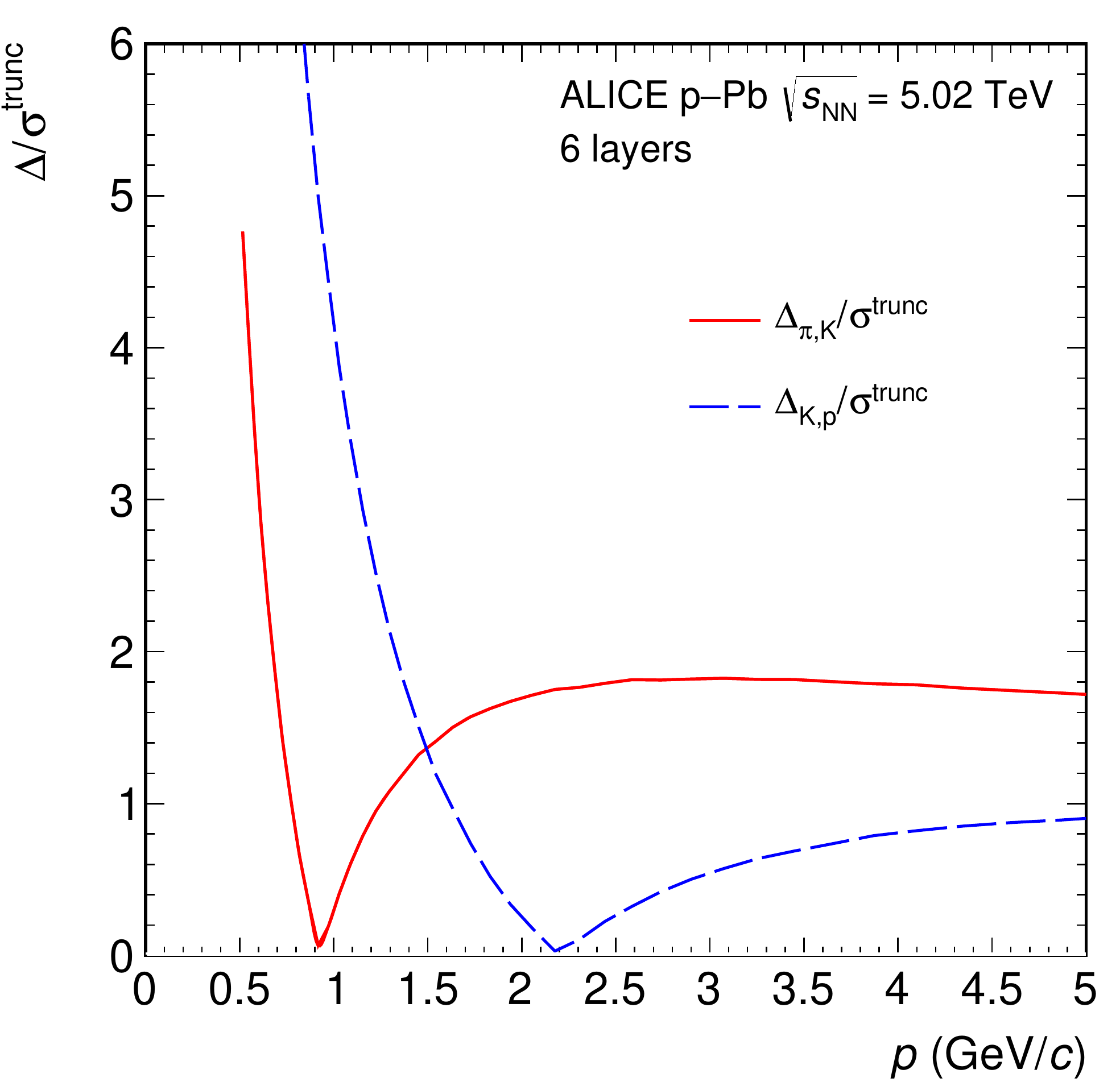} 
  \caption[]{Measured separation power ($\frac{\Delta}{\sigma^{\rm trunc}} = \frac{ S^{\rm trunc}_{ \pi \rm , K } - S^{\rm trunc}_{\rm K, p} }{ \sigma^{\rm trunc} } $) for $\pi$/K and K/p separation as a function of momentum.}
   \label{Figure_seppower}
\end{figure}

Figure~\ref{Figure_seppower} shows the pion-kaon and kaon-proton separation power as a function of momentum. The separation power is calculated as the distance between the expected truncated mean signal $S^{\rm trunc}$ of pions (kaons) and kaons (protons) divided by the resolution of the response: $\frac{\Delta}{\sigma^{\rm trunc}} = \frac{ S^{\rm trunc}_{ \pi \rm , K } - S^{\rm trunc}_{\rm K, p} }{ \sigma^{\rm trunc} }$. At low momenta an excellent separation power is achieved, at high momentum the separation power is about~2 for $\pi$/K and~1 for K/p.

\begin{figure}[bt]
   \centering
\includegraphics[width=0.5\textwidth]{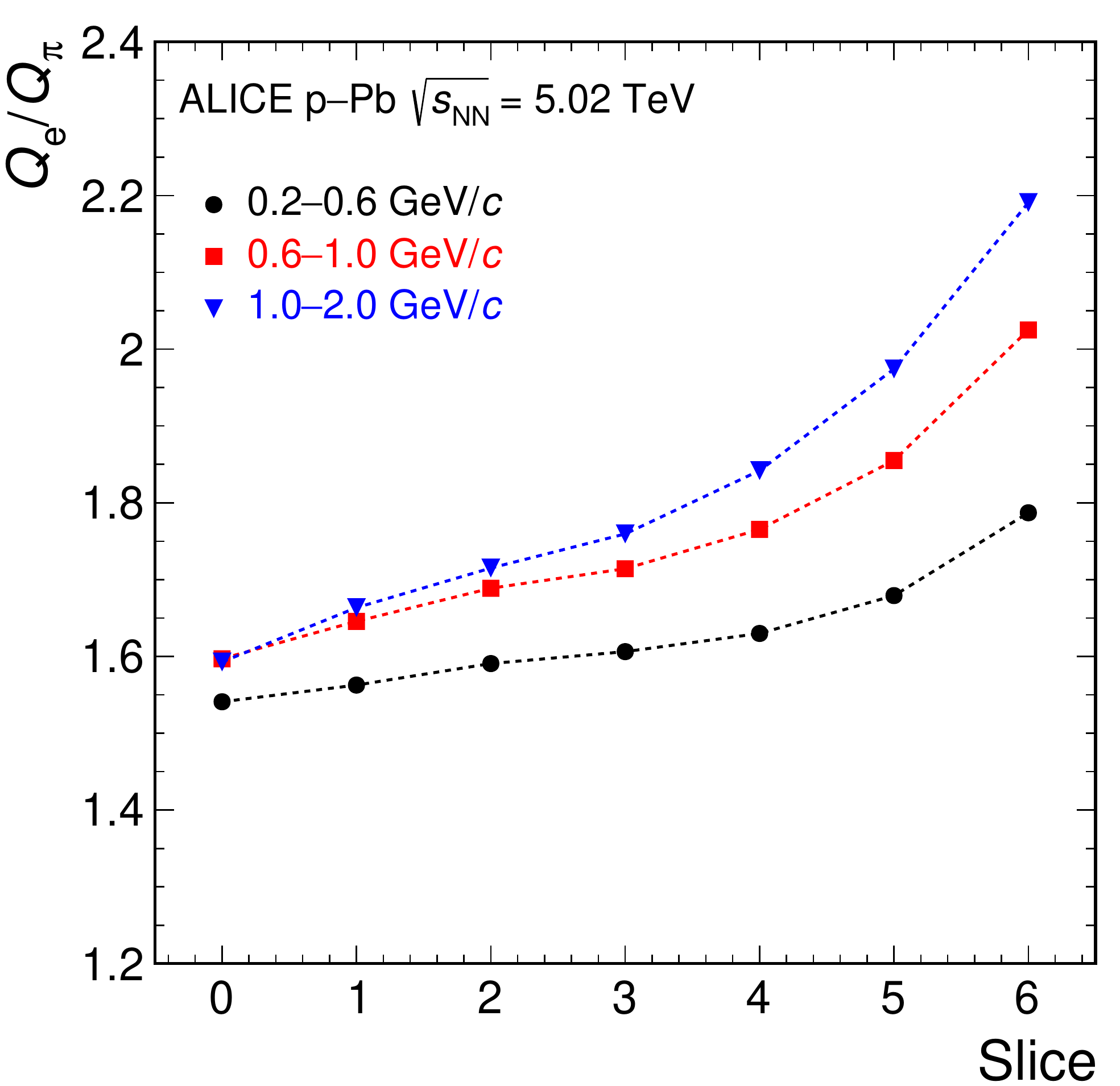} 
  \caption[]{Ratio of the average signal of electrons to that of pions 
  as a function of the depth in the detector (slice number; the lowest (highest) slice 
  number is farthest away from (closest to) the radiator).}
  \label{TRD_AvCharge}
\end{figure}

\subsection{Electron identification}

For the electron identification (eID), also the temporal evolution of the signal is used. For each TRD chamber the signal amplitudes of the clusters along a tracklet are redistributed into seven slices during the track reconstruction (see Section~\ref{Chaptertracking}). Each slice corresponds to about 5~mm of detector thickness for a track with normal incidence.
The ratio of the average signal for electrons and pions as a function of the slice number is shown in Fig.~\ref{TRD_AvCharge} for \ppb collisions at \sqrtsnn~=~5.02~TeV. At large slice numbers, i.e.\ long drift times, the TR contribution is visible because the TR photon is predominantly absorbed at the entrance of the drift region.

The eID performance is expressed in terms of the electron efficiency (the probability to correctly identify an electron) and the corresponding pion efficiency (the fraction of pions that are incorrectly identified as electrons). The inverse of the pion efficiency is the pion rejection factor.
The following methods are in use:
i) truncated mean (see previous section),
ii) a likelihood method with `dimensionality' (one-dimensional, LQ1D, corresponds to the total integrated charge~\cite{MFasel}, two-dimensional, LQ2D, for two charge bins~\cite{dlohner}, etc.),
iii) neural networks (NN)~\cite{mkroesen,aw-dok,Adler:2005rn}.

For the LQ2D method the signal is evaluated in two charge bins, i.e.\ the integrated signals of the first four slices and the last three slices are averaged. The latter sum contains most of the TR contribution. For the LQ3D method, the signals of the slices are combined as sums of the first three, the next two and the final two. Both the LQ7D and NN methods utilise 7~charge bins and thus benefit from the complete information contained in all 7~slices. While individual slices may be empty, the charge bins must contain a charge deposition. In physics analyses, this selection criterion does not introduce a loss of electrons when applying the LQ1D or the LQ2D methods, but causes a reduction in the number of electrons by about 40\% when the LQ7D method is used. The clean samples of electrons and pions described above are used to obtain references in momentum bins for particle identification. For each particle traversing the TRD, the likelihood values for electrons and pions, muons, kaons and protons are then calculated for each chamber via interpolation between adjacent momentum references. The global track particle identification is finally determined as the product of the single layer likelihood values.
In physics analyses, hadrons (e.g.\ pions) can be rejected with the TRD by applying either a cut on the likelihood or a pre-calculated momentum-dependent cut on the likelihood value for electrons. The latter provides a specified electron efficiency constant versus momentum. To cross-check the references and determine systematic uncertainties, electrons from photon conversions can be studied. 
In \pbpb collisions the mean of the charge deposit distributions shows a centrality (event multiplicity) dependence, of about 15\% comparing central and peripheral collisions~\cite{dlohner}, and therefore centrality-dependent references were introduced. 

The references can only be created after the relative gain calibration of the individual pads and the time-dependent gain calibration of the chambers as described in Section~\ref{Chaptercalib}. After this, the detector response is uniform across the acceptance and in time, and thus it can be studied in detail by combining 
all chambers and the full statistics of 1--2 months of data taking. 
Since the reference creation requires a large data sample, the reference distributions are only 
produced after the full physics reconstruction pass. This means that the reference creation can 
only be done later, during data analysis rather than already during reconstruction. 
The references for the truncated mean and the likelihood methods are stored for this purpose in the Offline Analysis Database (OADB) and read from there in the initialisation phase of the analysis tasks~\cite{ALIOFFLINE}.

\begin{figure}[bt]
\includegraphics[width=.47\textwidth]{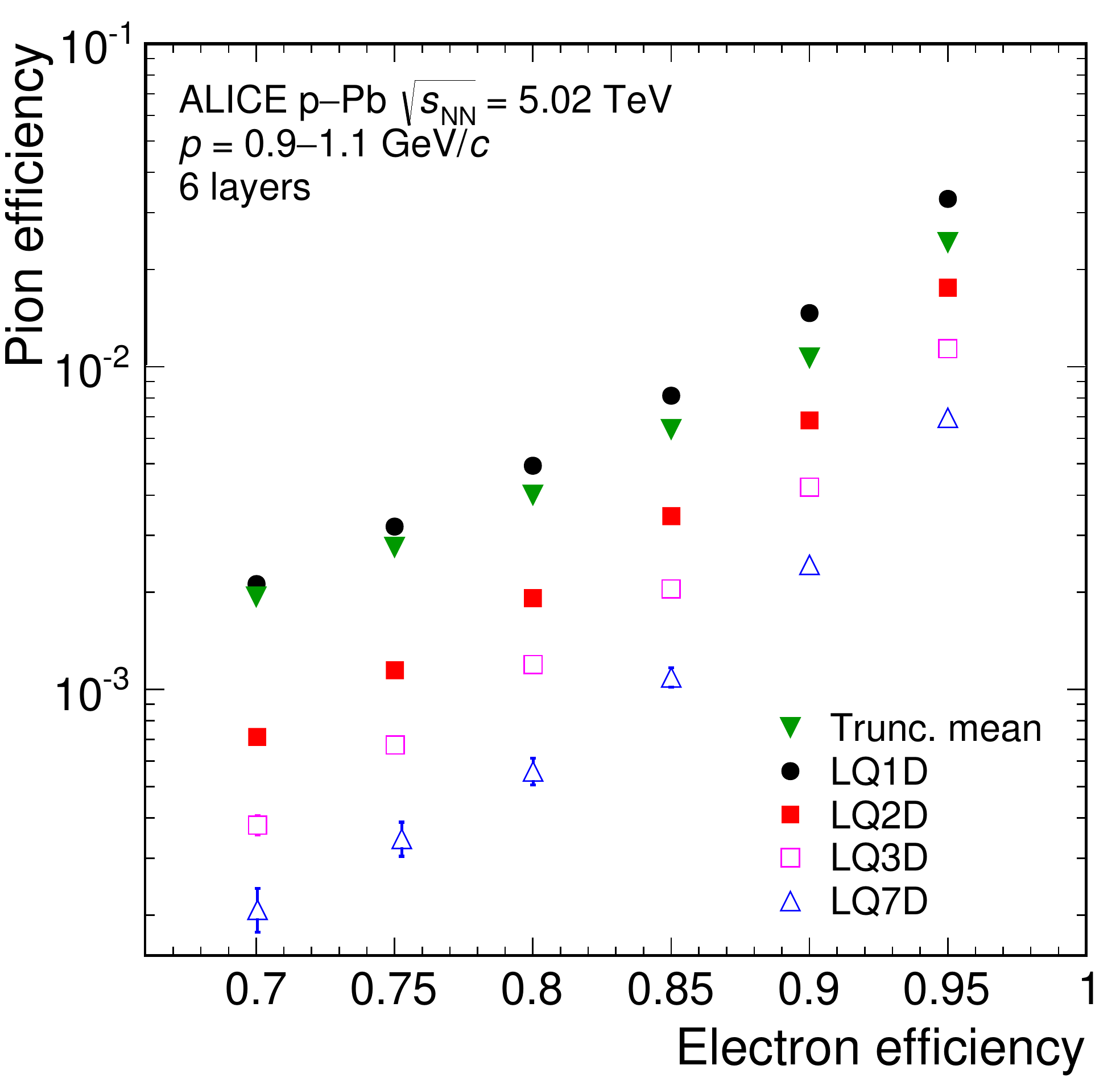} 
\hspace{\fill}
\includegraphics[width=.47\textwidth]{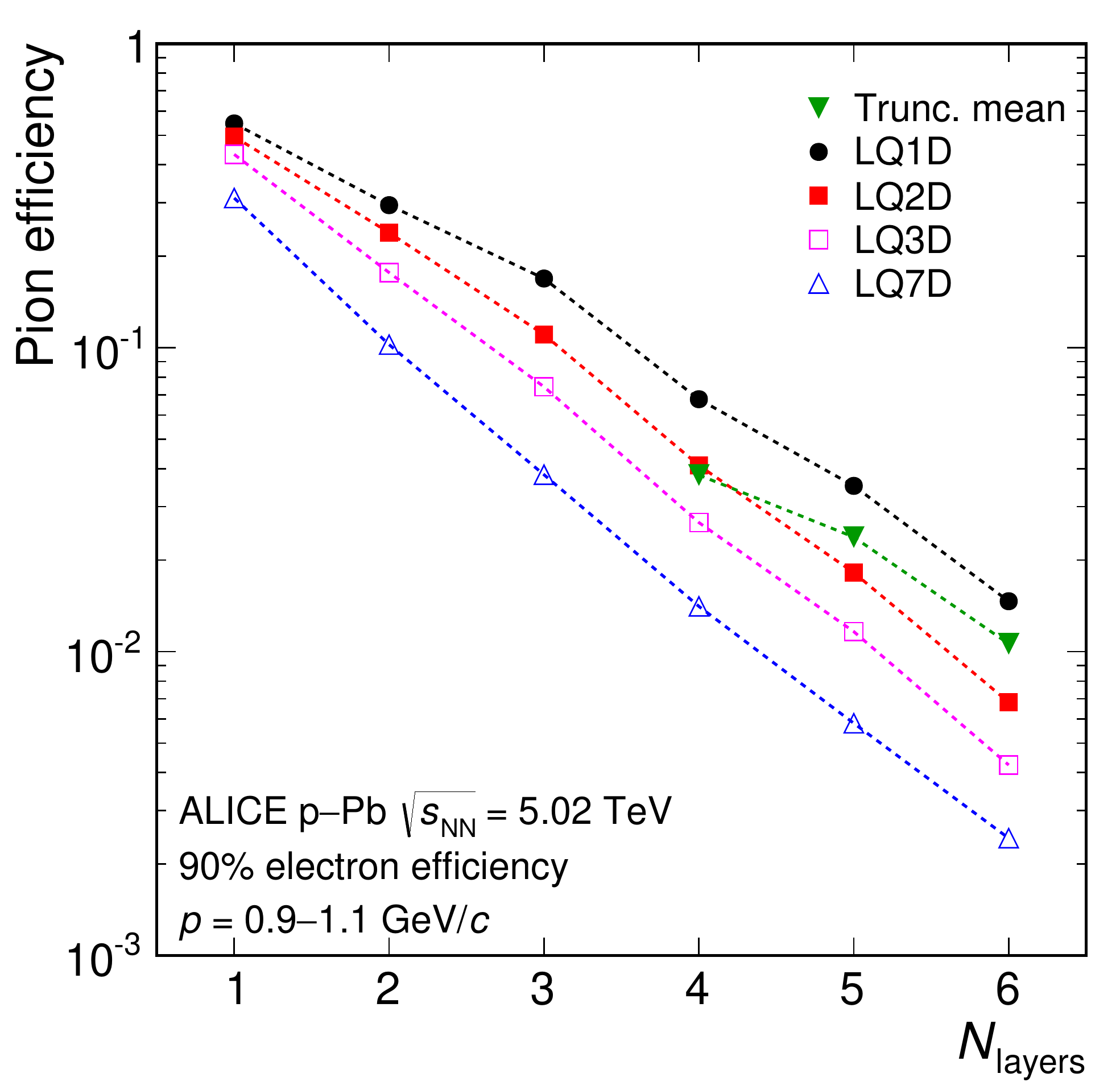} 
  \caption[]{Pion efficiency as a function of electron efficiency (left, 
for 6~detector layers) and as a function of the number of detector layers (right, for 90\% 
electron efficiency) for the various eID methods.
The results are compared for the momentum interval 0.9--1.1\gevc in \ppb collisions at \sqrtsnn~=~5.02~TeV. The results of the truncated mean method are only shown for a minimum of 4~tracklets, where the resolution is better than 18\% (see Fig.~\ref{Figure_truncmeanres}). }
   \label{TRD_PID}
\end{figure}

The pion efficiency for 1\gevc tracks is shown as a function of the electron efficiency and as a function of the number of detector layers providing signals for the various methods in Fig.~\ref{TRD_PID}. 
For all methods the pion rejection factor decreases as expected with decreasing number of contributing layers and 
a lower electron selection efficiency corresponds to a better pion rejection factor for all methods. 

A pion rejection factor of about~70 is obtained at a momentum of 1\gevc in \ppb collisions with the LQ1D method, the most simple identification algorithm. The LQ2D method yields a pion rejection factor far better than the design goal of 100 at 90\% electron efficiency found in test beams with prototypes~\cite{Bailhache:2006hs}. When using the temporal evolution of the signal even better performance is achieved, reaching a rejection of up to 410.

\begin{figure}[tb]
   \centering
\includegraphics[width=.5\textwidth]{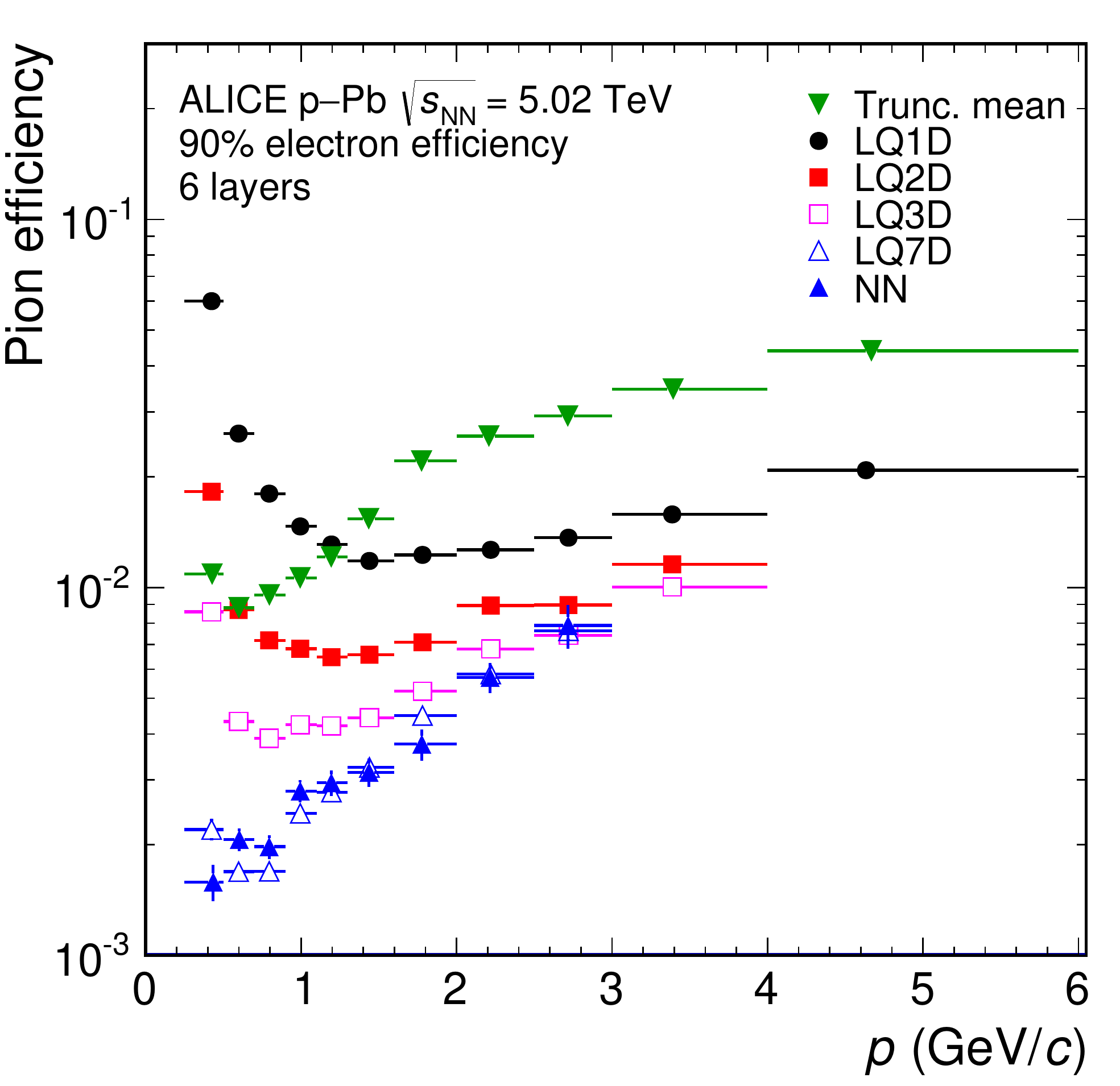} 
  \caption[]{Pion efficiency (for 90\% electron efficiency) as a function of momentum for the truncated mean, LQ1D, LQ2D, LQ3D, LQ7D and NN methods. The results are from \ppb collisions at \mbox{\sqrtsnn~=~5.02~TeV} and for tracks with signals in six layers.}
   \label{TRD_PIDvsmom}
\end{figure}

Figure~\ref{TRD_PIDvsmom} shows the momentum dependence of the pion efficiency for the different methods.
At low momenta, the pion rejection with the LQ1D method improves with increasing 
momentum because of the onset of the transition radiation. From 1--2\gevc upwards, the electron--pion separation power gradually decreases due to the saturation of the TR production and the relativistic rise of the specific energy loss of pions.
The other methods that make use of the temporal evolution of the signal provide substantial improvements, in particular for low and intermediate momenta. At high momenta (beyond 2\gevc), the limitation in statistics for the reference distributions is reflected in the rather modest improvements in the pion rejection in the muti-dimensional methods. The similar momentum-dependent shape of the likelihood methods is in parts due to the usage of the same data sample for reference creation. The best performance is achieved for the LQ7D and NN methods. However these methods are sensitive to a residual miscalibration of the drift velocity, while the truncated mean and LQ1D method are more robust against small miscalibration effects. 
At low momentum, where the energy loss dominates the signal, the truncated-mean method provides very good pion rejection. The rejection power of the method decreases at higher momenta, because the TR contribution, yielding higher charge deposits, is likely to be removed in the truncation~\cite{xlu}.

\begin{figure}[bt]
\centering
\includegraphics[width=0.6\textwidth]{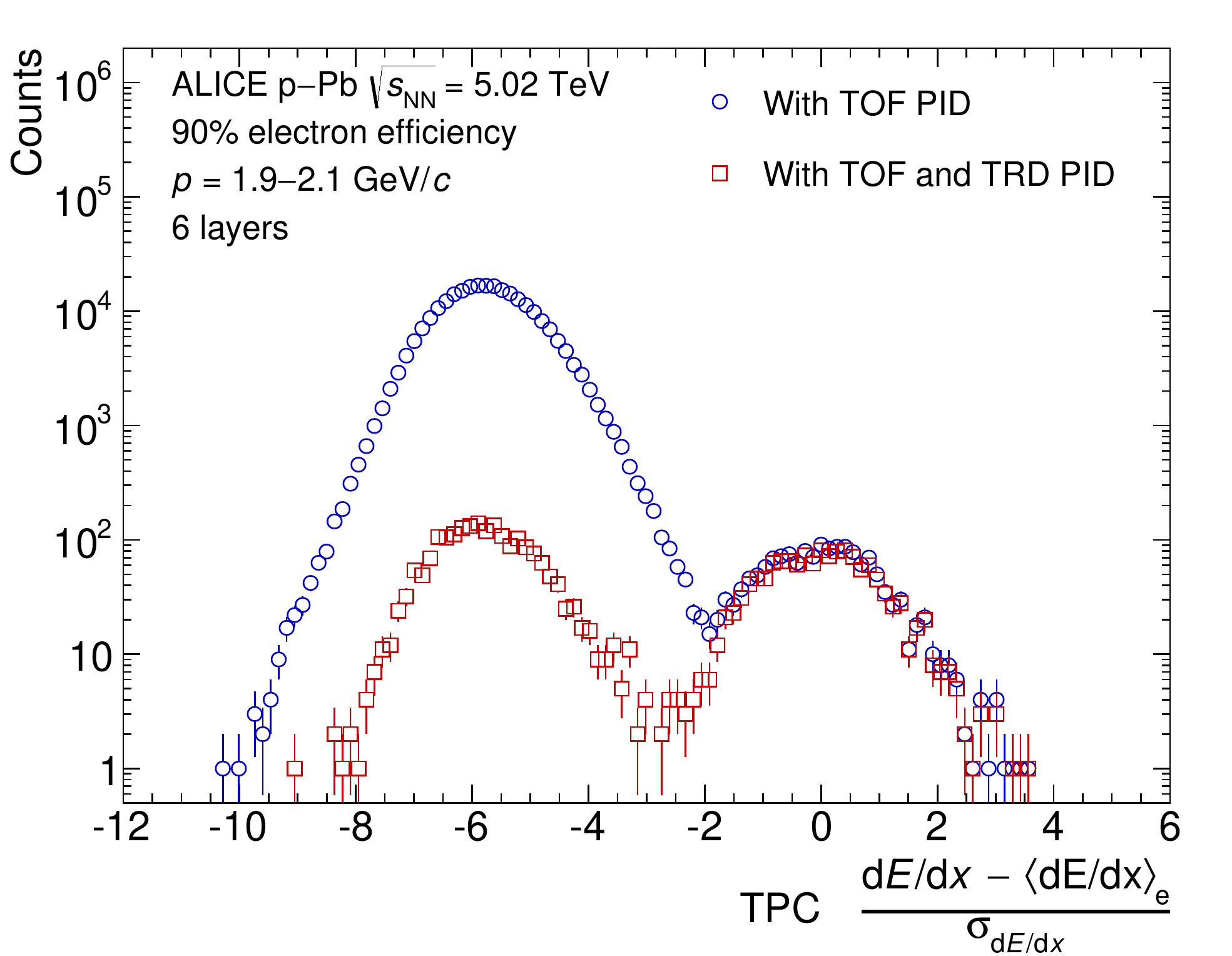}
\caption{\label{fig:combPIDelectrons}Difference in units of standard deviations between the measured TPC energy loss of a given track and the expected energy loss of an electron with TOF ($\pm 3 \sigma^\mathrm{TOF}_\mathrm{e}$) and TRD (90\% electron efficiency) electron identification. The distributions are shown for tracks with a momentum of 1.9--2.1\gevc within the TRD acceptance (6~layers in the TRD) in \ppb collisions at \sqrtsnn~=~5.02~TeV.}
\end{figure}

To visualise the strength of the TRD LQ2D electron identification method, the difference in units of standard deviations between the measured TPC energy loss of a given track and the expected energy loss of an electron for tracks with TOF and TOF+TRD particle identification is shown in Fig.~\ref{fig:combPIDelectrons}. 
The results are compared for tracks with a momentum of 1.9--2.1\gevc within the TRD acceptance. In this momentum interval electrons cannot be discriminated from pions using TOF-only electron identification. 
After applying the TRD electron identification with 90\% electron efficiency with the LQ2D method, hadrons are suppressed by about a factor of 130. The electron identification capabilities of the TRD thus allow selecting a very pure electron sample. This is important, e.g. for the measurement of electrons from heavy-flavour hadron decays.
Details on the usage of the electron identification for the latter measurement in pp collisions at $\sqrt{s}$~=~7~TeV can be found in~\cite{Abelev:2012xe}. 

In the Bayesian approach within ALICE~\cite{ALICEBayesPID}, where the identification capabilities of several detectors are combined, the TRD particle identification contributes with its estimate of the probability for a given particle to belong to a given species. For this purpose, transverse momentum dependent `propagation factors' for the priors, which represent the expected abundance of each particle species within the ITS and TPC acceptance, are calculated and stored in the analysis framework.

\section{Trigger}
\label{Chaptertrigger}

ALICE features a trigger system with three hardware levels and a
HLT farm~\cite{Abelev:2014ffa}. 
Apart from the contributions from the pretrigger system (see Section~\ref{sec:pt_ov}), the TRD contributes to physics triggers at level-1. These are based on tracks reconstructed
online in the GTU (see Section~\ref{sec:gtu}). The reconstruction is based on online tracklets (track segments corresponding to one read-out chamber) that are calculated locally in the FEE of each chamber. The local tracking in the FEE and the global online tracking in the GTU are discussed in the following.

As the trigger decision is based on individual tracks, a variety of
signatures can be implemented, only limited by the complexity of the
required calculations and the available time. In the following, the
triggers on cosmic-ray muons, electrons, light nuclei, and jets are discussed.

\begin{figure}[tb]
  \centering
  \includegraphics[width=.6\textwidth]{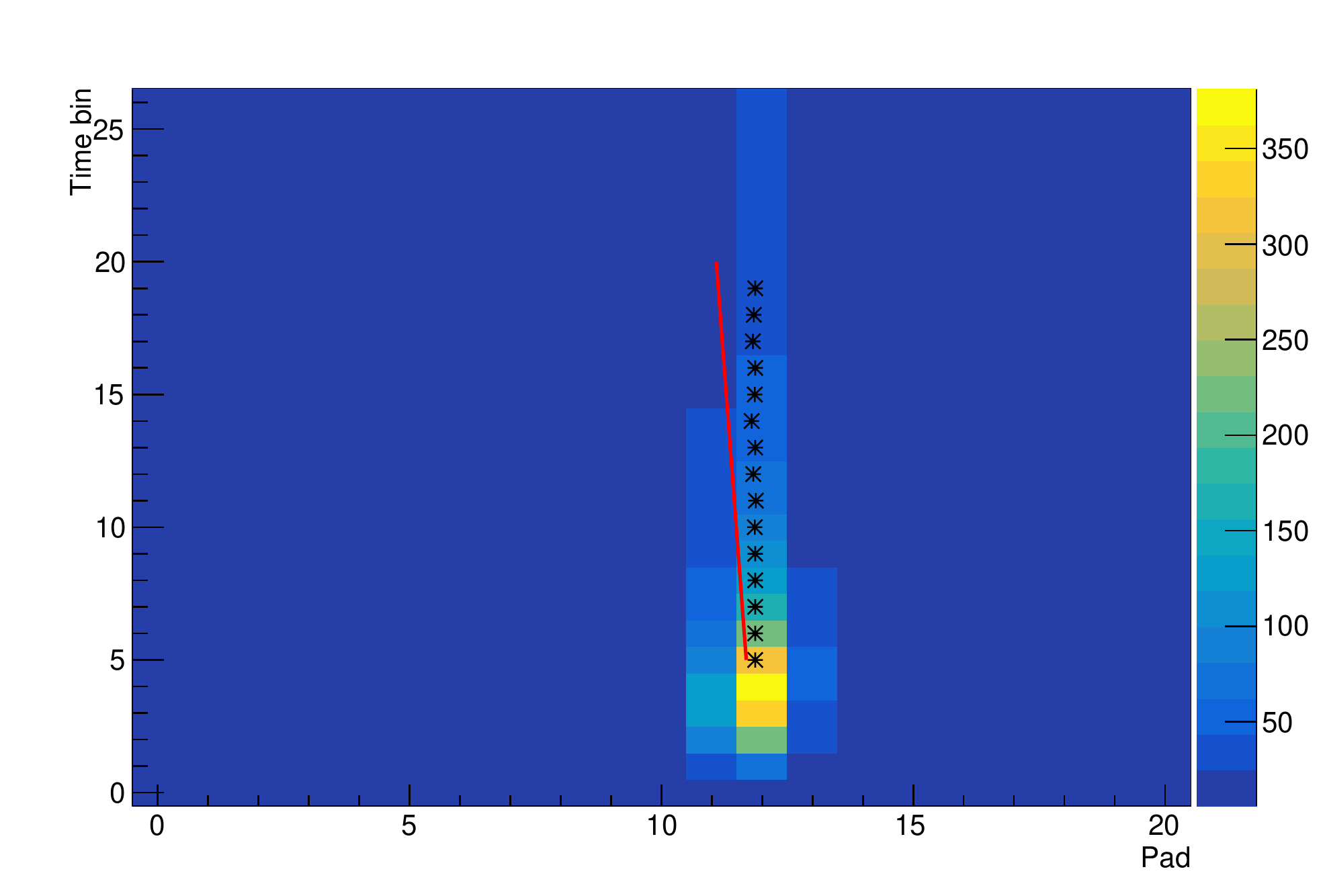}
  \caption{Example tracklet in one MCM. The ADC data for 26~time bins (100~ns each) from 21~channels are shown. The found clusters are marked as asterisks and the final tracklet, calculated as a straight line fit through the clusters, with Lorentz correction as a red line. 
}
 \label{fig:trkl_reco_evview}
\end{figure}

\subsection{Local online tracking}
\label{sec:local_online_tracking}

The local online tracking is carried out in parallel in the FEE (see
Section~\ref{sec:fee}). Each of the \num{65000}~MCMs processes data
from 21~pads, 3~of which are cross-fed from the neighbouring chips to
avoid inefficiencies at the borders of the chip (see
Fig.~\ref{fig:mcm_trkl}). For accurate online tracking, all relevant corrections and calibration
steps must be applied online. After appending two digits to avoid
rounding imprecisions, the digitised data are propagated through a
chain of filters. First, a pedestal filter is used to compensate for
variations in the baseline. A gain filter makes it
possible to correct for local gain variations, either caused by the
chamber or by the electronics. This equilibration is important for the
evaluation of the specific energy loss, which is used for online
particle identification. It uses correction factors derived from the
krypton calibration (see Section~\ref{sec:cal:kr}). A tail cancellation filter can be used to reduce the bias from ion tails of signals in preceding time bins. This improves the reconstruction of the radial cluster positions and of the deflection in the transverse plane.
The offline reconstruction takes the already applied
online corrections into account. For that purpose, all configuration
settings are stored in the OCDB and are, therefore, known during the
offline processing.

After the filtering, the data for one event are searched time bin-wise
for clusters by a hardware pre-processor. A cluster is found if the
charge on three adjacent pads exceeds a configurable threshold and the
center channel has the largest charge (see
Fig.~\ref{fig:trkl_reco_evview}). For each MCM and time bin,
transverse positions are calculated for up to six clusters. They are
used to calculate and store the (channel-wise) sums required for a linear
regression.

After the processing of all time bins, up to four channels with a
minimum number of found clusters are further processed (if more than
four channels exceed the threshold, the four of them with the largest number
of clusters are used). For the selected channels, a straight line fit
is computed from the pre-calculated sums. The fit results in information on the local transverse position $y$, the deflection in the bending plane $d_y$, the longitudinal position $z$,
and a PID value. The transverse position and deflection are calculated
from the fit, the longitudinal position is derived from the MCM
position, and the PID from a look-up table using the
accumulated charge as input.

\begin{figure}[tb]
  \centering
  \includegraphics[width=.6\textwidth]{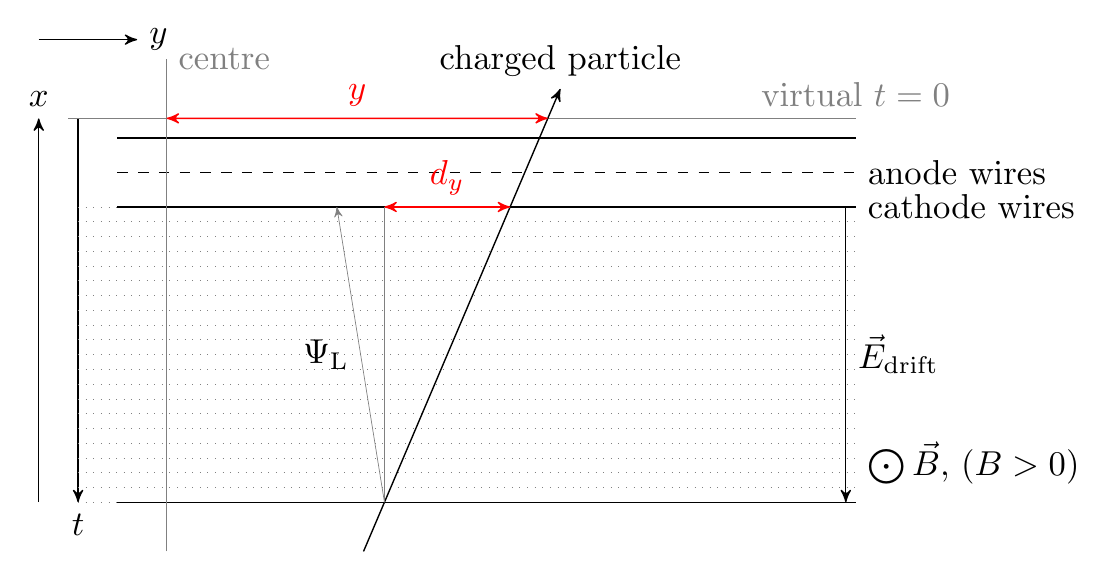}
  \caption{Sketch of the tracklet reconstruction. The tracklet reconstruction in the MCMs is performed in a
    local coordinate system. The tracklet comprises the information on
    $y$, $d_y$, $z$ and PID. The magnetic and electric field and the effect of the Lorentz angle ($\Psi_{\rm L}$) are indicated as well.}
  \label{fig:trkl_reco}
\end{figure}

The reconstructed values for $y$ and $d_y$ are corrected for 
systematic shifts caused by the Lorentz drift and the pad tilt. An example of a reconstructed tracklet is shown in
Fig.~\ref{fig:trkl_reco}. Eventually, the values (in fixed-point
representation) are packed into one 32-bit word per tracklet for
read-out.

\begin{figure}[tb]
  \centering
  \includegraphics[width=.6\textwidth]{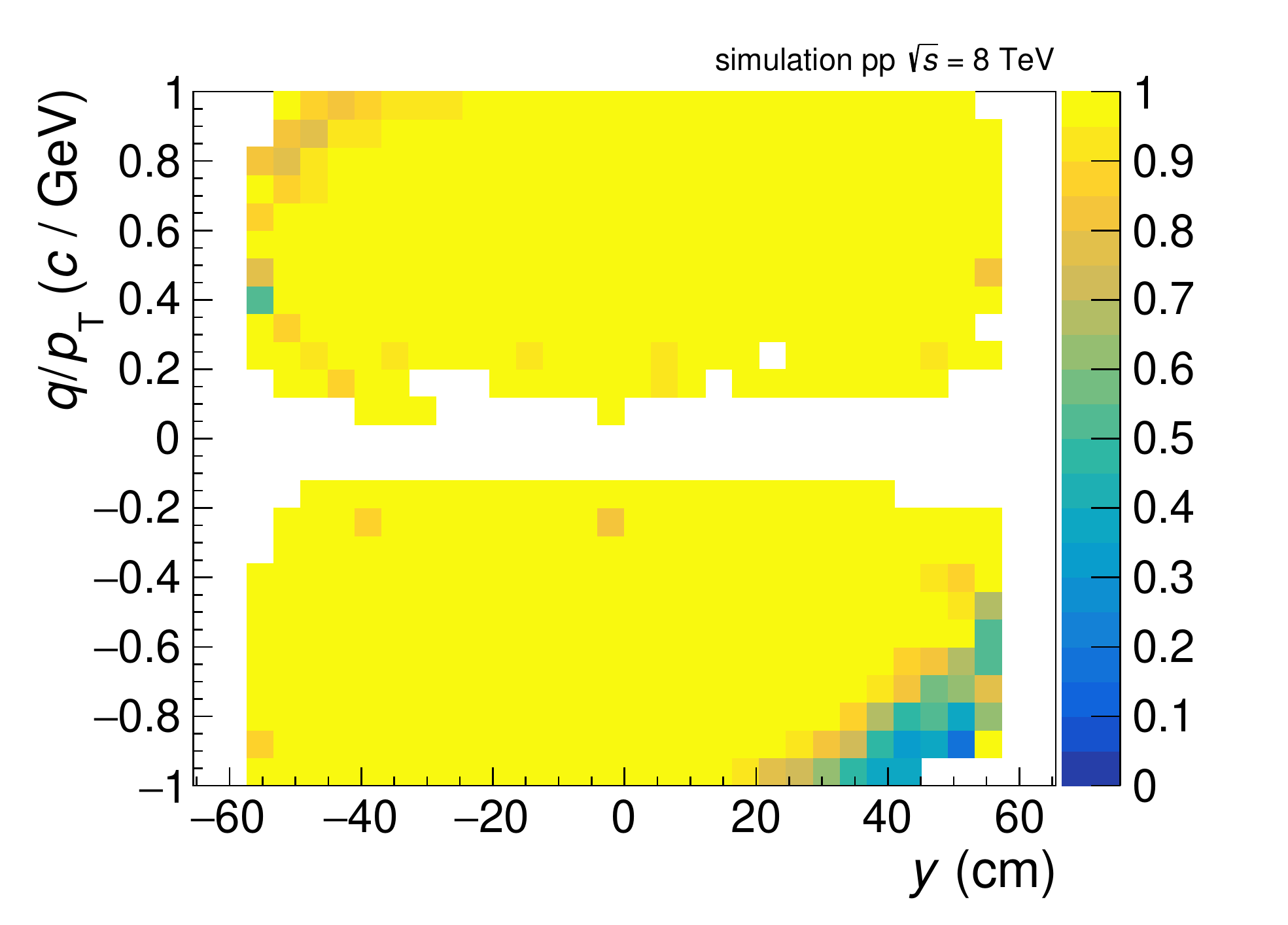}
  \caption{Reconstruction efficiencies for tracklets as
    a function of $y$ and $q/\pt$ for Monte Carlo simulations. The $z$-axis entries are zero-suppressed.
   }
  \label{fig:trkl_eff_mc}
\end{figure}

A realistic simulation of the local tracking was implemented in
the ALICE software framework and is used in Monte Carlo productions based on event
generators but can also be run on data recorded with the actual
detector. This allows cross-validating hardware and simulation,
and to study the effect of parameter changes on the tracklet
finding. Therefore, Monte Carlo simulations are well-suited to study the
performance of the online tracking algorithm with a given set of
configuration options since tracklets can be compared to track
references (track positions from Monte Carlo truth information). This allows tracklet efficiencies to be determined. An example is displayed in Fig.~\ref{fig:trkl_eff_mc}, which shows the efficiency of the
tracklet finding process for a typical set of parameters as a function
of $y$ and $q/\pt$. The efficiency drops for large $y$ and negative
$q/\pt$, where the asymmetry in $y$ is caused by a combination of the Lorentz correction
and the numerical range available for the deflection. The efficiency
is close to 100\% in the regime relevant for triggering. Furthermore,
shifts in $y$ and $d_y$ are calculated with respect to the expectation
from the Monte Carlo information. Besides a small systematic shift
because of the uncorrected misalignment, the distributions show widths
of about \SI{300}{\micro\metre} and \SI{1700}{\micro\metre} in $y$ and
$d_y$~\cite{Klein:2014rxa}, respectively.

\subsection{Global online tracking}

The global online tracking in the GTU operates stack-wise on the tracklets reconstructed and transmitted by the FEE. It is divided into a track matching
and a reconstruction stage. The algorithm used for the matching of the
tracklets is optimised for the high multiplicity environment of \pbpb
collisions~\cite{Cuveland:Diploma03}. It is implemented in the FPGAs
of the GTU (see Section~\ref{sec:gtu}) and operates in parallel on
subsets of tracklets that are compatible with a track in the $x$-$z$
plane. Groups of tracklets which fall into `roads' pointing to the
nominal primary vertex are pre-selected. The tracklets are propagated
to a virtual plane in the middle of the stack. Those which are close
enough on this plane are considered to belong to the same track. The
algorithm exploits a fixed read-out order of the tracklets to limit
the number of comparisons for the matching, meaning that a linear scaling
of the tracking time with the number of tracklets can be achieved.

\begin{figure}[bt]
  \centering
  \includegraphics[width=.5\textwidth]{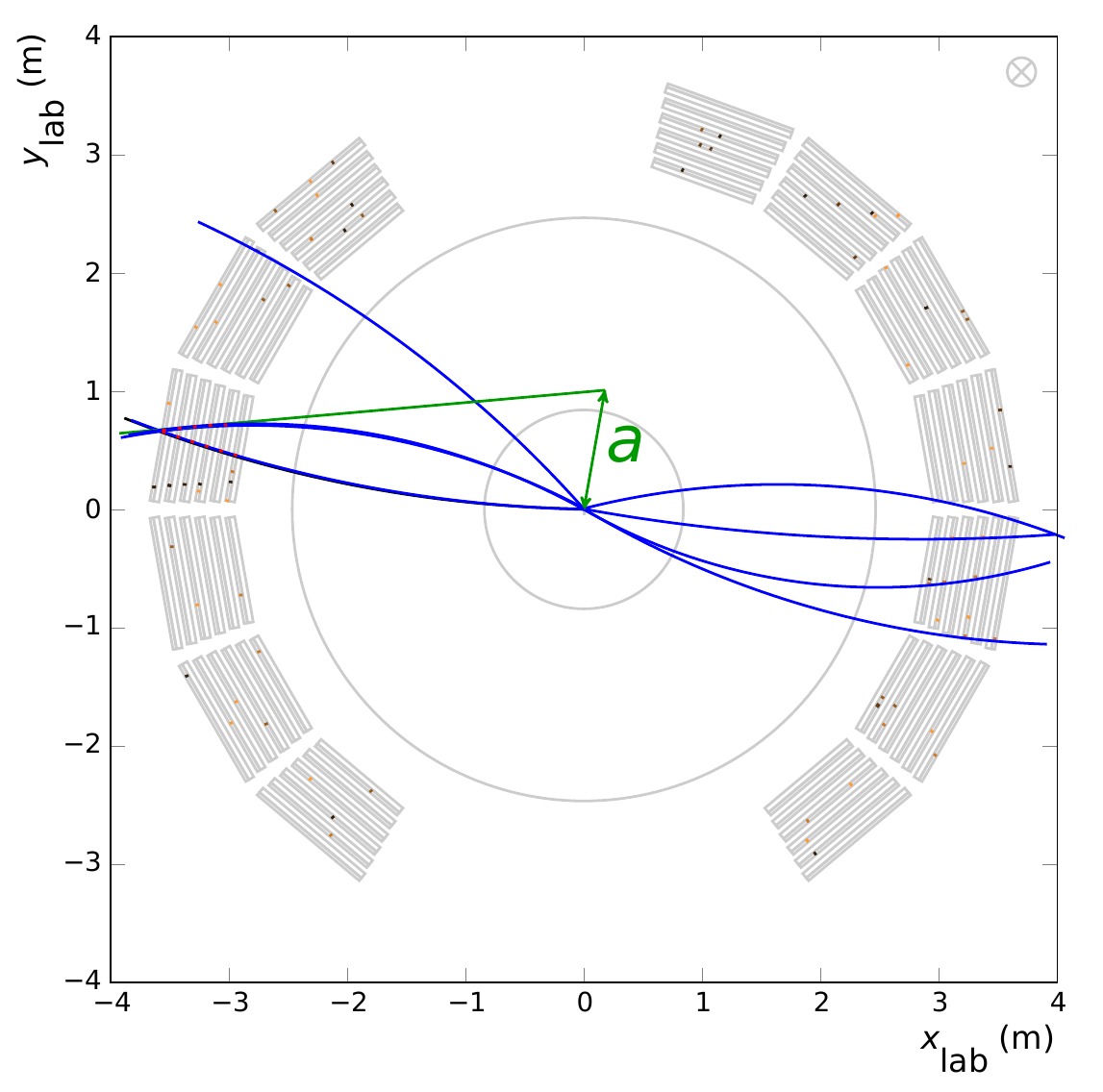}
  \caption{Event display showing the tracks available for the
  level-1 trigger from the online reconstruction (green) in comparison
  with helix fits to the contributing tracklets (blue). The offset $a$
  from the primary vertex used as measure for $1/\pt$ is shown as well. The colour coding of the tracklets (small boxes) is according to stacks.}
  \label{fig:gtu_evview}
\end{figure}

Global online tracks consist of at least four matching tracklets. The reconstruction stage uses the positions of the contributing tracklets to calculate a straight line fit (see Fig.~\ref{fig:gtu_evview}). The
computation is simplified by the use of pre-calculated and tabulated
coefficients, which depend on the layer mask. The approximation of a
straight line is adequate for the trigger-relevant tracks above
2\gevc. The transverse offset $a$ from the nominal vertex position
is then used to estimate the transverse momentum~\cite{Cuveland:Diploma03}. The PID value for
the track is calculated as the average over the contributing
tracklets. A precise simulation of all the tracking steps was
implemented and validated in AliRoot. It was used for systematic
studies of the tracking performance, see below.

\begin{figure}[tb]
  \centering
  \includegraphics[width=.7\textwidth]{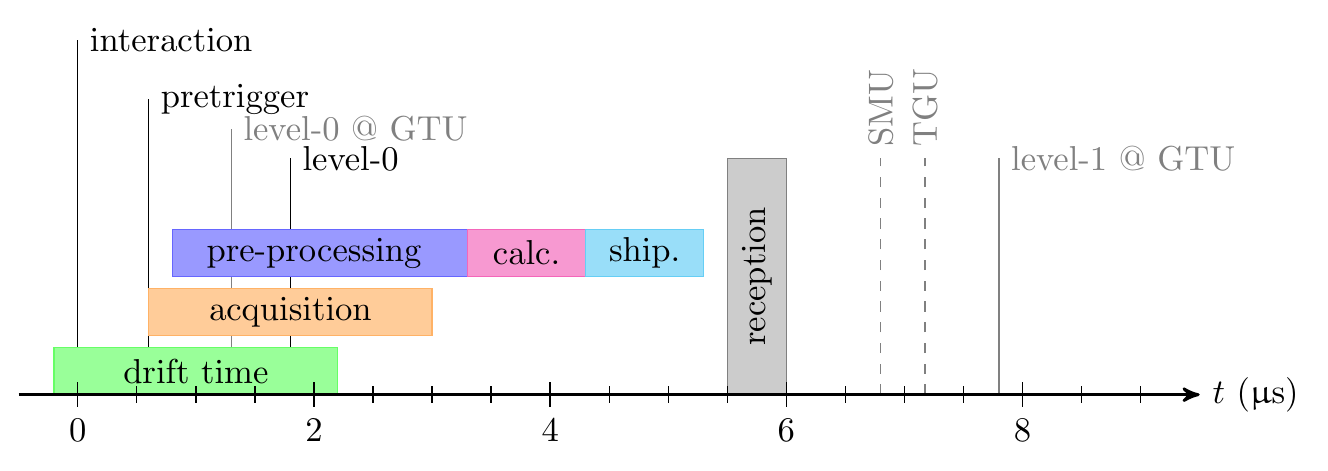}
  \caption{Timing of the various phases for the online tracking with respect to the interaction. 
  }
  \label{fig:trg_timing}
\end{figure}

Figure~\ref{fig:trg_timing} shows the timing of the online tracking together
with the constraints for the trigger contributions. Between
interactions, the FEE is in a sleep mode~\cite{Angelov:2006hu}. In this mode only the ADCs, the
digital filters, and the pipeline stages are active. The latter makes it
possible to process the data from the full drift time upon arrival of
a wake-up signal (see Section~\ref{sec:pt_ov}). The processing can be
aborted if it is not followed by a level-0 trigger. In this case, a
clear sequence is executed for resetting and putting the FEE back to
sleep mode. If a level-0 trigger was received, processing continues
and the tracklets are sent to the GTU. Here, the track matching and
reconstruction runs as the tracklets arrive. The tracks are used to
evaluate the trigger conditions (see next sections) until the
contribution for the level-1 trigger must be issued to the CTP
(about \SI{6}{\micro\second} after the level-0 trigger). The tracking
can continue beyond the contribution time for the trigger;
the resulting tracks are ignored for the decision but are available
for offline analysis (flagged as out-of-time). 

\begin{figure}[bth]
  \centering
  \includegraphics[width=.6\textwidth]{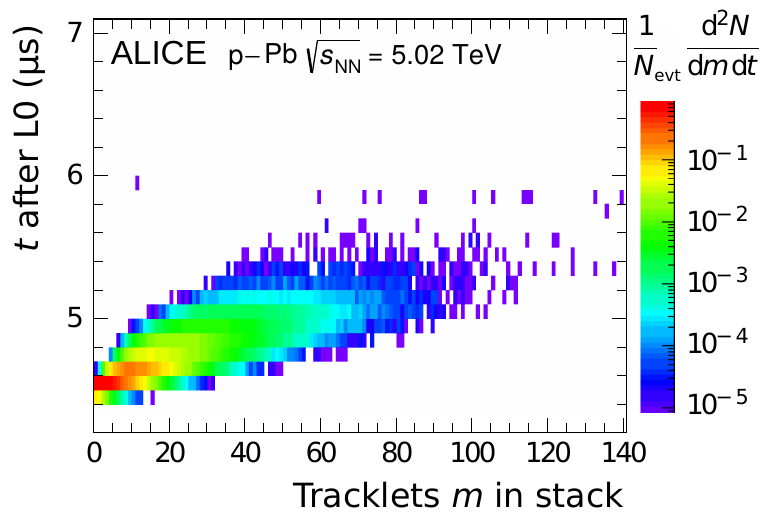}
  \caption{Dependence of the time required for the global online tracking on
    the tracklet multiplicity in a single stack.
    }
  \label{fig:trk_time_scaling}
\end{figure}

Figure~\ref{fig:trk_time_scaling} shows the
tracking time measured during data taking in p--Pb collisions. It shows the
expected linear scaling with the number of tracklets.

\begin{figure}[tb]
 \centering
  \includegraphics[width=.6\textwidth]{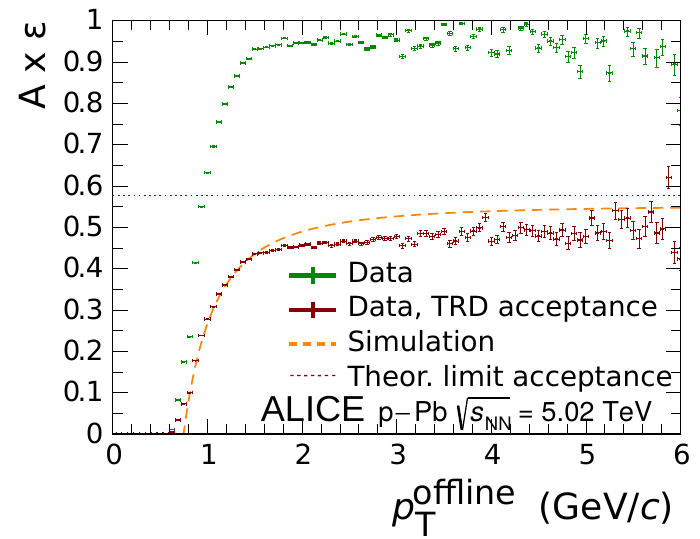}
  \caption{Acceptance times efficiency of the global online tracking for primary tracks (Data) and tracks in the detector acceptance (Data, TRD acceptance) as function of the transverse momentum of the global offline track (trigger threshold at
    2--3\gevc). The results of an ideal simulation, not considering non-operational parts of the real detector, are drawn for comparison. The dotted line shows the theoretical limit of the acceptance with $13$
    out of $18$ supermodules installed during the \ppb data taking period in \run{1}. }
  \label{fig:gtu_eff}
\end{figure}

The efficiency of the global online tracking is shown in
Fig.~\ref{fig:gtu_eff}. In order to separate the efficiency of the
online tracking from the acceptance and geometrical limitations, the
normalisation is done once for all primary tracks and once for those
which are findable, i.e.\ which have at least 4~tracklets assigned in
one stack in the offline tracking (TRD acceptance). The efficiency starts to rise
at about 0.6\gevc, reaches half of its asymptotic value at 1\gevc, and
saturates above about 1.5\gevc. Lower transverse momenta are not
relevant for the trigger operation and corresponding tracks are
suppressed at various stages. For comparison, the curve obtained from
an ideal Monte Carlo simulation shows slightly higher efficiencies.
The difference is caused by non-operational parts of the real detector (see Section~\ref{Chaptercomminbeam}) not being reflected in the ideal simulation.

\begin{figure}[tb]
 \centering
  \includegraphics[width=.6\textwidth]{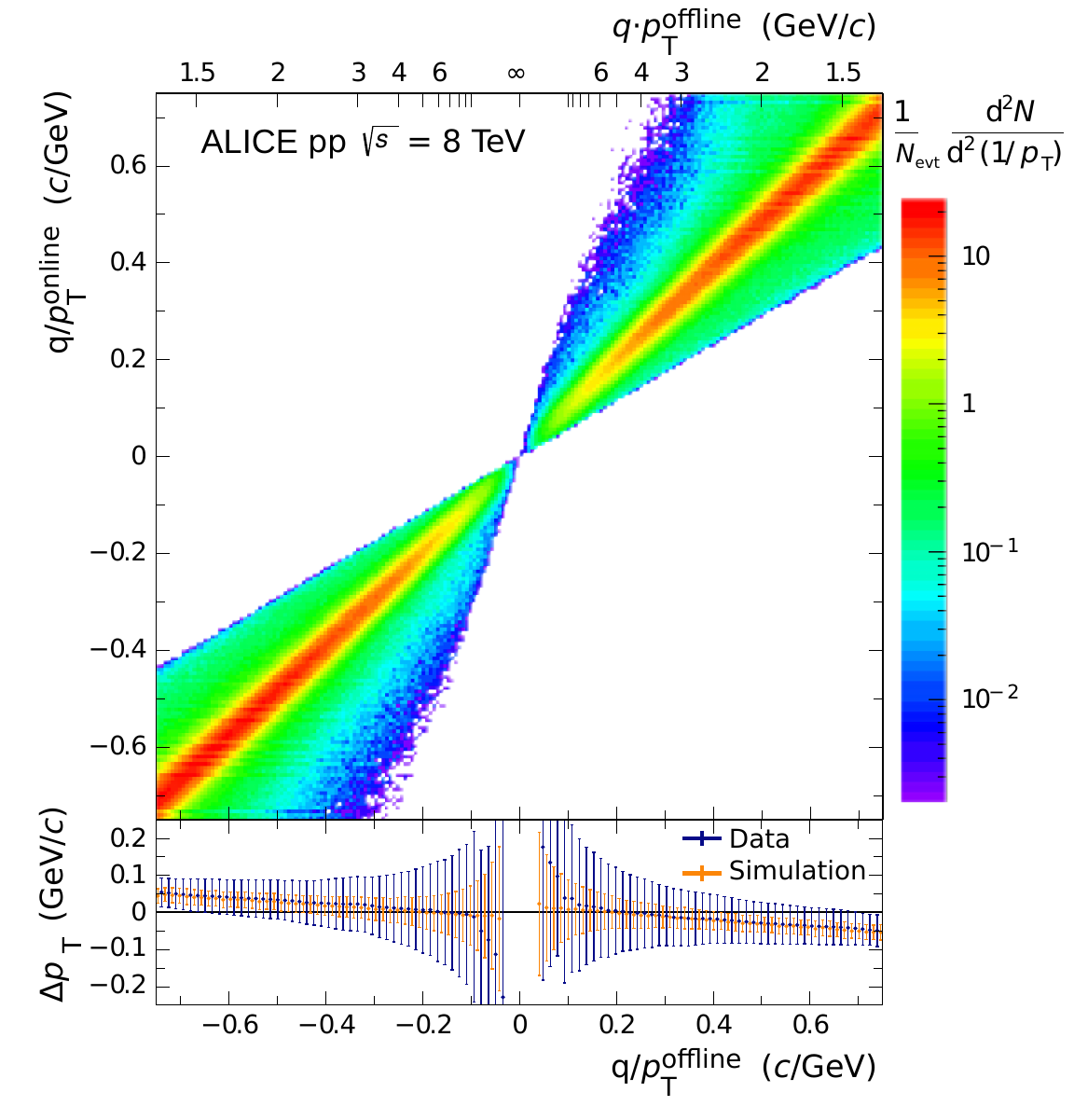}
  \caption{Top: Correlation between $1/p_\textrm{T}$ obtained from
    the online tracking and from a matched offline track for pp collisions at $\sqrt{s} = 8~\mathrm{TeV}$. Bottom: Difference (points) of the online and offline track \pt for data and simulation. The error bars indicate the corresponding width of the difference in $p_\textrm{T}$.
    }
  \label{fig:gtu_corr_a_pp}
\end{figure}

The correlation of the inverse transverse momentum from online and
offline tracking is established by matching global online tracks to global
offline tracks, reconstructed with ITS and TPC, based on a geometrical distance measure. An example for \pp
collisions at $\sqrt{s} = 8~\mathrm{TeV}$ is shown in Fig.~\ref{fig:gtu_corr_a_pp}. The online estimate correlates well with
the offline value in the transverse momentum range relevant for the
trigger thresholds, i.e.\ 2--3\gevc. The width of the correlation
corresponds to an online measured resolution of about 10\% for momenta
of $1.5-5\gevc$.

\begin{figure}[tb]
 \centering
  \includegraphics[width=.6\textwidth]{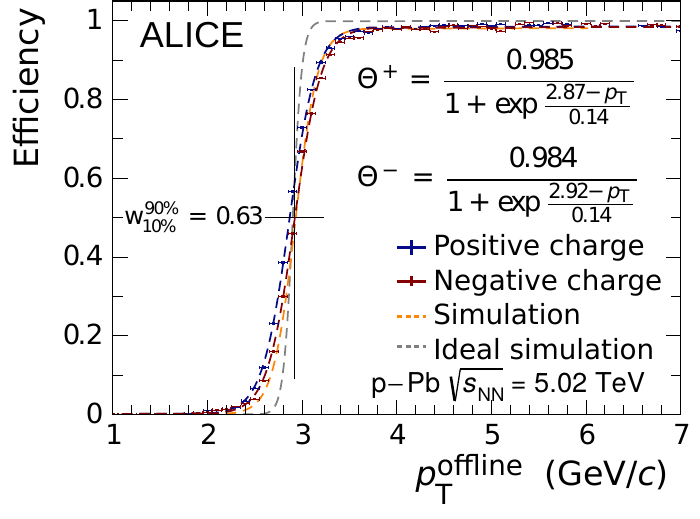}
  \caption{Turn-on curve of the trigger with a \pt threshold of 3\gevc for positively and negatively charged particles in comparison to the same variable computed in simulation with a realistic detector geometry (active channels). Also shown is the corresponding distribution for an ideal detector geometry (ideal simulation, not considering misalignment). The onset is characterised by a fit with a Fermi function.}
  \label{fig:gtu_cut_pt}
\end{figure}

The \pt resolution is crucial for the trigger since it
determines the sharpness of the threshold. It is shown in
Fig.~\ref{fig:gtu_cut_pt} for a \pt threshold of 3\gevc, where a width (10--90\%) of about 0.6\gevc is found. This is also well reproduced by simulations.

As a further development, the online tracking can benefit from taking the chamber alignment into account in the local tracking, and also by enabling the tail cancellation filter in the FEE. This will allow the use of tighter windows for the track matching and, thus, a reduction in combinatorial background while maintaining the same tracking efficiency. This is relevant for the online tracking in the high-multiplicity environment of Pb--Pb collisions. At the time of writing, these improvements are under development.

\subsection{Trigger on cosmic-ray muons}
\label{sec:cosmic_trg}

Cosmic-ray tracks are used for several purposes in the experiment,
e.g.\ for detector alignment after installation, and before
physics runs (see Section~\ref{Chapteralign}). Recording sufficient statistics requires a
good and clean trigger, in particular for tracks passing the
experiment horizontally, for which the rates are very low. Therefore,
the first level-1 trigger in ALICE was contributed by the TRD (even
before the LHC start-up) in order to select events containing tracks from
cosmic rays. It was operated on top of a level-0 trigger from TOF (TOF back-to-back coincidence). At
first, when the online tracking was still under commissioning, the
selection was based on coincident charge depositions in multiple
layers of any stack. Later, it used the full tracking
infrastructure with the condition requiring the presence of at least
one track in the event. This was sufficient to suppress the background
from the impure level-0 input from TOF.

\clearpage

\subsection{Trigger on jets}

Jets are commonly reconstructed by algorithms which cluster tracks
that are close in pseudorapidity and azimuth ($\eta$-$\varphi$ plane). The area covered by a
TRD stack roughly corresponds to that of a jet cone of radius $R =
0.2$. This allows the presence of several tracks above a
\pt threshold within one stack to be used as a signature for a high-\pt jet. The
TRD is only sensitive to the charged tracks of the jet, which is also
the part that is reconstructed using global offline tracking in the central
barrel detectors. 

\begin{figure}[tb]
  \centering
  \includegraphics[width=.6\textwidth]{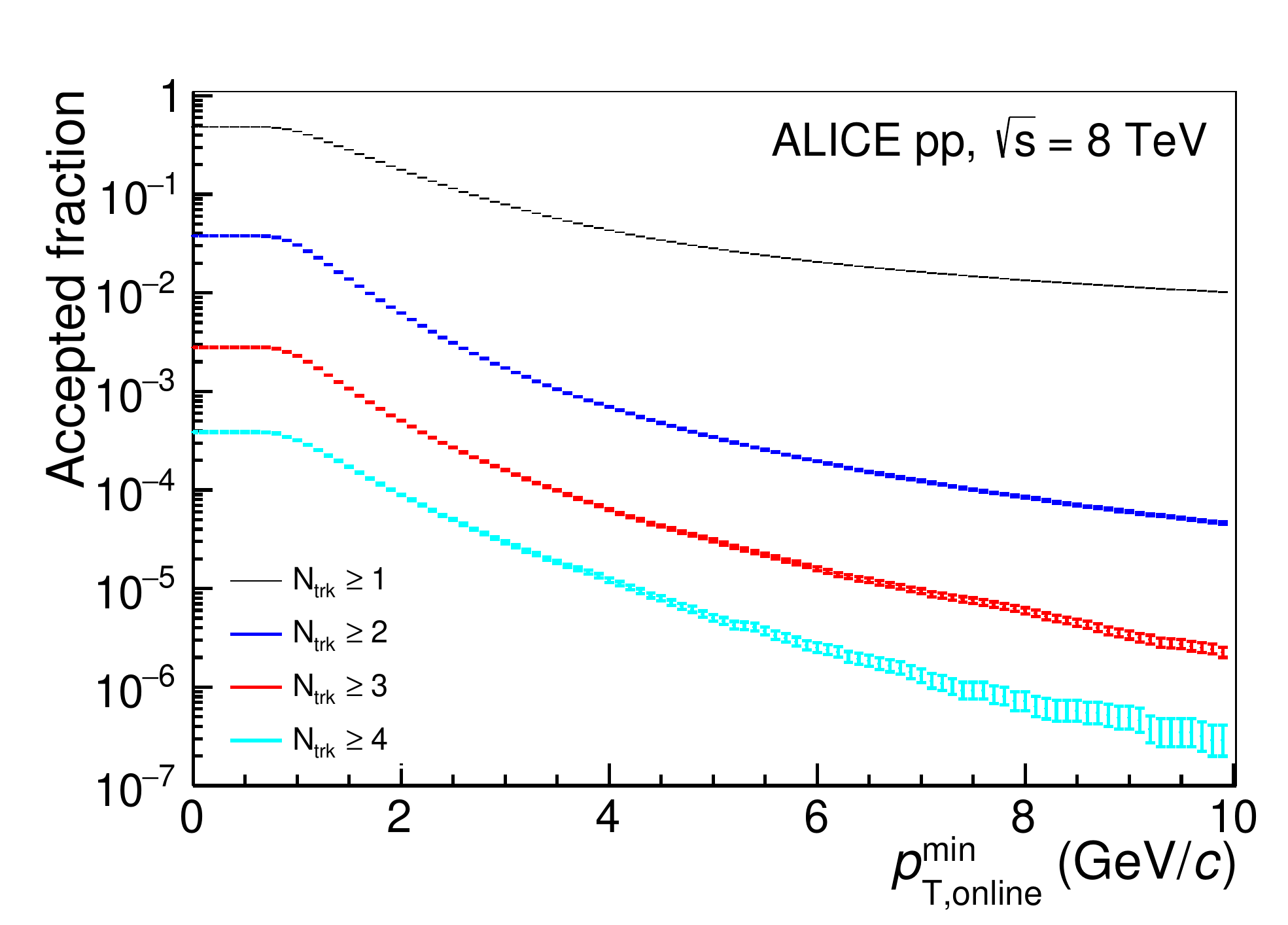}
  \caption{Rejection by the level-1 trigger for requiring
    1--4 tracks in any stack ($N_{\rm trk}$) above varying \pt thresholds for pp collisions at \sqrts~=~8~TeV~\cite{Klein:2014rxa}. The error bars indicate the statistical uncertainties. The distributions were obtained by counting the number of tracks in a stack above a given threshold and normalised by the number of sampled events. }
  \label{fig:trd_rej}
\end{figure}

In pp collisions at \sqrts~=~8~TeV and \ppb collisions at \sqrtsnn~=~5.02~TeV, the trigger sampled the anticipated integrated luminosity of about 200~nb$^{-1}$ and 1.4~nb$^{-1}$ in \run{1}, respectively.
Figure~\ref{fig:trd_rej} shows the rejection observed in \pp
collisions ($\sqrt{s} = 8~\mathrm{TeV}$) for the condition of a
certain number of global online tracks above a \pt threshold within any stack. As a
compromise between rejection and efficiency for the triggering on
jets, 3~tracks above 3\gevc were chosen as a trigger condition. This
results in a very good rejection, of about $1.5 \cdot 10^{-4}$.
The jet trigger was also used in \ppb collisions, where a good performance was achieved as well. However, the higher multiplicity reduces the rejection slightly.

\begin{figure}[bth]
  \centering
  \includegraphics[width=.47\textwidth]{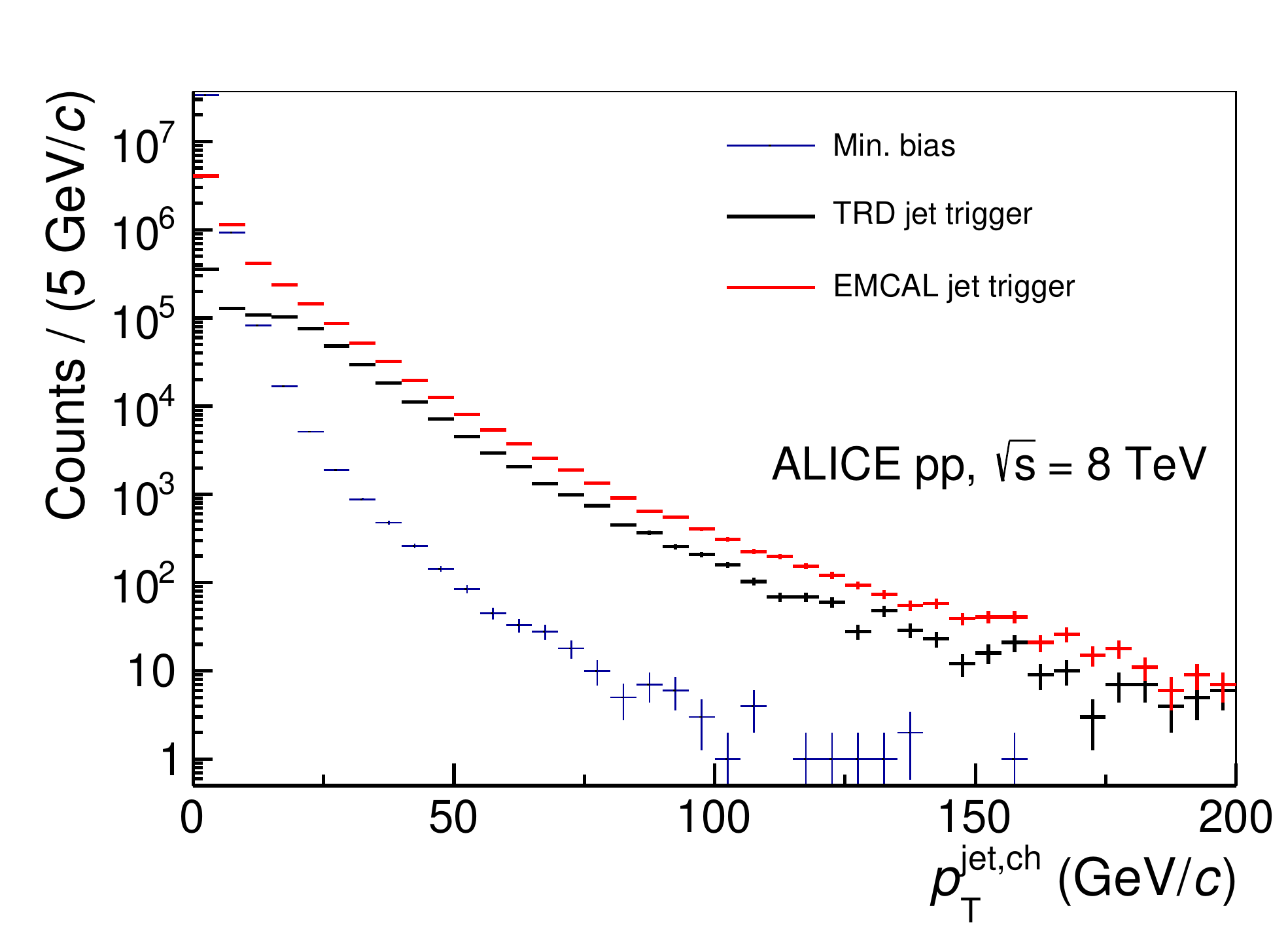}
  \hspace{\fill}
  \includegraphics[width=.47\textwidth]{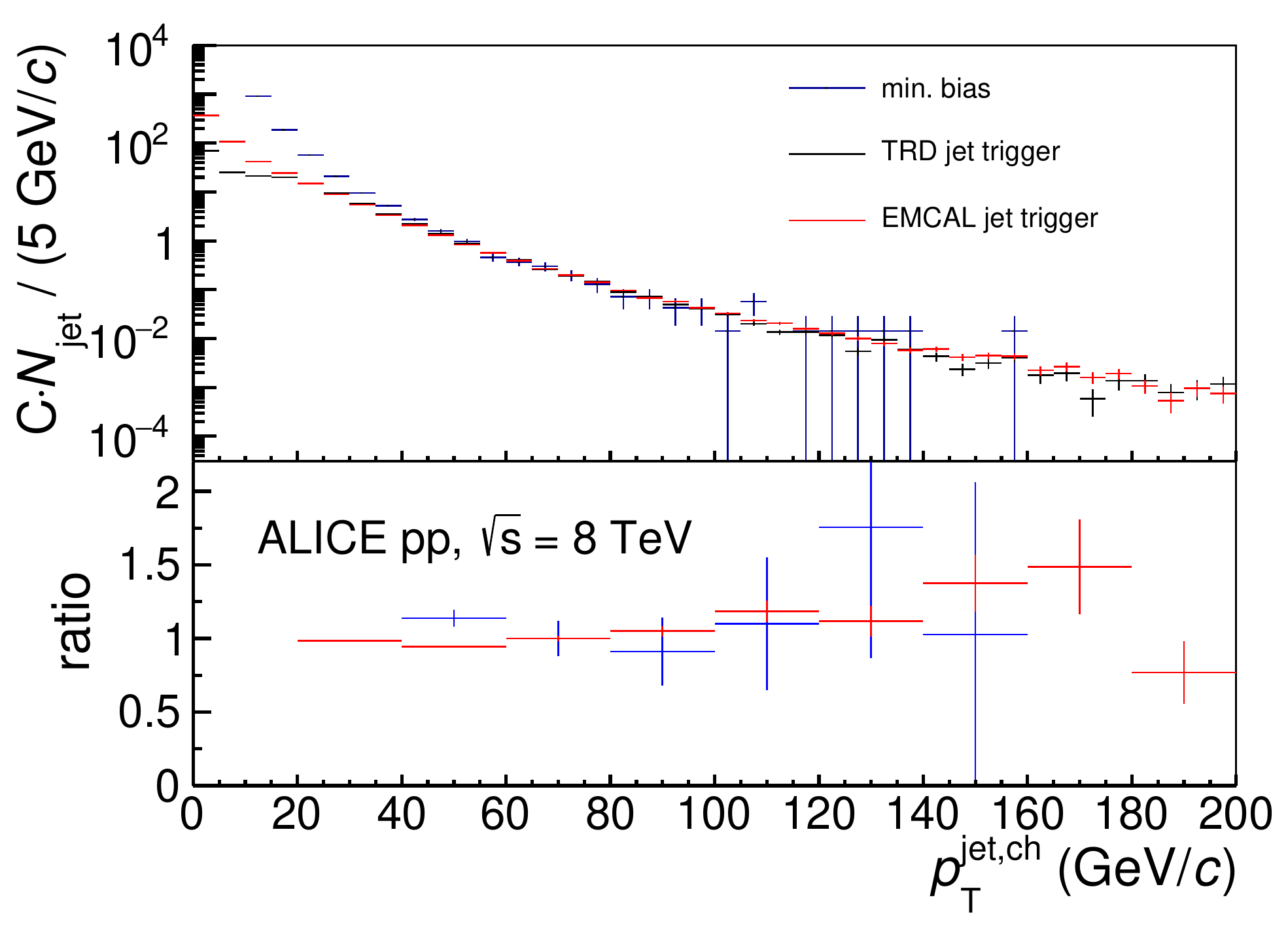}
  \caption{Left: \pt spectra of leading jets for the minimum-bias and triggered samples of pp collisions at \sqrts~=~8~TeV~\cite{Klein:2014rxa}. The leading jets are defined as the jets with the highest \pt in the event. Right: For comparison the \pt spectra were scaled to the same yield between 60 and 80\gevc. The spectra were re-binned to calculate the ratios.}
  \label{fig:trg_leadjet_pt}
\end{figure}

In Fig.~\ref{fig:trg_leadjet_pt} the jet \pt spectra from the
TRD-triggered data sample are shown. The jets were reconstructed using
the anti-kt jet finder from the Fastjet package~\cite{man:fastjet}
with a resolution parameter of $R = 0.4$. As
expected it extends to significantly larger jet \pt than the one from
the minimum-bias data sample. In order to judge the bias on the shape of
the spectrum, it is compared to an EMCal-triggered sample. At
sufficiently high \pt above about 50\gevc, the shapes of the spectra
agree.

\begin{figure}[bth]
  \centering
  \includegraphics[width=.6\textwidth]{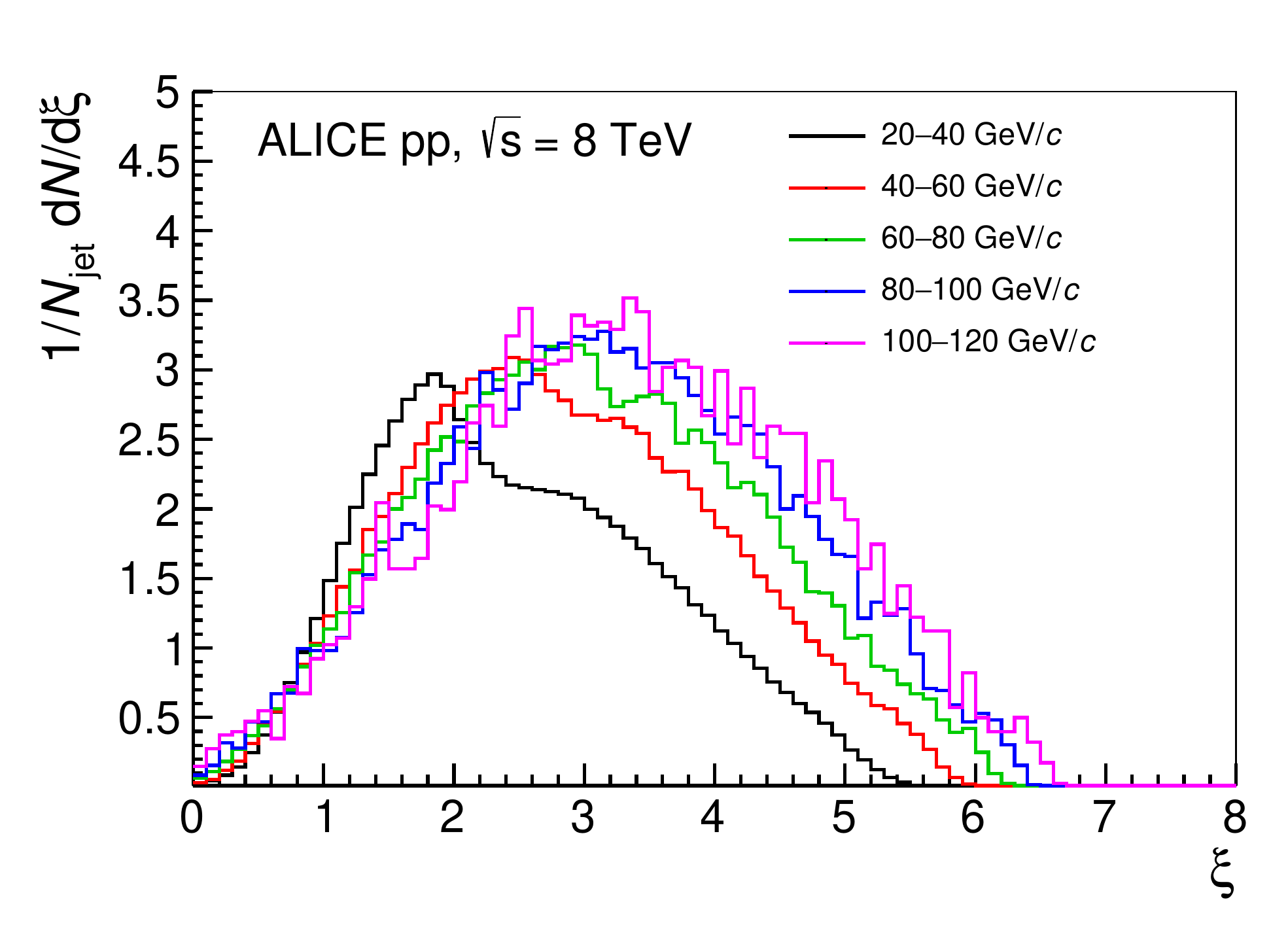}
  \caption{Fragmentation functions of leading jets from the
  TRD-triggered sample for jets in different \pt
  intervals in pp collisions at \sqrts~=~8~TeV~\cite{Klein:2014rxa}.  The leading jets are defined as the jets with the highest \pt in the event.
    }
  \label{fig:trg_leadjet_xi}
\end{figure}

To further judge the bias on the fragmentation, the raw fragmentation
function is shown as reconstructed from the jets in the TRD-triggered
data sample in Fig.~\ref{fig:trg_leadjet_xi}. The commonly used
variable $\xi$ is defined as
\begin{equation}
  \xi = - \log \frac{\pt^\mathrm{trk}}{\pt^\mathrm{jet}}   .
\end{equation}
For the lower jet \pt
intervals, a clear distortion can be seen at $\xi$ values corresponding to the
\pt threshold (in the given jet \pt interval). It disappears for higher
jet \pt, and agreement with fragmentation
functions obtained from an EMCal-triggered sample is found for jet
\pt above about 80\gevc~\cite{Klein:2014rxa}.

In order to improve the efficiency of the jet trigger, the counting of tracks can be extended over stack boundaries and, thus, avoid the acceptance gaps introduced between sectors and stacks. Corresponding studies are ongoing.

\subsection{Trigger on electrons}

During the tracklet reconstruction stage an electron likelihood is assigned to each tracklet allowing for an electron identification (see Section~\ref{sec:local_online_tracking}). It was calculated using a one-dimensional look-up table based on the total accumulated charge (the hardware also allows a two-dimensional LUT). The tracklet length is taken into account as a correction
factor applied to the charge, making the actual look-up table universal across the detector. The look-up table is created from reference
charge distributions of clean electron and pion samples obtained
through topological identification (see Section~\ref{Chapterpid}).

In order to select electrons at the trigger level, a combination of a
\pt threshold and a PID threshold can be used. The thresholds were optimised for
different physics cases. For electrons from semileptonic decays of
heavy-flavour hadrons, the goal was to extend the \pt reach at high values. Thus, a \pt~threshold of 3\gevc was chosen and the PID
threshold was adjusted to achieve a rejection of minimum-bias events
by a factor of about~100. For the measurement of quarkonia in the electron channel, a \pt~threshold of 2\gevc was chosen to cover most of the total cross-section. The PID
threshold was increased to achieve a similar rejection as for the
heavy-flavour trigger. Both triggers were used in \pp and \ppb (and Pb--p)
collisions and share a large fraction of the read-out bandwidth. For example in \ppb collisions at \mbox{\sqrtsnn = 5.02 TeV} recorded during \run{2}, about 45\% of the events of both electron triggers with late conversion rejection (see below) overlap.

The main background of the electron triggers is caused by the
conversion of photons in the detector material at large radii just in
front of or at the beginning of the TRD. The emerging electron-positron 
pairs look like high-\pt tracks and are likely to also be
identified as electrons as well. This background is suppressed by
requiring (in addition to the thresholds explained above) at least
five tracklets, one of which must be in the first layer. The background
can be further reduced by requiring that the online track can be
matched to a track in the TPC. However, this can not be done during
the online tracking, but only during the offline analysis or in the
HLT during data taking.

\begin{figure}[tb]
  \centering
  \includegraphics[width=.5\textwidth]{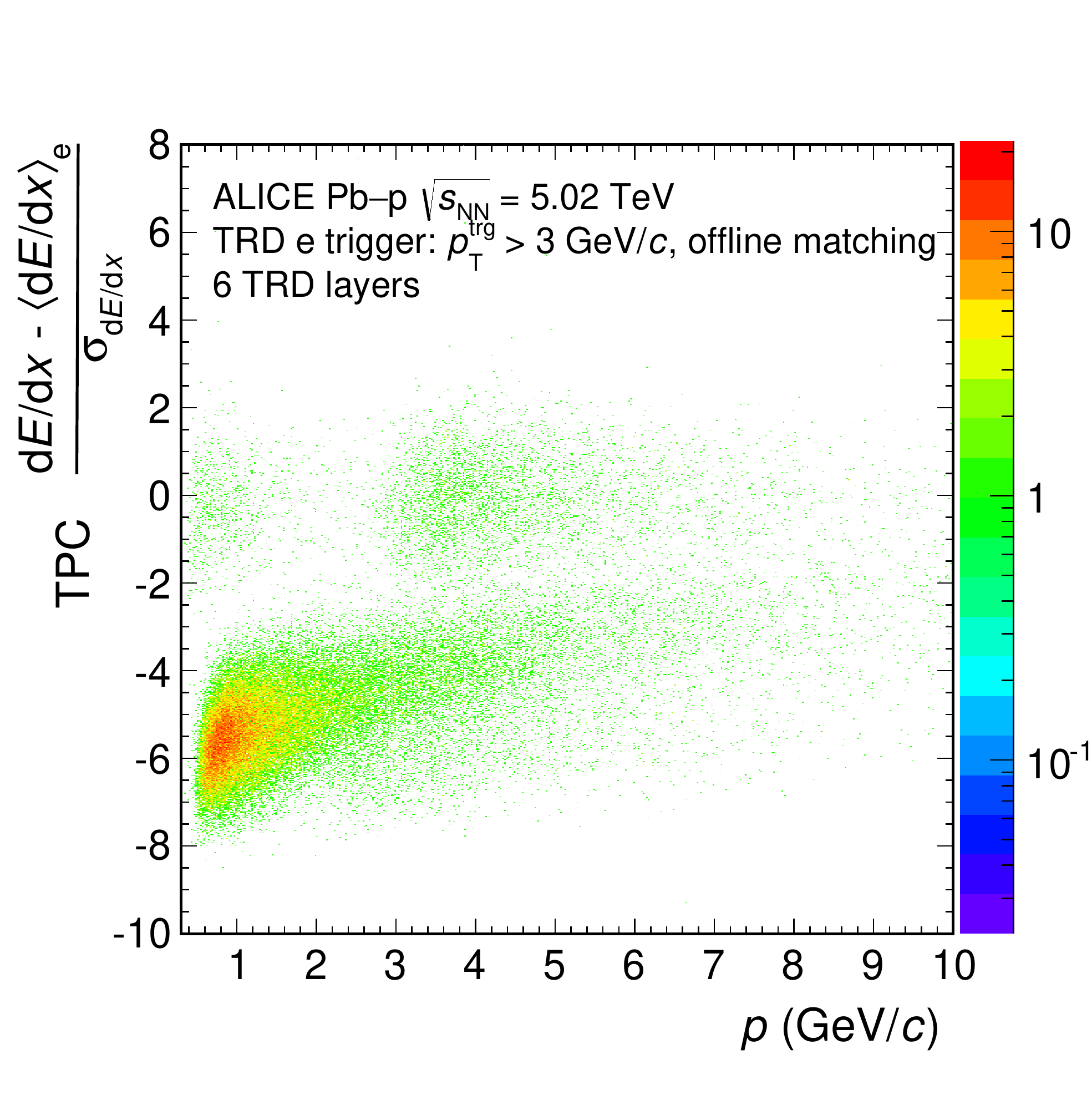}
  \caption{ $n_{\sigma^\mathrm{TPC}_\mathrm{e}}$ as a function of momentum for Pb--p collisions at \mbox{\sqrtsnn = 5.02 TeV} recorded with the electron trigger (\pt threshold at 3\gevc).
 Electrons from photon conversions in the detector material were rejected by
matching the online track with a track in the TPC.
 }
  \label{fig:nse_trgd}
\end{figure}

To judge the performance of the triggers, electron candidates are
identified using the signals from TPC, TOF, and TRD. For TPC and TOF
the selection is based on $n_{\sigma_\mathrm{e}}$, i.e. the deviation of the measured signal from the expected signal normalised to the expected resolution. Figure~\ref{fig:nse_trgd} shows the
distribution of this variable for the TPC as a function of
the track momentum $p$. The data sample was derived using an electron trigger with a \pt~threshold of 3\gevc and cleaned in the offline analysis by requiring matching with TPC tracks, i.e.\ rejecting
electrons from photon conversions. Above 3\gevc the enhancement of electrons is clearly
visible in the region around~$n_{\sigma^\mathrm{TPC}_\mathrm{e}} = 0$. 

\begin{figure}[tb]
  \centering
  \includegraphics[width=.5\textwidth]{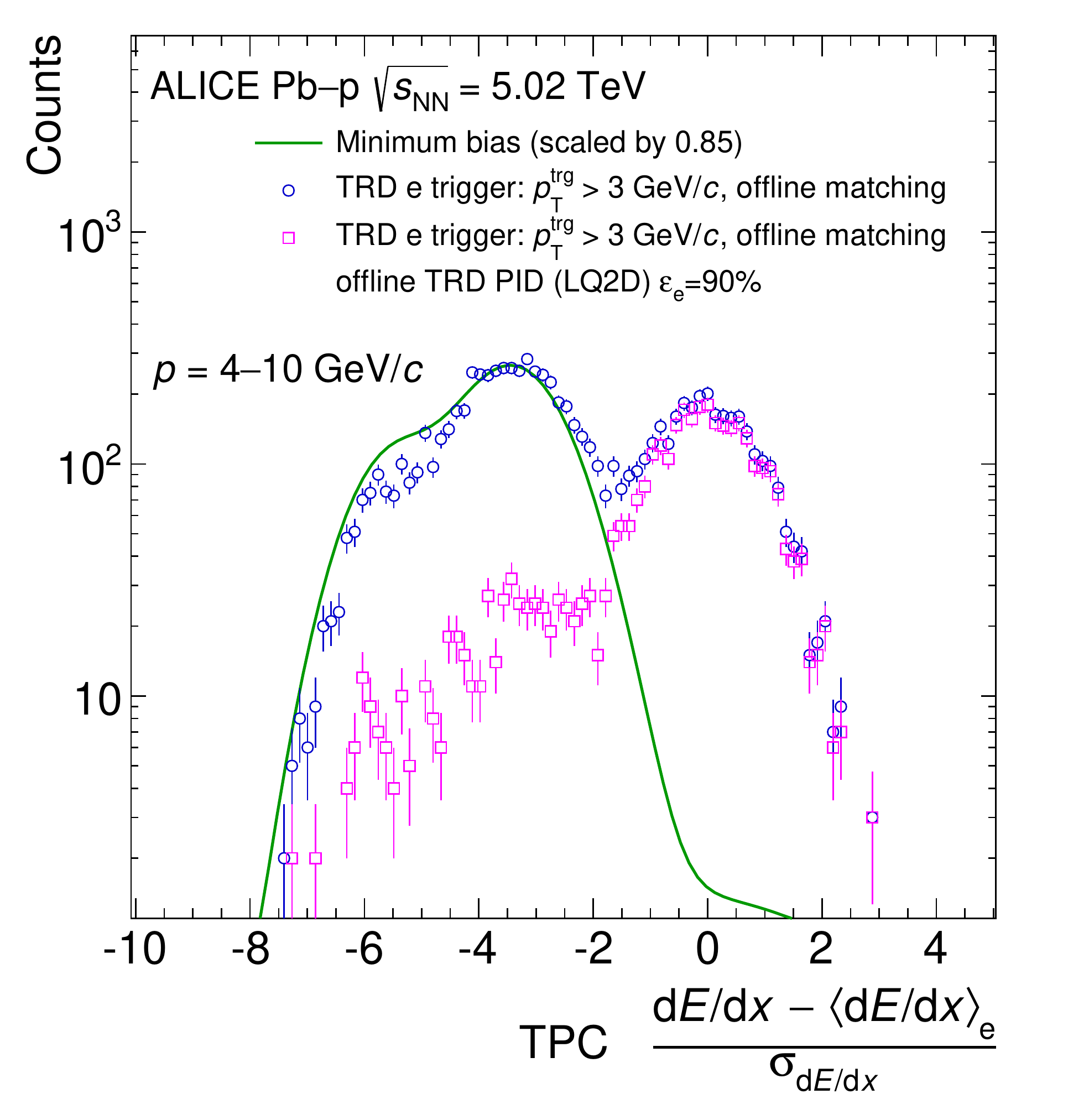}
  \caption{Electron selection for triggered data with and without the TRD offline PID (see Section~\ref{Chapterpid}) in Pb--p collisions at \mbox{\sqrtsnn = 5.02 TeV}. Electrons from photon conversions in the detector material were rejected by matching the online track with a track in the TPC. The corresponding distribution for minimum-bias data, scaled to the maximum of the distribution of the triggered data sample, is shown to visualise the TRD trigger capability to enhance electrons.}
  \label{fig:esel_trgd}
\end{figure}

\begin{figure}[tbh]
  \centering
  \includegraphics[width=.6\textwidth]{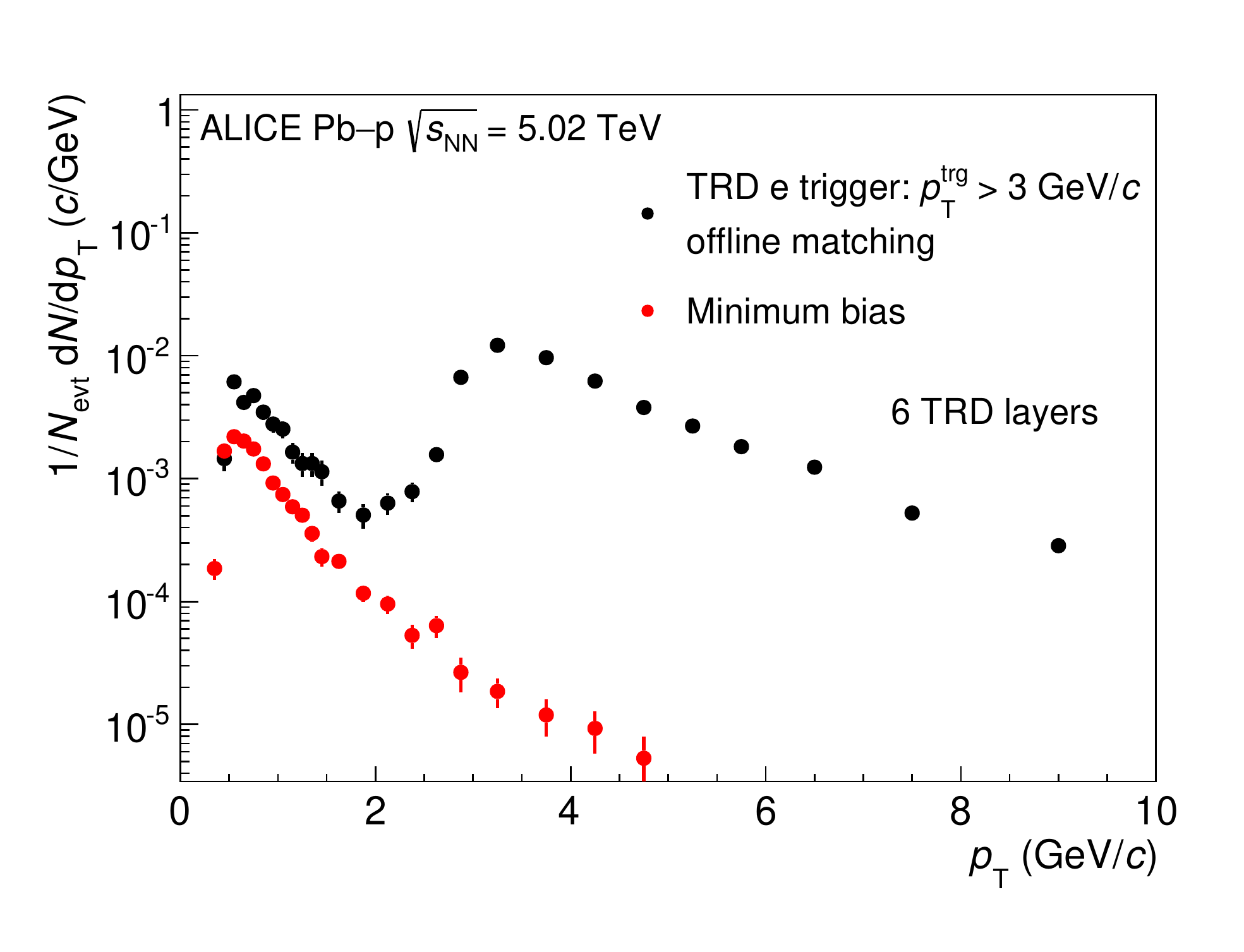}
  \caption{\pt spectra of identified electrons for the
  minimum-bias and TRD-triggered data sample of Pb--p collisions at \mbox{\sqrtsnn = 5.02 TeV}. For the result of the TRD-triggered sample, electrons from photon conversions in the detector material were rejected by matching
the online track with a track in the TPC.
   }
  \label{fig:ept_trgd}
\end{figure}

The enhancement due to the TRD electron trigger in comparison to the minimum-bias trigger is also clearly visible in Fig.~\ref{fig:esel_trgd}, which shows a projection of
$n_{\sigma^\mathrm{TPC}_\mathrm{e}}$ in a momentum interval for both data samples. A further suppression of hadrons can be achieved by exploiting the offline PID of the TRD (see
Section~\ref{Chapterpid}). Figure~\ref{fig:ept_trgd} shows the
\pt~spectra of electron candidates with 6~layers identified using the TPC and the TOF in the minimum-bias and triggered data sample. The expected onset at the trigger threshold of 3\gevc is observed for the triggered events and shows in comparison to the corresponding spectrum from minimum-bias collisions an enhancement of about 700.

\begin{figure}[tb]
 \centering
  \includegraphics[width=.5\textwidth]{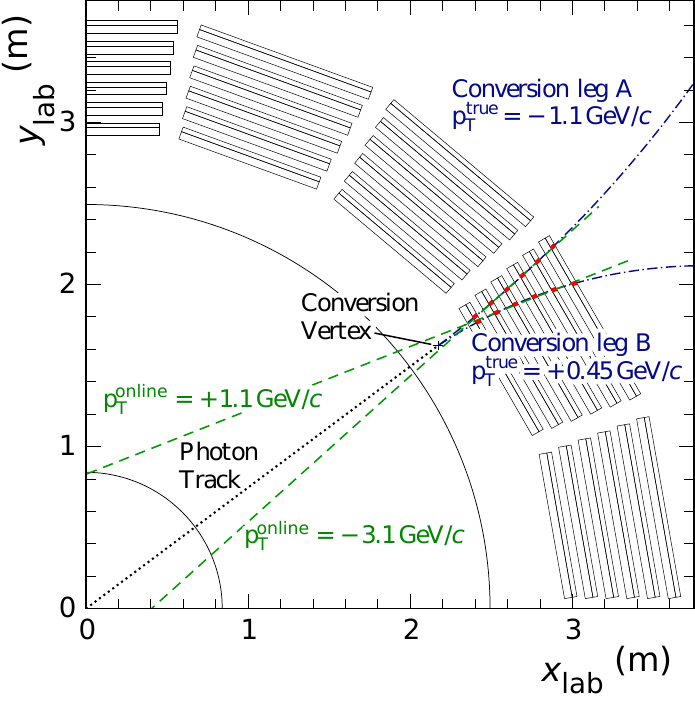}
  \caption{Photon converting into an e$^+$~e$^-$ pair at a large
    radius resembling a high-\pt track for the online tracking (green dashed line) since
    the offset to the primary vertex is small.}
  \label{fig:late_conv}
\end{figure}

The dominant background for the electron triggers, i.e.\ the conversion of photons at large radii close to the TRD entrance and in the first part of the TRD, was addressed before RUN 2. The \pt reconstruction in the online tracking assumes tracks originating from the primary vertex, which results in a too-high momentum for the electrons and positrons from `late conversions' as shown in Fig.~\ref{fig:late_conv}. An online rejection based on the calculation of the sagitta in the read-out chambers was implemented and validated. For a sagitta cut of $\Delta 1/p_{\rm T} = 0.2$~$c/\rm{GeV}$ an increased rejection of a factor of~7 at the same efficiency was achieved in pp collisions at $\sqrt{s}$~=~13~TeV~\cite{oschmidt}. For this selection criterion about 90\% of the late conversions are removed, while about 70\% of the good tracks are kept. This improvement allows only those tracks to be used for the electron trigger which are not tagged as late conversions. This setting was already successfully used in \run{2}.

\subsection{Trigger on nuclei}

A trigger on light nuclei was used for the first time in the
high-interaction \ppb and Pb--p data taking at \sqrtsnn~= 5.02~TeV in
2016. It exploits the much higher charge deposition from
multiply-charged particles. The trigger enhances mainly the statistics
of doubly-charged particles ($Z = 2$), i.e. $^3$He and $^4$He. The
trigger was operated with an estimated efficiency of about~30\% at a
rejection factor of about~600.

This trigger is also used in the \pp data taking at
13~TeV during \run{2} to significantly enhance the sample of light
nuclei. The trigger does not just enhance the sample of particles with
$Z=2$, but also of deuteron, triton and hypertriton (a bound state of
a proton, a neutron and a lambda hyperon, which decays weakly into a
$^3$He and a pion) nuclei. This will allow a precise determination of 
the mass and the lifetime of the latter.

\section{Summary}\label{Chaptersummary}

The physics objectives of the TRD together with the challenging LHC environment have led to an ambitious detector design. This required the development of a new chamber design with radiator and electronics.
After extensive tests of individual components and the full system, as well as commissioning with cosmic-ray tracks, the detector was ready for data taking with the first collisions provided by the LHC in 2009. During \run{1}, the original setup of 7~installed supermodules was further extended, reaching a maximum coverage of 13/18~in azimuth. The detector was completed in the \LS{1} before \run{2}. Since then it provides coverage of the full azimuthal acceptance of the central barrel. Read-out and trigger components were also upgraded. The developed gas system, services and infrastructure, read-out and electronics, and the Detector Control System allow the successful operation of the detector. The xenon-based gas mixture (over \SI{27}{\cubic\meter}) essential for the detection of the TR photons is re-circulated through the detector in order to reduce costs. To minimise the dead time and to cope with the read-out rates for heavy-ion data taking in~\run{2}, the data from the detector are processed in a highly parallelised read-out tree using a multi-event buffering technique, with link speeds to the DAQ of about 4~Gbit/s. Failsafe and reliable detector operation and its monitoring was achieved.
The resulting running efficiencies are about 100\% at read-out rates ranging from 100~Hz to 850~Hz in pp and p--Pb collisions, and up to 350~Hz in Pb--Pb collisions. 

Robust schemes for calibration, alignment and tracking were established. 
The TRD adds roughly \SI{70}{\centi\metre} to the lever arm of the other tracking detectors in ALICE. The $q/$\pt resolution of high transverse momentum tracks at 40\gevc is thus improved by about 40\%. In addition, the TRD increases the precision and efficiency of track matching of the detectors that lie behind it. Tracks anchored to the TRD are essential to correct the space charge distortions in the ALICE TPC. 

Several hadron and electron identification methods were developed. The electron identification performance is overall better than the design value. At 90\% electron efficiency, a pion rejection factor of about~70 is achieved at a momentum of 1\gevc for simple identification algorithms. When using the temporal evolution of the signal, a pion rejection factor of up to 410 is obtained. 

The complex and efficient design of the trigger allows the provision of triggers based on transverse momentum and electron identification in just about \SI{6}{\micro\second} after the level-0 trigger. This procedure successfully provides enriched samples of high-\pt electrons, light nuclei, and jets in pp and \ppb collisions. In pp collisions, e.g.\ at \sqrts~=~8~TeV, the jet trigger has efficiently sampled the foreseen integrated luminosity of about 200~nb$^{-1}$ during \run{1} with a constant rejection of around $1.5 \cdot 10^{-4}$. The TRD will contribute further to the physics output of the experiment in various areas, giving enriched samples of electrons, light nuclei and jets due to the trigger capabilities as well as its contributions to tracking and particle identification.

%
%

\newenvironment{acknowledgement}{\relax}{\relax}
\begin{acknowledgement}
\section*{Acknowledgements}
The ALICE Collaboration would like to acknowledge the significant contributions to the TRD project given by: A.~Anjam, W.~Amend, V.~Aprodu, H.~Baumeister, C.~Dinca, N.~Heine, T.~Herold, D.~Moisa, H.~Hinke, H.~H\"{o}bbel, V.~Kiworra$^{\ast}$, L.~Prodan, A.~Radu, W.~Verhoeven, A.~Wiesen\"{a}cker and the CERN gas group of EP-DT.


The ALICE Collaboration would like to thank all its engineers and technicians for their invaluable contributions to the construction of the experiment and the CERN accelerator teams for the outstanding performance of the LHC complex.
The ALICE Collaboration gratefully acknowledges the resources and support provided by all Grid centres and the Worldwide LHC Computing Grid (WLCG) collaboration.
The ALICE Collaboration acknowledges the following funding agencies for their support in building and running the ALICE detector:
A. I. Alikhanyan National Science Laboratory (Yerevan Physics Institute) Foundation (ANSL), State Committee of Science and World Federation of Scientists (WFS), Armenia;
Austrian Academy of Sciences and Nationalstiftung f\"{u}r Forschung, Technologie und Entwicklung, Austria;
Ministry of Communications and High Technologies, National Nuclear Research Center, Azerbaijan;
Conselho Nacional de Desenvolvimento Cient\'{\i}fico e Tecnol\'{o}gico (CNPq), Universidade Federal do Rio Grande do Sul (UFRGS), Financiadora de Estudos e Projetos (Finep) and Funda\c{c}\~{a}o de Amparo \`{a} Pesquisa do Estado de S\~{a}o Paulo (FAPESP), Brazil;
Ministry of Science \& Technology of China (MSTC), National Natural Science Foundation of China (NSFC) and Ministry of Education of China (MOEC) , China;
Ministry of Science, Education and Sport and Croatian Science Foundation, Croatia;
Ministry of Education, Youth and Sports of the Czech Republic, Czech Republic;
The Danish Council for Independent Research | Natural Sciences, the Carlsberg Foundation and Danish National Research Foundation (DNRF), Denmark;
Helsinki Institute of Physics (HIP), Finland;
Commissariat \`{a} l'Energie Atomique (CEA) and Institut National de Physique Nucl\'{e}aire et de Physique des Particules (IN2P3) and Centre National de la Recherche Scientifique (CNRS), France;
Bundesministerium f\"{u}r Bildung, Wissenschaft, Forschung und Technologie (BMBF) and GSI Helmholtzzentrum f\"{u}r Schwerionenforschung GmbH, Germany;
General Secretariat for Research and Technology, Ministry of Education, Research and Religions, Greece;
National Research, Development and Innovation Office, Hungary;
Department of Atomic Energy Government of India (DAE) and Council of Scientific and Industrial Research (CSIR), New Delhi, India;
Indonesian Institute of Science, Indonesia;
Centro Fermi - Museo Storico della Fisica e Centro Studi e Ricerche Enrico Fermi and Istituto Nazionale di Fisica Nucleare (INFN), Italy;
Institute for Innovative Science and Technology , Nagasaki Institute of Applied Science (IIST), Japan Society for the Promotion of Science (JSPS) KAKENHI and Japanese Ministry of Education, Culture, Sports, Science and Technology (MEXT), Japan;
Consejo Nacional de Ciencia (CONACYT) y Tecnolog\'{i}a, through Fondo de Cooperaci\'{o}n Internacional en Ciencia y Tecnolog\'{i}a (FONCICYT) and Direcci\'{o}n General de Asuntos del Personal Academico (DGAPA), Mexico;
Nederlandse Organisatie voor Wetenschappelijk Onderzoek (NWO), Netherlands;
The Research Council of Norway, Norway;
Commission on Science and Technology for Sustainable Development in the South (COMSATS), Pakistan;
Pontificia Universidad Cat\'{o}lica del Per\'{u}, Peru;
Ministry of Science and Higher Education and National Science Centre, Poland;
Korea Institute of Science and Technology Information and National Research Foundation of Korea (NRF), Republic of Korea;
Ministry of Education and Scientific Research, Institute of Atomic Physics and Romanian National Agency for Science, Technology and Innovation, Romania;
Joint Institute for Nuclear Research (JINR), Ministry of Education and Science of the Russian Federation and National Research Centre Kurchatov Institute, Russia;
Ministry of Education, Science, Research and Sport of the Slovak Republic, Slovakia;
National Research Foundation of South Africa, South Africa;
Centro de Aplicaciones Tecnol\'{o}gicas y Desarrollo Nuclear (CEADEN), Cubaenerg\'{\i}a, Cuba, Ministerio de Ciencia e Innovacion and Centro de Investigaciones Energ\'{e}ticas, Medioambientales y Tecnol\'{o}gicas (CIEMAT), Spain;
Swedish Research Council (VR) and Knut \& Alice Wallenberg Foundation (KAW), Sweden;
European Organization for Nuclear Research, Switzerland;
National Science and Technology Development Agency (NSDTA), Suranaree University of Technology (SUT) and Office of the Higher Education Commission under NRU project of Thailand, Thailand;
Turkish Atomic Energy Agency (TAEK), Turkey;
National Academy of  Sciences of Ukraine, Ukraine;
Science and Technology Facilities Council (STFC), United Kingdom;
National Science Foundation of the United States of America (NSF) and United States Department of Energy, Office of Nuclear Physics (DOE NP), United States of America.
\end{acknowledgement}

\bibliographystyle{utphys}
\bibliography{src/ref.bib}

\newpage
\appendix
\section{The ALICE Collaboration}
\label{app:collab}

\begingroup
\small
\begin{flushleft}
S.~Acharya\Irefn{org137}\And 
J.~Adam\Irefn{org96}\And 
D.~Adamov\'{a}\Irefn{org93}\And 
C.~Adler\Irefn{org103}\And 
J.~Adolfsson\Irefn{org32}\And 
M.M.~Aggarwal\Irefn{org98}\And 
G.~Aglieri Rinella\Irefn{org33}\And 
M.~Agnello\Irefn{org29}\And 
N.~Agrawal\Irefn{org46}\And 
Z.~Ahammed\Irefn{org137}
% ...
\And
\renewcommand\labelenumi{\textsuperscript{\theenumi}~}

\section*{Affiliation notes}
\renewcommand\theenumi{\roman{enumi}}
\begin{Authlist}
\item \Adef{org*}Deceased
\item \Adef{orgI}Dipartimento DET del Politecnico di Torino, Turin, Italy
\item \Adef{orgII}Georgia State University, Atlanta, Georgia, United States
\item \Adef{orgIII}Kirchhoff-Institut f\"{u}r Physik, Ruprecht-Karls-Universit\"{a}t Heidelberg, Heidelberg, Germany
\item \Adef{orgIV}Fachhochschule K\"{o}ln,  K\"{o}ln, Germany
\item \Adef{orgV}M.V. Lomonosov Moscow State University, D.V. Skobeltsyn Institute of Nuclear, Physics, Moscow, Russia
\item \Adef{orgVI}Department of Applied Physics, Aligarh Muslim University, Aligarh, India
\item \Adef{orgVII}Institute of Theoretical Physics, University of Wroclaw, Poland
\end{Authlist}

\section*{Collaboration Institutes}
\renewcommand\theenumi{\arabic{enumi}~}
\begin{Authlist}
\item \Idef{org1}A.I. Alikhanyan National Science Laboratory (Yerevan Physics Institute) Foundation, Yerevan, Armenia
\item \Idef{org2}Benem\'{e}rita Universidad Aut\'{o}noma de Puebla, Puebla, Mexico
\item \Idef{org3}Bogolyubov Institute for Theoretical Physics, Kiev, Ukraine
\item \Idef{org4}Bose Institute, Department of Physics  and Centre for Astroparticle Physics and Space Science (CAPSS), Kolkata, India
\item \Idef{org5}Budker Institute for Nuclear Physics, Novosibirsk, Russia
\item \Idef{org6}California Polytechnic State University, San Luis Obispo, California, United States
\item \Idef{org7}Central China Normal University, Wuhan, China
\item \Idef{org8}Centre de Calcul de l'IN2P3, Villeurbanne, Lyon, France
\item \Idef{org9}Centro de Aplicaciones Tecnol\'{o}gicas y Desarrollo Nuclear (CEADEN), Havana, Cuba
\item \Idef{org10}Centro de Investigaci\'{o}n y de Estudios Avanzados (CINVESTAV), Mexico City and M\'{e}rida, Mexico
\item \Idef{org11}Centro Fermi - Museo Storico della Fisica e Centro Studi e Ricerche ``Enrico Fermi', Rome, Italy
\item \Idef{org12}Chicago State University, Chicago, Illinois, United States
\item \Idef{org13}China Institute of Atomic Energy, Beijing, China
\item \Idef{org14}COMSATS Institute of Information Technology (CIIT), Islamabad, Pakistan
\item \Idef{org15}Department of Physics, Aligarh Muslim University, Aligarh, India
\item \Idef{org16}Department of Physics, Ohio State University, Columbus, Ohio, United States
\item \Idef{org17}Department of Physics, Pusan National University, Pusan, Republic of Korea
\item \Idef{org18}Department of Physics, Sejong University, Seoul, Republic of Korea
\item \Idef{org19}Department of Physics, University of Oslo, Oslo, Norway
\item \Idef{org20}Department of Physics and Technology, University of Bergen, Bergen, Norway
\item \Idef{org21}Dipartimento di Fisica dell'Universit\`{a} 'La Sapienza' and Sezione INFN, Rome, Italy
\item \Idef{org22}Dipartimento di Fisica dell'Universit\`{a} and Sezione INFN, Cagliari, Italy
\item \Idef{org23}Dipartimento di Fisica dell'Universit\`{a} and Sezione INFN, Trieste, Italy
\item \Idef{org24}Dipartimento di Fisica dell'Universit\`{a} and Sezione INFN, Turin, Italy
\item \Idef{org25}Dipartimento di Fisica e Astronomia dell'Universit\`{a} and Sezione INFN, Bologna, Italy
\item \Idef{org26}Dipartimento di Fisica e Astronomia dell'Universit\`{a} and Sezione INFN, Catania, Italy
\item \Idef{org27}Dipartimento di Fisica e Astronomia dell'Universit\`{a} and Sezione INFN, Padova, Italy
\item \Idef{org28}Dipartimento di Fisica `E.R.~Caianiello' dell'Universit\`{a} and Gruppo Collegato INFN, Salerno, Italy
\item \Idef{org29}Dipartimento DISAT del Politecnico and Sezione INFN, Turin, Italy
\item \Idef{org30}Dipartimento di Scienze e Innovazione Tecnologica dell'Universit\`{a} del Piemonte Orientale and INFN Sezione di Torino, Alessandria, Italy
\item \Idef{org31}Dipartimento Interateneo di Fisica `M.~Merlin' and Sezione INFN, Bari, Italy
\item \Idef{org32}Division of Experimental High Energy Physics, University of Lund, Lund, Sweden
\item \Idef{org33}European Organization for Nuclear Research (CERN), Geneva, Switzerland
\item \Idef{org34}Excellence Cluster Universe, Technische Universit\"{a}t M\"{u}nchen, Munich, Germany
\item \Idef{org35}Faculty of Engineering, Bergen University College, Bergen, Norway
\item \Idef{org36}Faculty of Mathematics, Physics and Informatics, Comenius University, Bratislava, Slovakia
\item \Idef{org37}Faculty of Nuclear Sciences and Physical Engineering, Czech Technical University in Prague, Prague, Czech Republic
\item \Idef{org38}Faculty of Science, P.J.~\v{S}af\'{a}rik University, Ko\v{s}ice, Slovakia
\item \Idef{org39}Faculty of Technology, Buskerud and Vestfold University College, Tonsberg, Norway
\item \Idef{org40}Frankfurt Institute for Advanced Studies, Johann Wolfgang Goethe-Universit\"{a}t Frankfurt, Frankfurt, Germany
\item \Idef{org41}Gangneung-Wonju National University, Gangneung, Republic of Korea
\item \Idef{org42}Gauhati University, Department of Physics, Guwahati, India
\item \Idef{org43}Helmholtz-Institut f\"{u}r Strahlen- und Kernphysik, Rheinische Friedrich-Wilhelms-Universit\"{a}t Bonn, Bonn, Germany
\item \Idef{org44}Helsinki Institute of Physics (HIP), Helsinki, Finland
\item \Idef{org45}Hiroshima University, Hiroshima, Japan
\item \Idef{org46}Indian Institute of Technology Bombay (IIT), Mumbai, India
\item \Idef{org47}Indian Institute of Technology Indore, Indore, India
\item \Idef{org48}Indonesian Institute of Sciences, Jakarta, Indonesia
\item \Idef{org49}INFN, Laboratori Nazionali di Frascati, Frascati, Italy
\item \Idef{org50}INFN, Laboratori Nazionali di Legnaro, Legnaro, Italy
\item \Idef{org51}INFN, Sezione di Bari, Bari, Italy
\item \Idef{org52}INFN, Sezione di Bologna, Bologna, Italy
\item \Idef{org53}INFN, Sezione di Cagliari, Cagliari, Italy
\item \Idef{org54}INFN, Sezione di Catania, Catania, Italy
\item \Idef{org55}INFN, Sezione di Padova, Padova, Italy
\item \Idef{org56}INFN, Sezione di Roma, Rome, Italy
\item \Idef{org57}INFN, Sezione di Torino, Turin, Italy
\item \Idef{org58}INFN, Sezione di Trieste, Trieste, Italy
\item \Idef{org59}Inha University, Incheon, Republic of Korea
\item \Idef{org60}Institut de Physique Nucl\'eaire d'Orsay (IPNO), Universit\'e Paris-Sud, CNRS-IN2P3, Orsay, France
\item \Idef{org61}Institute for Nuclear Research, Academy of Sciences, Moscow, Russia
\item \Idef{org62}Institute for Subatomic Physics of Utrecht University, Utrecht, Netherlands
\item \Idef{org63}Institute for Theoretical and Experimental Physics, Moscow, Russia
\item \Idef{org64}Institute of Experimental Physics, Slovak Academy of Sciences, Ko\v{s}ice, Slovakia
\item \Idef{org65}Institute of Physics, Academy of Sciences of the Czech Republic, Prague, Czech Republic
\item \Idef{org66}Institute of Physics, Bhubaneswar, India
\item \Idef{org67}Institute of Space Science (ISS), Bucharest, Romania
\item \Idef{org68}Institut f\"{u}r Informatik, Johann Wolfgang Goethe-Universit\"{a}t Frankfurt, Frankfurt, Germany
\item \Idef{org69}Institut f\"{u}r Kernphysik, Johann Wolfgang Goethe-Universit\"{a}t Frankfurt, Frankfurt, Germany
\item \Idef{org70}Institut f\"{u}r Kernphysik, Westf\"{a}lische Wilhelms-Universit\"{a}t M\"{u}nster, M\"{u}nster, Germany
\item \Idef{org71}Instituto de Ciencias Nucleares, Universidad Nacional Aut\'{o}noma de M\'{e}xico, Mexico City, Mexico
\item \Idef{org72}Instituto de F\'{i}sica, Universidade Federal do Rio Grande do Sul (UFRGS), Porto Alegre, Brazil
\item \Idef{org73}Instituto de F\'{\i}sica, Universidad Nacional Aut\'{o}noma de M\'{e}xico, Mexico City, Mexico
\item \Idef{org74}IRFU, CEA, Universit\'{e} Paris-Saclay, Saclay, France
\item \Idef{org75}iThemba LABS, National Research Foundation, Somerset West, South Africa
\item \Idef{org76}Joint Institute for Nuclear Research (JINR), Dubna, Russia
\item \Idef{org77}Konkuk University, Seoul, Republic of Korea
\item \Idef{org78}Korea Institute of Science and Technology Information, Daejeon, Republic of Korea
\item \Idef{org79}KTO Karatay University, Konya, Turkey
\item \Idef{org80}Laboratoire de Physique Subatomique et de Cosmologie, Universit\'{e} Grenoble-Alpes, CNRS-IN2P3, Grenoble, France
\item \Idef{org81}Lawrence Berkeley National Laboratory, Berkeley, California, United States
\item \Idef{org82}Moscow Engineering Physics Institute, Moscow, Russia
\item \Idef{org83}Nagasaki Institute of Applied Science, Nagasaki, Japan
\item \Idef{org84}National and Kapodistrian University of Athens, Physics Department, Athens, Greece
\item \Idef{org85}National Centre for Nuclear Studies, Warsaw, Poland
\item \Idef{org86}National Institute for Physics and Nuclear Engineering, Bucharest, Romania
\item \Idef{org87}National Institute of Science Education and Research, HBNI, Jatni, India
\item \Idef{org88}National Nuclear Research Center, Baku, Azerbaijan
\item \Idef{org89}National Research Centre Kurchatov Institute, Moscow, Russia
\item \Idef{org90}Niels Bohr Institute, University of Copenhagen, Copenhagen, Denmark
\item \Idef{org91}Nikhef, Nationaal instituut voor subatomaire fysica, Amsterdam, Netherlands
\item \Idef{org92}Nuclear Physics Group, STFC Daresbury Laboratory, Daresbury, United Kingdom
\item \Idef{org93}Nuclear Physics Institute, Academy of Sciences of the Czech Republic, \v{R}e\v{z} u Prahy, Czech Republic
\item \Idef{org94}Oak Ridge National Laboratory, Oak Ridge, Tennessee, United States
\item \Idef{org95}Petersburg Nuclear Physics Institute, Gatchina, Russia
\item \Idef{org96}Physics Department, Creighton University, Omaha, Nebraska, United States
\item \Idef{org97}Physics department, Faculty of science, University of Zagreb, Zagreb, Croatia
\item \Idef{org98}Physics Department, Panjab University, Chandigarh, India
\item \Idef{org99}Physics Department, University of Cape Town, Cape Town, South Africa
\item \Idef{org100}Physics Department, University of Jammu, Jammu, India
\item \Idef{org101}Physics Department, University of Rajasthan, Jaipur, India
\item \Idef{org102}Physikalisches Institut, Eberhard Karls Universit\"{a}t T\"{u}bingen, T\"{u}bingen, Germany
\item \Idef{org103}Physikalisches Institut, Ruprecht-Karls-Universit\"{a}t Heidelberg, Heidelberg, Germany
\item \Idef{org104}Physik Department, Technische Universit\"{a}t M\"{u}nchen, Munich, Germany
\item \Idef{org105}Purdue University, West Lafayette, Indiana, United States
\item \Idef{org106}Research Division and ExtreMe Matter Institute EMMI, GSI Helmholtzzentrum f\"ur Schwerionenforschung GmbH, Darmstadt, Germany
\item \Idef{org107}Rudjer Bo\v{s}kovi\'{c} Institute, Zagreb, Croatia
\item \Idef{org108}Russian Federal Nuclear Center (VNIIEF), Sarov, Russia
\item \Idef{org109}Saha Institute of Nuclear Physics, Kolkata, India
\item \Idef{org110}School of Physics and Astronomy, University of Birmingham, Birmingham, United Kingdom
\item \Idef{org111}Secci\'{o}n F\'{\i}sica, Departamento de Ciencias, Pontificia Universidad Cat\'{o}lica del Per\'{u}, Lima, Peru
\item \Idef{org112}SSC IHEP of NRC Kurchatov institute, Protvino, Russia
\item \Idef{org113}Stefan Meyer Institut f\"{u}r Subatomare Physik (SMI), Vienna, Austria
\item \Idef{org114}SUBATECH, IMT Atlantique, Universit\'{e} de Nantes, CNRS-IN2P3, Nantes, France
\item \Idef{org115}Suranaree University of Technology, Nakhon Ratchasima, Thailand
\item \Idef{org116}Technical University of Ko\v{s}ice, Ko\v{s}ice, Slovakia
\item \Idef{org117}Technical University of Split FESB, Split, Croatia
\item \Idef{org118}The Henryk Niewodniczanski Institute of Nuclear Physics, Polish Academy of Sciences, Cracow, Poland
\item \Idef{org119}The University of Texas at Austin, Physics Department, Austin, Texas, United States
\item \Idef{org120}Universidad Aut\'{o}noma de Sinaloa, Culiac\'{a}n, Mexico
\item \Idef{org121}Universidade de S\~{a}o Paulo (USP), S\~{a}o Paulo, Brazil
\item \Idef{org122}Universidade Estadual de Campinas (UNICAMP), Campinas, Brazil
\item \Idef{org123}Universidade Federal do ABC, Santo Andre, Brazil
\item \Idef{org124}University of Houston, Houston, Texas, United States
\item \Idef{org125}University of Jyv\"{a}skyl\"{a}, Jyv\"{a}skyl\"{a}, Finland
\item \Idef{org126}University of Liverpool, Liverpool, United Kingdom
\item \Idef{org127}University of Tennessee, Knoxville, Tennessee, United States
\item \Idef{org128}University of the Witwatersrand, Johannesburg, South Africa
\item \Idef{org129}University of Tokyo, Tokyo, Japan
\item \Idef{org130}University of Tsukuba, Tsukuba, Japan
\item \Idef{org131}Universit\'{e} Clermont Auvergne, CNRS--IN2P3, LPC, Clermont-Ferrand, France
\item \Idef{org132}Universit\'{e} de Lyon, Universit\'{e} Lyon 1, CNRS/IN2P3, IPN-Lyon, Villeurbanne, Lyon, France
\item \Idef{org133}Universit\'{e} de Strasbourg, CNRS, IPHC UMR 7178, F-67000 Strasbourg, France, Strasbourg, France
\item \Idef{org134}Universit\`{a} degli Studi di Pavia, Pavia, Italy
\item \Idef{org135}Universit\`{a} di Brescia, Brescia, Italy
\item \Idef{org136}V.~Fock Institute for Physics, St. Petersburg State University, St. Petersburg, Russia
\item \Idef{org137}Variable Energy Cyclotron Centre, Kolkata, India
\item \Idef{org138}Warsaw University of Technology, Warsaw, Poland
\item \Idef{org139}Wayne State University, Detroit, Michigan, United States
\item \Idef{org140}Wigner Research Centre for Physics, Hungarian Academy of Sciences, Budapest, Hungary
\item \Idef{org141}Yale University, New Haven, Connecticut, United States
\item \Idef{org142}Yonsei University, Seoul, Republic of Korea
\item \Idef{org143}Zentrum f\"{u}r Technologietransfer und Telekommunikation (ZTT), Fachhochschule Worms, Worms, Germany
\end{Authlist}
\endgroup
%
%
\end{document}